\documentclass[letterpaper,12pt,titlepage,oneside,final]{book}
 
\newcommand{\href}[1]{#1} 

\usepackage{ifthen}
\newboolean{PrintVersion}
\setboolean{PrintVersion}{true} 

\usepackage{amsmath,amssymb,amstext} 
\usepackage[pdftex]{graphicx} 
\usepackage{afterpage}

\usepackage{color}

\usepackage{tcolorbox}

\usepackage{collectbox}

\usepackage[utf8]{inputenc}
\usepackage[english]{babel}
\usepackage{subcaption}
\usepackage{bbold}

\newtheorem{theorem}{Theorem}[section]
\newtheorem{corollary}{Corollary}[theorem]
\newtheorem{lemma}[theorem]{Lemma}

\usepackage{sectsty}

\sectionfont{\fontsize{15}{15}\selectfont}

\subsectionfont{\fontsize{12}{15}\selectfont}

\usepackage{verbatim}
\usepackage{float}

\usepackage{csquotes}

\usepackage{bm}

\newcommand{\bi}{\mathbf}
\usepackage{titlesec} 
\usepackage{lipsum} 

\titleformat{\chapter}[display]
  {\bfseries\Large}
  {\filright\MakeUppercase{\chaptertitlename} \Large\thechapter}
  {1ex}
  {\titlerule\vspace{1ex}\filleft}
  [\vspace{1ex}\titlerule]

\usepackage[pdftex,letterpaper=true,pagebackref=false]{hyperref} 
\hypersetup{
    plainpages=false,       
    pdfpagelabels=true,     
    bookmarks=true,         
    unicode=false,          
    pdftoolbar=true,        
    pdfmenubar=true,        
    pdffitwindow=false,     
    pdfstartview={FitH},    
    pdftitle={Dissertation.V.Rokaj},    
    pdfnewwindow=true,      
    colorlinks=true,        
    linkcolor=black,         
    citecolor=blue,        
    filecolor=magenta,      
    urlcolor=cyan           
}
\ifthenelse{\boolean{PrintVersion}}{   
\hypersetup{	
    citecolor=black,%
    filecolor=black,%
    linkcolor=black,%
    urlcolor=black}
}{} 

\let\origdoublepage\cleardoublepage
\newcommand{\clearemptydoublepage}{%
  \clearpage{\pagestyle{empty}\origdoublepage}}
\let\cleardoublepage\clearemptydoublepage

\begin{document}

\pagestyle{empty}
\pagenumbering{roman}

\begin{titlepage}
        \begin{center}
        \vspace*{1.0cm}

        \Huge { Condensed Matter Systems in Cavity Quantum Electrodynamics }

        \vspace*{2.5cm}
        
        \normalsize Dissertation zur Erlangung des Doktorgrades\\
an der Fakultät für Mathematik, Informatik und
Naturwissenschaften,\\
Fachbereich Physik,
der Universität Hamburg

\vspace*{1.0cm}

        \normalsize
        vorgelegt von \\

        \vspace*{1.0cm}

        \Large
         Vasil Rokaj \\




        \vspace*{4.0cm}

       \large Hamburg\\ 
        19 October 2021 


        \end{center}
\end{titlepage}

\pagestyle{plain}
\setcounter{page}{2}

\cleardoublepage 
  
  \vspace*{\fill}
  
 \copyright\ Vasil Rokaj

  \noindent

  \bigskip
  
  \noindent

\cleardoublepage

\begin{align*}
&\text{Gutachter/innen der Dissertation:  }  &&\text{Prof.~Dr.~Angel Rubio}\\
&\text{}  &&\text{Prof.~Dr.~Klaus Sengstock}\\\\
&\text{Zusammensetzung der Prüfungskommission: }  &&\text{Prof.~Dr.~Angel Rubio}\\
&\text{}  &&\text{Prof.~Dr.~Roman Schnabel}\\
&\text{}  &&\text{Prof.~Dr.~Michael Potthoff}\\
&\text{}  &&\text{Dr.~Michael Ruggenthaler}\\
&\text{}  &&\text{Prof.~Dr.~Daniela Pfannkuche}\\\\
&\text{Vorsitzende/r der Prüfungskommission:  }  &&\text{Prof.~Dr.~Michael Potthoff}\\\\
&\text{Datum der Disputation:  }  &&\text{05.10.2021}\\\\
&\text{Vorsitzender Fach-Promotionsausschusses PHYSIK: }  &&\text{Prof.~Dr.~Wolfgang Hansen}\\\\
&\text{Leiter des Fachbereichs PHYSIK: }  &&\text{Prof.~Dr.~Günter H.~W.~Sigl}\\\\
&\text{Dekan der Fakultät MIN: }  &&\text{Prof.~Dr.~Heinrich Graener}\\\\
\end{align*}

\newpage

\afterpage{\null\newpage}

\cleardoublepage


\begin{center}\textbf{Abstract}\end{center}

Condensed matter physics and quantum electrodynamics (QED) have been long considered as distinct disciplines. This situation is changing rapidly by the progress in the field of cavity QED materials. Motivated by these advances we aim to bridge these fields by merging fundamental concepts coming from both sides. In the first part of the thesis we present how non-relativistic QED can be constructed and we discuss different forms of light-matter interaction in different gauges and that neglecting particular quadratic terms can lead to instabilities for the QED Hamiltonian. In the second part of the thesis we revisit the Sommerfeld model of the free electron gas in cavity QED and provide the exact analytic solution for this paradigmatic condensed matter system coupled to the cavity. We show that the cavity field modifies the optical conductivity of the electron gas and suppresses its Drude peak. Further, by constructing an effective field theory in the continuum of photon modes we show how the photon field leads to a many-body renormalization of the electron mass, which modifies the fermionic quasiparticle excitations of the Fermi liquid. In the last part of the thesis we show that translational symmetry for periodic materials in homogeneous magnetic fields can be restored by embedding the problem into QED. This leads to a generalization of Bloch's theory for electron-photon systems, that we named as QED-Bloch theory, which can be applied for the description of periodic materials in homogeneous magnetic fields and strongly coupled to the quantized cavity field. As a first application of our theory we consider Landau levels coupled to a cavity and we show that quasi-particle excitations between Landau levels and photons appear, called Landau polaritons. Further, for periodic materials in such setups, QED-Bloch theory predicts the emergence of novel fractal polaritonic energy spectra, which we name as fractal polaritons. The fractal polaritons are a polaritonic, QED analogue of the Hofstadter butterfly. In the limit of no cavity confinement, QED-Bloch theory recovers both the well-known Landau levels and the fractal spectrum of the Hofstadter butterfly, and can be applied for the description of periodic materials in strong magnetic fields.

\cleardoublepage

\begin{center}\textbf{Zusammenfassung}\end{center}

Die Festkörperphysik und die Quantenelektrodynamik (QED) werden gewöhnlicherweise als zwei getrennte Forschungsdisziplinen erachtet. Auf Grund von beachtlichen Fortschritten in der Hohlraum-QED Materialforschung ändert sich diese Sichtweise jedoch langsam. In dieser Arbeit wollen wir zur Verbindung dieser beiden Forschungsdisziplinen beitragen, indem wir fundamentale theoretische Konzepte der beiden Bereiche vereinheitlichen. Im ersten Teil dieser Dissertation wird als Grundlage die nicht-relativistische QED eingeführt und verschiedene Arten von Licht-Materie-Wechselwirkung und Eichtransformationen diskutiert. In diesem Kontext zeigen wir, dass die Vernachlässigung gewisser Wechselwirkungsterme zu unphysikalischen Instabilitäten in der QED führt. Im zweiten Teil präsentieren wir eine Erweiterung der Sommerfeld-Theorie des freien Elektronengases, welche eine paradigmatische Theorie der Festkörperphysik ist, für Hohlraum-QED. Die resultierende Hamilton-Gleichung des gekoppelten Licht-Materie-System ist analytisch lösbar und wir können zeigen, dass das Photonfeld die optische Leitfähigkeit des Elektronengases modifiziert sowie den Drude Peak verkleinert. Aufbauend auf den analytischen Lösungen wird eine effektive Quantenfeldtheorie konstruiert, welche zu einer Vielteilchen-Massenrenormierung führt und die fermionischen Anregungen der Quasi-Teilchen der Fermi-Flüssigkeits-Theorie modifiziert. Im letzten Teil der Arbeit stellen wir die durch homogene Magnetfelder gebrochene Translationsinvarianz periodischer Festkörper mittels der QED wieder her. Dies führt zu einer Verallgemeinerung der Bloch-Theorie, welche wir als QED-Bloch-Theorie bezeichnen. Mit Hilfe der QED-Bloch-Theorie können periodische Festkörper in homogenen Magnetfeldern beschrieben werden, die zugleich an das quantisierte Photonfeld koppeln. Als erste Anwendung unserer Theorie untersuchen wir Landau-Zustände, die stark an einem Photonfeld gekoppelt sind. Wir zeigen, dass auf Grund der Wechselwirkung der Landau-Zustände mit den Photonen neue Quasi-Teilchen (Landau-Polaritonen) entstehen. Die Theorie sagt voraus, dass für periodische Festkörper unter diesen Bedingungen fraktale polaritonische Energie-Spektra entstehen, die fraktalen Polaritonen. Diese stellen ein polaritonisches QED-Analogon zum ``Hofstadter-Butterfly" dar. Konsistenterweise erhält man im klassischen Grenzfall die wohlbekannten Landau-Zustände sowie das fraktale Spektrum vom Hofstadter-Butterfly. In diesem Fall dient die QED-Bloch-Theorie zur Beschreibung von periodisch angeordneten Materialien in starken Magnetfeldern.

\newpage


\begin{center}\textbf{Acknowledgements}\end{center}

First of all, I would like to thank my supervisors Angel Rubio and Michael Ruggenthaler for giving me the opportunity to do a PhD in a period where I was told that I am probably not good enough. They have both contributed enormously to my development and the research presented in this thesis. They were always open to hear my ideas and I have learned a tremendous lot from them. I feel privileged being taught how to do science close to them, Angel with his endless enthusiasm and far reaching intuition and Michael with his deep analytic insight and love for rigor.   

Then, I would like to thank Markus Penz with whom we have been working together for more than three years now. We had so much fun doing research together, looking at fractals and sometimes screaming from enthusiasm! Without his programming skills the work on the Hofstadter butterfly would not have been possible. Thank you for being a good friend as well.

I want to thank as well Michael Sentef for the collaboration we had but also for teaching me through his lectures and our discussions many topics and concepts of condensed matter physics and the attitude of a condensed matter theorist towards physics. 

I would also like to thank Florian Eich. With Florian we only worked together for a few months but his contribution was crucial for the work on the free electron gas in the cavity. He taught me how to think about electrons in the thermodynamic limit and I regret that we didn't interact more during our common stay at the Institute.

Further, I would like to thank Massimo Altarelli for the many discussions we had which proved extremely helpful to get things straight with several research problems in many occasions. Having to discuss with Massimo is like being a piece of metal in the hands of a blacksmith, he beats you hard but in the end you become sharper.

I want to thank also my office mates Davis Welakuh, Gabriel Topp, and Fabio Covito for making life in the office much more fun, with singing, playing kicker in coffee breaks and football on Fridays. I want to thank especially Davis with whom I also worked together, exchanged huge amounts of silly thoughts on QED and became close friends. Homie, you're the best!

Also I would like to thank our IMPRS coordinators Julia Quante and Neda Lotfiomran for their help with the PhD bureaucracy and for organizing courses, events and retreats for us. 

Last, I want to thank my partner in life Eirini. Without your love and support I would not have have made it.

\cleardoublepage



\textit{To my parents Miranda and Panagiotis}
\cleardoublepage

\renewcommand\contentsname{Table of Contents}
\tableofcontents
\cleardoublepage
\phantomsection




\pagenumbering{arabic}

\chapter{Introduction}

\begin{displayquote}
\footnotesize{The scientific enterprise is now largely involved in the creation of novelty---in the design of objects that never existed before and in the creation of conceptual frameworks to understand the complexity and novelty that can emerge from the \textit{known} foundations and ontologies.}
\end{displayquote}
\begin{flushright}
  \footnotesize{Silvan S.~Schweber\\
Physics, Community and the Crisis in Physical Theory~\cite{SchweberCrisis}}
\end{flushright}

Quantum electrodynamics (QED) is the cornerstone of our modern description for all phenomena involving the interaction between light and matter. This includes the interaction of atoms, molecules or many-body condensed matter systems with classical electromagnetic fields, lasers and even single photons. In the framework of QED both constituents are treated quantum-mechanically as dynamical entities, in the sense that light can shape matter and vice versa. Quantum electrodynamics, as it was formulated by Tomonaga, Schwinger, Feynman and Dyson~\cite{TomonagaNobel, FeynmanQED, SchwingerNobel, DysonQED, SchweberQEDHistory} is defined as a perturbative expansion of the scattering matrix (S-matrix), from which one can compute the probability of any scattering process occurring between charged particles and photons. This perturbative treatment relies on the fact that the light-matter interaction in free space is weak, as it is determined by the fine-structure constant $\alpha_{\textrm{fs}}=1/137$.

This paradigm of perturbative light-matter interactions is currently challenged by developments in the emerging field of cavity QED materials. Driven by the advances in fabricating nanostructures and nanomaterials, experimentalists are now able to couple atoms, molecules and 2D materials to the electromagnetic field in the strong and the ultrastrong coupling regime~\cite{ruggenthaler2017b, kockum2019ultrastrong}. In these novel regimes, light and matter lose their individual identity and form hybrid states called polaritons. These mixed quasiparticles can fundamentally alter the behavior and properties of materials~\cite{PolaritonPanorama}. 

In the last decade, a plethora of pathways have been explored, to achieve strong light-matter coupling, and several unprecedented phenomena involving polaritonic states have been observed. Quantum Hall systems under cavity confinement, in both the integer~\cite{Hagenmuller2010cyclotron, rokaj2019, Keller2020, ScalariScience, li2018} and the fractional~\cite{AtacPRL, SmolkaAtac} regime, have demonstrated ultrastrong coupling to the photon field and modifications of their transport properties~\cite{paravacini2019}. Light-matter interactions have been suggested to modify the electron-phonon coupling and the critical temperature of superconductors~\cite{SchlawinSuperconductivity, AtacSuperconductivity, sentef2018, Galitski} with the first experimental evidence already having appeared~\cite{A.Thomas2019}. Modifications of chemical properties and chemical reactions have been achieved through coupling to vacuum fields in polaritonic chemistry~\cite{ebbesen2016, hutchison2013, hutchison2012, orgiu2015, feist2017polaritonic, galego2016, flick2017, schafer2019modification}. Cavity control of excitons has been studied~\cite{LatiniRonca, ExcitonControl} and exciton-polariton condensation has been achieved~\cite{kasprzak2006, KeelingKenaCohen}. Further, the implications of coupling to chiral electromagnetic fields is currently investigated~\cite{ChiralCavities, ChiralPetersen67, PRAChiral, ChiralQuantumOptics}, and the possibility of cavity-induced ferroelectric phases has been proposed~\cite{LatiniFerro, Demlerferro}. 

Many of these fascinating phenomena are still poorly understood, and theoretical predictions depending on different kinds of modelling and approximations, are contradictory. Even the basic question of what is the correct Hamiltonian for the description of strongly coupled light-matter systems has been heavily debated~\cite{RouseGauges, DeBernardis2018, vukics2014, GalegoCasimir, DiStefano2019, HuoGauges, bernadrdis2018breakdown}. Special attention has been drawn to the question about the importance of the quadratic terms that appear in different gauges. Namely, the diamagnetic $\bi{A}^2$ term in the Coulomb gauge and the dipole self-energy in the length gauge~\cite{rokaj2017, schaeferquadratic}. We note that the debate over the importance of these terms, is closely related to the notorious quest over the superradiant phase transition~\cite{Lieb, CiutiSuperradiance, MazzaSuperradiance, MarquardtNoGO, Birula, AndolinaNoGo, ChirolliNoGO}.

\textbf{Fundamentals of Quantum Electrodynamics.}---To address properly this fundamental problem, we revisit the framework of non-relativistic QED, also known as Pauli-Fierz theory~\cite{spohn2004}. We present how the Pauli-Fierz Hamiltonian can be obtained as the non-relativistic limit of QED, and we show that in the Coulomb gauge the diamagnetic $\bi{A}^2$ term emerges, due the elimination of the positrons. We demonstrate that if the $\bi{A}^2$ term is omitted, then the Coulomb gauge Hamiltonian becomes unstable~\cite{rokaj2020}. In addition, in the so-called length-gauge form of the Pauli-Fierz Hamiltonian, a different quadratic term appears, the dipole-self energy. The importance of this term has been questioned and investigated~\cite{GalegoCasimir, NorahSelfPolarization, HuoPolarizedFock, KowalewskiControlling}. We show that without the dipole self-energy, the length-gauge Hamiltonian also becomes unstable and has no ground-state~\cite{rokaj2017, schaeferquadratic}. Further, if the dipole-self energy is ignored, the length-gauge Hamiltonian no longer respects translational symmetry.

It is important to emphasize that, throughout this work the concept of translational invariance will be used as a guiding principle and will be encountered multiple times, in different settings. This is not an arbitrary choice, but has to do with the fact that for solid state and condensed matter systems, translational symmetry is a defining property.

Furthermore, these intriguing phenomena coming from the field of cavity QED materials, that we mentioned, question the well-established concepts and methods of many-body physics and quantum optics. In many-body physics usually only the longitudinal photonic degrees of freedom are treated via the Coulomb interaction, while the transversal degrees of freedom of the photon field are neglected. On the other hand, in quantum optics, the photon field is treated in great detail but matter is described only by a few states. Both simplified pictures are far from providing a realistic and comprehensive account of these novel experiments in the field of cavity QED materials.

The state of the art in cavity QED materials calls for the development of new theoretical methods and tools, which will allow the description of condensed matter systems strongly coupled to the photon field. Motivated by all these developments, the aim of this thesis is exactly to pursue this direction and develop new theories and models for the emerging field of cavity QED materials.

\textbf{The Free Electron Gas in Cavity QED.}---To achieve this goal and provide the first (to the best of our knowledge) analytically solvable model for a condensed matter system coupled to the photon field, we revisit the paradigmatic Sommerfeld model of the free electron gas~\cite{Sommerfeld1928} in the framework of cavity QED~\cite{rokaj2020}. The defining properties of the Sommerfeld model are: (i) Coulomb interaction is neglected and (ii) continuous translational invariance. Making use of translational invariance and the fact that the momenta of the electrons are a good, conserved quantum number we manage to solve analytically this system in the long-wavelength limit, also known as dipole approximation, for an arbitrary, but finite amount of photon modes.

One of the most important outcomes of our analytic solution is that the quantized photon field mediates an all-to-all interaction between the electrons. This is very important from a conceptual point of view. Because our work, in addition to the concept of many-particle systems interacting via Coulomb forces, introduces the concept of many-body systems interacting via the quantized photon field and sets a new paradigm for many-body physics. 

In connection to the transport experiments performed for materials in cavities~\cite{paravacini2019, A.Thomas2019}, we show for the Sommerfeld model, that coupling to the cavity can modify the optical conductivity of the electron gas and can suppress the Drude peak and the DC conductivity of this system~\cite{rokaj2020}. This is an exciting and important result, because it makes apparent the fact that the quantized photon field can modify macroscopic properties of materials, like their conductivity.

Another great challenge for periodic solid state systems coupled to the electromagnetic field, is to go beyond the dipole approximation. The main problem in this setting is that the spatial variation of the vector potential of the electromagnetic field breaks the fundamental translational symmetry of the lattice. Thus, Bloch's theorem~\cite{Mermin} cannot be applied. This is an extremely important issue, as most of our understanding of solid state systems relies on Bloch theory, from which the band structure of materials can be obtained. We would like to emphasize, that this is not a merely theoretically driven question, but it is actually experimentally motivated, by the advent of Moir\'{e} materials~\cite{Moiremarvels, RubioMoire} which are currently probed under strong uniform magnetic fields. The enlarged Moir\'{e} unit cell enables to achieve large magnetic fluxes for the probe of the fractal spectrum of the Hofstadter butterfly~\cite{Hofstadter, DeanButterfly, WangButterfly, BarrierButterfly, BarrierButterfly}.

\textbf{Quantum Hall Systems in Cavity QED.}---Motivated by these advances, we go beyond the dipole approximation and we look into the setting of the quantum Hall effect~\cite{Klitzing}, in which the homogeneous magnetic field breaks translational symmetry~\cite{Landau}. This breaking of translational invariance has been a long standing problem for condensed matter physics, which we manage to solve by embedding it into non-relativistic QED~\cite{rokaj2019}. What we find is, that translational symmetry exists in the higher-dimensional, electronic plus photonic (polaritonic) configuration space. In this higher-dimensional space we generalize Bloch's theorem for coupled electron-photon systems. Thus, we name this framework as quantum electodynamical Bloch (QED-Bloch) theory~\cite{rokaj2019}. QED-Bloch theory allows for the construction of a polaritonic Bloch ansatz and the non-perturbative treatment of periodic materials in homogeneous magnetic fields, strongly coupled to the quantized photon field. 

As a first application of QED-Bloch theory we consider two-dimensional Landau levels under cavity confinement and we show that the cavity field leads to the emergence of mixed quasi-particle states between the Landau levels and the photons, known as Landau polaritons~\cite{rokaj2019, Hagenmuller2010cyclotron}. The Landau polaritons are currently of great interest, and have been measured experimentally~\cite{ScalariScience, paravacini2019, li2018}. Further, for two-dimensional periodic materials perpendicular to a magnetic field and under cavity confinement, QED-Bloch theory predicts the emergence of fractal polaritons, i.e., fractal polaritonic energy spectra~\cite{RokajButterfly2021}. This is a novel prediction of our theory, which has not been reported before, and constitutes a polaritonic analogue of the Hofstadter butterfly. In the limit of no quantized cavity field, QED-Bloch theory recovers the Landau levels~\cite{Landau} and the fractal spectrum of the Hofstadter butterfly~\cite{Hofstadter}, and thus provides a non-perturbative framework for the description of periodic materials in strong magnetic fields.

The Hofstadter butterfly has become now experimentally accessible with great accuracy in Moir\'{e} systems~\cite{DeanButterfly, WangButterfly, BarrierButterfly, ForsytheButterfly} and we believe our first-principles QED-Bloch framework, can help to understand these novel experiments. In addition, our prediction of the existence of fractal polaritonic spectra due to strong light-matter coupling, opens a new avenue for the exploration of fractal physics in the field of cavity QED materials and the probe of the fractional quantum Hall effects~\cite{TsuifractionalQHE, Laughlingfractional} with the use of cavity photons.

\section*{Outline \& Brief Summary}

The thesis is organized in three parts as follows.
 
\textit{Fundamentals of Quantum Electrodynamics.}---In the first part of the thesis we present how non-relativistic QED can be constructed. In chapter~\ref{Quantum Electrodynamics} we start from Maxwell's theory of electromagnetism, which we solve in free space, and we subsequently quantize. Then, we consider Dirac's theory for relativistic matter coupled to the photon field, and by taking the non-relativistic limit, we obtain the Pauli-Fierz theory. In chapter~\ref{Length Gauge QED} we look into the long-wavelength limit (dipole approximation) of the Pauli-Fierz theory. In the dipole approximation, the photon field is spatially homogeneous and respects translational invariance in the electronic configuration space, which is desirable for the description of condensed matter systems. In the dipole approximation there is a unitarily equivalent description known as the length gauge. In this gauge, the light-matter interaction depends on the dipole operator and a term depending on the quadrature of the dipole operator shows up, which is known as the dipole self-energy. As a consequence translational invariance in the electronic space is broken. However, as translational invariance is a physical property, it still exists in the full electronic plus photonic (polaritonic) space. We note that without the dipole self-energy the latter property does not hold. Further, we prove that without the dipole self-energy the Hamiltonian becomes unstable and has no ground-state. In addition, if the dipole self-energy is discarded, gauge invariance is broken and Maxwell's equations in matter are not satisfied.

\textit{The Free Electron Gas in Cavity QED.}---In the second part of the thesis, we revisit the Sommerfeld model of the free electron gas in the framework of cavity QED. In chapter~\ref{Free Electron Gas} we briefly review the Sommerfeld model and then in chapter~\ref{Free Electron Gas in cavity QED} we couple it to the quantized cavity field. Making use of translational invariance we provide the analytic solution for the free electron gas coupled to the cavity field. Then, in the thermodynamic limit we show that the hybrid  electron-photon ground state is a Fermi liquid dressed with (virtual) photons, and that without the diamagnetic $\bi{A}^2$ term the coupled system becomes unstable. In chapter~\ref{Cavity Responses} we perform linear response and we compute the optical conductivity of the free electron gas inside the cavity. We show that the cavity field modifies the conductive properties of the electron gas due to the emergence of plasmon-polariton resonances. Most importantly, the cavity field suppresses the DC conductivity and the Drude peak of the electron gas. Finally, in chapter~\ref{Effective QFT}, to go beyond the finite-mode approximation we construct an effective field theory in the continuum of electromagnetic modes. Exploiting this effective field theory we are able to show that the continuum of modes renormalizes the electron mass, modifies the quasi-particle excitations of the Fermi liquid and introduces dissipation into the system.

\textit{Quantum Hall Systems in Cavity QED.}---In the third part of the thesis we focus on quantum Hall systems confined inside cavities. In chapter~\ref{Landau Levels QHE} we briefly review how the quantization of the macroscopic Hall conductance can be described within the picture of non-interacting electrons in fully occupied Landau levels. Subsequently, in chapter~\ref{Bloch MTG} we present Bloch's theorem for periodic materials, we discuss that in quantum Hall systems, translational invariance is broken for periodic materials due to the magnetic field and we present how the magnetic translation group emerges in this setting. In chapter~\ref{QED Bloch theory} we demonstrate that the broken translational symmetry due to the magnetic field can be restored by embedding the problem into quantum electrodynamics. By doing so translational symmetry gets restored in the enlarged electronic plus photonic configuration space, in which we can use Bloch's theorem, and the framework is named quantum electrodynamical Bloch (QED-Bloch) theory. As a first application of QED-Bloch theory we consider two-dimensional Landau levels under cavity confinement and we show how the cavity modifies the Landau levels and that hybrid quasi-particle excitations emerge, known as Landau polaritons. As a further application, we describe two-dimensional periodic materials inside the cavity and perpendicular to a magnetic field, and we show that the energy spectrum of such electron-photon systems as a function of the light-matter coupling shows a novel polaritonic fractal pattern. In the limit of no quantized field we show that QED-Bloch theory recovers the well-known fractal spectrum of the Hofstadter butterfly.

\newpage

\section*{Publications}

The following articles have been published in the context of this thesis.

\begin{itemize}
   
 \item{V.~Rokaj, D.~M.~Welakuh, M.~Ruggenthaler and A.~Rubio, \href{https://iopscience.iop.org/article/10.1088/1361-6455/aa9c99/meta}{\textit{Light–Matter Interaction in the Long-Wavelength Limit: No Ground-State without Dipole Self-Energy}, J. Phys. B: At. Mol. Opt. Phys. \textbf{51}, 034005 (2018)}.}

\item{V.~Rokaj, M.~Penz, M.~A.~Sentef, M.~Ruggenthaler and A.~Rubio, \href{https://journals.aps.org/prl/abstract/10.1103/PhysRevLett.123.047202}{\textit{Quantum Electrodynamical Bloch Theory with Homogeneous Magnetic Fields}, Phys. Rev. Lett. \textbf{123}, 047202 (2019)}. (Highlighted as \textit{Editors' Suggestion})} 

\item{C. Sch\"{a}fer, M.~Ruggenthaler, V.~Rokaj, and A.~Rubio, \href{https://pubs.acs.org/doi/abs/10.1021/acsphotonics.9b01649}{\textit{Relevance of the Quadratic Diamagnetic and Self-Polarization Terms in Cavity Quantum Electrodynamics }, ACS Photonics \textbf{7}, 975-990 (2020)}.}

\item{V.~Rokaj, M.~Ruggenthaler, F.~G.~Eich and A.~Rubio,\href{https://arxiv.org/abs/2006.09236}{ \textit{The Free Electron Gas in Cavity Quantum Electrodynamics}, arXiv:2006.09236 [quant-ph] (2020)}. }

\item{V.~Rokaj, M.~Penz, M.~A.~Sentef, M.~Ruggenthaler and A.~Rubio, \href{https://arxiv.org/abs/2109.15075}{\textit{Polaritonic Hofstadter Butterfly and Cavity-Control of the Quantized Hall Conductance}, arXiv: 2109.15075 [cond-mat.mes-hall] (2021)}. }

\end{itemize}


\part{Fundamentals of Quantum Electrodynamics}

\chapter{Quantum Electrodynamics}\label{Quantum Electrodynamics}
\begin{displayquote}
\footnotesize{You have your quantum theater containing molecules, atoms or any other material, and you have the spectators which were the photons that you were using to see what the actors were doing. Now what we are trying to do is to add a new set of actors, the photons. Those create a new bigger play which has new stories to tell.}
\end{displayquote}
\begin{flushright}
  \footnotesize{Angel Rubio\\
Interview at Latest Thinking~\cite{Angel_Latest_Thinking}}
\end{flushright}

The aim of this chapter is to present how quantum electrodynamics is usually constructed. To be more specific we are primarily interested in deriving the non-relativistic limit of quantum electrodynamics which is also known as Pauli-Fierz theory~\cite{spohn2004}. However, as classical electromagnetism is inherently a relativistic theory (due to its Lorentz invariance) one cannot start directly from some kind of non-relativistic version of classical electromagnetism (if it exists) and then quantize this theory. Thus, for the construction of the Pauli-Fierz theory we will start from the classical theory of electromagnetism, we will solve Maxwell's equations in vacuum (absence of charges and currents) and then we will quantize the solution of the free electromagnetic field. On the other hand, for the description of quantum matter we will start from Dirac's relativistic quantum theory of the electron (and positron) which is described in terms of a 4-component spinor field~\cite{Diractheoryelectron, DiracQM}. Then, we will show how the Dirac spinor-field couples to the electromagnetic field and finally by taking the non-relativistic limit in which the positrons are eliminated from the theory, we will obtain the Pauli-Fierz Hamiltonian which describes non-relativistic (quantum) electrons coupled to the quantized electromagnetic field. We note that S.I. units are used throughout.

Before we continue we would like to mention that from this brief account one concludes that the primary ontological entity in quantum electrodynamics is the field, because both matter and radiation are described in terms of this concept. However, there is an important difference between these two fields which has been nicely captured by Peierls:
\begin{displayquote}
\footnotesize{The fact is that
there is ``the great difference between the wave field describing a particle and the
electromagnetic field describing radiation''. The electromagnetic field is something
measurable in principle as well as in practice, because a classical limit exists. However, the wave field representing the electron is never measurable, nor can one
obtain a classical description for such waves.}
\end{displayquote}
\begin{flushright}
  \footnotesize{R.~E.~Peierls\\
QED and the Men Who Made It~\cite{SchweberQEDHistory}}
\end{flushright}

\section{Classical Electromagnetism}\label{Classical Electromagnetism}

In the classical theory of electromagnetism, the electromagnetic field is described by two three-dimensional vector fields, namely the electric field $\bi{E}(\bi{r},t)$ and the magnetic field $\bi{B}(\bi{r},t)$. These fields are considered to extend (in principle) throughout the whole space like a fluid for example fills an empty vessel. Then one is interested in knowing how these fields spread in space and how they evolve in time. These properties would then determine the motion of charged particles. However, it is important to mention that because the charged particles are the source of radiation, they also produce electromagnetic fields (when they are accelerated) and change the overall electromagnetic field in space. These back-reaction effects in QED are captured nicely because both matter and light enter as dynamical entities that mutually influence each other.

Mathematically, the classical theory of electromagnetism is summed up in Maxwell's equations. In the presence of a charge density $\rho(\bi{r},t)$ and current density $\bi{j}(\bi{r},t)$ the electric and the magnetic field satisfy the equations~\cite{JacksonEM, Mandl}
\begin{eqnarray}
&&\nabla \cdot \bi{E}(\bi{r},t)=\frac{\rho(\bi{r},t)}{\epsilon_0},\label{div E}\\
&&\nabla \times \bi{E}(\bi{r},t)=-\frac{\partial \bi{B}(\bi{r},t)}{\partial t},\label{curl E}\\
&&\nabla \cdot \bi{B}(\bi{r},t)=0,\label{div B}\\
&&\nabla\times\bi{B}(\bi{r},t)=\mu_0 \bi{j}(\bi{r},t)+\mu_0\epsilon_0\frac{\partial \bi{E}(\bi{r},t)}{\partial t}, \label{curl B}
\end{eqnarray}
where $\mu_0$ is the vacuum magnetic permeability and $\epsilon_0$ is the vacuum electric permittivity. In free space (absence of charges and currents) the Maxwell equations are invariant under Lorentz transformations which means that Maxwell's theory is compatible with the special theory of relativity. Further, in free space both the electric and the magnetic field satisfy the wave equation, with both fields propagating through space at the speed of light $c$~\cite{JacksonEM}. 

Two out of the four Maxwell equations, the Eqs.~(\ref{curl E}) and (\ref{div B}), can be automatically satisfied by introducing a scalar and a vector potential $\phi(\bi{r},t)$ and $\bi{A}(\bi{r},t)$ respectively, and by defining the electric and magnetic fields as
\begin{eqnarray}\label{A and phi}
\bi{B}(\bi{r},t)=\nabla \times \bi{A}(\bi{r},t) \;\;\; \textrm{and} \;\;\; \bi{E}(\bi{r},t)=-\nabla\phi(\bi{r},t)-\frac{\partial \bi{A}(\bi{r},t)}{\partial t}.
\end{eqnarray}
From the above equations it is easy to check that the electric and the magnetic field stay invariant under the following transformations of the scalar and the vector potential
\begin{eqnarray}\label{gauge transformation A and phi}
\phi(\bi{r},t) \longrightarrow \phi^{\prime}(\bi{r},t)=\phi(\bi{r},t) +\frac{\partial f(\bi{r},t)}{\partial t} \;\;\; \textrm{and}\;\;\; \bi{A}(\bi{r},t)\longrightarrow \bi{A}^{\prime}(\bi{r},t)=\bi{A}(\bi{r},t)-\nabla f(\bi{r},t)\nonumber\\
\end{eqnarray}
where $f(\bi{r},t)$ is a twice differentiable scalar function. What this means is that the potentials $\{\phi(\bi{r},t),\bi{A}(\bi{r},t)\}$ and the potentials $\{\phi^{\prime}(\bi{r},t),\bi{A}^{\prime}(\bi{r},t)\}$ produce exactly the same electric and magnetic fields. This property of Maxwell's theory is called gauge invariance. This property is not a unique feature of the theory of electromagnetism but actually it is a feature shared by all fundamental theories of Nature, from Einstein's theory of gravitation to the strong nuclear interactions~\cite{Mandl}.

Substituting the expression for the electric and the magnetic field given in terms of the potentials~Eq.(\ref{A and phi}) into the remaining two Maxwell equations~(\ref{div E}) and (\ref{curl B}) we find that the scalar and the vector potential satisfy 
\begin{eqnarray}
&&-\nabla^2\phi(\bi{r},t)-\frac{\partial \left(\nabla \cdot \bi{A}(\bi{r},t)\right)}{\partial t}=\frac{\rho(\bi{r},t)}{\epsilon_0},\label{equation phi}\\
&&-\nabla^2\bi{A}(\bi{r},t)+\frac{1}{c^2}\frac{\partial^2\bi{A}(\bi{r},t)}{\partial t^2}+\nabla\left(\nabla \cdot \bi{A}(\bi{r},t)\right)= \mu_0\bi{j}(\bi{r},t)-\frac{1}{c^2}\frac{\partial \left(\nabla\phi(\bi{r},t)\right)}{\partial t}.\label{equation A}
\end{eqnarray}
To obtain the latter we also used the relation between the speed of light $c$, the vacuum permittivity $\epsilon_0$ and the vacuum permeability $\mu_0$, 
\begin{eqnarray}
\frac{1}{c^2}=\epsilon_0\mu_0.
\end{eqnarray}

\subsection{Coulomb Gauge}

The differential equations~(\ref{equation phi}) and (\ref{equation A}) for the scalar $\phi(\bi{r},t)$ and the vector $\bi{A}(\bi{r},t)$ potential are fairly complicated. However, upon an appropriate gauge choice they can be simplified considerably. In most cases the Coulomb gauge is used, in which the longitudinal degrees of freedom of the electromagnetic field are removed. We would like to mention that there are also other gauge choices like the Lorentz or the Feynman gauge in which relativistic covariance is manifest, but then one has to pay the price of treating all the degrees of freedom on equal footing~\cite{greiner1996}.  

In the Coulomb gauge the vector potential is chosen such that~\cite{JacksonEM, Mandl}
\begin{eqnarray}\label{Coulomb gauge}
\nabla \cdot \bi{A}(\bi{r},t)=0.
\end{eqnarray}
With this choice then the equations for the potentials take the simpler form
\begin{eqnarray}
&&-\nabla^2\phi(\bi{r},t)=\frac{\rho(\bi{r},t)}{\epsilon_0},\label{Poisson equation}\\
&&-\nabla^2\bi{A}(\bi{r},t)+\frac{1}{c^2}\frac{\partial^2\bi{A}(\bi{r},t)}{\partial t^2}= \mu_0\bi{j}(\bi{r},t)-\frac{1}{c^2}\frac{\partial \left(\nabla\phi(\bi{r},t)\right)}{\partial t}.
\end{eqnarray}
Further, the electric field can be decomposed into a purely longitudinal $\bi{E}^{||}(\bi{r},t)$ and purely transversal component $\bi{E}^{\perp}(\bi{r},t)$
\begin{eqnarray}
\bi{E}(\bi{r},t)=\bi{E}^{\perp}(\bi{r},t)+\bi{E}^{||}(\bi{r},t)
\end{eqnarray}
which are defined as
\begin{eqnarray}
\bi{E}^{\perp}(\bi{r},t)=-\frac{\partial \bi{A}(\bi{r},t)}{\partial t} \;\;\; \textrm{and}\;\;\; \bi{E}^{||}(\bi{r},t)=-\nabla\phi(\bi{r},t)\label{longitudinal n transversal}
\end{eqnarray}
and satisfy the conditions
\begin{eqnarray}
\nabla\cdot \bi{E}^{\perp}(\bi{r},t)=0 \;\;\; \textrm{and}\;\;\;\nabla\times\bi{E}^{||}(\bi{r},t)=0.
\end{eqnarray}

\subsection{The Free Electromagnetic Field}

We are now going to consider the case where the charge density and the current density are both zero, $\rho(\bi{r},t)=0$ and $\bi{j}(\bi{r},t)=0$, and the electromagnetic field simply propagates in empty space. In this case the equation for the scalar potential is
\begin{eqnarray}
\nabla^2\phi(\bi{r},t)=0.
\end{eqnarray}
Requiring also the scalar potential to vanish at infinity we find that the scalar potential has to be zero, $\phi(\bi{r},t)=0$~\cite{Mandl, greiner1996}. With this solution for $\phi(\bi{r},t)$ (and with $\bi{j}(\bi{r},t)=0$ ) we find that the vector potential satisfies the wave equation
\begin{eqnarray}\label{wave Eq A field}
-\nabla^2\bi{A}(\bi{r},t)+\frac{1}{c^2}\frac{\partial^2\bi{A}(\bi{r},t)}{\partial t^2}=0. 
\end{eqnarray}
To solve the above equation we consider the vector potential $\bi{A}(\bi{r},t)$ in a cubic box of length $L$ and volume $V=L^3$. Further, for the vector potential we employ periodic boundary conditions
\begin{eqnarray}
\bi{A}(0,y,z,t)=\bi{A}(L,y,z,t), \;\; \bi{A}(x,0,z,t)=\bi{A}(x,L,z,t) \;\; \textrm{and} \;\; \bi{A}(x,y,0,t)=\bi{A}(x,y,L,t).\nonumber
\end{eqnarray}
Then, it is straightforward to show that Eq.~(\ref{wave Eq A field}) admits plane wave solutions of the form
\begin{eqnarray}\label{plane wave solutions}
\bi{u}_{\bm{\kappa},\lambda}(\bi{r},t)=\bm{\varepsilon}_{\lambda}(\bm{\kappa})\frac{e^{\textrm{i}\left(\bm{\kappa}\cdot \bi{r}-\omega(\bm{\kappa})t\right)}}{\sqrt{V}},
\end{eqnarray}
where $\bm{\kappa}$ are the wave vectors, $\bm{\varepsilon}_{\lambda}(\bm{\kappa})$ are the polarization vectors and the frequency $\omega(\bm{\kappa})$ is a function of the wave vector 
\begin{eqnarray}\label{omega frequency}
\omega(\bm{\kappa})=c|\bm{\kappa}|.
\end{eqnarray}
Imposing the periodic boundary conditions on the plane wave solutions we find that the wave vectors must be of the form
\begin{eqnarray}
\bm{\kappa}=\frac{2\pi}{L}\bi{n} , \;\;\; \textrm{where}\;\;\; \bi{n} \in \mathbb{Z}^3.
\end{eqnarray}
Moreover, from the fact that the vector potential is chosen to be in the Coulomb gauge~(\ref{Coulomb gauge}) we find that polarization vectors must be orthogonal to the wave vectors
\begin{eqnarray}\label{trasversality}
\bm{\varepsilon}_{\lambda}(\bm{\kappa})\cdot \bm{\kappa}=0 \;\;\; \forall \;\bm{\kappa}. 
\end{eqnarray}
This implies that the vector potential $\bi{A}(\bi{r},t)$ in the Coulomb gauge is transverse and that it has only two independent polarization vectors $\bm{\varepsilon}_1(\bm{\kappa})$ and $\bm{\varepsilon}_2(\bm{\kappa})$, which can be chosen mutually perpendicular
\begin{eqnarray}\label{orthonormality}
\bm{\varepsilon}_{\lambda}(\bm{\kappa})\cdot \bm{\varepsilon}_{\lambda^{\prime}}(\bm{\kappa})=\delta_{\lambda\lambda^{\prime}} \;\; \textrm{with}\;\; \lambda,\lambda^{\prime}=1,2.
\end{eqnarray}
An explicit form for the polarization vectors, such that the conditions of transversality and orthonormality in Eqs.~(\ref{trasversality}) and (\ref{orthonormality}) are satisfied is
\begin{eqnarray}\label{polarization vectors}
\bm{\varepsilon}_{1}(\bm{\kappa})=\frac{1}{|\bm{\kappa}|}\left(\sqrt{\kappa^2_y+\kappa^2_z}, \frac{-\kappa_x\kappa_y}{\sqrt{\kappa^2_y+\kappa^2_z}}, \frac{-\kappa_x\kappa_z}{\sqrt{\kappa^2_y+\kappa^2_z}}\right)\;\; \textrm{and}\;\; \bm{\varepsilon}_2(\bm{\kappa})=\frac{1}{\sqrt{\kappa^2_y+\kappa^2_z}}\left(0,\kappa_z,-\kappa_y\right).\nonumber\\
\end{eqnarray}

The plane waves of $\bi{u}_{\bm{\kappa},\lambda}(\bi{r},t)$ of Eq.~(\ref{wave Eq A field}) form a complete set of transverse orthonormal vector fields and consequently the generic solution for the vector potential $\bi{A}(\bi{r},t)$ will be given as a Fourier series
\begin{eqnarray}\label{vector potential}
\bi{A}(\bi{r},t)=\sum_{\bm{\kappa}, \lambda} \frac{\bm{\varepsilon}_{\lambda}(\bm{\kappa})}{\sqrt{V}}\left[\alpha_{\bm{\kappa},\lambda}e^{\textrm{i}\left(\bm{\kappa}\cdot \bi{r}-\omega(\bm{\kappa})t\right)}+\alpha^*_{\bm{\kappa},\lambda} e^{-\textrm{i}\left(\bm{\kappa}\cdot \bi{r}-\omega(\bm{\kappa})t\right)}\right],
\end{eqnarray}
where $\alpha_{\bm{\kappa},\lambda}$ and its complex conjugate $\alpha^*_{\bm{\kappa},\lambda}$ are scalars. Having found the expression for the vector potential $\bi{A}(\bi{r},t)$, using Eq.~(\ref{A and phi}) we can also obtain the expressions for the electric and the magnetic field in free space
\begin{eqnarray}
&&\bi{E}^{\perp}(\bi{r},t)=\sum_{\bm{\kappa}, \lambda} \frac{ \textrm{i}\omega(\bm{\kappa}) \bm{\varepsilon}_{\lambda}(\bm{\kappa})}{\sqrt{V}}\left[\alpha_{\bm{\kappa},\lambda}e^{\textrm{i}\left(\bm{\kappa}\cdot \bi{r}-\omega(\bm{\kappa})t\right)}-\alpha^*_{\bm{\kappa},\lambda} e^{-\textrm{i}\left(\bm{\kappa}\cdot \bi{r}-\omega(\bm{\kappa})t\right)}\right],\label{electric field}\\
&&\bi{B}(\bi{r},t)= \sum_{\bm{\kappa}, \lambda} \frac{\textrm{i}\bm{\kappa} \times \bm{\varepsilon}_{\lambda}(\bm{\kappa})}{\sqrt{V}}\left[\alpha_{\bm{\kappa},\lambda}e^{\textrm{i}\left(\bm{\kappa}\cdot \bi{r}-\omega(\bm{\kappa})t\right)}-\alpha^*_{\bm{\kappa},\lambda} e^{-\textrm{i}\left(\bm{\kappa}\cdot \bi{r}-\omega(\bm{\kappa})t\right)}\right]\label{magnetic field}.
\end{eqnarray}

\subsection{The Coulomb Potential}\label{Coulomb Potential}
What we described so far is the general solution of Maxwell's equations in free space. Let us consider now the situation where there is also a charge density $\rho(\bi{r},t)$ distributed in space. In this case the scalar potential $\phi(\bi{r},t)$ satisfies the Poisson equation~(\ref{Poisson equation}). The Poisson equation can be solved with the use of Green's functions~\cite{JacksonEM}. To do so one searches for a Green's function $G(\bi{r},\bi{r}^{\prime})$ satisfying the equation
\begin{eqnarray}
\nabla^2G(\bi{r},\bi{r}^{\prime})=\delta\left(\bi{r}-\bi{r}^{\prime}\right).
\end{eqnarray}
A solution to the above equation in three spatial dimensions is\footnote{It is important to mention that the Green's function for the Poisson equation here is obtained for the scalar potential $\phi(\bi{r},t)$ considered in infinite space. This might seem inconsistent to the way we treated the vector potential $\bi{A}(\bi{r},t)$, in a finite space of volume $V$ with periodic boundary conditions, but the two treatments become mathematically the same upon taking the volume of space to infinity $V\rightarrow\infty$.}
\begin{eqnarray}
G(\bi{r},\bi{r}^{\prime})=-\frac{1}{4\pi|\bi{r}-\bi{r}^{\prime}|}.
\end{eqnarray}
Then it is straightforward to show that the scalar potential 
\begin{eqnarray}
\phi(\bi{r},t)=\frac{-1}{\epsilon_0}\int d^3r^{\prime}G(\bi{r},\bi{r}^{\prime})\rho(\bi{r}^{\prime},t)=\int d^3r^{\prime}\frac{\rho(\bi{r}^{\prime},t)}{ 4\pi\epsilon_0|\bi{r}-\bi{r}^{\prime}|}\label{general phi}
\end{eqnarray}
given in terms of the Green's function, satisfies the Poisson equation~(\ref{Poisson equation}). From the above result we understand that the scalar potential $\phi(\bi{r},t)$ is completely determined by the charge density $\rho(\bi{r},t)$. This means that it has no independent degrees of freedom. This is a consequence of the fact that in Eq.~(\ref{Poisson equation}) no time-derivatives of $\phi(\bi{r},t)$ show up. This implies that the scalar potential is not dynamical. We would like also to emphasize that in the case of a non-zero charge distribution $\rho(\bi{r},t)$, the longitudinal component of the electric field is non-trivial.

Finally, we would like to highlight that for a charge distribution of point particles with charges $q_i$
\begin{eqnarray}
\rho(\bi{r},t)=\sum_{i}q_i\delta(\bi{r}-\bi{r}_i(t))
\end{eqnarray}
we obtain the standard Coulomb potential
\begin{eqnarray}
\phi_{\textrm{C}}(\bi{r},t)=\sum_i\frac{q_i}{4\pi \epsilon_0|\bi{r}-\bi{r}_i(t)|}\label{Coulomb phi}.
\end{eqnarray}

\section{Quantization of the Electromagnetic Field}\label{EM field Quantization}

Having found the generic solution of Maxwell's equations for the electromagnetic field in the absence of the charges and currents, our aim now is to proceed with the quantization of the electromagnetic field. To do so we will follow the standard quantization procedure known as canonical quantization~\cite{DiracQM}.

In canonical quantization one identifies the canonical variables of the theory and the conjugate momenta and then promotes the classical Poisson bracket into a commutator. In Maxwell's theory the canonical variable (or canonical field) is the vector potential $\bi{A}(\bi{r},t)$, since as it was demonstrated, in free space the electric and the magnetic field can be derived from it. The conjugate field to the vector potential is the transverse electric field $\bi{E}^{\perp}(\bi{r},t)$. In principle to show that the electric field is the conjugate field to the vector potential one needs to go into the framework of Lagrangian field theory~\cite{Mandl}. However, this is beyond the scope of this section and our purposes. An intuitive way to understand that this is indeed true is that in analogy to the momentum which is (up to a constant) the time derivative of the position (canonical variable) in classical mechanics, the electric field is also the time derivative (up to a minus) of the vector potential (see Eq.~(\ref{A and phi})) which is the canonical field.

Then, the vector potential and the electric field are promoted into field operators
\begin{eqnarray}
\{\bi{A}(\bi{r},t), \bi{E}^{\perp}(\bi{r},t)\} \longrightarrow \{\hat{\bi{A}}(\bi{r},t), \hat{\bi{E}}^{\perp}(\bi{r},t)\}
\end{eqnarray}
which must satisfy the equal time commutation relations~\cite{Mandl, Rene2019, buhmann2013dispersionI}
\begin{eqnarray}\label{canonical quantization}
&&[\hat{A_i}(\bi{r},t),\epsilon_0 \hat{E}^{\perp}_j(\bi{r}^{\prime},t)]=-\textrm{i}\hbar\delta^{\perp}_{ij}(\bi{r}-\bi{r}^{\prime})\;\;\; \textrm{and}\;\;\\
&& [\hat{A}_i(\bi{r},t), \hat{A}_j(\bi{r}^{\prime},t)]=[\hat{E}^{\perp}_i(\bi{r},t), \hat{E}^{\perp}_j(\bi{r}^{\prime},t)]=0\nonumber,
\end{eqnarray}
where $\hat{A}_i(\bi{r},t)$ and $\hat{E}_j(\bi{r},t)$ are the components of the vector potential and the electric field respectively and $\delta^{\perp}_{ij}(\bi{r}-\bi{r}^{\prime})$ is the transverse delta distribution~\cite{greiner1996}
\begin{eqnarray}\label{transverse delta}
\delta^{\perp}_{ij}(\bi{r}-\bi{r}^{\prime})=\left(\delta_{ij}-\frac{\partial_i\partial_j}{\nabla^2}\right)\delta(\bi{r}-\bi{r}^{\prime}) \;\;\; \textrm{with}\;\;\; i,j=1,2,3.
\end{eqnarray}
We note that the transverse delta distribution is necessary for the quantization of the electromagnetic field in the Coulomb gauge. If a simple delta distribution had been used, then the quantization conditions~(\ref{canonical quantization}) would actually violate the Coulomb gauge~\cite{greiner1996}.

The quantization of the electromagnetic field has the consequence that the coefficients $\alpha_{\bm{\kappa},\lambda}$ and their complex conjugate $\alpha^*_{\bm{\kappa},\lambda}$ are also now promoted into operators
\begin{eqnarray}
\{\alpha_{\bm{\kappa},\lambda}, \alpha^{*}_{\bm{\kappa},\lambda}\} \longrightarrow \{\hat{\alpha}_{\bm{\kappa},\lambda}, \hat{\alpha}^{\dagger}_{\bm{\kappa},\lambda}\}. 
\end{eqnarray}
To find now the commutation relations between the operators $\hat{\alpha}_{\bm{\kappa},\lambda}$ and $\hat{\alpha}^{\dagger}_{\bm{\kappa},\lambda}$ we need the expression for $\hat{\alpha}_{\bm{\kappa},\lambda}$ and $\hat{\alpha}^{\dagger}_{\bm{\kappa},\lambda}$ in terms of the vector potential $\hat{\bi{A}}(\bi{r},t)$ and the transverse electric field $\hat{\bi{E}}^{\perp}(\bi{r},t)$. For this we multiply the expression for the vector potential and the electric field in Eqs~(\ref{vector potential}) and (\ref{electric field}) with the plane wave 
\begin{eqnarray}
\frac{1}{\sqrt{V}}e^{-\textrm{i}(\bm{\kappa}^{\prime}\cdot\bi{r}-\omega(\bm{\kappa}^{\prime})t)}
\end{eqnarray}
and we integrate over $\bi{r}$
\begin{eqnarray}\label{integral over A n E}
&&\frac{1}{\sqrt{V}}\int d^3r e^{-\textrm{i}(\bm{\kappa}^{\prime}\cdot\bi{r}-\omega(\bm{\kappa}^{\prime})t)} \hat{\bi{A}}(\bi{r},t)= \sum_{\lambda}\bm{\varepsilon}_{\lambda}(\bm{\kappa}^{\prime})\hat{\alpha}_{\bm{\kappa}^{\prime},\lambda} +e^{2\textrm{i}\omega(\bm{\kappa}^{\prime})t}\sum_{\lambda}\bm{\varepsilon}_{\lambda}(-\bm{\kappa}^{\prime}) \hat{\alpha}^{\dagger}_{-\bm{\kappa}^{\prime},\lambda}\;\; \textrm{and}\nonumber\\
&&\frac{1}{\sqrt{V}}\int d^3r e^{-\textrm{i}(\bm{\kappa}^{\prime}\cdot\bi{r}-\omega(\bm{\kappa}^{\prime})t)} \hat{\bi{E}}^{\perp}(\bi{r},t)= \textrm{i}\omega(\bm{\kappa}^{\prime})\left[\sum_{\lambda}\bm{\varepsilon}_{\lambda}(\bm{\kappa}^{\prime})\hat{\alpha}_{\bm{\kappa}^{\prime},\lambda} -e^{2\textrm{i}\omega(\bm{\kappa}^{\prime})t}\sum_{\lambda}\bm{\varepsilon}_{\lambda}(-\bm{\kappa}^{\prime}) \hat{\alpha}^{\dagger}_{-\bm{\kappa}^{\prime},\lambda}\right].\nonumber\\
\end{eqnarray}
To obtain the expressions above, the completeness relation of the plane waves was used
\begin{eqnarray}\label{plane wave completeness}
\frac{1}{V}\int_{V}d^3r e^{\textrm{i}(\bm{\kappa}-\bm{\kappa}^{\prime})\cdot\bi{r}}=\delta_{\bm{\kappa}\bm{\kappa}^{\prime}}
\end{eqnarray}
and a summation over the momenta  was performed. Solving the system of linear equations in~(\ref{integral over A n E}) we find for $\hat{\alpha}_{\bm{\kappa},\lambda}$
\begin{eqnarray}
\sum_{\lambda}\bm{\varepsilon}_{\lambda}(\bm{\kappa}) \hat{\alpha}_{\bm{\kappa},\lambda}=\frac{1}{2\sqrt{V}}\int d^3r\left[\hat{\bi{A}}(\bi{r},t)+\frac{1}{\textrm{i}\omega(\bm{\kappa})}\hat{\bi{E}}^{\perp}(\bi{r},t)\right]e^{-\textrm{i}(\bm{\kappa}\cdot\bi{r}-\omega(\bm{\kappa})t)}
\end{eqnarray}
where the index $\bm{\kappa}^{\prime}$ was exchanged with $\bm{\kappa}$ for simplicity. Further, we multiply the equation above with the polarization $\bm{\varepsilon}_{\lambda}(\bm{\kappa})$, we use the orthonormality relation for the polarization~(\ref{orthonormality}) and then $\hat{\alpha}_{\bm{\kappa},\lambda}$ is 
\begin{eqnarray}\label{alpha operator}
\hat{\alpha}_{\bm{\kappa},\lambda}=\frac{1}{2\sqrt{V}}\int d^3re^{-\textrm{i}(\bm{\kappa}\cdot\bi{r}-\omega(\bm{\kappa})t)}\bm{\varepsilon}_{\lambda}(\bm{\kappa})\cdot\left[\hat{\bi{A}}(\bi{r},t)+\frac{1}{\textrm{i}\omega(\bm{\kappa})}\hat{\bi{E}}^{\perp}(\bi{r},t)\right].
\end{eqnarray}
Analogously, one finds the expression for $\hat{\alpha}^{\dagger}_{\bm{\kappa},\lambda}$
\begin{eqnarray}\label{alpha dagger operator}
\hat{\alpha}^{\dagger}_{\bm{\kappa},\lambda}=\frac{1}{2\sqrt{V}}\int d^3re^{\textrm{i}(\bm{\kappa}\cdot\bi{r}-\omega(\bm{\kappa})t)}\bm{\varepsilon}_{\lambda}(\bm{\kappa})\cdot\left[\hat{\bi{A}}(\bi{r},t)-\frac{1}{\textrm{i}\omega(\bm{\kappa})}\hat{\bi{E}}^{\perp}(\bi{r},t)\right].
\end{eqnarray}
From the expression for $\hat{\alpha}_{\bm{\kappa},\lambda}$ and $\hat{\alpha}^{\dagger}_{\bm{\kappa},\lambda}$ one can compute straightforwardly their commutation relations
\begin{eqnarray}
&&[\hat{\alpha}_{\bm{\kappa},\lambda}, \hat{\alpha}^{\dagger}_{\bm{\kappa}^{\prime},\lambda^{\prime}}] = \frac{1}{4V} \iint d^3r d^3r^{\prime} e^{-\textrm{i}(\bm{\kappa}\cdot\bi{r}-\omega(\bm{\kappa})t)}e^{\textrm{i}(\bm{\kappa}^{\prime}\cdot\bi{r}^{\prime}-\omega(\bm{\kappa}^{\prime})t)}\times \\ &&\Bigg[\frac{-1}{\textrm{i}\omega(\bm{\kappa}^{\prime})}\left[\bm{\varepsilon}_{\lambda}(\bm{\kappa}) \cdot\hat{\bi{A}}(\bi{r},t),\bm{\varepsilon}_{\lambda^{\prime}}(\bm{\kappa}^{\prime})\cdot\hat{\bi{E}}^{\perp}(\bi{r}^{\prime},t)\right]
+\frac{1}{\textrm{i}\omega(\bm{\kappa})}\left[\bm{\varepsilon}_{\lambda}(\bm{\kappa}) \cdot\hat{\bi{E}}^{\perp}(\bi{r},t),\bm{\varepsilon}_{\lambda^{\prime}}(\bm{\kappa}^{\prime})\cdot\hat{\bi{A}}(\bi{r}^{\prime},t)\right]\Bigg]\nonumber
\end{eqnarray}
We use the commutation relation between the electric field and the vector potential given in Eq.~(\ref{canonical quantization}) and the definition for the transverse delta distribution (\ref{transverse delta}) and we have
\begin{eqnarray}
[\hat{\alpha}_{\bm{\kappa},\lambda}, \hat{\alpha}^{\dagger}_{\bm{\kappa}^{\prime},\lambda^{\prime}}] &=& \frac{\hbar}{4\epsilon_0V} \iint d^3r d^3r^{\prime} e^{-\textrm{i}(\bm{\kappa}\cdot\bi{r}-\omega(\bm{\kappa})t)}e^{\textrm{i}(\bm{\kappa}^{\prime}\cdot\bi{r}^{\prime}-\omega(\bm{\kappa}^{\prime})t)} \times \\
&\times&\varepsilon^{i}_{\lambda}(\bm{\kappa})\varepsilon^{j}_{\lambda^{\prime}}(\bm{\kappa}^{\prime})\left(\delta_{ij}-\frac{\partial_i\partial_j}{\nabla^2}\right)\delta(\bi{r}-\bi{r}^{\prime})\left[\frac{1}{\omega(\bm{\kappa})}+\frac{1}{\omega(\bm{\kappa}^{\prime})}\right].\nonumber
\end{eqnarray}
For the term containing the partial derivatives $\partial_i\partial_j$ we perform an integration by parts (in which the boundary term is set to zero), and this term vanishes because it becomes proportional to $\bm{\varepsilon}_{\lambda}(\bm{\kappa})\cdot \bm{\kappa}$ which is equal to zero, because the polarization vectors are orthogonal to the wave vectors. Then, we use the properties of the Kronecker-$\delta$ and we also integrate over $\bi{r}^{\prime}$ 
\begin{eqnarray}
[\hat{\alpha}_{\bm{\kappa},\lambda}, \hat{\alpha}^{\dagger}_{\bm{\kappa}^{\prime},\lambda^{\prime}}] = \frac{\hbar}{4\epsilon_0 V} \int d^3r e^{\textrm{i}(\bm{\kappa}^{\prime}-\bm{\kappa})\cdot\bi{r}}e^{\textrm{i}t(\omega(\bm{\kappa})-\omega(\bm{\kappa}^{\prime}))}\bm{\varepsilon}_{\lambda}(\bm{\kappa})\cdot\bm{\varepsilon}_{\lambda^{\prime}}(\bm{\kappa}^{\prime})\left(\frac{1}{\omega(\bm{\kappa}^{\prime})}+\frac{1}{\omega(\bm{\kappa})}\right).
\end{eqnarray}
Finally, integrating over $\bi{r}$, using the completeness relation Eq.~(\ref{plane wave completeness}) and the orthonormality of the polarization vectors~(\ref{orthonormality}) one obtains the commutation  relations for $\hat{\alpha}_{\bm{\kappa},\lambda}$ and $\hat{\alpha}^{\dagger}_{\bm{\kappa},\lambda}$
\begin{eqnarray}\label{bosonic algebra}
[\hat{\alpha}_{\bm{\kappa},\lambda}, \hat{\alpha}^{\dagger}_{\bm{\kappa}^{\prime},\lambda^{\prime}}]=\frac{\hbar}{2\epsilon_0\omega(\bm{\kappa})}\delta_{\bm{\kappa},\bm{\kappa}^{\prime}}\delta_{\lambda, \lambda^{\prime}}.
\end{eqnarray}
Moreover, it is trivial to show that 
\begin{eqnarray}
[\hat{\alpha}_{\bm{\kappa},\lambda}, \hat{\alpha}_{\bm{\kappa}^{\prime},\lambda^{\prime}}]=[\hat{\alpha}^{\dagger}_{\bm{\kappa},\lambda}, \hat{\alpha}^{\dagger}_{\bm{\kappa}^{\prime},\lambda^{\prime}}]=0.
\end{eqnarray}
To normalize the commutations relations between $\hat{\alpha}_{\bm{\kappa},\lambda}$ and $\hat{\alpha}^{\dagger}_{\bm{\kappa},\lambda}$ we introduce a new set of scaled operators $\hat{a}_{\bm{\kappa},\lambda}$ and $\hat{a}^{\dagger}_{\bm{\kappa},\lambda}$ defined as
\begin{eqnarray}
\hat{a}_{\bm{\kappa},\lambda}=\sqrt{\frac{2\epsilon_0\omega(\bm{\kappa})}{\hbar}} \hat{\alpha}_{\bm{\kappa},\lambda} \;\;\; \textrm{and}\;\;\; \hat{a}^{\dagger}_{\bm{\kappa},\lambda}=\sqrt{\frac{2\epsilon_0\omega(\bm{\kappa})}{\hbar}} \hat{\alpha}^{\dagger}_{\bm{\kappa},\lambda}.
\end{eqnarray}
The operators $\hat{a}_{\bm{\kappa},\lambda}$ and $\hat{a}^{\dagger}_{\bm{\kappa},\lambda}$ then satisfy the normalized commutation relations
\begin{eqnarray}
[\hat{a}_{\bm{\kappa},\lambda},\hat{a}^{\dagger}_{\bm{\kappa}^{\prime},\lambda^{\prime}}]=\delta_{\bm{\kappa}\bm{\kappa}^{\prime}}\delta_{\lambda\lambda^{\prime}}\;\;\; \textrm{and}\;\;\; [\hat{a}_{\bm{\kappa},\lambda},\hat{a}_{\bm{\kappa}^{\prime},\lambda^{\prime}}]=[\hat{a}^{\dagger}_{\bm{\kappa},\lambda},\hat{a}^{\dagger}_{\bm{\kappa}^{\prime},\lambda^{\prime}}]=0.
\end{eqnarray}
In terms of the new operators $\hat{a}_{\bm{\kappa},\lambda}$ and $\hat{a}^{\dagger}_{\bm{\kappa},\lambda}$ the quantized vector potential, and the quantized electric and magnetic fields are
\begin{eqnarray}
&&\hat{\bi{A}}(\bi{r},t)=\sum_{\bm{\kappa}, \lambda} \sqrt{\frac{\hbar}{2\epsilon_0V\omega(\bm{\kappa})}}\bm{\varepsilon}_{\lambda}(\bm{\kappa})\left[\hat{a}_{\bm{\kappa},\lambda}e^{\textrm{i}\left(\bm{\kappa}\cdot \bi{r}-\omega(\bm{\kappa})t\right)}+\hat{a}^{\dagger}_{\bm{\kappa},\lambda} e^{-\textrm{i}\left(\bm{\kappa}\cdot \bi{r}-\omega(\bm{\kappa})t\right)}\right],\label{quantum A field}\\
&&\hat{\bi{E}}^{\perp}(\bi{r},t)=\sum_{\bm{\kappa}, \lambda} \sqrt{\frac{ \hbar \omega(\bm{\kappa}) }{2\epsilon_0V}}\textrm{i}\bm{\varepsilon}_{\lambda}(\bm{\kappa})\left[\hat{a}_{\bm{\kappa},\lambda}e^{\textrm{i}\left(\bm{\kappa}\cdot \bi{r}-\omega(\bm{\kappa})t\right)}-\hat{a}^{\dagger}_{\bm{\kappa},\lambda} e^{-\textrm{i}\left(\bm{\kappa}\cdot \bi{r}-\omega(\bm{\kappa})t\right)}\right],\label{quantum electric field}\\
&&\hat{\bi{B}}(\bi{r},t)= \sum_{\bm{\kappa}, \lambda} \sqrt{\frac{\hbar}{2\epsilon_0V\omega(\bm{\kappa})}}\textrm{i}\bm{\kappa} \times \bm{\varepsilon}_{\lambda}(\bm{\kappa})\left[\hat{a}_{\bm{\kappa},\lambda}e^{\textrm{i}\left(\bm{\kappa}\cdot \bi{r}-\omega(\bm{\kappa})t\right)}-\hat{a}^{\dagger}_{\bm{\kappa},\lambda} e^{-\textrm{i}\left(\bm{\kappa}\cdot \bi{r}-\omega(\bm{\kappa})t\right)}\right]\label{quantum magnetic field}.
\end{eqnarray}
At first glance this last step in which the new, scaled operators $\hat{a}_{\bm{\kappa},\lambda}$ and $\hat{a}^{\dagger}_{\bm{\kappa},\lambda}$ were introduced might seem trivial and without any physical significance. However, it is important to highlight that this is not entirely true. Because by introducing the new set of operators, the expressions for the vector potential and for the electric and magnetic fields changed in such a way that Planck's constant $\hbar$ showed up. This is a consequence of the quantization conditions of the electromagnetic field in Eq.~(\ref{canonical quantization}) and makes manifest the fact that the electromagnetic fields being considered now are quantum mechanical. 

\subsection{Hamiltonian of the Electromagnetic Field}

For a complete quantum mechanical description of the photon field also the Hamiltonian operator describing the theory is necessary. From classical electromagnetic theory we know that the energy of the electromagnetic field is
\begin{eqnarray}\label{transeversal energy}
H^{\perp}_{\textrm{\tiny{\textrm{EM}}}}=\frac{\epsilon_0}{2}\int_{V} d^3r\left[\big(\bi{E}^{\perp}\big)^2(\bi{r},t)+c^2\bi{B}^2(\bi{r},t)\right].
\end{eqnarray}
To promote the classical Hamiltonian $H$ into a quantum Hamiltonian operator $\hat{H}$ we replace the classical electric and magnetic fields by their quantized counterparts defined in Eqs.~(\ref{quantum electric field}) and (\ref{quantum magnetic field})
\begin{eqnarray}
\hat{H}^{\perp}_{\textrm{\tiny{\textrm{EM}}}}=\frac{\epsilon_0}{2}\int_{V} d^3r\left[\left(\hat{\bi{E}}^{\perp}\right)^2(\bi{r},t)+c^2\hat{\bi{B}}^2(\bi{r},t)\right].
\end{eqnarray}
We substitute the expression for the electric and the magnetic field given in Eqs.~(\ref{quantum electric field}) and (\ref{quantum magnetic field}) and after some tedious and laborious amount of algebra, we obtain the final expression for the quantized Hamiltonian describing the photon field~\cite{Mandl, buhmann2013dispersionI, greiner1996}
\begin{eqnarray}\label{EM Hamiltonian}
\hat{H}^{\perp}_{\textrm{\tiny{EM}}}=\sum_{\bm{\kappa},\lambda}\hbar\omega(\bm{\kappa})\left(\hat{a}^{\dagger}_{\bm{\kappa},\lambda}\hat{a}_{\bm{\kappa},\lambda}+\frac{1}{2}\right).
\end{eqnarray}
As we see the quantized Hamiltonian is now written completely in terms of the operators $\hat{a}^{\dagger}_{\bm{\kappa},\lambda}, \hat{a}_{\bm{\kappa},\lambda}$. Comparing the expression for $\hat{H}$ to the Hamiltonian of a harmonic oscillator~\cite{GriffithsQM} we see that $\hat{H}^{\perp}_{\tiny{\textrm{EM}}}$ is described actually by an infinite sum of harmonic oscillators. Moreover, the operators $\hat{a}^{\dagger}_{\bm{\kappa},\lambda}, \hat{a}_{\bm{\kappa},\lambda}$ satisfy the standard bosonic algebra of the harmonic oscillator. Due to this analogy we call $\hat{a}^{\dagger}_{\bm{\kappa},\lambda}$ and $\hat{a}_{\bm{\kappa},\lambda}$ creation and annihilation operators respectively. This algebraic similarity to the quantum harmonic oscillator will help us to construct the photonic Hilbert space and to establish the concept of photons.

\subsection{Photons \& Photonic Hilbert Space}

The concept of photon dates back to the beginning of the twentieth century and to the early days of quantum mechanics. In 1905 Einstein~\cite{PhotoelectricEinstein} introduced the concept of the photon as the quantum of the electromagnetic field, to explain the photoelectric effect. Throughout this section we are dealing with the quantum theory of the electromagnetic field, but so far the notion of photons has been completely absent and we have restricted our discussion purely to fields, classical and quantum. Thus naturally the question arises: where are the photons in this theory?

To answer this question one needs to take a closer look into the Hilbert space of the Hamiltonian $\hat{H}^{\perp}_{\textrm{\tiny{EM}}}$ which describes the photon field. The operators $\hat{a}^{\dagger}_{\bm{\kappa},\lambda}, \hat{a}_{\bm{\kappa},\lambda}$ satisfy the bosonic commutation relations of Eq.~(\ref{bosonic algebra}). Using this bosonic algebra it is possible to construct the complete photonic Hilbert space. 

First, we define the operator $\hat{N}_{\bm{\kappa},\lambda}$, which is usually known as the number density operator or mode occupation,
\begin{eqnarray}
\hat{N}_{\bm{\kappa},\lambda}=\hat{a}^{\dagger}_{\bm{\kappa},\lambda}\hat{a}_{\bm{\kappa},\lambda}.
\end{eqnarray}
Then, using the bosonic algebra~(\ref{bosonic algebra}) we can straightforwardly compute the commutation relations between $\hat{N}_{\bm{\kappa},\lambda}$ and the annihilation and creation operators
\begin{eqnarray}\label{number operator algebra}
[\hat{N}_{\bm{\kappa},\lambda},\hat{a}_{\bm{\kappa},\lambda}]=-\hat{a}_{\bm{\kappa},\lambda} \;\;\;\; \textrm{and}\;\;\;\; [\hat{N}_{\bm{\kappa},\lambda},\hat{a}^{\dagger}_{\bm{\kappa},\lambda}]=\hat{a}^{\dagger}_{\bm{\kappa},\lambda}.
\end{eqnarray}
Further, we assume that the operator $\hat{N}_{\bm{\kappa},\lambda}$ has a normalized eigenstate $|n\rangle$ with eigenvalue $n$
\begin{eqnarray}
\hat{N}_{\bm{\kappa},\lambda}|n\rangle_{\bm{\kappa},\lambda}=n|n\rangle_{\bm{\kappa},\lambda}
\end{eqnarray}
this is a fairly generic assumption which we are allowed to make. It is easy to show that the eigenvalues of the number density operator are positive. To show this we consider the expectation value of $\hat{N}_{\bm{\kappa},\lambda}$ with respect to its eigenstate $|n\rangle_{\bm{\kappa},\lambda}$
\begin{eqnarray}
{}_{\lambda, \bm{\kappa}}\langle n|\hat{N}_{\bm{\kappa},\lambda}|n\rangle_{\bm{\kappa},\lambda}=n\; {}_{\lambda, \bm{\kappa}}\langle n|n\rangle_{\bm{\kappa},\lambda}=n,
\end{eqnarray}
where we used the fact that the state $|n\rangle_{\bm{\kappa},\lambda}$ is normalized to 1. Using the definition for $\hat{N}_{\bm{\kappa},\lambda}$ we find that the expectation value ${}_{\lambda, \bm{\kappa}}\langle n|\hat{N}_{\bm{\kappa},\lambda}|n\rangle_{\bm{\kappa},\lambda}$ is actually equal to the norm of the state $|s\rangle=\hat{a}_{\bm{\kappa},\lambda}|n\rangle_{\bm{\kappa},\lambda}$,
\begin{eqnarray}\label{positive definite eigenvalues}
n={}_{\lambda, \bm{\kappa}}\langle n|\hat{N}_{\bm{\kappa},\lambda}|n\rangle_{\bm{\kappa},\lambda}= \underbrace{{}_{\lambda, \bm{\kappa}}\langle n|\hat{a}^{\dagger}_{\bm{\kappa},\lambda}}_{\langle s|}\underbrace{\hat{a}_{\bm{\kappa},\lambda}|n\rangle_{\bm{\kappa},\lambda}}_{|s\rangle}=\langle s|s\rangle \geq 0
\end{eqnarray}
which of course on a Hilbert space is positive. This implies that the eigenvalues of $\hat{N}_{\bm{\kappa},\lambda}$ are strictly positive.

Moreover, by applying the creation operator $\hat{a}^{\dagger}_{\bm{\kappa},\lambda}$ on the state $|n\rangle_{\bm{\kappa},\lambda}$ we can construct more eigenstates of the operator $\hat{N}_{\bm{\kappa},\lambda}$ with higher eigenvalues. Being more precise we find that the state $\hat{a}^{\dagger}_{\bm{\kappa},\lambda}|n\rangle_{\bm{\kappa},\lambda}$ is an eigenstate of the number operator $\hat{N}_{\bm{\kappa},\lambda}$
\begin{eqnarray}
\hat{N}_{\bm{\kappa},\lambda}\left(\hat{a}^{\dagger}_{\bm{\kappa},\lambda}|n\rangle_{\bm{\kappa},\lambda}\right)=(n+1)\hat{a}^{\dagger}_{\bm{\kappa},\lambda}|n\rangle_{\bm{\kappa},\lambda}
\end{eqnarray}
with eigenvalue $n+1$. We can define the normalized eigenstate with eigenvalue $n+1$ as
\begin{eqnarray}
|n+1\rangle_{\bm{\kappa},\lambda}=\frac{\hat{a}^{\dagger}_{\bm{\kappa},\lambda}}{\sqrt{n+1}}|n\rangle_{\bm{\kappa},\lambda}
\end{eqnarray}
By applying multiple times the creation operator on $|n\rangle_{\bm{\kappa},\lambda}$ we can construct even higher eigenstates of the operator $\hat{N}_{\bm{\kappa},\lambda}$
\begin{eqnarray}
\hat{N}_{\bm{\kappa},\lambda}\left(\left(\hat{a}^{\dagger}_{\bm{\kappa},\lambda}\right)^m|n\rangle_{\bm{\kappa},\lambda}\right)=(n+m)\left(\hat{a}^{\dagger}_{\bm{\kappa},\lambda}\right)^m|n\rangle_{\bm{\kappa},\lambda}.
\end{eqnarray}
Then, from the above equation we can write the normalized eigenstate of $\hat{N}_{\bm{\kappa},\lambda}$ with eigenvalue $n+m$ as
\begin{eqnarray}
|n+m\rangle_{\bm{\kappa},\lambda}= \frac{1}{c_m}\left(\hat{a}^{\dagger}_{\bm{\kappa},\lambda}\right)^m|n\rangle_{\bm{\kappa},\lambda}\;\;\; \textrm{where}\;\;\; c_m=\prod^m_{i=1}\sqrt{n+i}.
\end{eqnarray}
In addition, if we apply the annihilation operator on the state $|n\rangle_{\bm{\kappa},\lambda}$ we obtain eigenstates of the number operator $\hat{N}_{\bm{\kappa},\lambda}$ but with lower eigenvalues,
\begin{eqnarray}
\hat{N}_{\bm{\kappa},\lambda}\left(\hat{a}_{\bm{\kappa},\lambda}|n\rangle\right)=(n-1)\hat{a}_{\bm{\kappa},\lambda}|n\rangle_{\bm{\kappa},\lambda}
\end{eqnarray}
and by applying the annihilation operator multiple times on $|n\rangle_{\bm{\kappa},\lambda}$ we have
\begin{eqnarray}
\hat{N}_{\bm{\kappa},\lambda}\left(\left(\hat{a}_{\bm{\kappa},\lambda}\right)^l|n\rangle_{\bm{\kappa},\lambda}\right)=(n-l)\left(\hat{a}_{\bm{\kappa},\lambda}\right)^l|n\rangle_{\bm{\kappa},\lambda},
\end{eqnarray}
and as we did before we also label the above states by their eigenvalue with respect to $\hat{N}_{\bm{\kappa},\lambda}$
\begin{eqnarray}
|n-l\rangle_{\bm{\kappa},\lambda}=\frac{1}{d_l} \left(\hat{a}_{\bm{\kappa},\lambda}\right)^l|n\rangle_{\bm{\kappa},\lambda}\;\;\; \textrm{where}\;\;\; d_l=\prod^{l-1}_{i=0}\sqrt{n-i}.
\end{eqnarray}
Obviously, for $l=n$ we have the $0_{\textrm{th}}$ state $|0\rangle_{\bm{\kappa},\lambda}$ for which the operator $\hat{N}_{\bm{\kappa},\lambda}$ has zero as an eigenvalue. From the fact that the eigenvalues of $\hat{N}_{\bm{\kappa},\lambda}$ are positive we find that the state $|-1\rangle_{\bm{\kappa},\lambda}=\hat{a}_{\bm{\kappa},\lambda}|0\rangle_{\bm{\kappa},\lambda}$ cannot exist, because if it existed it would yield a negative eigenvalue for the number density operator. Thus, we conclude that the state $|0\rangle_{\bm{\kappa},\lambda}$ is the lowest eigenstate of the operator $\hat{N}_{\bm{\kappa},\lambda}$ and that when we apply the operator $\hat{a}_{\bm{\kappa},\lambda}$ on $|0\rangle_{\bm{\kappa},\lambda}$, the state gets annihilated,
\begin{eqnarray}\label{photon vacuum}
\hat{a}_{\bm{\kappa},\lambda}|0\rangle_{\bm{\kappa},\lambda}=0.
\end{eqnarray}
The state $|0\rangle_{\bm{\kappa},\lambda}$ is the ground-state of the mode with momentum $\bm{\kappa}$ and polarization $\lambda$. Then, we can define the ground-state of the full photon field as the tensor product over the ground-states of all photon-modes
\begin{eqnarray}
|0\rangle =\bigotimes_{\bm{\kappa},\lambda} |0\rangle_{\bm{\kappa},\lambda}.
\end{eqnarray}
The state above is also known as the electromagnetic vacuum or simply as the vacuum state. This state bears this name because it contains no excitations of the photon field, i.e., no photons. To understand this point first we need to define the photon number operator. The photon number operator $\hat{N}_{\textrm{ph}}$ is defined as the sum of the mode occupation $\hat{N}_{\bm{\kappa},\lambda}$ over all the modes and polarizations of the electromagnetic field
\begin{eqnarray}
\hat{N}_{\textrm{ph}}=\sum_{\bm{\kappa},\lambda}\hat{N}_{\bm{\kappa},\lambda}.
\end{eqnarray}
If we compute now the expectation value of the photon number operator with respect to the vacuum state $|0\rangle$ we find that the vacuum state has a zero amount of photons 
\begin{eqnarray}
\langle 0|\hat{N}_{\textrm{ph}}|0\rangle=0.
\end{eqnarray}
\textit{Zero-Point Energy.}---Although the vacuum state of the electromagnetic field has no photons it carries a significant amount of energy. This energy of the vacuum it is not just significant but comes out to be actually infinite. To see this we have to compute the expectation value of the Hamiltonian of the electromagnetic field $\hat{H}^{\perp}_{\tiny{\textrm{EM}}}$ with respect to the vacuum state
\begin{eqnarray}
E_{\textrm{vac}}=\langle 0|\hat{H}^{\perp}_{\tiny{\textrm{EM}}}|0\rangle= \sum_{\bm{\kappa}}\hbar \omega(\bm{\kappa}).
\end{eqnarray}
We note that in the above expression we summed over the two polarizations which give exactly the same contribution. Because the sum above runs through an infinite amount of momenta $\bm{\kappa}$ and the photon frequency is a linear function of the norm of the photon momenta, $\omega(\bm{\kappa})=c|\bm{\kappa}|$, it is quite clear that the energy of the vacuum diverges $E_{\textrm{vac}}\rightarrow \infty$~\cite{spohn2004, Weinberg, Srednicki}. This infinite amount of energy is a consequence of the uncertainty principle and the zero-point or vacuum fluctuations of the electromagnetic field. This energy is analogous to the zero-point energy of the quantum harmonic oscillator~\cite{GriffithsQM, DiracQM} with the only difference that here we have an infinite amount of harmonic oscillators due to the infinite degrees of freedom (photon-modes) of the electromagnetic field.

However, this superficial divergence is not a great difficulty for the theory because merely the vacuum energy of the electromagnetic field defines the reference with respect to which all the energies in QED have to be computed. So what we mean by that is that whenever we compute the energy of a given state in QED (in free space) we need to subtract from the result the energy of the vacuum, in order to obtain a well-defined, finite answer.

It is important to highlight, that in free space the zero-point energy does not have any physical meaning and can be discarded. However, If one is interested in the change of energy of the electromagnetic field due to a pair of perfectly conducting metallic plates, the zero-point energy needs to be handled properly. The zero-point energy in this case leads to the emergence of macroscopic Casimir~\cite{Casimir:1948dh} and Casimir-Polder forces~\cite{casimir1948influence} which have measurable effects. We will look into these forces more closely in chapter~\ref{Effective QFT}.

\textit{Photons}.---Having defined the state of the photon field which contains no photons we would like also to give the state of the electromagnetic field that contains one or multiple photons. For a state to contain photons the expectation value of the photon number operator $\hat{N}_{\textrm{ph}}$ must be non-zero. To construct such a state we apply the creation operator $\hat{a}_{\bm{\kappa},\lambda}$ on the vacuum state of the photon field
\begin{eqnarray}
\hat{a}^{\dagger}_{\bm{\kappa},\lambda}|0\rangle.
\end{eqnarray}
Then, to check for photons we compute the expectation value of the number operator and we find that this state contains one photon
\begin{eqnarray}
\langle 0| \hat{a}_{\bm{\kappa},\lambda} \hat{N}_{\textrm{ph}}\hat{a}^{\dagger}_{\bm{\kappa},\lambda}|0\rangle=1.
\end{eqnarray}
Thus, the state $\hat{a}^{\dagger}_{\bm{\kappa},\lambda}|0\rangle\equiv |1_p\rangle_{\bm{\kappa},\lambda}$ is interpreted as the state which contains 1 photon~\footnote{We note that the notation $1_p$ stands for 1 photon.} with momentum $\bm{\kappa}$ and polarization $\lambda$. Further, the 1-photon state $|1_p\rangle_{\bm{\kappa},\lambda}$ is an eigenstate of the of the Hamiltonian of the electromagnetic field $\hat{H}^{\perp}_{\tiny{\textrm{EM}}}$ with eigenenergy $\hbar\omega(\bm{\kappa})$
\begin{eqnarray}
\hat{H}^{\perp}_{\textrm{\tiny{EM}}} |1_p\rangle_{\bm{\kappa},\lambda}=\hbar \omega(\bm{\kappa})|1_p\rangle_{\bm{\kappa},\lambda}.
\end{eqnarray}
To obtain the above result the infinite vacuum energy of the photon field was discarded. From the latter result, wee see that the 1-photon state which is the first excited state of the photon field, has energy $\hbar \omega(\bm{\kappa})$ as proposed by Einstein in 1905~\cite{PhotoelectricEinstein}. From this whole discussion it becomes clear that in QED the concept of the photon is not fundamental, but it is a derived concept, described as the first excited state of the quantized electromagnetic field. Finally, for completeness we would like to to mention that the state containing $n$ photons can be constructed by applying $n$ times the creation operator on the vacuum state
\begin{eqnarray}
|n_p\rangle_{\bm{\kappa},\lambda}=\frac{\left(\hat{a}^{\dagger}_{\bm{\kappa},\lambda}\right)^n}{\sqrt{n!}}|0\rangle.
\end{eqnarray}
The $n$-photon ($n_p$) state is also an eigenstate of $\hat{H}^{\perp}_{\tiny{\textrm{EM}}}$ with eigenenergy
\begin{eqnarray}
\hat{H}^{\perp}_{\textrm{\tiny{EM}}} |n_p\rangle_{\bm{\kappa},\lambda}=\hbar \omega(\bm{\kappa}) n  |n_p\rangle_{\bm{\kappa},\lambda}.
\end{eqnarray}

\subsection{Coulomb Energy}
The Hamiltonian $\hat{H}^{\perp}_{\textrm{\tiny{EM}}}$ in Eq.~(\ref{EM Hamiltonian}) describes the energy of the transeversal degrees of freedom of the free electromagnetic field. However, as we showed in section~\ref{Coulomb Potential} in the presence of charges there is also a contribution from the scalar potential $\phi(\bi{r},t)$ of Eq.~(\ref{general phi}) which is responsible for the longitudinal component of the electric field $\bi{E}^{||}(\bi{r},t)=-\nabla\phi(\bi{r},t)$. Then the energy contribution from the longitudinal part of the electric field is~\cite{Mandl} 
\begin{eqnarray}
H^{||}_{\textrm{\textrm{EM}}}=\frac{\epsilon_0}{2}\int \left(\bi{E}^{||}\right)^2(\bi{r},t)\; d^3r = \frac{\epsilon_0}{2}\int \nabla \phi(\bi{r},t) \cdot \nabla \phi(\bi{r},t) d^3r.
\end{eqnarray}
After performing a partial integration, and making use of the Poisson equation~(\ref{Poisson equation}) and the expression for $\phi(\bi{r},t)$ given by Eq.~(\ref{general phi}) we obtain
\begin{eqnarray}
H^{||}_{\textrm{\textrm{EM}}}=-\frac{\epsilon_0}{2}\int \phi(\bi{r},t)\nabla^2\phi(\bi{r},t) d^3r=\frac{1}{2} \iint \frac{\rho(\bi{r}^{\prime},t)\rho(\bi{r},t)}{ 4\pi\epsilon_0|\bi{r}-\bi{r}^{\prime}|}d^3r d^3r^{\prime}.
\end{eqnarray}
Finally, for a charge distribution made of point charges $\rho(\bi{r},t)=\sum_iq_i \delta(\bi{r}-\bi{r}_i)$ we find that the energy contribution due to the longitudinal degrees of freedom is the standard Coulomb potential energy~\cite{JacksonEM}
\begin{eqnarray}
W_{\textrm{C}}(|\bi{r}_i-\bi{r}_j|)\equiv H^{||}_{\tiny{\textrm{EM}}}=\frac{1}{2}\sum_{i \neq j}\frac{q_iq_j}{4\pi\epsilon_0|\bi{r}_i-\bi{r}_j|}.\label{Coulomb energy}
\end{eqnarray}
Lastly, we note that the longitudinal component of the electric field as it is purely determined by the charge distribution $\rho(\bi{r},t)$ is not subject to quantization. Thus, the energy due to the longitudinal component can be simply added to the Hamiltonian of the electromagnetic field without the need of performing a quantization procedure.

\section{Relativistic Quantum Matter: Dirac Equation}

\begin{displayquote}
\footnotesize{Until that time [the introduction of the Dirac equation] I had the impression that in quantum theory we had come back into the harbor, into the port. Dirac's paper (on the spinning electron) threw us out into the open sea again.}
\end{displayquote}
\begin{flushright}
  \footnotesize{Werner Heisenberg on the Dirac equation\\
QED and the Men Who Made It~\cite{SchweberQEDHistory}}
\end{flushright}

The aim of this section is to discuss how quantum matter can be coupled to the electromagnetic field. In section~\ref{Classical Electromagnetism} the classical theory of electromagnetism was introduced. There it was mentioned that although Maxwell's theory is called classical (since it is not quantized) it is actually a relativistic theory as it is invariant under Lorentz transformations. On this ground, and because electrons are the source of radiation, Dirac wanted to establish a relativistic and quantum mechanical theory of the electron. This he accomplished in his seminal paper~\cite{Diractheoryelectron} in which he introduced the equation which now bears his name. The Dirac equation reads as follows
\begin{eqnarray}
\textrm{i}\hbar\gamma^{\mu}\partial_{\mu} \Psi_{\tiny{\mathcal{D}}}(\bi{r},t)=m_{\textrm{e}}c\Psi_{\tiny{\mathcal{D}}}(\bi{r},t),
\end{eqnarray}
where $\Psi_{\tiny{\mathcal{D}}}(\bi{r},t)$ is a $4$-component spinor field describing the relativistic electron.
\begin{eqnarray}
\Psi_{\tiny{\mathcal{D}}}(\bi{r},t)=\left(\begin{tabular}{c}
		$\psi_1(\bi{r},t)$ \\
	    $\psi_2(\bi{r},t)$  \\
	    $\psi_3(\bi{r},t)$\\
	    $\psi_4(\bi{r},t)$
	\end{tabular}\right)\equiv \left(\begin{tabular}{c}
		$\Psi(\bi{r},t)$ \\
	    $\Phi(\bi{r},t)$  \\
	    \end{tabular}\right)
\end{eqnarray}
Here we have adopted the standard relativistic conventions in which $\mu=0,1,2,3$ and $\partial_{\mu}=(\partial_0,\partial_i)=(\partial/c\partial t, \nabla)$~\cite{Scharf}. Further, the matrices $\gamma^{\mu}$ are $4\times 4$ matrices satisfying the Clifford algebra
\begin{eqnarray}
\gamma^{\mu}\gamma^{\nu}+\gamma^{\nu}\gamma^{\mu}=2\eta^{\mu\nu} \;\;\; \textrm{with}\;\;\; \eta^{\mu\nu}=\textrm{diag}(1,-1,-1,-1).
\end{eqnarray}
A particular representation for the matrices $\gamma^{\mu}$ which will prove later useful is the Dirac representation
\begin{eqnarray}
\gamma^{0}=\left(\begin{tabular}{c c}
		$\mathbb{1}$ & $0$ \\
	    $0$ & $-\mathbb{1}$ \\
	    \end{tabular}\right)\;\;\; \textrm{and} \;\;\; \gamma^{i}= \left(\begin{tabular}{c c}
		$0$ & $\sigma^{i}$ \\
	    $-\sigma^{i}$ & $0$ \\
	    \end{tabular}\right)\;\;\; \textrm{with}\;\;\; i=1,2,3.
\end{eqnarray}
The matrices $\sigma^{i}$ are the well-known Pauli matrices~\cite{GriffithsQM}. Having defined everything then it is straightforward to check that the Dirac equation is invariant under Lorentz transformations and thus compatible with special relativity. Moreover, the Dirac equation is invariant under the phase transformation
\begin{eqnarray}
\Psi_{\tiny{\mathcal{D}}}(\bi{r},t)\rightarrow \Psi^{\prime}_{\tiny{\mathcal{D}}}(\bi{r},t)=e^{\textrm{i}\chi}\Psi_{\tiny{\mathcal{D}}}(\bi{r},t)
\end{eqnarray}
with $\chi$ being a real number. This transformation is an element of the $U(1)$ unitary group, and this particular symmetry is known as global $U(1)$ gauge invariance. This property of the Dirac equation is exploited in order to couple the electron to the electromagnetic field. The construction goes as follows.

Let us promote first the real number $\chi$ into a function $\chi(\bi{r},t)$. Then the following local $U(1)$ gauge transformation can be performed 
\begin{eqnarray}
\Psi_{\tiny{\mathcal{D}}}(\bi{r},t)\rightarrow \Psi^{\prime}_{\tiny{\mathcal{D}}}(\bi{r},t)=e^{\textrm{i}\chi(\bi{r},t)}\Psi_{\tiny{\mathcal{D}}}(\bi{r},t).
\end{eqnarray}
Of course the Dirac equation is not invariant under this local gauge transformation
\begin{eqnarray}
\textrm{i}\hbar\gamma^{\mu}\partial_{\mu} \Psi^{\prime}_{\tiny{\mathcal{D}}}(\bi{r},t)=e^{\textrm{i}\chi(\bi{r},t)}\textrm{i}\hbar\gamma^{\mu}\partial_{\mu} \Psi_{\tiny{\mathcal{D}}}(\bi{r},t)-\hbar\gamma^{\mu}\partial_{\mu}\chi(\bi{r},t)\Psi_{\tiny{\mathcal{D}}}(\bi{r},t).
\end{eqnarray}
To compensate the second term showing up in the above equation, the electromagnetic four potential $A_{\mu}(\bi{r},t)=(\phi(\bi{r},t)/c,\bi{A}(\bi{r},t))$ is added to the Dirac equation, where $\phi(\bi{r},t)$ and $\bi{A}(\bi{r},t)$ are the scalar and the vector potential of the electromagnetic field. Then, the Dirac equation for the electron coupled to the four potential $A_{\mu}(\bi{r},t)$ of an external electromagnetic field is
\begin{eqnarray}\label{Dirac A field}
 \gamma^{\mu}\Big(\textrm{i}\hbar\partial_{\mu}-eA_{\mu}(\bi{r},t)\Big)\Psi_{\tiny{\mathcal{D}}}(\bi{r},t)=m_{\textrm{e}}c\Psi_{\tiny{\mathcal{D}}}(\bi{r},t).
\end{eqnarray}
The above equation is invariant under the local $U(1)$ gauge transformation, which involves both the Dirac field and the electromagnetic four potential
\begin{eqnarray}
&&\Psi_{\tiny{\mathcal{D}}}(\bi{r},t)\rightarrow \Psi^{\prime}_{\tiny{\mathcal{D}}}(\bi{r},t)=e^{\textrm{i}\chi(\bi{r},t)}\Psi_{\tiny{\mathcal{D}}}(\bi{r},t) \;\;\; \\
&& A_{\mu}(\bi{r},t)\rightarrow A^{\prime}_{\mu}(\bi{r},t)=A_{\mu}(\bi{r},t)-\frac{\hbar}{e}\partial_{\mu}\chi(\bi{r},t).
\end{eqnarray}
This is how the quantum mechanical description of a relativistic electron coupled to an external electromagnetic field is constructed~\cite{Mandl, Scharf}.

\section{Non-Relativistic Quantum Electrodynamics}

As we already stated the main focus of this thesis lies on investigating non-relativistic phenomena in the framework of quantum electrodynamics. This is because we are interested in the low energy limit of QED in which pair (electron-positron) creation cannot occur. In the previous section we showed how relativistic electrons can be coupled in quantum mechanics to the electromagnetic field. Here, the aim is to show how a non-relativistic version of quantum electrodynamics can be constructed by considering the non-relativistic limit of the Dirac equation. In addition the quantum nature of the electromagnetic field will be added, which was absent in the previous section. 

\subsection{Non-Relativistic Limit of the Dirac Equation}
Let us first show how the Dirac equation looks in the non-relativistic limit. To do so first one multiplies Eq.~(\ref{Dirac A field}) with $c\gamma^0$ and then splits $\partial_{\mu}$ into its temporal and spatial components
\begin{eqnarray}\label{Dirac Hamiltonian}
\textrm{i}\hbar\frac{\partial \Psi_{\tiny{\mathcal{D}}}(\bi{r},t)}{\partial t}= \Big[\bi{S}\cdot\big(-\textrm{i}\hbar\nabla-e\bi{A}(\bi{r},t)\big)+e\phi(\bi{r},t) +m_{\textrm{e}}c^2 \gamma^{0} \Big]\Psi_{\tiny{\mathcal{D}}}(\bi{r},t),
\end{eqnarray}
where $\bi{S}$ is
\begin{eqnarray}
\bi{S}=\left(\begin{tabular}{c c}
		$0$ & $\bm{\sigma}$ \\
	    $\bm{\sigma}$ & $0$ 
	    \end{tabular}\right) \;\;\;\; \textrm{and}\;\;\;\; \bm{\sigma}=(\sigma^1,\sigma^2,\sigma^3).
\end{eqnarray}
The benefit of separating the temporal part from the spatial part is that we can identify the Hamiltonian in the Dirac theory, from analogy to the Schr\"{o}dinger equation $\hat{H}\psi=\textrm{i}\hbar\partial_t\psi$. Thus, we find that the Hamiltonian in the Dirac theory which generates the time evolution of the spinor field $\Psi_{\tiny{\mathcal{D}}}(\bi{r},t)$ is
\begin{eqnarray}
\hat{H}_{\tiny{\mathcal{D}}}=\bi{S}\cdot\big(-\textrm{i}\hbar\nabla-e\bi{A}(\bi{r},t)\big)+e\phi(\bi{r},t) +m_{\textrm{e}}c^2 \gamma^{0}.
\end{eqnarray}
As a next step the following ansatz is introduced for the four-component spinor field~\cite{Scharf}
\begin{eqnarray}
\Psi_{\tiny{\mathcal{D}}}(\bi{r},t)=\left(\begin{tabular}{c c}
		$\Psi(\bi{r},t)$ \\
	    $\Phi(\bi{r},t)$ 
	    \end{tabular}\right)e^{-\frac{\textrm{i}m_{\textrm{e}}c^2}{\hbar}t}.
\end{eqnarray}
Substituting the above ansatz into the Dirac equation in~(\ref{Dirac Hamiltonian}) one obtains the following set of coupled equations for the components of the spinor field
\begin{eqnarray}
&&\textrm{i}\hbar\frac{\partial \Psi(\bi{r},t)}{\partial t}=e\phi(\bi{r},t)\Psi(\bi{r},t)+c\bm{\sigma}\cdot\big(-\textrm{i}\hbar\nabla-e\bi{A}(\bi{r},t)\big)\Phi(\bi{r},t),\label{Diff eq Psi} \\
&&\textrm{i}\hbar\frac{\partial \Phi(\bi{r},t)}{\partial t}=\left(e\phi(\bi{r},t)-2m_{\textrm{e}}c^2\right)\Phi(\bi{r},t)+c\bm{\sigma}\cdot\big(-\textrm{i}\hbar\nabla-e\bi{A}(\bi{r},t)\big)\Psi(\bi{r},t).\label{Diff eq Phi}
\end{eqnarray}
The next step is to take the non-relativistic limit. To do so first we divide Eq.(\ref{Diff eq Phi}) by $2m_{\textrm{e}}c^2$. Then the non-relativistic limit is considered by taking the limit where the speed of light is infinite, $c\rightarrow \infty$. To leading order in $1/c$ one finds
\begin{eqnarray}
\Phi(\bi{r},t)=\frac{1}{2m_{\textrm{e}}c}\bm{\sigma}\cdot\big(-\textrm{i}\hbar \nabla-e\bi{A}(\bi{r},t)\big)\Psi(\bi{r},t)+\mathcal{O}(1/c^2).
\end{eqnarray}
Substituting the above equation into Eq.~(\ref{Diff eq Psi}) we find
\begin{eqnarray}
\textrm{i}\hbar\frac{\partial \Psi(\bi{r},t)}{\partial t}=\frac{1}{2m_{\textrm{e}}}\left[\bm{\sigma}\cdot\big(-\textrm{i}\hbar\nabla-e\bi{A}(\bi{r},t)\big)\right]^2\Psi(\bi{r},t)+e\phi(\bi{r},t)\Psi(\bi{r},t).
\end{eqnarray}
Lastly, after expanding the square bracket we obtain
\begin{eqnarray}\label{NR Dirac}
\textrm{i}\hbar\frac{\partial \Psi(\bi{r},t)}{\partial t}=\frac{1}{2m_{\textrm{e}}}\big(-\textrm{i}\hbar\nabla-e\bi{A}(\bi{r},t)\big)^2\Psi(\bi{r},t)+e\phi(\bi{r},t)\Psi(\bi{r},t)-\frac{e\hbar}{2m_{\textrm{e}}}\bm{\sigma}\cdot\bi{B}(\bi{r},t)\Psi(\bi{r},t).\nonumber\\
\end{eqnarray}
The above equation describes a slowly moving, non-relativistic electron coupled to an external classical electromagnetic field. As it can be seen the above equation is actually the Schr\"{o}dinger equation with an additional coupling term to the magnetic field $\bi{B}(\bi{r},t)$. This additional term signifies that due to the spin $\bm{\sigma}$ the non-relativistic electron has acquired the magnetic moment $\bm{\mu}=\mu_{\textrm{B}}\bm{\sigma}$, where $\mu_{\textrm{B}}=e\hbar/2m_{\textrm{e}}$ is the Bohr magneton. From Eq.~(\ref{NR Dirac}) we conclude that the Hamiltonian describing a non-relativistic electron interacting with a classical electromagnetic field is
\begin{eqnarray}
\hat{H}_{\tiny{\textrm{NR}\mathcal{D}}}=\frac{1}{2m_{\textrm{e}}}\big(-\textrm{i}\hbar\nabla-e\bi{A}(\bi{r},t)\big)^2+e\phi(\bi{r},t)-\frac{e\hbar}{2m_{\textrm{e}}}\bm{\sigma}\cdot\bi{B}(\bi{r},t).
\end{eqnarray}
Before we continue we would like to mention that the above result could also be derived by simply exchaning the non-relativistic covariant momentum $\big(-\textrm{i}\hbar\nabla-e\bi{A}(\bi{r},t)\big)^2$ with $\left[\bm{\sigma}\cdot\big(-\textrm{i}\hbar\nabla-e\bi{A}(\bi{r},t)\big)\right]^2$~\cite{spohn2004}. In this way the Pauli-Stern-Gerlach term $\sigma\cdot \bi{B}$ would be included. However, this would be an ad hoc construction for the inclusion of spin. Dirac on the other side, was able to recover the Pauli-Stern-Gerlach term and show how spin emerges, by merging special relativity with quantum theory, and then taking the non-relativistic limit of his theory. The explanation of spin as a consequence of combining special relativity and quantum mechanics was one of the early successes of Dirac's theory~\cite{DiracQM, Diractheoryelectron}.

\subsection{Many-Electron Non-Relativistic QED Hamiltonian}

So far we introduced the Hamiltonian for the transversal and the longitudinal components of the electromagnetic field $\hat{H}^{\perp}_{\tiny{\textrm{EM}}}$ in Eq.~(\ref{EM Hamiltonian}) and $H^{||}_{\tiny{\textrm{EM}}}\equiv W_{\textrm{C}}(\bi{r}_i-\bi{r}_j)$ in Eq.~(\ref{Coulomb energy}) respectively. Further, we presented the Hamiltonian for non-relativistic matter coupled to the electromagnetic field $\hat{H}_{\tiny{\textrm{NR}\mathcal{D}}}$ in Eq.~(\ref{NR Dirac}). 

Now by collecting all the different elements of quantum electrodynamics we would like to give the general Hamiltonian describing $N$ interacting electrons coupled to the quantized photon field $\hat{\bi{A}}(\bi{r},t)$ and to an external classical electromagnetic field $\bi{A}_{\textrm{ext}}(\bi{r},t)$, in the presence also of scalar potentials $v_{\textrm{ext}}(\bi{r})\; (\equiv e\phi_{\textrm{ext}}(\bi{r}))$ representing the clamped nuclei of an atom or the ions in the periodic potential of a solid. This Hamiltonian is known as the Pauli-Fierz Hamiltonian and reads as~\cite{spohn2004, cohen1997photons, faisal1987}
\begin{eqnarray}\label{Pauli Fierz Hamiltonian}
\hat{H}_{\textrm{PF}}&=&\hat{H}_{\tiny{\textrm{NR}\mathcal{D}}}+\hat{V}_{\textrm{ext}}+\hat{H}^{\perp}_{\tiny{\textrm{EM}}}+H^{||}_{\tiny{\textrm{EM}}}=\nonumber\\
&=&\frac{1}{2m_{\textrm{e}}}\sum^{N}_{j=1}\Big[\bm{\sigma}_j \cdot \big(-\textrm{i}\hbar\nabla_j-e\hat{\bi{A}}(\bi{r}_j,t)-e\bi{A}_{\textrm{ext}}(\bi{r}_j,t)\big)\Big]^2+\sum^N_{j=1}v_{\textrm{ext}}(\bi{r}_j)\nonumber\\
&+&\frac{1}{2}\sum_{i \neq j}\frac{e^2}{4\pi\epsilon_0|\bi{r}_i-\bi{r}_j|}+\sum_{\bm{\kappa},\lambda}\hbar\omega(\bm{\kappa})\left(\hat{a}^{\dagger}_{\bm{\kappa},\lambda}\hat{a}_{\bm{\kappa},\lambda}+\frac{1}{2}\right).
\end{eqnarray}
We would like to mention that the Pauli-Fierz Hamiltonian can be straightforwardly generalized for the interaction of electrons with positively charged particles like the nuclei of atoms. This implies that the Pauli-Fierz theory should be able to describe all non-relativistic phenomena occurring between light and matter at low energies. In the recent years there has been a great amount of effort to make the description all these phenomena tractable, in a unified theory of light and matter mainly in the framework of quantum electrodynamical density functional theory~\cite{ruggenthaler2014, ruggenthaler2017b, FlickElectron-Nuclear, TokatlyPRL}.

It is clear that high-energy processes in QED like electron-positron pair production cannot be described within the Pauli-Fierz Hamiltonian. For the description of such phenomena the full relativistic treatment of QED becomes necessary. This is one particular limit in which the Pauli-Fierz theory is known to become inadequate. However, it is not absolutely clear to this day what is the exact range of applicability of the Pauli-Fierz theory. Primarily, this has to do with the fact that there are no exact, non-perturbative solutions for any real physical system, like for example the hydrogen atom, coupled to the full photon field. Thus, no exact or complete comparison to experiment can be claimed for the Pauli-Fierz theory. In addition, QED it is known to be plagued by divergences due to the infinite amount of photonic degrees of freedom~\cite{spohn2004, Weinberg, Srednicki, Mandl}. Of course by introducing a non-relativistic cutoff $\Lambda$ for the photon momenta this problem is resolved and non-relativistic QED becomes finite~\cite{spohn2004, Liebstability}. However, the problem still persists in the sense that it is not absolutely clear what is the exact value of the ultraviolet cutoff $\Lambda$. We will touch upon some of these fundamental issues of non-relativistic QED in chapter~\ref{Effective QFT} for the simple case of a gas of free electrons coupled to full continuum of electromagnetic modes. It is important to highlight that although ultraviolet divergences are present in non-relativistic QED, infrared divergences do not occur due to the elimination of the the positrons from the theory and the appearance of the diamagnetic $\bi{A}^2$ term in the Pauli-Fierz Hamiltonian.

In contrast to the ultraviolet behavior of the Pauli-Fierz Hamiltonian, which remains an open problem, important physically desired mathematical properties have been established rigorously for the Pauli-Fierz theory. It has been proven that the Pauli-Fierz Hamiltonian is a self-adjoint operator~\cite{HiroshimaSelf-adjoint} which implies that this Hamiltonian generates a unitary time evolution on an appropriately chosen Hilbert space representing the physical states of the theory~\cite{spohn2004}. In addition, it has been shown that the Pauli-Fierz Hamiltonian has a ground-state under fairly generic conditions~\cite{HiroshimaGSQEDI, HiroshimaGSQEDII}. The existence of a ground-state implies that the variational principle can be applied, which is a cornerstone for electronic structure methods. These properties are very important for a sound physical theory and provide a certain amount of confidence that the Pauli-Fierz theory is a proper framework for the description of the interaction of non-relativistic matter with light.



\chapter{QED in the Length Gauge \& the Dipole Self-Energy}\label{Length Gauge QED}

\begin{displayquote}
\footnotesize{Cavity quantum electrodynamics can be defined, in a nutshell, as the
physics of a spin and an oscillator in interaction.}
\end{displayquote}
\begin{flushright}
 \footnotesize{S.~Haroche \& J.~M. Raimond\\
Exploring the Quantum~\cite{ExploringQuantum}}
\end{flushright}

In the previous chapter we showed how non-relativistic quantum electrodynamics is constructed and we derived the Pauli-Fierz Hamiltonian. Assuming a form factor for the photon field that suppresses infinitely high photon momenta, and allowing for external fields of Kato type, the Pauli-Fierz Hamiltonian is bounded from below and thus obeys a variational principle for ground-states~\cite{spohn2004}. Further, the Pauli-Fierz Hamiltonian was constructed by taking the non-relativistic limit of QED and is consequently gauge invariant, which means that physical observables do not depend on the gauge choice. These two properties are fundamental and very much desired for a quantum theory of light-matter interactions and constitute the primary reasons for choosing to work with the Pauli-Fierz theory.

However, in cavity QED much more simplified descriptions of light-matter interactions are usually employed. In many cases the paradigmatic few-level models of quantum optics, like the Rabi, the Jaynes-Cummings~\cite{shore1993, Braak, cohen1997photons} or the Dicke model~\cite{dicke1954} are used to describe such cavity QED setups. These models are the cornerstone of our understanding of light-matter interactions, but due to the fact that they treat matter with the use of only a few quantum states, it has been shown by several different research groups~\cite{bernadrdis2018breakdown, HuoGauges, Dickegaugeinvariance, DiStefano2019, RouseGauges, Stokes2019} that these models actually break gauge invariance in the ultrastrong coupling regime~\cite{kockum2019ultrastrong} and that their predictions depend significantly on the gauge choice. These gauge ambiguities arise primarily due to truncation of the electronic Hilbert space and show up also for more complicated coupled light-matter systems, like tight-binding models coupled to the photon field~\cite{Eckstein2020, SchiroGaugeTB}. Furthermore, in the field of polaritonic chemistry and cavity QED materials there is an ongoing discussion on whether the diamagnetic $\bi{A}^2$ term needs to be included in the Pauli-Fierz Hamiltonian in the Coulomb gauge~\cite{vukics2014, HagenmullerSPT, ChirolliNoGO} and respectively whether the dipole self energy in the length gauge needs to be taken into account~\cite{GalegoCasimir, rokaj2017, schaeferquadratic}.  

The aim of this chapter is to present how the Pauli-Fierz Hamiltonian in the length gauge is derived and to demonstrate the necessity of the dipole self energy by showing that fundamental properties of the Pauli-Fierz Hamiltonian, like stability, gauge invariance and translational symmetry are broken, if the dipole self energy is not taken into account~\cite{rokaj2017}.

\section{The Length Gauge Hamiltonian}

In the field of cavity QED, where atomic or molecular systems interact with the photons of an optical cavity, a simplified version of the Pauli-Fierz Hamiltonian is used. For such systems, the spatial extension of the matter system is much smaller than the wavelength of the relevant photon modes. Due to this, we can neglect the spatial variation of the electromagnetic field $e^{\pm \textrm{i} (\bm{\kappa}\cdot\bi{r}-\omega(\bm{\kappa})t)}\approx 1$. This approximation is known by different names: it is either called the long-wavelength or optical limit as well as dipole approximation~\cite{cohen1997photons, faisal1987, spohn2004}. In the dipole approximation the spatial dependence of the quantized vector potential is not taken into account and the the Pauli-Fierz Hamiltonian is 
\begin{eqnarray}
\hat{H}&=&\frac{1}{2m_{\textrm{e}}}\sum^{N}_{j=1}\big(\textrm{i}\hbar\nabla_j+e\hat{\bi{A}}\big)^2+\sum^N_{j=1}v_{\textrm{ext}}(\bi{r}_j)\nonumber\\
&+&\frac{1}{2}\sum_{i \neq j}\frac{e^2}{4\pi\epsilon_0|\bi{r}_i-\bi{r}_j|}+\sum_{\bm{\kappa},\lambda}\hbar\omega(\bm{\kappa})\left(\hat{a}^{\dagger}_{\bm{\kappa},\lambda}\hat{a}_{\bm{\kappa},\lambda}+\frac{1}{2}\right).
\end{eqnarray}
In the Hamiltonian above also the Stern-Gerlach term $\bm{\sigma}\cdot\hat{\bi{B}}$ has been neglected and the external classical vector potential has been taken to zero, $\bi{A}_{\textrm{ext}}=0$. Further, in the dipole approximation the vector potential, the electric field and the magnetic field become spatially uniform
\begin{eqnarray}
&&\hat{\bi{A}}=\sum_{\bm{\kappa}, \lambda} \sqrt{\frac{\hbar}{2\epsilon_0V\omega(\bm{\kappa})}}\bm{\varepsilon}_{\lambda}(\bm{\kappa})\left[\hat{a}_{\bm{\kappa},\lambda}+\hat{a}^{\dagger}_{\bm{\kappa},\lambda} \right],\label{dipole A field}\\
&&\hat{\bi{E}}=\sum_{\bm{\kappa}, \lambda} \sqrt{\frac{ \hbar \omega(\bm{\kappa}) }{2\epsilon_0V}}\textrm{i}\bm{\varepsilon}_{\lambda}(\bm{\kappa})\left[\hat{a}_{\bm{\kappa},\lambda}-\hat{a}^{\dagger}_{\bm{\kappa},\lambda} \right],\label{dipole electric field}\\
&&\hat{\bi{B}}= \sum_{\bm{\kappa}, \lambda} \sqrt{\frac{\hbar}{2\epsilon_0V\omega(\bm{\kappa})}}\textrm{i}\bm{\kappa} \times \bm{\varepsilon}_{\lambda}(\bm{\kappa})\left[\hat{a}_{\bm{\kappa},\lambda}-\hat{a}^{\dagger}_{\bm{\kappa},\lambda} \right]\label{dipole magnetic field}.
\end{eqnarray}
Before we continue, we would like to emphasize that the dipole approximation has been proven very successful in the field of cavity QED as it has allowed for the description of a wide range of different experimental set-ups~\cite{kockum2019ultrastrong, ruggenthaler2017b}. Also the paradigmatic few-level models of quantum optics like the Rabi, the Jaynes-Cummings~\cite{shore1993, garraway2011} and the Dicke model~\cite{dicke1954} are all derived in the dipole approximation. Furthermore, it is important to mention that the dipole approximation is important also for solid state and condensed matter systems in which translational invariance is of fundamental importance. The Pauli-Fierz Hamiltonian for dipolar electromagnetic fields and for $v_{\textrm{ext}}(\bi{r})=0$ is invariant under translations in the electronic configuration space $\bi{r} \rightarrow \bi{r}+\bi{a}$. This is an important property for homogeneous systems, like for example the jellium model~\cite{Mermin, Vignale}. Also, in the case of a periodic external potential $v_{\textrm{ext}}(\bi{r})=v_{\textrm{ext}}(\bi{r}+\bi{R}_{\bi{n}})$, is invariant under Bravais lattice translations $\bi{r}\rightarrow \bi{r}+\bi{R}_{\bi{n}}$, which implies that Bloch's theorem can be applied for the description of periodic solids~\cite{Mermin, Callaway}

The basic principle that was employed for the construction of QED is gauge invariance. The fact that QED is a gauge theory implies that depending on the gauge choice the Hamiltonian describing QED takes a different form. However, we would like to emphasize that all observables in QED are invariant with respect to the gauge choice. In chapter~\ref{Quantum Electrodynamics} we worked in the Coulomb gauge, which is one of the most frequently employed gauges. However, in the long-wavelength limit or dipole approximation, there is another gauge that is commonly used, the so-called length gauge~\cite{rokaj2017, flick2017, ruggenthaler2014}, also known as multipolar or Power-Zienau-Woolley gauge~\cite{babiker1983derivation, woolley1980gauge}. So let us demonstrate the steps in order to go from the Coulomb gauge Hamiltonian to the length gauge Hamiltonian. To do so, first we need to define the annihilation and creation operators of the photon field $\hat{a}_{\bm{\kappa},\lambda}, \hat{a}^{\dagger}_{\bm{\kappa},\lambda}$ with respect to the displacement coordinates $q_{\bm{\kappa},\lambda}$ and their conjugate momenta $\partial/\partial q_{\bm{\kappa},\lambda}$
\begin{equation}\label{q coordinate and momenta}
	\hat{a}_{\bm{\kappa},\lambda}=\frac{1}{\sqrt{2}}\left(q_{\bm{\kappa},\lambda}+\frac{\partial}{\partial q_{\bm{\kappa},\lambda}}\right)\qquad \textrm{and} \qquad \hat{a}^{\dagger}_{\bm{\kappa},\lambda}=\frac{1}{\sqrt{2}}\left(q_{\bm{\kappa},\lambda}-\frac{\partial}{\partial q_{\bm{\kappa},\lambda}}\right).
\end{equation}
Substituting these expressions for the annihilation and the creation operators, the transverse part of the electromagnetic Hamiltonian $\hat{H}^{\perp}_{\tiny{\textrm{EM}}}$ becomes
\begin{eqnarray}
\hat{H}^{\perp}_{\tiny{\textrm{EM}}}=\sum_{\bm{\kappa},\lambda}\hbar\omega(\bm{\kappa})\left(\hat{a}^{\dagger}_{\bm{\kappa},\lambda}\hat{a}_{\bm{\kappa},\lambda}+\frac{1}{2}\right)=\sum_{\bm{\kappa},\lambda}\frac{\hbar\omega(\bm{\kappa})}{2}\left(-\frac{\partial^2}{\partial q^2_{\bm{\kappa},\lambda}} +q^2_{\bm{\kappa},\lambda}\right).
\end{eqnarray}
Further, with respect to the displacement coordinates and their momenta the quantized vector potential is 
\begin{eqnarray}\label{A field in Qs}
\hat{\bi{A}}=\sqrt{\frac{\hbar}{\epsilon_0V}}\sum_{\bm{\kappa}, \lambda} \frac{\bm{\varepsilon}_{\lambda}(\bm{\kappa})}{\omega(\bm{\kappa})}q_{\bm{\kappa},\lambda}.
\end{eqnarray}
Subsequently, the QED Hamiltonian after expanding the covariant kinetic term takes the next form
\begin{eqnarray}\label{Velocity Hamiltonian}
	\hat{H}&=&\frac{1}{2m}\sum\limits^{N}_{j=1}\left(-\hbar^2\mathbf{\nabla}^2_{j} +2\textrm{i}e\hbar\hat{\mathbf{A}}\cdot\mathbf{\nabla}_j +e^2\hat{\mathbf{A}}^2\right)+\frac{1}{4\pi\epsilon_0}\sum\limits^{N}_{j< k}\frac{e^2}{|\mathbf{r}_j-\mathbf{r}_k|}\nonumber\\
	&+&\sum\limits^{N}_{j=1}v_{ext}(\mathbf{r}_{j})+\sum_{\bm{\kappa},\lambda}\frac{\hbar\omega(\bm{\kappa})}{2}\left(-\frac{\partial^2}{\partial q^2_{\bm{\kappa},\lambda}} +q^2_{\bm{\kappa},\lambda}\right).
\end{eqnarray}

The first step to obtain the length gauge Hamiltonian is to perform the following unitary transformation~\cite{faisal1987, spohn2004, cohen1997photons} 
\begin{equation}\label{Length unitary transf}
\hat{H}_{L}' = \hat{U}^{\dagger} \hat{H} \hat{U},\;\;\; \textrm{where}\;\;\;  \hat{U}=\exp\left(\frac{\textrm{i}e}{\hbar}\hat{\mathbf{A}}\cdot \mathbf{R}\right),
\end{equation}
where $\bi{R}=\sum\limits^{N}_{i=1}\bi{r}_i$ is the full dipole operator, the sum over the positions of all the charged particles. The individual terms in the Hamiltonian~(\ref{Velocity Hamiltonian}) transform as
\begin{eqnarray}\label{eq2.8}
2\textrm{i}\hbar e\hat{\mathbf{A}}\cdot \mathbf{\nabla}_j & \longrightarrow&  2\textrm{i}\hbar e\hat{U}^{\dagger}\hat{\mathbf{A}}\cdot \mathbf{\nabla}_j \hat{U}= 2i\hbar e\hat{\mathbf{A}}\cdot \mathbf{\nabla}_j -2e^2\hat{\mathbf{A}}^{2}\nonumber\\
	-\hbar^2\mathbf{\nabla}^2_j & \longrightarrow & -\hbar^2 \hat{U}^{\dagger}\mathbf{\nabla}^2_j  \hat{U}=-\hbar^2 \mathbf{\nabla}^2_j -2\textrm{i}\hbar e\hat{\mathbf{A}}\cdot \mathbf{\nabla}_j +e^2\hat{\mathbf{A}}^{2}\\
	e^2\hat{\mathbf{A}}^{2} & \longrightarrow & e^2\hat{U}^{\dagger}\hat{\mathbf{A}}^2\hat{U}=e^2\hat{\mathbf{A}}^{2}\nonumber\\
	-\frac{\partial^2}{\partial q^2_{\bm{\kappa},\lambda}} & \longrightarrow &  -\hat{U}^{\dagger}\frac{\partial^2}{\partial q^2_{\bm{\kappa},\lambda}}\; \hat{U}=-\frac{\partial^2}{\partial q^2_{\bm{\kappa},\lambda}}-\textrm{i}\frac{2e\bm{\varepsilon}_{\lambda}(\bm{\kappa})\cdot\mathbf{R}}{\sqrt{\hbar\epsilon_0V\omega(\bm{\kappa})}}\frac{\partial}{\partial q_{\bm{\kappa},\lambda}} +\left(\frac{e\bm{\varepsilon}_{\lambda}(\bm{\kappa})\cdot \mathbf{R}}{\sqrt{\hbar\epsilon_0 V\omega(\bm{\kappa})}}\right)^2 \nonumber.
\end{eqnarray}

The rest of the terms in the Hamiltonian of Eq.~(\ref{Velocity Hamiltonian}) are invariant under this transformation because they commute with the operator $\hat{U}$. The Hamiltonian after the unitary transformation takes the form
\begin{eqnarray}\label{eq2.9}
	\hat{H}^{'}_L&=&-\frac{\hbar^2}{2m}\sum\limits^{N}_{i=1}\bi{\nabla}^2_{i}+\frac{1}{4\pi\epsilon_0}\sum\limits^{N}_{i< j}\frac{e^2}{|\bi{r}_i-\bi{r}_j|}+\sum\limits^{N}_{i=1}v_{ext}(\bi{r}_{i})\\
	&+&\sum\limits_{\bm{\kappa},\lambda}\frac{\hbar\omega(\bm{\kappa})}{2}\left[-\frac{\partial^2}{\partial q^2_{\bm{\kappa},\lambda}}+q^2_{\bm{\kappa},\lambda} -2\textrm{i}e\frac{\bm{\varepsilon}_{\lambda}(\bm{\kappa})\cdot\bi{R}}{\sqrt{\hbar\epsilon_0V\omega(\bm{\kappa})}}\frac{\partial}{\partial q_{\bm{\kappa},\lambda}} +\left(\frac{e\bm{\varepsilon}_{\lambda}(\bm{\kappa})\cdot\bi{R}}{\sqrt{\hbar\epsilon_0V\omega(\bm{\kappa})}}\right)^2\right].\nonumber
\end{eqnarray}
In the Hamiltonian above we see that the diamagnetic $\bi{A}^2$ term has been eliminated due to the unitary transformation. However, due to this transformation we see that we have now a new electronic harmonic potential $\left(\bm{\varepsilon}_{\lambda}(\bm{\kappa})\cdot\bi{R}\right)^2$ showing up. This term is known as the dipole self-energy~\cite{faisal1987, rokaj2017}. As it is clear from Eqs.~(\ref{eq2.8}), this electron-electron interaction does not come from the transformation of the $\mathbf{A}^{2}$ term. This means that the dipole self-energy and the diamagnetic $\bi{A}^2$ term are clearly not equivalent. 

Further, to get into the length gauge Hamiltonian we also have to perform a canonical variable transformation that swaps the displacement coordinates of the photon modes with their respective conjugate momenta~\cite{ruggenthaler2014}  
\begin{eqnarray}\label{eq2.10}
	i\frac{\partial}{\partial q_{\bm{\kappa},\lambda} } \longrightarrow p_{\bm{\kappa},\lambda}\qquad \textrm{and}\qquad q_{\bm{\kappa},\lambda} \longrightarrow -i\frac{\partial}{\partial p_{\bm{\kappa},\lambda}}.
\end{eqnarray}
The transformation above is merely a Fourier transform with respect to the mode photonic coordinates $q_{\bm{\kappa},\lambda}$ of the full wave-function 
\begin{equation}\label{eq2.11}
		\Psi'(..., q_{\bm{\kappa},\lambda},...)\longrightarrow \Psi(...,p_{\bm{\kappa},\lambda},...)=\frac{1}{\sqrt{2\pi}}\int^{\infty}_{-\infty} e^{- i q_{\bm{\kappa},\lambda}p_{\bm{\kappa},\lambda}}\Psi'(...,q_{\bm{\kappa},\lambda},...)\textrm{d} q_{\bm{\kappa},\lambda}.
\end{equation}
This variable transformation leaves the commutation relations unchanged. The Hamiltonian in the length gauge then is
\begin{eqnarray}\label{Lenght Hamiltonian}
	\hat{H}_L&=&-\frac{\hbar^2}{2m}\sum\limits^{N}_{i=1}\bi{\nabla}^2_i+\frac{1}{4\pi \epsilon_0} \sum\limits^{N}_{i< j}\frac{e^2}{|\bi{r}_i-\bi{r}_j|}+\sum\limits^{N}_{i=1}v_{ext}(\bi{r}_{i})\nonumber\\
	&+&\sum\limits_{\bm{\kappa},\lambda}\left[ -\frac{\hbar\omega(\bm{\kappa})}{2}\frac{\partial^2}{\partial p^2_{\bm{\kappa},\lambda}}+\frac{\hbar\omega(\bm{\kappa})}{2}\left(p_{\bm{\kappa},\lambda}-\frac{e\bm{\varepsilon}_{\lambda}(\bm{\kappa})\cdot \bi{R}}{\sqrt{\hbar\epsilon_0V\omega(\bm{\kappa})}}\right)^2\right] .
\end{eqnarray}
The length gauge Hamiltonian above contains the explicit bilinear electron-photon interaction 
\begin{equation}\label{Bilinear interaction}
	\hat{V}_{int} = -\sum_{\bm{\kappa},\lambda}e \sqrt{\frac{\hbar\omega(\bm{\kappa})}{\epsilon_0V}} \bm{\varepsilon}_{\lambda}(\bm{\kappa})\cdot\bi{R} \;p_{\bm{\kappa},\lambda},
\end{equation}
as well as the dipole self-energy~\cite{cohen1997photons, spohn2004, ruggenthaler2014}
\begin{equation}\label{Dipole self energy}
	\hat{\varepsilon}_{dip}  = \sum_{\bm{\kappa},\lambda} \frac{\hbar\omega(\bm{\kappa})}{2}\left(\frac{e\bm{\varepsilon}_{\lambda}(\bm{\kappa})\cdot\bi{R}}{\sqrt{\hbar\epsilon_0V\omega(\bm{\kappa})}}\right)^{2}. 
\end{equation}
These terms, and particularly the dipole self-energy, in $\hat{H}_L$ arise because the length-gauge transformation~(\ref{Length unitary transf}) mixes the matter and the photonic degrees of freedom. More specifically, the coordinate $p_{\bm{\kappa},\lambda}$ does not correspond anymore to a purely photonic degree of freedom, but rather to the electromagnetic displacement field which is the sum of the electric field and the polarization field due to matter~\cite{JacksonEM}. We will look into this point in more detail in section~\ref{Maxwell matter}.

\section{Translational Inavariance}

An alternative way to understand how the length gauge transformation mixes the matter and the photonic degrees freedom is by looking into translational invariance. 

To do so, we will consider the simple case where the external scalar potential is zero, $v_{ext}(\bi{r})=0$. In this case the Coulomb gauge Hamiltonian of Eq.~(\ref{Velocity Hamiltonian}) is invariant under translations in the electronic configuration space $\bi{r}\rightarrow \bi{r}+\bi{a}$, where $\bi{a}$ is an arbitrary vector. This means that the electronic translation operator
\begin{equation}
    \hat{T}(\bi{a})=\exp\left(\frac{\textrm{i}}{\hbar}\sum_{j=1}^{N}\bi{a}\cdot \hat{\bi{p}}_j\right)=\exp\left(\sum_{j=1}^{N} \bi{a}\cdot\nabla_j\right),
\end{equation}
commutes with the Hamiltonian $\hat{H}$ of Eq.~(\ref{Velocity Hamiltonian})
\begin{equation}\label{eq2.21}
   [\hat{H},\hat{T}(\bi{a})]=0.
\end{equation}
However, it is quite clear that the length-gauge Hamiltonian $\hat{H}_L$ does not commute with the electronic translation operator $\hat{T}(\bi{a})$, because of the the interaction term $\hat{V}_{int}$ (\ref{Bilinear interaction}) and the dipole self-energy $\hat{\varepsilon}_{dip}$ (\ref{Dipole self energy}) which have a linear and quadratic, dependence respectively, on the total dipole $\bi{R}=\sum_j\bi{r}_j$ of the electrons.

Since the Coulomb gauge Hamiltonian (in the dipole approximation) is invariant under translations $\hat{H} = \hat{T}(\bi{a}) \hat{H} \hat{T}^{\dagger}(\bi{a})$ by using the unitary transformation~(\ref{Length unitary transf}) we find that the length gauge Hamiltonian is invariant under the following transformation
\begin{equation}\label{eq2.25}
    \hat{H}_{L}'=\hat{U}^{\dagger}\hat{T}(\bi{a})\hat{U}\hat{H}_{L}'\hat{U}^{\dagger}\hat{T}^{\dagger}(\bi{a})\hat{U}.
\end{equation}
Thus, in the length gauge the translation operator is transformed as well via $\hat{T}_{L}'(\bi{a})=\hat{U}^{\dagger}\hat{T}(\bi{a})\hat{U}$. With the help of the Baker-Hausdorff-Campbell formula can be written as
\begin{equation}\label{eq2.26a}
    \hat{T}_{L}'(\bi{a})=\hat{U}^{\dagger}\hat{T}(\bi{a})\hat{U} = \exp\left[\frac{\textrm{i}}{\hbar} \sum_{j=1}^{N}\bi{a}\cdot \left(\hat{\bi{p}}_j  + e\hat{\mathbf{A}}\right)\right].
\end{equation}
After performing also the Fourier transformation~(\ref{eq2.10}) we find 
\begin{eqnarray}\label{eq2.26b}
    \hat{T}_{L}(\bi{a})& =& \exp\left[\frac{\textrm{i}}{\hbar} \sum_{j=1}^{N}\bi{a}\cdot \left(\hat{\bi{p}}_j  + e\sum^{M}_{\bm{\kappa},\lambda}\frac{\sqrt{\hbar} \bm{\varepsilon}_{\lambda}(\bm{\kappa})}{\sqrt{\epsilon_0V\omega(\bm{\kappa})}}\left(-\textrm{i} \frac{\partial}{\partial p_{\bm{\kappa},\lambda}}\right)\right)\right]\nonumber\\
    &=&\exp\left[\frac{\textrm{i}}{\hbar} \sum_{j=1}^{N}\bi{a}\cdot \hat{\bi{p}}_j + \frac{\textrm{i}}{\hbar}\sum^{M}_{\bm{\kappa},\lambda}d_{\bm{\kappa},\lambda}\left(-\textrm{i}\hbar\frac{\partial}{\partial p_{\bm{\kappa},\lambda}}\right)\right],
\end{eqnarray}
where $d_{\bm{\kappa},\lambda}= eN\bm{\varepsilon}_{\lambda}(\bm{\kappa})\cdot \bi{a}/\sqrt{\hbar\epsilon_0 V\omega(\bm{\kappa})}$. Thus, the original translation, restricted on the electronic subspace becomes a generalized translation in the full polaritonic (electronic plus photonic) configuration space of dimension $3N+M$ such that
\begin{equation}\label{polariton translations}
 (\bi{r}_j,p_{\bm{\kappa},\lambda})\longrightarrow \left(\bi{r}_j+\bi{a},p_{\bm{\kappa},\lambda}+ d_{\bm{\kappa},\lambda}\right).
\end{equation}
That this is the case can be straightforwardly checked by performing the above combined electronic plus photonic translation to the length gauge Hamiltonian of Eq.~(\ref{Lenght Hamiltonian}). By doing so we find that the crucial term  
\begin{equation}\label{symmetric term}
 \left(p_{\bm{\kappa},\lambda}-\frac{e\bm{\varepsilon}_{\lambda}(\bm{\kappa})\cdot \bi{R}}{\sqrt{\hbar\epsilon_0V\omega(\bm{\kappa})}}\right)^2
\end{equation}
is indeed invariant. This means that the simplistic interpretation of $\bi{r}$ corresponding to matter and $p_{\bm{\kappa},\lambda}$ to the photonic degrees of freedom is not valid. Both are mixtures of matter and photons and consequently polaritonic in nature~\cite{rokaj2017}. It is important to note that as can be understood from Eq.~(\ref{symmetric term}) translational symmetry would not be respected if the the quadratic dipole-self energy term $\hat{\varepsilon}_{dip}\sim \left(\bm{\varepsilon}_{\lambda}(\bm{\kappa})\cdot \bi{R}\right)^2$ was not present in the length gauge Hamiltonian. The fact that translational symmetry in the length gauge exists in the full electronic plus photonic configuration space, implies that in this gauge, Bloch's theorem can be generalized and applied in the polaritonic space. Such an approach has been explored in the framework of quantum electrodynamical Bloch theory~\cite{rokaj2019}. We will look into this framework in detail in chapter~\ref{QED Bloch theory}.

\section{Photon Number in the Length Gauge}

The fact that the length gauge transformation mixes light and matter in such a way that $\bi{r}$ and $p_{\bm{\kappa},\lambda}$ can no longer be interpreted as purely electronic and photonic operators respectively, can also be understood by looking into the photonic observables. In QED one of the most fundamental photonic observables is the number of photons contained in a state. The photon number operator $\hat{N}_{\textrm{ph}}$ is defined as~\cite{cohen1997photons, spohn2004}
\begin{eqnarray}
\hat{N}_{\textrm{ph}}=\sum_{\bm{\kappa},\lambda}\hat{a}^{\dagger}_{\bm{\kappa},\lambda}\hat{a}_{\bm{\kappa},\lambda},
\end{eqnarray}
where $\hat{a}_{\bm{\kappa},\lambda}$ and $\hat{a}^{\dagger}_{\bm{\kappa},\lambda}$ are respectively the annihilation and creation operators of the photon field. The annihilation and creation operators with respect to the displacement coordinates $q_{\bm{\kappa},\lambda}$ and their conjugate momenta $\partial/\partial q_{\bm{\kappa},\lambda}$ are given by Eq.~(\ref{q coordinate and momenta}). Then in terms of $q_{\bm{\kappa},\lambda}$ and $\partial/\partial q_{\bm{\kappa},\lambda}$ the photon number operator is
\begin{eqnarray}
\hat{N}_{\textrm{ph}}=\sum_{\bm{\kappa},\lambda}\left(-\frac{1}{2}\frac{\partial^2}{\partial q^2_{\bm{\kappa},\lambda}}+\frac{1}{2}q^2_{\bm{\kappa},\lambda}-\frac{1}{2}\right).
\end{eqnarray}
By performing the length gauge transformation on the photon number operator we can obtain its expression in the length gauge. First we perform the unitary transformation of Eq.~(\ref{Length unitary transf}) and we have for $\hat{N}_{\textrm{ph}}$
\begin{eqnarray}
\hat{U}^{\dagger}\hat{N}_{\textrm{ph}} \hat{U}=\sum_{\bm{\kappa},\lambda}\left[ -\frac{1}{2}\frac{\partial^2}{\partial q^2_{\bm{\kappa},\lambda}}-\textrm{i}\frac{e\bm{\varepsilon}_{\lambda}(\bm{\kappa})\cdot\mathbf{R}}{\sqrt{\hbar\epsilon_0V\omega(\bm{\kappa})}}\frac{\partial}{\partial q_{\bm{\kappa},\lambda}} +\frac{1}{2}\left(\frac{e\bm{\varepsilon}_{\lambda}(\bm{\kappa})\cdot \mathbf{R}}{\sqrt{\hbar\epsilon_0 V\omega(\bm{\kappa})}}\right)^2 +\frac{1}{2}q^2_{\bm{\kappa},\lambda}-\frac{1}{2}\right].\nonumber\\
\end{eqnarray}
To obtain the above result we made use of Eq.~(\ref{eq2.8}) for the transformation of the kinetic term of the photon modes $\partial^2/\partial q^2_{\bm{\kappa},\lambda}$. In addition, by performing also the swap between the coordinates and their relative momenta described in Eq.~(\ref{eq2.10}) we obtain the final expression for the photon number operator in the length gauge~\cite{schaeferquadratic}
\begin{eqnarray}
\hat{N}_{\textrm{ph}}=\sum\limits_{\bm{\kappa},\lambda}\left[ -\frac{1}{2}\frac{\partial^2}{\partial p^2_{\bm{\kappa},\lambda}}+\frac{1}{2}\left(p_{\bm{\kappa},\lambda}-\frac{e\bm{\varepsilon}_{\lambda}(\bm{\kappa})\cdot \bi{R}}{\sqrt{\hbar\epsilon_0V\omega(\bm{\kappa})}}\right)^2-\frac{1}{2}\right].
\end{eqnarray}
From the expression above we see clearly that the photon number operator in length gauge depends also on the full dipole operator $\bi{R}=\sum_{j}\bi{r}_j$, which typically is understood as an observable related to matter. Consequently, here we see that due to the mixing between the electronic and the photonic degrees of freedom, even the most fundamental photonic observable, the photon number operator, bears an influence coming from the electrons. Finally, we would like emphasize that in many cases it is wrongly assumed that the photon number operator in the length gauge it is simply given by the expression
\begin{eqnarray}
\hat{N}^{\prime}_{\textrm{ph}}= \sum\limits_{\bm{\kappa},\lambda}\left[ -\frac{1}{2}\frac{\partial^2}{\partial p^2_{\bm{\kappa},\lambda}}+\frac{1}{2}p_{\bm{\kappa},\lambda} -\frac{1}{2}\right].
\end{eqnarray}
As it was shown explicitly in~\cite{schaeferquadratic} the expectation values of these two operators $\hat{N}_{\textrm{ph}}$ and $\hat{N}^{\prime}_{\textrm{ph}}$ differ significantly and omitting the $\bi{R}$-dependent terms in the definition of the photon number operator can lead to completely different results about the amount of photons contained in the eigenstates of an electron-photon system. Finally, we would also like to mention that preserving the $\bi{R}$-dependent terms, i.e., the bilinear term $\bm{\varepsilon}_{\lambda}(\bm{\kappa})\cdot \bi{R}p_{\bm{\kappa},\lambda}$ and the quadratic dipole sel-energy $\hat{\varepsilon}_{dip}\sim \left(\bm{\varepsilon}_{\lambda}(\bm{\kappa})\cdot \bi{R}\right)^2$, in the expression for $\hat{N}_{\textrm{ph}}$, guarantees that the results obtained for the photon occupations are gauge invariant~\cite{schaeferquadratic}. 

\section{No Ground-State without the Dipole Self-Energy}\label{No GS without DSE}

Now we enter one of the main points that we aim to address in this section, namely the stability of the length-gauge Hamiltonian and the question of whether this Hamiltonian is bounded from below and has a ground-state. The properties of having a Hamiltonian which is bounded from below and has a stable ground-state are of fundamental importance in order to employ the variational principle and extend ground-state density-functional theory to non-relativistic QED in the dipole approximation~\cite{ruggenthaler2014, ruggenthaler2015, flick2017, TokatlyPRL}. 

But before we proceed, we would like to make more precise what we mean by a ground-state for the electron-photon system. It has been proven that in most cases where the bare electronic Hamiltonian (not being coupled to the photons) has a ground-state, the same holds also for the minimal-coupling Hamiltonian~\cite{spohn2004}. Having a ground-state $\Psi_{gs}$ means that we cannot find any other state $\Psi$ in the self-adjoint domain of the Hamiltonian that has an energy smaller than the energy $E_{gs}$ corresponding to the ground-state $\Psi_{gs}$. This means that the Hamiltonian is bounded from below, i.e., for every state $\Psi$ in the domain of the Hamiltonian holds $\langle \Psi|\hat{H}|\Psi\rangle \geq E_{gs}$. This property is true for both the bare electronic Hamiltonian and the minimal-coupling Hamiltonian in the Coulomb gauge, and it has been shown for a broad class of external potentials, e.g., $v_{ext}(\bi{r})\in L^2(\mathbb{R}^3)+L^{\infty}(\mathbb{R}^{3})$~\cite{spohn2004, Blanchard, Teschl, Kato}.

In addition, the length gauge Hamiltonian, being merely a unitarily equivalent form of the minimal-coupling Hamiltonian in the Coulomb gauge, will also be bounded from below and will have a stable ground-state. This guarantees that the variational principle is intact and that density-functional methods are applicable to the description of interacting electron-photon systems~\cite{ruggenthaler2014, ruggenthaler2015}. But for this very important and fundamental property to be satisfied, all terms appearing in the length-gauge Hamiltonian need to be included and to be taken into account.

However, in many cases the dipole self-energy $\hat{\varepsilon}_{dip}$ that arises in the length-gauge picture is ignored and eliminated from the length-gauge Hamiltonian. This is done based on the argument that the dipole self-energy depends on the quantization volume of the electromagnetic field, and therefore for the interaction of photons with a single atom or molecule one may take the limit $V\rightarrow \infty$ and in this case $\hat{\varepsilon}_{dip}\rightarrow 0$~\cite{faisal1987}. Due to this argument the dipole self-energy is supposed to be important only in the thermodynamic limit where the number of atoms or molecules interacting with the photon field becomes exceedingly large, $N\rightarrow \infty$. Another reason for which the dipole self-energy is frequently omitted in fields like cavity and circuit QED~\cite{grynberg2010} is because for the paradigmatic two-level models of quantum optics, like the Rabi and Jaynes-Cummings model, the dipole self-energy contributes only a constant energy offset~\cite{rokaj2017, ruggenthaler2014}.

Here, in order to investigate the impact of the dipole self-energy for the spectral properties of the length-gauge Hamiltonian, we will consider what happens if we ignore this harmonic self-interaction. For simplicity, in this section, we will consider the case where we have only one electron in a binding Coulombic potential interacting with one mode of the electromagnetic field. The general case of $N$ interacting electrons coupled to $M$ modes of the photon field can be treated analogously and we present it in detail in appendix~\ref{Many Mode DSE}.

In this simple case the Hamiltonian~(\ref{Lenght Hamiltonian}) takes the  form
\begin{eqnarray}\label{eq3.5}
	\hat{H}_{L}=-\frac{\hbar^2}{2m_{\textrm{e}}}\bi{\nabla}^2 -\frac{\hbar\omega}{2}\frac{\partial^2}{\partial p^2}+\frac{\hbar\omega}{2}\left(p-\frac{e\bm{\varepsilon}\cdot \bi{r}}{\sqrt{\hbar\epsilon_0V\omega}}\right)^2+v_{ext}(\bi{r}).
\end{eqnarray}
This Hamiltonian describes a single electron coupled to a single mode of a high-Q cavity, which means that we do not take into account dissipation. Further, we assume that the electron can escape from the cavity and thus we consider the electron in full space $\mathbb{R}^3$, like in the uncoupled case. The length gauge Hamiltonian without the dipole self-energy $\hat{H}^{'}=\hat{H}_{L}-\hat{\varepsilon}_{dip}$ is
\begin{equation}\label{eq3.7}
	\hat{H}^{'}=-\frac{\hbar^2}{2m}\bi{\nabla}^2 -\frac{\hbar\omega}{2}\frac{\partial^2}{\partial p^2} +\frac{\hbar\omega}{2}p^2-\left(\bm{\lambda}\cdot\bi{r}\right) p+v_{ext}(\bi{r})\;\; \textrm{where}\;\;\bm{\lambda}=e\sqrt{\frac{\hbar\omega}{\epsilon_0V}}\bm{\varepsilon}.
\end{equation}
As we already discussed extensively, the Hamiltonian $\hat{H}_{L}$ which includes the dipole self-energy is bounded from below. The question that arises now is whether the Hamiltonian $\hat{H}'$, without the dipole self-energy, is bounded from below as well?

\begin{figure}[h]
\begin{center}
   \includegraphics[width=3in, height=2.5in]{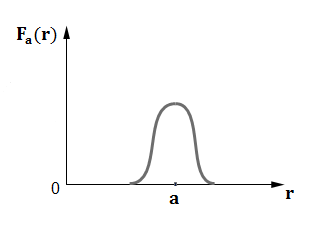}
\caption{\label{fig:two}Schematic representation in one-dimension of the electronic wavefunction $F_a(\bi{r})$. The wavefunction $F_a(\bi{r})$ (and all its derivatives) are non-zero only in the unit ball $|\bi{r}-\bi{a}|<1$.} 
\end{center}
\end{figure}

To answer this question, we will consider a trial wavefunction and we will compute the energy of this wavefunction with respect to $\hat{H}^{'}$. For the photonic part of the wavefunction we choose
\begin{equation}\label{eq3.8}
	\Phi(p)=\frac{1}{\sqrt{2}}\left[\phi_1(p)+\phi_{2}(p)\right],
\end{equation}
where the functions $\phi_1(p)$ and $\phi_{2}(p)$ are the first and the second excited state respectively, of the standard harmonic oscillator~\cite{GriffithsQM, Teschl}. For the electronic part we choose the following localized wavefunction
\begin{eqnarray}\label{eq3.10}
	F_{a}(\bi{r})=
	\begin{cases}
		\mathcal{N}\exp[-\frac{1}{1-|\bi{r}-\bi{a}|^2}],\qquad \textrm{if} \qquad|\bi{r}-\bi{a}|<1 \\
		 \qquad0, \hspace{2,8cm} \textrm{if}\qquad|\bi{r}-\bi{a}|\geq 1 \\
		\textrm{where} \qquad\bi{a}=a\bi{w}, \qquad a\in \mathbb{R}
	\end{cases}
\end{eqnarray}
where $\bi{w}$ is a non-zero vector and $a$ an arbitrary parameter. The wavefunction depicted in Fig.~\ref{fig:two} is non-zero in the unit ball $|\bi{r}-\bi{a}|<1$, is normalized, with $\mathcal{N}$ being its normalization constant, and is infinitely many times differentiable. We could have chosen any other well-behaved function but we fix this one for definiteness. Thus, the complete electron-photon wavefunction is the tensor product between $F_{a}(\bi{r})$ and $\Phi(p)$,
\begin{equation}\label{eq3.11}
	\Psi=F_{a}(\bi{r})\otimes\Phi(p).
\end{equation}
It is important to mention that the wavefunction $\Psi$ lies in the domain of $\hat{H}_L$ and of $\hat{H}^{\prime}$ because $\langle \hat{H}_L \Psi|\hat{H}_L\Psi\rangle < \infty$ and $\langle \hat{H}^{\prime} \Psi|\hat{H}^{\prime}\Psi\rangle < \infty$. The energy with respect to $\hat{H}'$ consists of four different terms
\begin{eqnarray}\label{eq3.13}
\langle \Psi|\hat{H}^{'}|\Psi\rangle= -\frac{\hbar^2}{2m}\langle F_{a}|\bi{\nabla}^2| F_{a}\rangle+\langle \Phi |\hat{H}_p |\Phi\rangle+\langle F_{a}|\hat{V}_{ext}| F_{a}\rangle+ \langle \Psi|\hat{V}_{int}|\Psi\rangle.
\end{eqnarray}
First we have the kinetic energy of the electron, the second term is the photon-energy, the third is the potential energy and the last one is the contribution due to the bilinear interaction between the electron and the photons. For the kinetic term of the electron we have
\begin{eqnarray}\label{eq3.14}
	-\frac{\hbar^2}{2m}\langle F_{a}| \bi{\nabla}^2| F_{a}\rangle=-\frac{\hbar^2\mathcal{N}^2}{2m}\int\limits_{|\bi{r}-\bi{a}|<1} e^{-\frac{1}{1-|\bi{r}-\bi{a}|^2}} \bi{\nabla}^2 \left(e^{-\frac{1}{1-|\bi{r}-\bi{a}|^2}}\right) d^3r.
\end{eqnarray}
Because the momentum operator is invariant under translations, we can perform the translation $\bi{r} \rightarrow \bi{r}+\bi{a}$ without changing the operator. Then, we have
\begin{eqnarray}\label{eq3.17}
-\frac{\hbar^2}{2m}\langle F_{a}|\bi{\nabla}^2| F_{a}\rangle&=&-\frac{\hbar^2}{2m}\langle F_{0}|\bi{\nabla}^2| F_{0}\rangle=\\
&=&-\frac{\hbar^2|\mathcal{N}|^2}{2m}\int\limits_{|\bi{r}|<1} e^{-\frac{1}{1-|\bi{r}|^2}} \bi{\nabla}^2 \left(e^{-\frac{1}{1-|\bi{r}|^2}}\right)d^3r=T<\infty.\nonumber
\end{eqnarray}
The result of integration above is finite because the integral is performed over the unit ball $|\bi{r}|<1$ and because all derivatives of $F_{a}(\bi{r})$ are finite. Then, for the energy of the photons we have
\begin{eqnarray}\label{eq3.18}
	\langle\Phi|\hat{H}_p| \Phi\rangle=\frac{1}{2}(E_1+E_2) \qquad\textrm{where}\qquad E_n=\hbar\omega\left(n+\frac{1}{2}\right)\qquad \forall \; n \in \mathbb{N}\;.
\end{eqnarray}
We note that $E_n$ are the energy levels of the harmonic oscillator~\cite{GriffithsQM}. Next we need to compute the energy of the electron due to the external scalar potential $v_{\textrm{ext}}$. The external potential is chosen to be an attractive, binding potential which consequently means that the contribution of $v_{ext}$ will be negative. We note that this is not some random choice but it is the standard choice for describing atomic and molecular systems,
\begin{equation}\label{eq3.19}
	\langle F_{a}|\hat{V}_{ext}| F_{a}\rangle=-V_{a} \qquad \textrm{where}\qquad V_{a}\geq0.
\end{equation}
Finally, we compute the energy of the bilinear interaction
\begin{eqnarray}\label{eq3.20}
\langle\Psi|\hat{V}_{int}| \Psi\rangle=-\langle\Phi| p |\Phi\rangle \langle F_{a}|\left( \bm{\lambda}\cdot \bi{r}\right)| F_{a}\rangle 
\end{eqnarray}
that with $\langle \Phi| p |\Phi\rangle=1$ becomes 
\begin{equation}
    \langle \Psi|\hat{V}_{int}| \Psi\rangle=-\bm{\lambda}\cdot\langle F_{a}|\bi{r}|F_{a}\rangle=-|\mathcal{N}|^2\int \limits_{|\bi{r}-\bi{a}|<1}\bm{\lambda}\cdot \bi{r}e^{-\frac{2}{1-|\bi{r}-\bi{a}|^2}} d^3r.
\end{equation}
We perform once more the translation $\bi{r}\rightarrow \bi{r}+\bi{a}$ and the integral takes the form
\begin{eqnarray}\label{eq3.21}
 \langle \Psi|\hat{V}_{int}| \Psi\rangle = -|\mathcal{N}|^2\int \limits_{|\bi{r}|<1}\bm{\lambda}\cdot \bi{r}\;e^{-\frac{2}{1-|\bi{r}|^2}}d^3r -\bm{\lambda}\cdot\bi{a}\;\langle F_0|F_0\rangle=-\bm{\lambda}\cdot\bi{a}=-a 
\end{eqnarray}
where we have now chosen $\bi{w} = \bm{\lambda}/|\bm{\lambda}|^2$. The first integral in the above equation is zero, because we integrate an odd function over a symmetric integration volume. Then, we see that the contribution of the bilinear interaction is proportional to $-a$. Summing all four contributions, we obtain the result for the total energy 
\begin{equation}\label{eq3.22}
	\langle \Psi|\hat{H}^{'}|\Psi\rangle= T+\frac{1}{2}(E_1+E_2)-V_a-a \leq T+\frac{1}{2}(E_1+E_2)- a  \sim -a\;.
\end{equation}
From the expression for the energy it becomes clear that the Hamiltonian $\hat{H}^{\prime}$, which does not include $\hat{\varepsilon}_{dip}$, is unbounded from below, because the parameter $a$ can be chosen arbitrarily ($F_{a}$ can be moved further and further away from the origin) and we can therefore lower the energy of $\hat{H}^{'}$ indefinitely. Thus, we come to the conclusion that without the dipole self-energy the length gauge Hamiltonian becomes unbounded from below and has no ground-state~\cite{rokaj2017}. We would like to note that this result is not so surprising as we subtracted a harmonic potential from the Hamiltonian~(\ref{eq3.5}) and despite claims in literature this term is dominant and cannot be discarded. 

\textit{Comments on Assumptions.}---Finally, we would like to comment on the choice of considering the particles to be in full space while the photon field was quantized in a box with periodic boundary conditions (see section~\ref{EM field Quantization}), as well as considering a purely negative external potential $v_{ext}(\bi{r})$. First of all, we want to emphasize that allowing for different lengths in the different directions of the quantization volume is straightforward, and in this way we can model not only free space but also a planar cavity. Further, it would be not a problem to use other boundary conditions, like zero boundary conditions in the $z$-direction and periodic ones in the $(x,y)$-plane and in the very end take the limit to infinity for the open directions~\cite{spohn2004}. But all these considerations become superfluous in the long-wavelength limit (or dipole approximation), where the spatial profile of the electromagnetic modes is not taken into account. Further, the explicit spatial form of the photon field does not change the harmonic nature of the dipole self-energy, and for that reason we employed periodic boundary conditions. However, for the investigation of the ground-state in the length gauge Hamiltonian it makes a difference if we enclose the particles in a finite volume. In this case it would not be possible to lower the energy of $\hat{H}^{'}$ indefinitely. But what would happen in this case is that we would find a ground-state that is localized at the edge of the quantization box. This was demonstrated exactly in a numeric fashion in~\cite{schaeferquadratic}. Such a wavefunction however would not be a physical ground-state of an atom because the electrons of the atom would not be localised around the nucleus. Further, in this case we would also have a maximally allowed box length for a given atomic or molecular system which would just cut the region in which the dipole self-energy becomes dominant. Keeping the dipole self-energy in the length-gauge Hamiltonian has the advantage of allowing for a treatment independent of the box-size, which is definitely a physically desirable property that yields physically acceptable localised ground-states. Lastly, we would like to mention that in the case of a high-Q cavity, the cavity mirrors can be modeled as barrier for the matter-particles by adding a very large repulsive potential at the assumed positions of the mirrors. But even with the inclusion of such a potential barrier would not invalidate our proof that the length gauge Hamiltonian without the dipole self-energy is unbounded from below, because the potentials we consider here are in the Banach space $L^{2}(\mathbb{R}^3)+L^{\infty}(\mathbb{R}^3)$ and in the limit $|\bi{r}| \rightarrow \infty$ only the bounded part of the potential survives. Thus, by shifting $F_{a}$ arbitrarily, only a contribution proportional to the limiting constant $v_{ext}(\bi{r}) \rightarrow v_{ext}^{\infty}$ contributes. As a consequence, such a positive potential in $L^{2}(\mathbb{R}^3)+L^{\infty}(\mathbb{R}^3)$ cannot compensate the linear decrease in energy due to the bilinear interaction $\hat{V}_{int}$ between the photons and the charged particles.

\section{Maxwell's Equations in Matter}\label{Maxwell matter}

Clearly, as we showed in the previous section, despite the claims in the literature, the dipole self-energy is a very important term for the length-gauge Hamiltonian and neglecting this term dramatically changes the properties of the interacting light-matter system. Without the dipole self-energy a ground-state becomes impossible, and even for arbitrarily small but finite coupling to the photon field the light-matter system decays and no stable ground-state exists. Therefore, the dipole self-energy is necessary for describing the static properties of the combined matter-photon system.

Besides this extremely important effect, there is also another crucial point for which the dipole self-energy cannot be neglected which has to do with the fact that Maxwell's equations for the electric field are not satisfied if $\hat{\varepsilon}_{dip}$ is omitted from $\hat{H}_L$~\cite{rokaj2017}. So let us see how this actually happens.

As we stated in the previous sections, a peculiarity of the length gauge is that it mixes the electronic and the photonic degrees of freedom. This fact can be understood also from the expression of the electric field in the length-gauge picture. From the definition of the vector potential operator~(\ref{A field in Qs}) and performing the transformation on the photonic coordinates defined in Eq.~(\ref{eq2.10}) we find that the $\bi{A}$-field in the length gauge is 
\begin{equation}\label{eq4.10}
\hat{\bi{A}} = \sum_{\bm{\kappa},\lambda} \hat{\bi{A}}_{\bm{\kappa},\lambda}, \qquad \textrm{where} \qquad \hat{\bi{A}}_{\bm{\kappa},\lambda} = -\textrm{i} \sqrt{\frac{\hbar}{\epsilon_0V\omega(\bm{\kappa})}}\bm{\varepsilon}_{\lambda}(\bm{\kappa})\frac{\partial}{\partial p_{\bm{\kappa},\lambda}}.
\end{equation}
Then, from the Heisenberg equations of motion~\cite{GriffithsQM} we can obtain the expression for the electric field
\begin{equation}\label{eq4.11}
\hat{\bi{E}}=-\frac{d \hat{\bi{A}}}{d t}=-\frac{\textrm{i}}{\hbar}[\hat{H}_{L},\hat{\bi{A}}]=\sum_{\bm{\kappa},\lambda} \hat{\bi{E}}_{\bm{\kappa},\lambda} \;\; \textrm{with} \;\; \hat{\bi{E}}_{\bm{\kappa},\lambda} = \sqrt{\frac{\hbar\omega(\bm{\kappa})}{\epsilon_0V}}\bm{\varepsilon}_{\lambda}(\bm{\kappa})\left(p_{\bm{\kappa},\lambda}-\frac{\bm{\varepsilon}_{\lambda}(\bm{\kappa})\cdot \bi{R}}{\sqrt{\epsilon_0V\hbar\omega(\bm{\kappa})}}\right).
\end{equation}
Further, by defining the polarization operator $\hat{\bi{P}}$ as 
\begin{equation}\label{eq4.12}
\hat{\bi{P}}=\epsilon_0\sum_{\bm{\kappa},\lambda}\bm{\varepsilon}_{\lambda}(\bm{\kappa}) \frac{e\bm{\varepsilon}_{\lambda}(\bm{\kappa})\cdot \bi{R}}{\epsilon_0V},
\end{equation}
we find that the photonic coordinates $p_{\bm{\kappa},\lambda}$ actually correspond to the displacement field $\hat{\bi{D}}$
\begin{equation}\label{eq4.13}
\hat{\bi{D}}=\epsilon_0\sum_{\bm{\kappa},\lambda}\sqrt{\frac{\hbar\omega(\bm{\kappa})}{\epsilon_0V}}\bm{\varepsilon}_{\lambda}(\bm{\kappa})p_{\bm{\kappa},\lambda}
\end{equation}
of the Maxwell equations in matter, i.e., $\hat{\bi{D}}= \epsilon_0\hat{\bi{E}}+ \hat{\bi{P}}$~\cite{JacksonEM}. We note that this result would be true even if we discarded the dipole self-energy from the length-gauge Hamiltonian. The consequences of omitting the dipole self-energy term only show up in higher derivatives of the vector potential and lead to a violation of the equations of motion. 

To see how this happens we will compute the equations of motion for the $\bi{A}$-field and the $\bi{E}$-field, with and without the dipole self-energy. Let us start with the case where $\hat{\varepsilon}_{dip}$ is present. Firstly, we compute the time-derivative for the electric field which relates to the second time-derivative of the vector potential $\dot{\bi{E}}=-\ddot{\bi{A}}$, and we obtain
\begin{equation}\label{eq4.14}
\frac{d^2}{d t^2}\hat{\bi{A}} + \sum_{\bm{\kappa},\lambda}\omega^2(\bm{\kappa}) \hat{\bi{A}}_{\bm{\kappa},\lambda} =\frac{\textrm{i}\hbar e}{\epsilon_0Vm_{\textrm{e}}}\sum_{\bm{\kappa},\lambda}\bm{\varepsilon}_{\lambda}(\bm{\kappa})\sum^{N}_{i=1}\bm{\varepsilon}_{\lambda}(\bm{\kappa})\cdot\nabla_{i} .
\end{equation}
The equation above is the mode resolved inhomogeneous Maxwell equation, with the inhomogeneity coming from the presence of the paramagnetic current operator on the right-hand side~\cite{JacksonEM}. For the computation of the equation of motion for the electric field we make the choice $v_{ext}(\bi{r}) = 0$ to simplify the further analysis. We would like to emphasize that this choice corresponds to the paradigmatic system of the homogeneous electron gas, also known as the jellium model~\cite{Mermin, Vignale}. Then, for the equation of motion for the electric field we have~\cite{rokaj2017}
\begin{equation}\label{eq4.15}
\frac{d^2}{d t^2} \hat{\bi{E}}=-\sum_{\bm{\kappa},\lambda}\omega^2(\bm{\kappa})\hat{\bi{E}}_{\bm{\kappa},\lambda}-\frac{Ne^{2} }{\epsilon_0Vm_{\textrm{e}}}\hat{\bi{E}}=-\sum_{\bm{\kappa},\lambda}\left(\omega^2(\bm{\kappa})+\omega^2_p\right)\hat{\bi{E}}_{\bm{\kappa},\lambda}.
\end{equation}
The equation of motion of the electric field is the well-known mode resolved Maxwell equation. Moreover, we see that due to the interaction with matter there is an additional term contributing to the oscillation frequencies of the of the electric field. The bare photon frequencies $\omega(\bm{\kappa})$ get dressed by the contribution $\omega_p$ coming from matter
\begin{equation}\label{eq4.16}
\omega_p=\sqrt{\frac{e^{2} N}{\epsilon_0Vm_{\textrm{e}}}}=\sqrt{\frac{e^2n_{\textrm{e}}}{\epsilon_0 m_{\textrm{e}} }},
\end{equation}
where $n_{\textrm{e}}=N/V$ is the electron density. The total frequencies of the electric field therefore are
\begin{equation}\label{eq4.17}
\widetilde{\omega}^{2}(\bm{\kappa}) = \omega^{2}(\bm{\kappa}) + \omega_{p}^{2}. 
\end{equation}
This change in the frequency spectrum of the electric field is a diamagnetic shift due to the interactions with matter and depends on the full electron density $n_{\textrm{e}}$ via the plasma frequency $\omega_p$~\cite{rokaj2017, rokaj2020}. This diamagnetic contribution has been observed experimentally in resonant matter-photon systems in the ultra-strong coupling regime~\cite{todorov2010, todorov2012, TodorovPRX2014}.

However, if we ignore the dipole self-energy and compute the equations of motion for the electric field in the case where $v_{\textrm{ext}}(\bi{r})=0$ we find that the equations of motion differ significantly~\cite{rokaj2017}
\begin{equation}\label{eq4.18}
\frac{d^2}{d t^2}\hat{\bi{E}}+\sum_{\bm{\kappa},\lambda}\omega^2(\bm{\kappa})\hat{\bi{E}}_{\bm{\kappa},\lambda}=-\omega^2_p\hat{\mathbf{D}}.
\end{equation}
As it is clear from the above result, neglecting the dipole self-energy leads to a wrong description of the electromagnetic field, because on the right-hand side of equation~(\ref{eq4.18}) we do not have the electric field $\hat{\bi{E}}$ but the displacement field $\hat{\bi{D}}$. This implies that if we neglect the dipole self-energy the electric field does not satisfy the mode-resolved Maxwell equation and we get a completely wrong description for the electromagnetic field coupled to matter. Therefore, the dipole self-energy cannot be ignored and must always be included in order to have a complete and consistent physical description of an interacting photon-matter system~\cite{rokaj2017}.

\part{The Free Electron Gas in Cavity QED}

\chapter{The Free Electron Gas}\label{Free Electron Gas}

The model of the free electron gas introduced by Sommerfeld in 1928~\cite{Sommerfeld1928} is a paradigmatic model for solid state and condensed matter physics. Originally it was introduced for the description of thermal and conduction properties of metals. Since then it has served as one of the fundamental models for understanding and describing materials. Further, with the inclusion of the electron-electron interactions and the positive ion charges in terms of a background medium, the free electron gas was transformed into the homogeneous electron gas, also known as the jellium model~\cite{Mermin, Vignale}, which with the advent of density functional theory (DFT) and the local density approximation (LDA)~\cite{HohenbergKohn} has become one of the most useful and successful computational tools for physics, chemistry and materials science~\cite{RubioReview}. Also within Landau's Fermi liquid theory~\cite{LandauFermiLiquid}, the free electron gas was used as the fundamental building block~\cite{Nozieres}. In addition, the free electron gas in the presence of strong homogeneous magnetic fields has also proven to be extremely important for the description of the integer and the fractional quantum Hall effects~\cite{Klitzing, LaughlinPRB, Laughlingfractional, TsuifractionalQHE}. Due to the importance of the free electron gas for condensed matter physics and its wide applicability, in what follows we focus on this system.

\section{Free Electrons in a ``Periodic'' Box}

The Sommerfeld model of the free electron gas consists of $N$ non-interacting electrons confined in a cube whose sides are of length $L$ and volume $V=L^3$. Because the electrons do not interact with one another the Hamiltonian describing this system is the sum of the kinetic energy operators of all the particles
\begin{eqnarray}\label{free electron H}
\hat{H}=\sum^N_{j=1}\hat{H}_{j}=\frac{-\hbar^2}{2m_{\textrm{e}}}\sum^N_{j=1}\nabla^2_j.
\end{eqnarray}
This implies that we can find the full set of eigenstates of the system by solving the single-particle Hamiltonian $\hat{H}_j$ and then construct the many-body eigenstates from the single-particle ones. To describe the fact that the electrons are confined within the cube of volume $V$ we need to impose boundary conditions on the wavefunctions of the electrons. The natural choice would be the wavefunctions to satisfy zero boundary conditions which make clear that the electrons do not escape outside of the material. Zero boundary conditions though lead to solutions which are standing waves, which are not convenient to work with. Another possibility is to impose periodic boundary conditions~\cite{Mermin}, which as it was proven by Lebowitz and Lieb in~\cite{Lieb_Boundary_Conditions}, do not affect the bulk properties of a system. Thus, we choose to impose periodic boundary conditions which are more convenient to work with from a mathematical point of view. Such boundary conditions imply that the wavefunctions describing our system need to satisfy
\begin{eqnarray}\label{Bounday Conditions}
\phi(x+L,y,z)=\phi(x,y+L,z)=\phi(x,y,z+L)=\phi(x,y,z).
\end{eqnarray}
With such boundary conditions and because the Hamiltonian commutes with the momentum operator, $[\hat{H},\nabla]=0$, the single-particle eigenfunctions are plane waves of the form
\begin{eqnarray}\label{plane wave}
\phi_{\bi{k}}(\bi{r})=\frac{e^{\textrm{i}\bi{k}\cdot\bi{r}}}{\sqrt{V}} \;\;\; \textrm{with} \;\;\; \bi{k}=\frac{2\pi \bi{n}}{L} \;\;\; \textrm{and}\;\;\; \bi{n} \in \mathbb{Z}^3.
\end{eqnarray}
The quantum number $\bi{k}$ multiplied by $\hbar$ is the momentum $\bi{p}=\hbar \bi{k}$ of the free particle, which due to translational invariance of the system, is a conserved quantity. The eigenenergy of the eigenfunction $\phi_{\bi{k}}(\bi{r})$ is 
\begin{eqnarray}
E_{\bi{k}}=\frac{\hbar^2\bi{k}^2}{2m_{\textrm{e}}}.
\end{eqnarray}
From the single-particle eigenfunctions we can construct the many-body eigenstates of the system. Because electrons are fermions the many-body eigenstates must be antisymmetric under the exchange of any two electrons. To satisfy the fermionic statistics we use a Slater determinant~\cite{Mermin}.
\begin{eqnarray}\label{Slater}
    \Psi_{\bi{K}}(\bi{r}_1\sigma_1,..,\bi{r}_N\sigma_N)=\frac{1}{\sqrt{N!}}\;\begin{tabular}{|c c c c|}
		$\phi_{\mathbf{k}_1}(\bi{r}_1\sigma_1)$ &$ \phi_{\mathbf{k}_1}(\bi{r}_2\sigma_2)$ & $\cdot\cdot\cdot$ & $\phi_{\bi{k}_{1}} (\bi{r}_N\sigma_N)$\\
		$\phi_{\mathbf{k}_2}(\bi{r}_1\sigma_1)$ & $\phi_{\mathbf{k}_2}(\bi{r}_2\sigma_2)$ & $\cdot\cdot\cdot$ & $\phi_{\bi{k}_{2}} (\bi{r}_N\sigma_N)$\\
		$\cdot$ & $\cdot$ & $\cdot\cdot\cdot$& $\cdot$\\
		$\cdot$ & $\cdot$ & $\cdot\cdot\cdot$& $\cdot$\\
		$\phi_{\mathbf{k}_N}(\bi{r}_1\sigma_1)$ & $\phi_{\mathbf{k}_N}(\bi{r}_2\sigma_2)$ & $\cdot\cdot\cdot$ & $\phi_{\bi{k}_{N}} (\bi{r}_N\sigma_N)$
	\end{tabular}
\end{eqnarray}
We note that $\bi{K}=\sum_j\bi{k}_j$ is the collective momentum of the electrons and we introduced it here in order to indicate the fact that each many-body eigenstate is defined with respect to a particular distribution of the electrons in $\bi{k}$-space and depends on the collective momentum of the electron gas. Thus, the collective momentum $\bi{K}$ is used as a label to distinguish different many-body states of the electron gas. The energy $E_{\bi{K}}$ of the Slater determinant $\Phi_{\bi{K}}(\bi{r}_1,..,\bi{r}_N)$ is then the sum of the single-particle energies
\begin{eqnarray}
E_{\bi{K}}=\sum^N_{j=1}E_{\bi{k}_j}=\sum^N_{j=1}\frac{\hbar^2\bi{k}^2_j}{2m_{\textrm{e}}}.
\end{eqnarray}

\section{Ground State Energy in the Thermodynamic Limit}

Our aim now is to compute the ground-state energy of the free electron gas in the so-called thermodynamic limit, in which the number of particles and the volume of the system tend to infinity but in such a way that the electron density $n_{\textrm{e}}=N/V$ stays fixed~\cite{Mermin, Vignale}. To achieve this, first we need to find what is actually the ground-state of the system in the thermodynamic limit. 

In the previous section we found that the full set of eigenstates of the system is described by Slater determinants $\Psi_{\bi{K}}(\bi{r}_1,..,\bi{r}_N)$ of Eq.~(\ref{Slater}). But as we already mentioned these states are defined by a distribution in $\bi{k}$-space. Thus, to find the ground state of the system we need to find the optimal distribution for the electrons in $\bi{k}$-space. 

The allowed values for $\bi{k}$ are discrete as we can see from Eq.~(\ref{plane wave}). This implies that the allowed $\bi{k}$ form a cubic grid in $\bi{k}$-space. From this fact one might expect intuitively that the ground state distribution for small number of particles should be a cube since all three directions $k_x,k_y$ and $k_z$ are equivalent. However, here we are interested in very large amount of particles and very large $\bi{k}$-space regions which eventually become so regular that the ground of distribution in $\bi{k}$-space can be regarded to be a sphere. This sphere is known as the Fermi sphere~\cite{Mermin}. The fact that the ground state distribution has to be a sphere can be understood from the fact that the Hamiltonian of Eq.~(\ref{free electron H}) is spherically symmetric.

The volume $\Omega_{\mathcal{S}}$ of the Fermi sphere is defined by its radius $k_\textrm{F}$ which is the highest occupied wave vector in $\bi{k}$-space and it is known as the Fermi wave vector. Then, the volume $\Omega_{\mathcal{S}}$ is   
\begin{eqnarray}
\Omega_{\mathcal{S}}=\frac{4\pi k^3_\textrm{F}}{3}.
\end{eqnarray}
We note that using the Fermi wave vector we can also define the highest occupied energy in the system as $E_\textrm{F}=\hbar^2k^2_\textrm{F}/2m_{\textrm{e}}$ which is known as the Fermi energy. Having found the volume of the distribution in $\bi{k}$-space, in terms of $k_{\textrm{F}}$, we can also find the number of allowed states in $\bi{k}$-space as the quotient between $\Omega_{\mathcal{S}}$ and the volume occupied by a single state in $\bi{k}$-space, which is $(2\pi/L)^3$. We therefore conclude that number of allowed states contained in the Fermi sphere $\Omega_{\mathcal{S}}$ is
\begin{eqnarray}
\#\textrm{states}=\frac{\Omega_S}{(2\pi/L)^3}=\frac{\Omega_{\mathcal{S}}V}{8\pi^3},
\end{eqnarray}
where $V$ is the volume of the material in real space. Since each wave vector $\bi{k}$ can be occupied by two electrons, due to the spin degeneracy, we find that the number of electrons that can be accommodated in the Fermi sphere is
\begin{eqnarray}
N=2\#\textrm{states}=\frac{\Omega_{\mathcal{S}}V}{4\pi^3}.
\end{eqnarray}
As a consequence the electron density in the system as a function of the Fermi wave vector is 
\begin{eqnarray}\label{electron density}
n_{\textrm{e}}=\frac{N}{V}=\frac{k^3_{\textrm{F}}}{3\pi^2}.
\end{eqnarray}

Having defined the ground-state distribution in $\bi{k}$-space and the electron density, in the thermodynamic limit, we can attempt to compute the ground-state energy density of the system. The total energy of the system inside the Fermi sphere for doubly occupied wave vectors is
\begin{eqnarray}
E_{gs}=2\sum_{|\bi{k}|<k_F}\frac{\hbar^2}{2m_{\textrm{e}}}\bi{k}^2.
\end{eqnarray}
The summation of a smooth function as the one appearing above can be done as follows~\cite{Mermin}: The volume of $\bi{k}$-space per allowed $\bi{k}$ value is $\Delta\bi{k}=8\pi^3/V$ and consequently we can write the sum of the previous equation as
\begin{eqnarray}
\sum_{|\bi{k}|<k_{F}}\bi{k}^2=\frac{V}{8\pi^3} \sum_{|\bi{k}|<k_{F}}\bi{k}^2\Delta\bi{k}. 
\end{eqnarray}
In the limit where the volume of the system approaches infinity ($V\rightarrow \infty$) the measure $\Delta\bi{k}$ goes to zero ($\Delta\bi{k}\rightarrow 0$) and the sum approaches an integral. Thus, the energy per volume in the limit where $V\rightarrow \infty$ is 
\begin{eqnarray}
\mathcal{E}_{gs} \equiv \lim_{V\to\infty}\frac{E_{gs}}{V}=\frac{\hbar^2}{8\pi^3 m_{\textrm{e}}}\iiint\limits_{|\bi{k}|<k_{F}} d^3k \bi{k}^2.
\end{eqnarray}
Going into spherical coordinates in $\bi{k}$-space and performing the integral, we find the energy density as a function of the Fermi momentum 
\begin{eqnarray}\label{energy density kF}
\mathcal{E}_{gs}(k_\textrm{F})=\frac{\hbar^2k^5_\textrm{F}}{10\pi^2 m_{\textrm{e}}}.
\end{eqnarray}
Further, using the relation for the electron density $n_\textrm{e}$ as a function of the Fermi momentum in Eq.~(\ref{electron density}) we can find the energy density $\mathcal{E}_{gs}$ as a function of the electron density
\begin{eqnarray}
\mathcal{E}_{gs}(n)=\frac{3\hbar^2(3\pi)^{2/3}}{10m_{\textrm{e}}}n^{5/3}_{\textrm{e}}.
\end{eqnarray}
Such expressions of the energy density in terms of the electron density are very important for the construction of functional approximations in density functional theory, like for example in the case of the LDA functional~\cite{HohenbergKohn}. Moreover, we note that from the expression of the energy density one can compute also other kinds of thermodynamic properties of interest like the energy per particle, the pressure exerted by the electron gas or the compressibility of the system~\cite{Mermin}.

\section{Density of States}

Another very important quantity for solid state systems is the density of states. The density of states describes the number of states that are available in a system in a particular energy neighborhood, and is essential for determining the carrier concentrations and energy distributions of carriers.  Furthermore, the density of states is connected to the transport properties of materials~\cite{Mermin, Vignale}. Generally the density of states $\mathcal{D}(E)$ of a system of volume $V$ and with $N$ countable energy levels is defined as 
\begin{eqnarray}\label{DOS Def}
\mathcal{D}(E)=\frac{2}{V}\sum^N_{i=1}\delta(E-E_i)
\end{eqnarray}
where $E_i$ are the energy levels of the system and the prefactor $2$ appears due to the spin degeneracy. In our case we have a system which in the thermodynamic limit has a continuous spectrum. Following the procedure we described before, on how to perform summations in the thermodynamic limit, the density of states in the limit $V\rightarrow \infty$ and for a continuous spectrum becomes
\begin{eqnarray}
\mathcal{D}(E)=\lim_{V \to \infty}\frac{2}{V}\sum_{\bi{k}}\delta(E-E_{\bi{k}})=\frac{2}{(2\pi)^3}\iiint d^3k \delta(E-E_{\bi{k}}).
\end{eqnarray}
The energy spectrum of the free electrons is $E_{\bi{k}}=\hbar^2\bi{k}^2/2m_{\textrm{e}}$. Substituting this energy expression and going into spherical coordinates $(k_x,k_y,k_z)\longrightarrow (k_r,k_{\theta},k_{\phi})$ we obtain
\begin{eqnarray}
\mathcal{D}(E)=\frac{2}{(2\pi)^3}\iint \sin k_{\theta} dk_{\theta}dk_{\phi} \int dk_r k^2_r \delta\left(E-\frac{\hbar^2k^2_r}{2m_{\textrm{e}}}\right).
\end{eqnarray}
We integrate over the angle coordinates 
\begin{eqnarray}
\mathcal{D}(E)=\frac{8\pi}{(2\pi)^3}\int dk_r k^2_r \delta\left(E-\frac{\hbar^2k^2_r}{2m_{\textrm{e}}}\right).
\end{eqnarray}
For the last integral over $k_r$ we introduce the variable $s$
\begin{eqnarray}
s=\frac{\hbar^2 k^2_r}{2m_{\textrm{e}}}\;\;\; \textrm{with}\;\;\; ds=\frac{\hbar^2 k_r dk_r}{m_{\textrm{e}}} 
\end{eqnarray}
and the integral for $\mathcal{D}(E)$ transforms as
\begin{eqnarray}
\mathcal{D}(E)=\frac{\sqrt{2}}{\pi^2}\left(\frac{m_{\textrm{e}}}{\hbar^2}\right)^{3/2} \int ds \sqrt{s} \delta(E-s),
\end{eqnarray}
and after integrating over the variable $s$ we obtain the expression for the density of states
\begin{eqnarray}
\mathcal{D}(E)=\frac{\sqrt{2}}{\pi^2}\left(\frac{m_{\textrm{e}}}{\hbar^2}\right)^{3/2} \sqrt{E} \;\;\; \textrm{for}\;\;\; E<E_F.
\end{eqnarray}
We note that the expression above for the density of states holds for energies smaller than the Fermi energy and that for energies above the Fermi energy the density of states is zero~\cite{Mermin}.

\chapter{The Free Electron Gas in Cavity QED}\label{Free Electron Gas in cavity QED}

\begin{displayquote}
\footnotesize{I remember Fermi used to ask ``Where is the hydrogen atom of this problem?'' 
Where, in what domain, will we find a simple system with a relatively simple law for its description,
which will be the forerunner or the test of a real theory?}
\end{displayquote}
\begin{flushright}
  \footnotesize{Murray Gell-Mann\\
Particles and Principles~\cite{Gell-Mann}}
\end{flushright}

Most of our understanding of light-matter interactions is based on finite-system (or few-level) models coming from the field of quantum optics, like the Rabi~\cite{Braak}, Jaynes-Cummings~\cite{shore1993} or the Dicke model~\cite{dicke1954}. These models have been proven to be extremely successful for the description of coupled light-matter systems and particularly for the field of cavity QED. However, during the last decade these fundamental models are being challenged by developments in the emerging field of cavity QED materials~\cite{ruggenthaler2017b}, in which also extended solid state systems are coupled to the quantized cavity-field.

It is well-known that extended solid state or condensed matter systems behave very much differently than finite-system models. This brings into question whether the few-level models of quantum optics can be used for the description of macroscopic systems, like materials, strongly coupled to the quantized field of a cavity. Consequently, it becomes desirable to have a an example of an exactly solvable extended system strongly coupled to the quantized electromagnetic field of a cavity, analogously to the Rabi and the Dicke model. 
  
For that purpose, in this chapter our aim is to revisit the paradigmatic Sommerfeld model~\cite{Sommerfeld1928} of the free electron gas in the framework of cavity QED.  

\section{Electron Gas in Cavity QED}\label{2DEG in QED}

\begin{figure}[h]
\begin{center}
\includegraphics[scale=0.8]{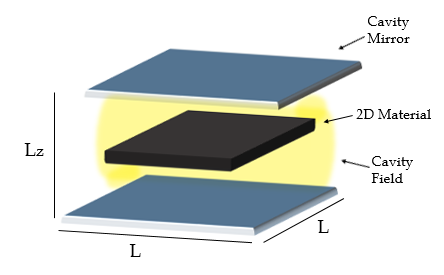}
\caption{\label{HEG_Cavity} Schematic depiction of a 2D material confined inside a cavity. The cavity mirrors are of length $L$ and area $S=L^2$. The area of the material is also $S$, while the distance between the mirrors of the cavity is $L_z$. }    
\end{center}
\end{figure}

Here we are interested in the two-dimensional free electron gas (2DEG) confined inside a cavity, as depicted in Fig.~\ref{HEG_Cavity}. For that purpose we consider the Pauli-Fierz Hamiltonian of Eq.~(\ref{Pauli Fierz Hamiltonian}) in the absence of any external potential, $v_{\textrm{ext}}(\bi{r})=0$, and we neglect the Coulomb interaction as in the original Sommerfeld model~\cite{Sommerfeld1928}. Further, we neglect the Pauli (Stern-Gerlach) term $\hat{\bm{\sigma}}\cdot\hat{ \bi{B}}(\bi{r})$ and we consider the vector potential of the electromagnetic field in the long-wavelength limit. Under these assumptions the Hamiltonian for the 2DEG in the cavity is
\begin{eqnarray}\label{Pauli-Fierz} 
\hat{H}=\frac{1}{2m_{\textrm{e}}}\sum\limits^{N}_{j=1}\left(\textrm{i}\hbar \mathbf{\nabla}_{j}+e \hat{\mathbf{A}}\right)^2+\sum\limits_{\bm{\kappa},\lambda}\hbar\omega(\bm{\kappa})\left(\hat{a}^{\dagger}_{\bm{\kappa},\lambda}\hat{a}_{\bm{\kappa},\lambda}+\frac{1}{2}\right) .
\end{eqnarray}
We note that for our system the long-wavelength limit or dipole approximation is respected and justified, because we are considering a 2D material (effectively of zero thickness) confined in the cavity, as shown in Fig.~\ref{HEG_Cavity}. Since we restrict our considerations to two dimensions, the momentum operator has only two components $\nabla=(\partial_x,\partial_y)$. Moreover, the 2DEG in the $(x,y)$ plane is assumed to be macroscopic and consequently the electrons can be described with the use of periodic boundary conditions, as in the original Sommerfeld model~\cite{Sommerfeld1928}. 

In cavity QED it is very typical to employ the single-mode approximation in which the cavity is considered to have a particular mode which is on resonance with the matter system and the rest of the electromagnetic modes are neglected. As a starting point we will also take the approximation where the field consists of a single mode with frequency $\omega$ but we will keep both polarizations of the field. As we will see later, the single-mode case will be used for the construction of an effective quantum field theory in which the full 2D continuum of electromagnetic modes interacts with the 2DEG. The polarization vectors of the quantized field are chosen to be in the $(x,y)$ plane such that the mode interacts with the 2DEG. For the polarizations to be orthogonal we choose $\bm{\varepsilon}_1=\mathbf{e}_x$ and $\bm{\varepsilon}_2=\mathbf{e}_y$. We note that these two polarization vectors with respect to the general definition for the polarization vectors in Eq.~(\ref{polarization vectors}) correspond to $\kappa_x=\kappa_y=0$. This implies that the photon field in the $(x,y)$ plane carries no momentum, and thus translational symmetry is preserved in the $(x,y)$ plane, in accordance with the dipole-approximation. Also, it is important to note that the analytic solution for the single-mode case that we will present in what follows, can also be generalized for an arbitrary, finite amount of modes. This is presented in detail in appendix~\ref{Mode-Mode Interactions}.

With these extra assumptions enforced, the Pauli-Fierz Hamiltonian of Eq.~(\ref{Pauli-Fierz}), after expanding the covariant kinetic energy, reads
\begin{eqnarray}\label{single mode Hamiltonian}
\hat{H}=\sum\limits^{N}_{j=1}\left[-\frac{\hbar^2}{2m_{\textrm{e}}}\nabla^2_j +\frac{\textrm{i}e\hbar}{m_{\textrm{e}}} \hat{\mathbf{A}}\cdot\nabla_j\right]+\underbrace{ \frac{Ne^2}{2m_{\textrm{e}}} \hat{\mathbf{A}}^2+\sum\limits^{2}_{\lambda=1}\hbar\omega\left(\hat{a}^{\dagger}_{\lambda}\hat{a}_{\lambda}+\frac{1}{2}\right)}_{\hat{H}_p}.
\end{eqnarray}
and the quantized vector potential takes the simple form
\begin{eqnarray}\label{AinDipole}
\hat{\bi{A}}=\left(\frac{\hbar}{\epsilon_0V}\right)^{\frac{1}{2}}\sum^{2}_{\lambda=1}\frac{\bm{\varepsilon}_{\lambda}}{\sqrt{2\omega}}\left( \hat{a}_{\lambda}+\hat{a}^{\dagger}_{\lambda}\right).
\end{eqnarray}

In Eq.~(\ref{single mode Hamiltonian}) there exists a purely photonic part $\hat{H}_p$ which depends only on the photonic operators $\hat{a}^{\dagger}_{\lambda},\hat{a}_{\lambda}$. Inserting the expression for $\hat{\mathbf{A}}$ given by Eq.~(\ref{AinDipole}) and introducing the diamagnetic shift $\omega_p$
\begin{eqnarray}\label{plasma frequency}
    \omega_p=\sqrt{\frac{e^2 N}{m_{\textrm{e}}\epsilon_0V}}=\sqrt{\frac{e^2n_{\textrm{e}}}{m_{\textrm{e}}\epsilon_0}},
\end{eqnarray}
the purely photonic part $\hat{H}_p$ reads
\begin{eqnarray}\label{photonicpart}
\hat{H}_p=\sum^{2}_{\lambda=1}\left[\hbar\omega\left(\hat{a}^{\dagger}_{\lambda}\hat{a}_{\lambda}+\frac{1}{2}\right)+\frac{\hbar \omega^2_p}{4\omega}\left(\hat{a}_{\lambda}+\hat{a}^{\dagger}_{\lambda}\right)^2\right].
\end{eqnarray}
The diamagnetic shift $\omega_p$ is a consequence of having the $\bi{A}^2$ in the Pauli-Fierz Hamiltonian and shows up due to the collective coupling of the full electron density $n_{\textrm{e}}=N/V$ to the quantized cavity-field~\cite{rokaj2017, rokaj2019, todorov2010, todorov2012, faisal1987}. This implies that $\omega_p=\sqrt{e^2 n_{\textrm{e}}/m_{\textrm{e}}\epsilon_0}$ is the plasma frequency inside the cavity. We would like to emphasize that the 3D electron density $n_{\textrm{e}}=N/V$ is defined via the 2D electron density of the material $n_{2\textrm{D}}=N/S$ and the distance between the cavity mirrors $L_z$ as $n_{\textrm{e}}=n_{2\textrm{D}}/L_z$.

The photonic part $\hat{H}_p$ can be diagonalized by introducing a new set of bosonic annihilation and creation operators $\hat{b}^{\dagger}_{\lambda},\hat{b}_{\lambda}$
\begin{eqnarray}\label{boperators}
\hat{b}_{\lambda}&=&\frac{1}{2\sqrt{\omega\widetilde{\omega}}}\left[\hat{a}_{\lambda}\left(\widetilde{\omega}+\omega\right)+\hat{a}^{\dagger}_{\lambda}\left(\widetilde{\omega}-\omega\right)\right]\\
\hat{b}^{\dagger}_{\lambda}&=&\frac{1}{2\sqrt{\omega\widetilde{\omega}}} \left[\hat{a}_{\lambda}\left(\widetilde{\omega}-\omega\right)+\hat{a}^{\dagger}_{\lambda}\left(\widetilde{\omega}+\omega\right)\right].\nonumber
\end{eqnarray}
where the frequency $\widetilde{\omega}$ is defined as
\begin{eqnarray}\label{plasmon polariton}
    \widetilde{\omega}=\sqrt{\omega^2+\omega^2_p}.
\end{eqnarray}
The above frequency is a dressed frequency which depends on the cavity frequency $\omega$ and the diamagnetic shift (or plasma frequency) $\omega_p$. Thus, $\widetilde{\omega}$ should be interpreted as a plasmon-polariton frequency. As we will see later in section~\ref{Cavity Responses}, the dressed frequency corresponds to a plasmon-polariton excitation of the light-matter hybrid system.  The operators $\hat{b}_{\lambda},\hat{b}^{\dagger}_{\lambda}$ satisfy bosonic commutation relations $[\hat{b}_{\lambda},\hat{b}^{\dagger}_{\lambda^{\prime}}]=\delta_{\lambda,\lambda^{\prime}}$ for $\lambda,\lambda^{\prime}=1,2$. In terms of the operators $\hat{b}_{\lambda},\hat{b}^{\dagger}_{\lambda}$ the photonic part $\hat{H}_p$ is equal to the sum of two non-interacting harmonic oscillators
\begin{equation}\label{Hpinb}
    \hat{H}_{p}=\sum^{2}_{\lambda=1}\hbar\widetilde{\omega}\left(\hat{b}^{\dagger}_{\lambda}\hat{b}_{\lambda}+\frac{1}{2}\right)
\end{equation}
and the vector potential $\hat{\bi{A}}$ is
\begin{eqnarray}\label{Ainb}
\hat{\bi{A}}=\left(\frac{\hbar}{\epsilon_0V}\right)^{\frac{1}{2}}\sum^{2}_{\lambda=1}\frac{\bm{\varepsilon}_{\lambda}}{\sqrt{2\widetilde{\omega}}}\left( \hat{b}_{\lambda}+\hat{b}^{\dagger}_{\lambda}\right).
\end{eqnarray}
From the above expression we see that the quantized vector potential $\hat{\bi{A}}$ got renormalized and now depends on the plasmon-polariton $\widetilde{\omega}$~\cite{rokaj2019}. Substituting back into Eq.~(\ref{single mode Hamiltonian}) the expressions for $\hat{H}_p$ and $\hat{\bi{A}}$ given by Eqs.~(\ref{Hpinb}) and~(\ref{Ainb}) respectively, and introducing the parameter $g_0$
\begin{eqnarray}\label{g coupling}
g_0=\frac{e\hbar}{m_{\textrm{e}}}\left(\frac{\hbar}{\epsilon_0V2\widetilde{\omega}}\right)^{\frac{1}{2}},
\end{eqnarray}
we obtain the following expression for the the Hamiltonian of Eq.~(\ref{single mode Hamiltonian})
\begin{eqnarray}\label{H in bs}
\hat{H}=-\frac{\hbar^2}{2m_{\textrm{e}}}\sum^N_{j=1}\nabla^2_j+\sum^2_{\lambda=1}\hbar\widetilde{\omega}\left(\hat{b}^{\dagger}_{\lambda}\hat{b}_{\lambda}+\frac{1}{2}\right)+\textrm{i}g_0\sum^2_{\lambda=1}\left( \hat{b}_{\lambda}+\hat{b}^{\dagger}_{\lambda}\right)\bm{\varepsilon}_{\lambda}\cdot \sum^N_{j=1}\nabla_j.
\end{eqnarray}
We note that $g_0$ in Eq.~(\ref{g coupling}) can be interpreted as the single-particle light-matter coupling constant. The Hamiltonian is invariant under translations in the electronic space, because it includes only the electronic momentum operators. Thus, $\hat{H}$ commutes with $\nabla$, $[\hat{H},\nabla]$=0, and the two operators share eigenfunctions. For the electronic wavefunctions we employ periodic boundary conditions. Thus, the eigenfunctions of the momentum operator $\nabla$ and the Hamiltonian are plane waves of the form~\cite{Sommerfeld1928, Mermin}
\begin{eqnarray}\label{single electron}
    \phi_{\mathbf{k}_j}(\mathbf{r}_j)=\frac{e^{\textrm{i}\mathbf{k}_j\cdot \mathbf{r}_j} }{\sqrt{S}} \;\; \textrm{with} \;\; 1\leq j \leq N ,
\end{eqnarray}
where $\mathbf{k}_j=2\pi (n^x_{j}/L,n^y_{j}/L)$ are the momenta of the electrons, with $\mathbf{n}_j=\left(n^x_j,n^y_j\right)\in \mathbb{Z}^2$, and $S=L^2$ is the area of the 2D material inside the cavity. The plane waves in Eq.~(\ref{single electron}) are the single-particle eigenfunctions. To construct the many-body eigenfunctions which satisfy the proper fermionic statistics we will use a Slater determinant built out of the single-particle eigenfunctions of Eq.~(\ref{single electron}) 
\begin{eqnarray}\label{Slater determinant}
    \Phi_{\bi{K}}(\bi{r}_1\sigma_1,..,\bi{r}_N\sigma_N)=\frac{1}{\sqrt{N!}}\;\begin{tabular}{|c c c c|}
		$\phi_{\mathbf{k}_1}(\bi{r}_1\sigma_1)$ &$ \phi_{\mathbf{k}_1}(\bi{r}_2\sigma_2)$ & $\cdot\cdot\cdot$ & $\phi_{\bi{k}_{1}} (\bi{r}_N\sigma_N)$\\
		$\phi_{\mathbf{k}_2}(\bi{r}_1\sigma_1)$ & $\phi_{\mathbf{k}_2}(\bi{r}_2\sigma_2)$ & $\cdot\cdot\cdot$ & $\phi_{\bi{k}_{2}} (\bi{r}_N\sigma_N)$\\
		$\cdot$ & $\cdot$ & $\cdot\cdot\cdot$& $\cdot$\\
		$\cdot$ & $\cdot$ & $\cdot\cdot\cdot$& $\cdot$\\
		$\phi_{\mathbf{k}_N}(\bi{r}_1\sigma_1)$ & $\phi_{\mathbf{k}_N}(\bi{r}_2\sigma_2)$ & $\cdot\cdot\cdot$ & $\phi_{\bi{k}_{N}} (\bi{r}_N\sigma_N)$
	\end{tabular} \nonumber\\
\end{eqnarray}
where $\bi{K}=\sum_j\bi{k}_j$ is the collective momentum of the electrons and we introduced it to indicate the fact that the ground state and the excited states of the system depend on the distribution of the electrons in $\bi{k}$-space and particularly on the collective momentum of the electron gas. As we did in the previous chapter we denote the Slater determinant as $\Phi_{\bi{K}}\equiv \Phi_{\bi{K}}(\bi{r}_1\sigma_1,..,\bi{r}_N\sigma_N)$. We apply now the Hamiltonian $\hat{H}$ of Eq.~(\ref{H in bs}) on the eigenfunction $\Phi_{\bi{K}}$ and we have
\begin{eqnarray}\label{H on Psik}
    \hat{H}\Phi_{\bi{K}}=\Bigg [\sum^2_{\lambda=1}\left[\hbar\widetilde{\omega}\left(\hat{b}^{\dagger}_{\lambda}\hat{b}_{\lambda}+\frac{1}{2}\right)-g_0\left( \hat{b}_{\lambda}+\hat{b}^{\dagger}_{\lambda}\right)\bm{\varepsilon}_{\lambda}\cdot\bi{K}\right]+ \sum^N_{j=1}\frac{\hbar^2\bi{k}^2_j}{2m_{\textrm{e}}}\Bigg ] \Phi_{\bi{K}} \;\; \textrm{with}\;\; \bi{K}=\sum^N_{j=1}\bi{k}_j.\nonumber\\
\end{eqnarray}
As a next step we define another set of bosonic operators $\hat{c}^{\dagger}_{\lambda},\hat{c}_{\lambda}$
\begin{eqnarray}\label{c operators}
    \hat{c}_{\lambda}=\hat{b}_{\lambda}-\frac{g_0\bm{\varepsilon}_{\lambda}\cdot \bi{K}}{\hbar\widetilde{\omega}} \;\;\; \textrm{and}\;\;\; \hat{c}^{\dagger}_{\lambda}=\hat{b}^{\dagger}_{\lambda}-\frac{g_0\bm{\varepsilon}_{\lambda}\cdot \bi{K}}{\hbar\widetilde{\omega}},
\end{eqnarray}
and the operator $\hat{H}\Phi_{\bi{K}}$ of Eq.~(\ref{H on Psik}) takes the form 
\begin{eqnarray}\label{Projection H}
    \hat{H}\Phi_{\bi{K}}=\Bigg[ \sum^2_{\lambda=1}\left[\hbar\widetilde{\omega}\left(\hat{c}^{\dagger}_{\lambda}\hat{c}_{\lambda}+\frac{1}{2}\right)-\frac{g^2_0}{\hbar\widetilde{\omega}}\left(\bm{\varepsilon}_{\lambda}\cdot \mathbf{K}\right)^2\right]+\frac{\hbar^2}{2m_{\textrm{e}}}\sum\limits^{N}_{j=1}\mathbf{k}^2_j\Bigg] \Phi_{\bi{K}}.
\end{eqnarray}
 We note that also the operators $\hat{c}^{\dagger}_{\lambda},\hat{c}_{\lambda}$ satisfy bosonic cummutation relations $[\hat{c},\hat{c}^{\dagger}_{\lambda^{\prime}}]=\delta_{\lambda\lambda^{\prime}}$ for $\lambda, \lambda^{\prime}=1,2$. For the operator $\hat{H}_{\lambda}=\hbar\widetilde{\omega}\left(\hat{c}^{\dagger}_{\lambda}\hat{c}_{\lambda}+1/2\right)$ which has the form of a harmonic oscillator it is well-known that the complete set of eigenstates is~\cite{GriffithsQM}
\begin{eqnarray}\label{c eigenstates}
    |n_{\lambda},\bm{\varepsilon}_{\lambda}\cdot\bi{K}\rangle_{\lambda}=\frac{(\hat{c}^{\dagger}_{\lambda})^{n_{\lambda}}}{\sqrt{n_{\lambda}!}}|0,\bm{\varepsilon}_{\lambda}\cdot\mathbf{K}\rangle_{\lambda} \;\; \textrm{with}\;\; n_{\lambda}\in\mathbb{Z}, \lambda=1,2\nonumber\\
\end{eqnarray}
where $|0,\bm{\varepsilon}_{\lambda}\cdot\mathbf{K}\rangle_{\lambda}$ is the ground-state of $\hat{H}_{\lambda}$, which gets annihilated by $\hat{c}_{\lambda}$~\cite{GriffithsQM}. Further, the eigenvalues of $\hat{H}_{\lambda}$ are $\hbar\widetilde{\omega}(n_{\lambda}+1/2)$. The operator $\hat{H}\Phi_{\bi{K}}$ in Eq.~(\ref{Projection H}) in terms of the bosonic operators $\hat{c}^{\dagger}_{\lambda},\hat{c}_{\lambda}$ contains only the sum over $\hat{H}_{\lambda}$, as a consequence by applying $\hat{H}\Phi_{\bi{K}}$ on the states $\prod_{\lambda}|n_{\lambda},\bm{\varepsilon}_{\lambda}\cdot\bi{K}\rangle_{\lambda}$ we find
\begin{eqnarray}\label{eigenvalue eigenstate Equation}
    \hat{H}\Phi_{\bi{K}} \prod^2_{\lambda=1}|n_{\lambda},\bm{\varepsilon}_{\lambda}\cdot\bi{K}\rangle_{\lambda}=\left[ \sum^2_{\lambda=1}\left[\hbar\widetilde{\omega}\left(n_{\lambda}+\frac{1}{2}\right)-\frac{g^2_0\left(\bm{\varepsilon}_{\lambda}\cdot \mathbf{K}\right)^2}{\hbar\widetilde{\omega}}\right]+\sum\limits^{N}_{j=1}\frac{\hbar^2\mathbf{k}^2_j}{2m_{\textrm{e}}}\right]\Phi_{\bi{K}}  \prod^2_{\lambda=1}|n_{\lambda},\bm{\varepsilon}_{\lambda}\cdot\bi{K}\rangle_{\lambda}.\nonumber\\
    \end{eqnarray}
From the equation above we conclude that the complete set of eigenstates of the Hamiltonian in Eq.~(\ref{single mode Hamiltonian}) describing the 2DEG coupled to the cavity field is
\begin{eqnarray}\label{eigenstates}
   \Phi_{\bi{K}} \prod^2_{\lambda=1}|n_{\lambda},\bm{\varepsilon}_{\lambda}\cdot\bi{K}\rangle_{\lambda} \;\; \textrm{with}\;\; \lambda=1,2\;\; \textrm{and}\;\;  \bi{K}=\sum^N_{j=1}\bi{k}_j
\end{eqnarray}
and the full eigenspectrum reads as
\begin{eqnarray}\label{eigenspectrum}
   E_{n_{\lambda},\bi{k}}=\sum^2_{\lambda=1}\left[\hbar\widetilde{\omega}\left(n_{\lambda}+\frac{1}{2}\right)-\frac{\gamma}{N}\frac{\left(\bm{\varepsilon}_{\lambda}\cdot \hbar\bi{K}\right)^2}{2m_{\textrm{e}}}\right]+\sum\limits^{N}_{j=1}\frac{\hbar^2\mathbf{k}^2_j}{2m_{\textrm{e}}}.
\end{eqnarray}
In the above expression for the energy spectrum there exists also a negative term proportional to the square of collective momentum of all the electrons in the 2DEG, $\sim \left(\bm{\varepsilon}_{\lambda}\cdot \hbar\bi{K}\right)^2$. This particular term is an all-to-all interaction between the electrons, mediated by the photon field. We call it all-to-all because the momentum of each individual electron couples to the momenta of all the other electrons. 

Moreover, we would like to mention that to obtain the expression of Eq.~(\ref{eigenspectrum}) we substituted the single-particle coupling constant $g_0$ given by Eq.~(\ref{g coupling}) in Eq.~(\ref{eigenvalue eigenstate Equation}) and introduced the parameter $\gamma$
\begin{eqnarray}\label{collective coupling}
   \gamma=\frac{2m_{\textrm{e}}N}{\hbar^2}\frac{g^2_0}{\hbar\widetilde{\omega}}=\frac{\omega^2_p}{\widetilde{\omega}^2}=\frac{\omega^2_p}{\omega^2+\omega^2_p}\leq 1.
\end{eqnarray}
The parameter $\gamma$ must be interpreted as the collective coupling constant of the 2DEG to the cavity field. The collective coupling $\gamma$ depends on the cavity frequency and the electron density in the cavity $n_{\textit{e}}$ via the dressed frequency $\omega_p$ defined in Eq.~(\ref{plasma frequency}). This means that the more charges in the system, the stronger the light-matter coupling. Further, we emphasize that the collective coupling $\gamma$ is dimensionless and most importantly $\gamma$ has an upper bound and cannot exceed one. This upper bound, as we will later see in section~\ref{Instability and A2}, guarantees the stability of the electron-photon system.

\section{Ground State in the Thermodynamic Limit}\label{Ground State}

\begin{displayquote}
\footnotesize{In this limit, a large collection $N \rightarrow \infty$ of objects can behave completely differently---have different symmetry from anything that the separate objects can themselves exhibit.}
\end{displayquote}
\begin{flushright}
  \footnotesize{Philip Anderson about the thermodynamic limit\\
More and Different~\cite{MoreandDifferent}}
\end{flushright}

Having diagonalized the Hamiltonian $\hat{H}$ of Eq.~(\ref{single mode Hamiltonian}) for $N$ free electrons coupled to a single mode, we want now to find the ground state of this many-body system in the thermodynamic (or large $N$) limit. To do so, we need to minimize the energy of the many-body spectrum given by Eq.~(\ref{eigenspectrum}) in the limit where the number of electrons $N$ and the area of the materials $S=L^2$ become arbitrarily large and approach the thermodynamic limit. This procedure needs to be performed in such a way that the 2D electron density $n_{2\textrm{D}}=N/S$ stays fixed. 

\begin{figure}[h]
\begin{center}
\includegraphics[height=7cm,width=9cm]{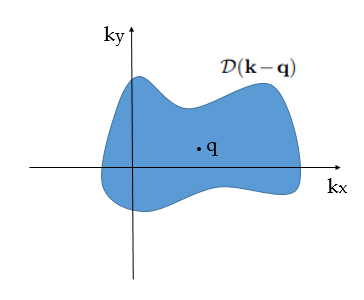}
\caption{\label{Distribution Kspace}Illustration of a generic distribution $\mathcal{D}(\bi{k}-\bi{q})$ in $\bi{k}$-space. The shape $\mathcal{D}$ as well as the origin $\bi{q}$ of the distribution are arbitrary. To find the ground-state distribution in $\bi{k}$-space we have to minimize the energy density of the system with respect to both the shape $\mathcal{D}$ and the origin $\bi{q}$. }    
\end{center}
\end{figure}

To define properly the electron density we need to compute the number of allowed states in a region of $\bi{k}$-space of volume $\Omega_{\mathcal{D}}$. The volume $\Omega_{\mathcal{D}}$ is defined with respect to a distribution $\mathcal{D}$ which has particular shape in $\bi{k}$-space. The number of allowed states in the volume $\Omega_{\mathcal{D}}$ is 
\begin{eqnarray}
\#\textrm{states}=\frac{\Omega_{\mathcal{D}}}{(2\pi)^2/L^2}=\frac{\Omega_{\mathcal{D}}S}{(2\pi)^2}.
\end{eqnarray}
 The volume $\Omega_{\mathcal{D}}$ with respect to the generic distribution $\mathcal{D}(\mathbf{k}-\mathbf{q})$  whose origin $\mathbf{q}$ is an arbitrary point in $\bi{k}$-space (see Fig.~\ref{Distribution Kspace}) is defined as
\begin{eqnarray}
\Omega_{\mathcal{D}}=\iint \limits_{-\infty}^{+\infty} \mathcal{D}(\mathbf{k}-\mathbf{q}) d^2k=\iint \limits_{-\infty}^{+\infty} \mathcal{D}(\mathbf{u}) d^2u,
\end{eqnarray}
where we performed the shift $\bi{u}=\bi{k}-\bi{q}$. With the use of $\Omega_{\mathcal{D}}$ we can now define the 2D electron density. The number of electrons $N$ we can accommodate within the volume $\Omega_{\mathcal{D}}$ is 2 times (due to the spin degeneracy) the number of allowed states $N=2\#\textrm{states}=2\Omega_{\mathcal{D}}S/(2\pi)^2.$ Thus, with find that the 2D electron density is
\begin{eqnarray}\label{density}
n_{2\textrm{D}}=\frac{N}{S}=\frac{2\Omega_{\mathcal{D}}}{(2\pi)^2}.
\end{eqnarray}
As we already stated, our aim is to find the ground state of the system. From Eq.~(\ref{eigenspectrum}) it is clear that the energy minimizes for $n_{\lambda}=0$ for both $\lambda=1,2$. Thus, the photonic contribution to the ground state energy is $E_{\textrm{p}}=\hbar\widetilde{\omega}$, which is constant and independent of the momenta of the electrons. As a consequence, the photonic contribution can be neglected for finding the ground-state distribution of the electrons in $\bi{k}$-space. Thus, the ground-state energy of the system as a function of the momenta $\bi{k}_j$ is
\begin{eqnarray}\label{E of k}
   E_{\bi{k}}=\frac{\hbar^2}{2m_{\textrm{e}}}\left[\sum\limits^{N}_{j=1}\bi{k}^2_j-\frac{\gamma}{N}\sum^2_{\lambda=1}\left(\bm{\varepsilon}_{\lambda}\cdot \bi{K}\right)^2\right],
\end{eqnarray}
In the expression above we have two contributions: (i) a positive one, which is the sum over the individual kinetic energies of all the electrons and we denote by $T$, and (ii) a negative one which is minus the square of the collective momentum $\bi{K}=\sum_j\bi{k}_j$ of the electrons. 

To find the ground-state distribution in $\bi{k}$-space we need to minimize the energy density $E_{\bi{k}}/S$ with respect to the distribution $\mathcal{D}(\bi{k}-\bi{q})$. In the thermodynamic limit, the number of particles $N$ and the area $S$ of the 2D material tend to infinity and the sums in the expression for the energy density $E_{\bi{k}}/S$ turn into integrals~\cite{Mermin}. Thus, for the kinetic energy per area $T/S$ in the large $N,S$ limit with doubly occupied momenta we have~\cite{Mermin}.
\begin{eqnarray}\label{Kinetic energy}
\frac{T}{S}=\frac{\hbar^2}{2m_{\textrm{e}}}\lim_{S\to\infty}\frac{2}{S}\sum_{\mathbf{k}}\bi{k}^2=\frac{\hbar^2}{2m_{\textrm{e}}}\frac{2}{(2\pi)^2}\iint \limits^{+\infty}_{-\infty} d^2k \mathcal{D}(\mathbf{k}-\mathbf{q})\mathbf{k}^2\nonumber\\
\end{eqnarray}
We perform the transformation $\bi{u}=\bi{k}-\bi{q}$ and we obtain 
\begin{eqnarray}
\frac{T}{S}=\frac{\hbar^2}{2m_{\textrm{e}}}\left[\underbrace{\frac{2}{(2\pi)^2}\iint \limits^{+\infty}_{-\infty} d^2u \mathcal{D}(\mathbf{u})\mathbf{u}^2}_{t_{\mathcal{D}}}+2\mathbf{q}\underbrace{\frac{2}{(2\pi)^2}\iint \limits^{+\infty}_{-\infty} d^2u \mathcal{D}(\mathbf{u})\mathbf{u}}_{\mathbf{K}_{\mathcal{D}}}+\mathbf{q}^2\underbrace{\frac{2}{(2\pi)^2}\iint \limits^{+\infty}_{-\infty} d^2u \mathcal{D}(\mathbf{u})}_{n_{2\textrm{D}}}\right]\nonumber\\
\end{eqnarray}
Thus, the kinetic energy per area $\tau[\mathcal{D}]=T/S$ is
\begin{eqnarray}\label{TperV}
\tau[\mathcal{D}]=\frac{\hbar^2}{2m_{\textrm{e}}}\left(t_{\mathcal{D}}+2\mathbf{q}\cdot\mathbf{K}_{\mathcal{D}}+\mathbf{q}^2n_{2\textrm{D}}\right).
\end{eqnarray}
Before we continue, let us comment and give an interpretation of each term appearing above. The term $t_{\mathcal{D}}$ is the standard kinetic energy of the electron gas with respect to the distribution centered at zero $\mathcal{D}(\mathbf{u})$~\cite{Mermin}. The term $\mathbf{K}_{\mathcal{D}}$ is the collective momentum of the electrons with respect to $\mathcal{D}(\mathbf{u})$, and $\mathbf{q}^2 n_{2\textrm{D}}$ is the kinetic energy due to the arbitrary origin of the distribution, $\mathcal{D}(\mathbf{k}-\mathbf{q})$ (see Fig.~\ref{Distribution Kspace}). This last term depends on the 2D density $n_{2\textrm{D}}$ and the origin of the distribution $\bi{q}$, but it is independent of the shape of the distribution $\mathcal{D}$. 

Let us continue by computing the second term appearing in Eq.~(\ref{E of k}), $\bm{\varepsilon}_{\lambda}\cdot \bi{K}$. The collective momentum per area $\bi{K}/S$, with doubly occupied momenta, in the thermodynamic limit is
\begin{eqnarray}
\frac{\bi{K}}{S}&=&\frac{2}{(2\pi)^2}\iint \limits^{+\infty}_{-\infty} d^2k \mathcal{D}(\mathbf{k}-\mathbf{q})\mathbf{k}.
\end{eqnarray}
Performing again the shift $\bi{u}=\bi{k}-\bi{q}$ we find
\begin{eqnarray}
\frac{\bi{K}}{S}=\frac{2}{(2\pi)^2}\iint \limits^{+\infty}_{-\infty} d^2u \mathcal{D}(\mathbf{u})\mathbf{u}+\mathbf{q}\frac{2}{(2\pi)^2}\iint \limits^{+\infty}_{-\infty} d^2u \mathcal{D}(\mathbf{u})=\mathbf{K}_{\mathcal{D}}+\mathbf{q}n_{2\textrm{D}}.
\end{eqnarray}
Taking now the inner product of the above quantity with the polarization vectors $\bm{\varepsilon}_{\lambda}\cdot\bi{K}/S$, squaring it and then multiplying by $S$, we find the following result for the second term in Eq.~(\ref{E of k}) 
\begin{eqnarray}\label{negative term}
 \frac{\left(\bm{\varepsilon}_{\lambda}\cdot\mathbf{K}\right)^2}{S}=S\left(\bm{\varepsilon}_{\lambda}\cdot\mathbf{K}_{\mathcal{D}}+\bm{\varepsilon}_{\lambda}\cdot\mathbf{q}n_{2\textrm{D}}\right)^2.
  \end{eqnarray}
 Adding the two contributions which we found in Eqs.~(\ref{TperV}) and~(\ref{negative term}) we obtain the expression for the energy density as a function of the shape of the distribution $\mathcal{D}$ and the origin $\bi{q}$
 \begin{eqnarray}\label{energy density}
  \mathcal{E}[\mathcal{D}]\equiv\frac{E_{\bi{k}}}{S}=\frac{\hbar^2}{2m_{\textrm{e}}}\left[t_{\mathcal{D}}+2\mathbf{q}\cdot\mathbf{K}_{\mathcal{D}}+\mathbf{q}^2n_{2\textrm{D}}-\frac{\gamma}{n_{2\textrm{D}}}\sum^2_{\lambda=1}\left(\bm{\varepsilon}_{\lambda}\cdot\mathbf{K}_{\mathcal{D}}+\bm{\varepsilon}_{\lambda}\cdot\mathbf{q}n_{2\textrm{D}}\right)^2\right].
 \end{eqnarray}
 The energy density first has to be minimized with respect to the origin of the distribution $\bi{q}=(q_x,q_y)$. For that we compute the derivative of the energy density $ \mathcal{E}[\mathcal{D}]$ with respect to both components of the vector $\bi{q}=(q_x,q_y)$, 
 \begin{eqnarray}\label{minimization over q}
  &&\frac{\partial \mathcal{E}[\mathcal{D}]}{\partial q_x}=\frac{\hbar^2}{2m_{\textrm{e}}}2(1-\gamma)\left(K^x_{\mathcal{D}}+q_xn_{2\textrm{D}}\right)=0, \nonumber\\
  &&\frac{\partial \mathcal{E}[\mathcal{D}]}{\partial q_y}=\frac{\hbar^2}{2m_{\textrm{e}}}2(1-\gamma)\left(K^y_{\mathcal{D}}+q_yn_{2\textrm{D}}\right)=0.
 \end{eqnarray}
 From the above set of equations we find the optimal vector $\bi{q}_0$ that minimizes the energy density
 \begin{eqnarray}\label{optimal q}
  \bi{q}_0=-\frac{\bi{K_{\mathcal{D}}}}{n_{2\textrm{D}}},
 \end{eqnarray}
 where we see surprisingly that the optimal vector $\bi{q}_0$ is independent of the coupling constant $\gamma$. Substituting the optimal vector $\bi{q}_0$ into the Eq.~(\ref{energy density}) we find for the energy density
 \begin{eqnarray}\label{optimal energy density}
  \mathcal{E}[\mathcal{D}]|_{\bi{q_0}}=\frac{\hbar^2}{2m_{\textrm{e}}}\left[t_{\mathcal{D}}-\frac{\bi{K}^2_{\mathcal{D}}}{n_{2\textrm{D}}}\right].
 \end{eqnarray}
 Having optimized the energy density with respect to the origin $\bi{q}$ of the distribution $\mathcal{D}(\bi{k}-\bi{q})$ the remaining task is to optimize the energy with respect to the shape $\mathcal{D}$ of the  distribution in momentum space. In general to perform such a minimization of the energy functional $\mathcal{E}[\mathcal{D}]$ it is a cumbersome task. Thus, in order to find the optimal shape of the distribution we will use some physical intuition.
 
The energy density $\mathcal{E}[\mathcal{D}]$ and the optimal origin $\bi{q}_0$, given by Eqs.~(\ref{optimal energy density}) and (\ref{optimal q}) respectively, are both independent of the coupling constant $\gamma$. This indicates that the ground-state and the ground-state energy of the electrons in the thermodynamic limit is independent of the coupling to the cavity mode. Guided by this observation let us compare the energy density in Eq.~(\ref{optimal energy density}) with the energy density of the original Sommerfeld model~\cite{Mermin} not coupled to a cavity mode. 
 
 \subsection{Comparison to Free Electron Gas Not Coupled to a Cavity}
\begin{figure}[h]
 \begin{center}
    \includegraphics[height=7cm,width=9cm]{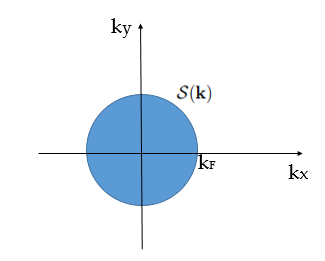}
\caption{\label{Fermi Sphere}Schematic representation of the ground-state distribution of the 2D free electron model not coupled to a cavity. The ground state distribution is the 2D Fermi sphere $\mathcal{S}(\bi{k})$ with radius $|\bi{k}_{\textrm{F}}|$ (Fermi momentum). For the 2DEG coupled to the cavity, the ground-state distribution in $\bi{k}$-space is also the 2D Fermi sphere $\mathcal{S}(\bi{k})$ with radius $|\bi{k}_{\textrm{F}}|$.} 
 \end{center}
\end{figure}

 In the original free electron model introduced by Sommerfeld, the energy of the system is the sum over the kinetic energies of all the electrons~\cite{Sommerfeld1928,Mermin}
 \begin{eqnarray}\label{eg no coupling}
  E^{nc}_{\bi{k}}=\frac{\hbar^2}{2m_{\textrm{e}}}\sum^N_{j=1}\bi{k}_j.
 \end{eqnarray}
 Due to rotational symmetry, the ground-state $\bi{k}$-space distribution is the standard 2D Fermi sphere $\mathcal{S}(\bi{k})$~\cite{Mermin} as shown in Fig.~\ref{Fermi Sphere}. 
 
But let us forget for a moment the fact that we know that the ground state distribution in $\bi{k}$-space is the 2D Fermi sphere, and consider again a generic distribution in $\bi{k}$-space $\mathcal{D}(\bi{k}-\bi{q})$ with an arbitrary origin $\bi{q}$ as the one in Fig.~\ref{Distribution Kspace}. For such a generic distribution the ground state energy density, which we already computed it in Eq.~(\ref{TperV}), is
\begin{eqnarray}\label{energy density no coupling}
 \mathcal{E}^{nc}[\mathcal{D}]=\frac{\hbar^2}{2m_{\textrm{e}}}\frac{2}{(2\pi)^2}\iint \limits^{+\infty}_{-\infty} d^2k \mathcal{D}(\mathbf{k}-\mathbf{q})\mathbf{k}^2=\frac{\hbar^2}{2m_{\textrm{e}}}\left(t_{\mathcal{D}}+2\mathbf{q}\cdot\mathbf{K}_{\mathcal{D}}+\mathbf{q}^2n_{2\textrm{D}}\right).
\end{eqnarray}
Minimizing the above energy density with respect to the origin $\bi{q}$ of the distribution we find the optimal origin $\bi{q}_0$
\begin{eqnarray}
 \frac{\partial \mathcal{E}^{nc}[\mathcal{D}]}{\partial \bi{q}}=\frac{\hbar^2}{2m_{\textrm{e}}}2\left(\bi{K}_{\mathcal{D}}+\bi{q}n_{2\textrm{D}}\right)=0 \Longrightarrow \bi{q}_0=-\frac{\bi{K}_{\mathcal{D}}}{n_{2\textrm{D}}}
\end{eqnarray}
 which is the same with the one we found in Eq.~(\ref{optimal q}) for the electron gas coupled to the cavity mode. Substituting $\bi{q}_0$ into the expression for the energy density $\mathcal{E}^{nc}[\mathcal{D}]$ of the uncoupled electron gas in Eq.~(\ref{energy density no coupling}) we obtain
\begin{eqnarray}\label{optimal energy no coupling}
 \mathcal{E}^{nc}[\mathcal{D}]|_{\bi{q_0}}=\frac{\hbar^2}{2m_{\textrm{e}}}\left[t_{\mathcal{D}}-\frac{\bi{K}^2_{\mathcal{D}}}{n_{2\textrm{D}}}\right].
\end{eqnarray}
Comparing the energy density above, for the electron gas not coupled to the cavity $\mathcal{E}^{nc}[\mathcal{D}]|_{\bi{q_0}}$, to the energy density $\mathcal{E}[\mathcal{D}]|_{\bi{q_0}}$ in Eq.~(\ref{optimal energy density}) of the electron gas coupled to the cavity mode, we see that they are exactly the same 
\begin{eqnarray}
 \mathcal{E}[\mathcal{D}]|_{\bi{q_0}}=\mathcal{E}^{nc}[\mathcal{D}]|_{\bi{q_0}}.
\end{eqnarray}
This means that both energy functionals, the coupled and the uncoupled, get minimized exactly by the same distribution in $\bi{k}$-space. As we already mentioned for the uncoupled free electrons gas, the shape of the ground-state distribution in $\bi{k}$-space is the 2D Fermi sphere $\mathcal{S}$. For a sphere the collective momentum is zero, $\bi{K}_{\mathcal{S}}=0$, due to the parity symmetry of the Fermi sphere, $\bi{k}\rightarrow -\bi{k}$. As consequence the optimal origin of the sphere is also zero $\bi{q}_0=0$. Thus, we find that for the coupled system the ground-state momentum distribution is also the 2D the Fermi sphere $\mathcal{S}(\bi{k})$ centered at zero, as depicted in Fig.~\ref{Fermi Sphere}.   

Since the collective momentum over the Fermi sphere is zero, $\bi{K}=\sum_j\bi{k}_j=0$, the ground-state of the electron gas coupled to the cavity is 
\begin{eqnarray}\label{Thermodynamic GS}
  |\Psi_{gs}\rangle=|\Phi_{0}\rangle \otimes  \prod^2_{\lambda=1}|0,0\rangle_{\lambda},
\end{eqnarray}
where $|\Phi_0\rangle$ is the Slater determinant given by Eq.~(\ref{Slater determinant}) with zero collective momentum $\bi{K}=0$. The fact that the ground state distribution of the electrons in $\bi{k}$-space is the Fermi sphere implies that the 2DEG coupled to the cavity field is a Fermi liquid~\cite{Baym, LandauFermiLiquid}.

Moreover, for the 2D Fermi sphere, the ground state energy density is equal to the kinetic energy density $t_{\mathcal{D}}$, whose expression as a function of the Fermi momentum $\bi{k}_{\textrm{F}}$ is
\begin{eqnarray}
 \mathcal{E}[\mathcal{S}]=\frac{\hbar^2}{2m_{\textrm{e}}}\frac{2}{(2\pi)^2}\iint \limits^{+\infty}_{-\infty} d^2k \mathcal{S}(\mathbf{k})\mathbf{k}^2=\frac{\hbar^2k^4_{\textrm{F}}}{16\pi m_{\textrm{e}}}.
\end{eqnarray}
We note that a discrepancy shows up between the energy density of the electrons and of the energy density of the photon field. The contribution of the photon field to the ground-state energy density as we can deduce from Eq.~(\ref{eigenspectrum}) is $E_p/S=\hbar\widetilde{\omega}/S$, and in the thermodynamic limit, where the area of the material $S$ goes macroscopic, $E_p/S=\hbar\widetilde{\omega}/S$ is miniscule and strictly speaking goes to zero. On the other hand the energy density, of the electronic sector is finite. This implies that only the electron gas contributes to the ground state energy density of the interacting electron-photon system in the cavity. This hints towards the fact that for the photons and electrons to contribute equally on the ground state a continuum of modes for the photon field need to be taken into account.

Lastly, from the fact that the electronic ground-state is the standard Fermi sphere and that the energy density of the cavity field in the thermodynamic limit is zero, one might conclude that the ground-state of the hybrid electron-photon system is trivial and there are no quantum fluctuation effects. However, this is not true. In order to classify completely the electron-photon ground-state we need to look also at the photon occupation.

\subsection{Ground State Photon Occupation}

The photon number operator is defined as
\begin{eqnarray}
\hat{N}_{\textrm{ph}}=\sum^2_{\lambda=1}\hat{a}^{\dagger}_{\lambda}\hat{a}_{\lambda}.    
\end{eqnarray}
To calculate the ground-state photon occupation we need to express the photon number operator in terms of the bosonic operators $\hat{c}^{\dagger}_{\lambda}, \hat{c}_{\lambda}$ defined in Eq.~(\ref{c operators}). Using Eqs.~(\ref{boperators}) and~(\ref{c operators}) we find for the photon number operator 
\begin{eqnarray}
   \hat{N}_{\textrm{ph}}&=&\sum^2_{\lambda=1} \frac{1}{4\omega\widetilde{\omega}}\Bigg [ \left(\omega^2-\widetilde{\omega}^2\right)\left(\hat{c}_{\lambda}+\frac{g\bm{\varepsilon}_{\lambda}\cdot \bi{K}}{\hbar\widetilde{\omega}}\right)^2 +\left(\omega-\widetilde{\omega}\right)^2\left(\hat{c}_{\lambda}+\frac{g\bm{\varepsilon}_{\lambda}\cdot \bi{K}}{\hbar\widetilde{\omega}}\right)\left(\hat{c}^{\dagger}_{\lambda}+\frac{g\bm{\varepsilon}_{\lambda}\cdot \bi{K}}{\hbar\widetilde{\omega}}\right)\nonumber\\
   &+&\left(\omega+\widetilde{\omega}\right)^2\left(\hat{c}_{\lambda}+\frac{g\bm{\varepsilon}_{\lambda}\cdot \bi{K}}{\hbar\widetilde{\omega}}\right)\left(\hat{c}^{\dagger}_{\lambda}+\frac{g\bm{\varepsilon}_{\lambda}\cdot \bi{K}}{\hbar\widetilde{\omega}}\right)+\left(\omega^2-\widetilde{\omega}^2\right)\left(\hat{c}^{\dagger}_{\lambda}+\frac{g\bm{\varepsilon}_{\lambda}\cdot \bi{K}}{\hbar\widetilde{\omega}}\right)^2\Bigg].
\end{eqnarray}
In the ground-state the collective momentum is zero, $\bi{K}=0$. Moreover, from all the terms appearing above only the term that first creates and then destroys a bosonic excitation $\hat{c}_{\lambda}\hat{c}^{\dagger}_{\lambda}$ gives a non-zero contribution. As a consequence we find that the ground-state photon occupation is
\begin{eqnarray}
    \langle \hat{N}_{\textrm{ph}} \rangle_{gs}\equiv \langle \Psi_{gs}|\hat{N}_{\textrm{ph}}|\Psi_{gs}\rangle=\frac{\left(\widetilde{\omega}-\omega\right)^2}{2\omega\widetilde{\omega}}.
\end{eqnarray}
The result above shows that the photon occupation is non-zero. This implies that there are virtual photons in the ground-state of the hybrid electron-photon system. This phenomenon has also been reported for dissipative systems~\cite{DeLiberato2017}. From the fact that the ground-state of the 2DEG in the cavity contains photons we conclude that there are quantum fluctuations of the photon field in the ground-state due to the electron-photon coupling. Thus, our system is not a trivial Fermi liquid, but rather it is a Fermi liquid dressed with photons.

Moreover, the ground-state photon occupation exhibits an interesting behavior with respect to the electron density. For electron densities small enough for the diamagnetic shift $\omega_p=\sqrt{e^2 n_{\textrm{e}}/m_{\textrm{e}}\epsilon_0}$ to be much smaller than the cavity frequency, $\omega_p\ll\omega$, the dressed frequency $\widetilde{\omega}=\sqrt{\omega^2_p+\omega^2}$ is approximately equal to the cavity frequency, $\widetilde{\omega} \approx \omega$. In this case the ground-state photon occupation is zero, $ \langle \hat{N}_{\textrm{ph}} \rangle_{gs}=0$. This recovers nicely the correct decoupling limit.

However, for large electronic densities such that $\omega_p \gg \omega$, the dressed frequency is $\widetilde{\omega} \approx \omega_p$ and the numerator in the expression for $\langle \hat{N}_{\textrm{ph}} \rangle_{gs}$ is approximately $\omega^2_p$. In this case, we find that the ground-state photon occupation has a square root dependence on the electronic density
\begin{eqnarray}
\langle \hat{N}_{\textrm{ph}} \rangle_{gs} \sim \sqrt{n_{\textrm{e}}}.
\end{eqnarray}
This shows that the number of photons in the ground-state increases by adding more electrons to the system. This behavior of the photon occupation might be related to the superradiant phase transition~\cite{Lieb} and could potentially provide some insights on how to achieve this correlated phase between light and matter, which still remains elusive.

\section{Critical Coupling, Instability \& the Diamagnetic $\bi{A}^2$ term}\label{Instability and A2}

Up to here we examined rigorously and in full generality the behavior of the free electron gas coupled to the cavity, in the regime where the cavity mode $\omega$ is non-zero and the collective coupling constant $\gamma$, defined in Eq.~(\ref{collective coupling}), is smaller than one. But naturally, the following question arises: what happens in the limit where the cavity frequency approaches zero, $\omega\rightarrow 0$, and the collective coupling reaches its maximum value $\gamma\rightarrow 1$? 

\begin{figure}[H]
 \begin{center}
\includegraphics[height=7cm,width=9cm]{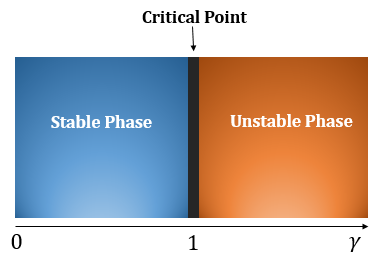}
\caption{\label{Phase Diagram} Phase diagram for the 2DEG coupled to the cavity. The system is stable and has a ground state for $\gamma<1$. At the critical coupling $\gamma_c=1$ the ground state is infinitely degenerate. Beyond the critical coupling $\gamma > 1$ the system is unstable and has no ground-state.}     
 \end{center}
\end{figure}

We will refer to the maximum value of the coupling constant $\gamma$ as critical coupling, $\gamma_c=1$. As we will see, an interesting transition happens for the system at the critical coupling, from a stable phase to an unstable phase, as it is summarized by the phase diagram in Fig.~\ref{Phase Diagram}.  

\subsection{Critical Coupling and Infinite Degeneracy}

At the critical coupling $\gamma_c=1$ the energy density $\mathcal{E}[\mathcal{D}]$ given by Eq.~(\ref{energy density}) becomes independent of the origin $\bi{q}$
\begin{eqnarray}\label{critical energy density}
  \mathcal{E}[\mathcal{D}]|_{\gamma_c}=\frac{\hbar^2}{2m_{\textrm{e}}}\left[t_{\mathcal{D}}-\frac{\bi{K}^2_{\mathcal{D}}}{n_{2\textrm{D}}}\right].
\end{eqnarray}
This means that the ground state of the system is not unique. Moreover, Eq.~(\ref{optimal q}) from which we determined the optimal value for the vector $\bi{q}$ is trivially satisfied and all $\bi{q}$ are possible.

\begin{figure}[H]
 \begin{center}
\includegraphics[height=6cm,width=7.5cm]{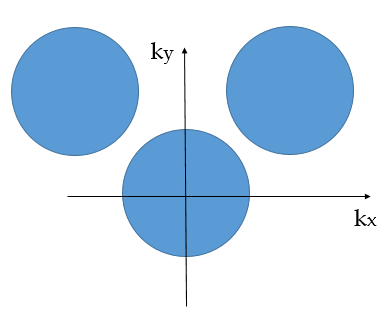}
\caption{\label{Degeneracy}Schematic depiction of the infinite degeneracy for $\gamma_c=1$. All two-dimensional Fermi spheres of arbitrary center have exactly the same ground-state energy. } \end{center}
\end{figure}

In the previous section we showed that the energy density of Eq.~(\ref{critical energy density}) minimizes for the 2D sphere $\mathcal{S}(\bi{k}-\bi{q})$. However, the energy density $\mathcal{E}[\mathcal{D}]|_{\gamma_c}$ at the critical coupling is independent of the origin $\bi{q}$ and the optimal $\bi{q}$ cannot be determined from Eq.~(\ref{optimal q}). This means that all spheres of the form $\mathcal{S}(\bi{k}-\bi{q})$ are energetically degenerate and have the same ground-state energy density. This is also depicted in Fig.~\ref{Degeneracy}. As a consequence, the ground-state $\bi{k}$-space distribution it is not unique but rather at the critical coupling $\gamma_c=1$ it is infinitely degenerate with respect to origin of the $\bi{k}$-space distribution of the electrons. 

We note that such an infinite degeneracy appears also for a 2D electron gas in the presence of perpendicular, homogeneous magnetic field. For such systems we have the Landau levels exhibiting exactly this behavior~\cite{Landau}. This infinite degeneracy is directly connected to the quantum Hall effect and the quantization of the macroscopic Hall conductance~\cite{Klitzing}. The link between QED and the quantum Hall effect has also been investigated recently in the framework of quantum electrodynamical Bloch theory~\cite{rokaj2019}.

We would also like to mention that the fact that all spheres $\mathcal{S}(\bi{k}-\bi{q})$ of arbitrary origin $\bi{q}$ are energetically degenerate means that the ground-state of our system is invariant under translations in $\bi{k}$-space. This implies that it is invariant under Galilean boosts.

\subsection{No Ground State Beyond the Critical Coupling}\label{No Ground State}

To complete our investigation, we will consider now also the case where the collective coupling constant becomes larger than the critical coupling $\gamma_c=1$. From its definition in Eq.~(\ref{collective coupling}) the collective coupling $\gamma$ is not allowed to exceed the critical coupling, but exploring this scenario will provide further physical insight why this should not happen. 
\begin{figure}[h]
 \begin{center}
\includegraphics[height=6cm,width=7.5cm]{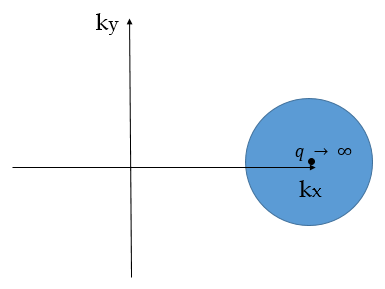}
\caption{\label{Instability}For $\gamma>1$ by shifting the 2D Fermi sphere arbitrarily far in $\bi{k}$-space the energy of the system can be lowered indefinitely. This implies that the energy is unbounded from below and that the system is unstable.}     
 \end{center}
\end{figure}

Without loss of generality, we simplify our consideration to the case where the photon field has a single polarization vector $\bm{\varepsilon}_1=\bi{e}_x$ and $\bm{\varepsilon}_2=0$. In this case the energy density $\mathcal{E}[\mathcal{D}]$ given by Eq.~(\ref{energy density}) as a function of the $x$-component of $\bi{q}=(q_x,q_y)$ is
\begin{eqnarray}
  \mathcal{E}(q_x)=\frac{\hbar^2}{2m_{\textrm{e}}}(1-\gamma)\left(2q_xK^x_{\mathcal{D}}+q^2_xn_{2\textrm{D}}\right).
\end{eqnarray}
 In the equation above we have neglected all terms in Eq.~(\ref{energy density}) independent of $q_x$. For $\gamma>1$ the energy density above is unbounded from below and has no minimum because $1-\gamma<0$ is negative and taking the limit for $q_x$ to infinity the energy density goes to minus infinity
 \begin{eqnarray}\label{negative divergence}
  \lim_{q_x \to \infty}\mathcal{E}(q_x)=\frac{\hbar^2}{2m_{\textrm{e}}} \lim_{q_x \to \infty}(1-\gamma)\left(2q_xK^x_{\mathcal{D}}+q^2_xn_{\textrm{e}}\right)= -\infty.\nonumber\\
\end{eqnarray}
 This shows that the 2DEG coupled to the cavity mode for $\gamma>1$ has no ground state and the system in this case is unstable, because by shifting further and further the origin $\bi{q}$ of the $\bi{k}$-space distribution (see also Fig.\ref{Instability}), the energy density can be lowered indefinitely\footnote{We would like to highlight that this argument is similar to the one for the lack of ground state in the length gauge when the dipole self-energy is omitted. In the length gauge the energy can be lowered indefinitely by shifting the electronic wavefunction further and further in real space~\cite{rokaj2017}.}. Thus, we come to the conclusion that the upper bound of the collective coupling $\gamma$ in Eq.~(\ref{collective coupling}) guarantees the stability of the interacting electron-photon system. 
 
\subsection{No-Go Theorem and the $\bi{A}^2$ Term}
In this section now we would like to investigate the importance of the often neglected~\cite{vukics2014} diamagnetic $\bi{A}^2$ term for our system. The influence of the diamagnetic term has been studied theoretically in multiple publications~\cite{schaeferquadratic,vukics2014, bernadrdis2018breakdown, DiStefano2019} and its influence has also been experimentally measured~\cite{li2018}. Moreover, the elimination of the diamagnetic $\bi{A}^2$ term is known to be responsible for the notorious superradiant phase transition of the Dicke model~\cite{dicke1954}. The superradiant phase transition was firstly predicted by Hepp and Lieb~\cite{Lieb} for the Dicke model in the thermodynamic limit and soon after derived in a different way by Wang and Hioe~\cite{Wang}. However, the existence of the superradiant phase was challenged by a no-go theorem~\cite{Birula} which demonstrated that the superradiant phase transition in atomic systems appeared completely due to the elimination of the $\bi{A}^2$ term. More recently, another demonstration of a superradiant phase transition was predicted for circuit QED set-ups~\cite{CiutiSuperradiance}, which was challenged again by another no-go theorem applicable also to circuit QED systems~\cite{MarquardtNoGO}. The debate over the existence of the superradiant phase transition is still ongoing, with new demonstrations emerging from the field of cavity QED materials~\cite{HagenmullerSPT,MazzaSuperradiance} accompanied though by the respective no-go theorems~\cite{ChirolliNoGO, AndolinaNoGo}. 

To examine the importance of the diamagnetic $\bi{A}^2$ term for our system, we will study the 2DEG coupled to the cavity field in the absence of the $\bi{A}^2$ term. From the Hamiltonian $\hat{H}$ in Eq.~(\ref{single mode Hamiltonian}) we can obtain straightforwardly the Hamiltonian $\hat{H}^{\prime}$ for the 2DEG coupled to the cavity mode when the $\bi{A}^2$ term is neglected  $\hat{H}^{\prime}=\hat{H}-Ne^2\hat{\bi{A}}^2/2m_{\textrm{e}}$ 
 \begin{eqnarray}\label{noA2}
	\hat{H}^{\prime}&=&\sum\limits^{N}_{j=1}\left[-\frac{\hbar^2\nabla^2_j}{2m_{\textrm{e}}} +\frac{\textrm{i}e\hbar}{m_{\textrm{e}}} \hat{\mathbf{A}}\cdot\nabla_j\right]+\sum^2_{\lambda=1}\hbar\omega \left[\hat{a}^{\dagger}_{\lambda}\hat{a}_{\lambda}+\frac{1}{2}\right].\nonumber\\
\end{eqnarray}
As we explained in section~\ref{2DEG in QED}, our system is translationally invariant in the electronic space. Consequently, the many-body electronic eigenfunction is the Slater determinant of plane waves given by Eq.~(\ref{Slater determinant}). Introducing the coupling parameter
\begin{eqnarray}\label{g no a2}
g^{\prime}_0=\frac{e\hbar}{m_{\textrm{e}}}\left(\frac{\hbar}{2\epsilon_0 \omega V}\right)^{1/2},
\end{eqnarray}
applying the Hamiltonian $\hat{H}^{\prime}$ on the Slater determinant $\Phi_{\bi{K}}$ and substituting the definition for the quantized vector potential $\hat{\bi{A}}$ given by Eq.~(\ref{AinDipole}) we find
 \begin{eqnarray}\label{H no A2 on Psik}
    \hat{H}^{\prime}\Phi_{\bi{K}}=\Bigg\{\sum^2_{\lambda=1}\left[\hbar\omega\left(\hat{a}^{\dagger}_{\lambda}\hat{a}_{\lambda}+\frac{1}{2}\right)-g^{\prime}_0\left( \hat{a}_{\lambda}+\hat{a}^{\dagger}_{\lambda}\right)\bm{\varepsilon}_{\lambda}\cdot\bi{K}\right]+ \frac{\hbar^2}{2m_{\textrm{e}}}\sum^N_{j=1}\bi{k}^2_j\Bigg\} \Phi_{\bi{K}}.
\end{eqnarray}
The Hamiltonian $\hat{H}^{\prime}$ has precisely the same form with $\hat{H}$ in Eq.~(\ref{H on Psik}). Following the same procedure for diagonalizing $\hat{H}$, which we demonstrated in detail in section~\ref{2DEG in QED}, we can diagonalize the Hamiltonian $\hat{H}^{\prime}$ as well, and we find that the full eigenspectrum is
\begin{eqnarray}\label{eigenspectrum no A2}
   E_{n_{\lambda},\bi{k}}=\sum^2_{\lambda=1}\left[\hbar\omega\left(n_{\lambda}+\frac{1}{2}\right)-\frac{\gamma^{\prime}}{N}\frac{\hbar^2\left(\bm{\varepsilon}_{\lambda}\cdot \mathbf{K}\right)^2}{2m_{\textrm{e}}}\right]+\sum\limits^{N}_{j=1}\frac{\hbar^2\mathbf{k}^2_j}{2m_{\textrm{e}}}\nonumber\\
\end{eqnarray}
where we substituted the parameter $g^{\prime}_0$ of Eq.~(\ref{g no a2}) and we introduced the coupling constant $\gamma^{\prime}$
\begin{eqnarray}
   \gamma^{\prime}=\frac{2m_{\textrm{e}}N}{\hbar^2}\frac{(g^{\prime}_0)^2}{\hbar\omega}=\frac{\omega^2_p}{\omega^2}
\end{eqnarray}
in analogy to the collective coupling $\gamma$ given by Eq.~(\ref{collective coupling}). The dressed frequency $\widetilde{\omega}$ does not appear anymore, neither in the coupling $\gamma^{\prime}$ nor in the energy spectrum~(\ref{eigenspectrum no A2}), because the vector potential and the energy of the mode do not get renormalized by the $\bi{A}^2$ term. Comparing the spectrum of Eq.~(\ref{eigenspectrum no A2}) for the Hamiltonian $\hat{H}^{\prime}$, to the spectrum in Eq.~(\ref{eigenspectrum}) derived for the Hamiltonian of Eq.~(\ref{single mode Hamiltonian}) which included the $\bi{A}^2$ term, we see that they are the same, up to replacing $\widetilde{\omega}$ with $\omega$ and $\gamma$ with $\gamma^{\prime}$. The last one is a crucial difference, because the coupling $\gamma^{\prime}$ has no upper bound and can be arbitrarily large, as $\omega_p$ can be larger than $\omega$. In section~\ref{No Ground State} we showed that the spectrum in Eq.~(\ref{eigenspectrum no A2}), has no ground-state if the coupling constant becomes larger than one. Obviously, for large densities the diamagnetic shift $\omega_p$ can be larger than the cavity frequency $\omega$ and $\gamma^{\prime}$ can exceed the critical coupling one. As a consequence, the Hamiltonian $\hat{H}^{\prime}$ will not have a ground-state. 

This proves that eliminating the diamagnetic $\bi{A}^2$ term, is a no-go situation for the 2DEG coupled to the cavity field. Thus, for the proper description of such an extended solid-state system the diamagnetic $\bi{A}^2$ term is absolutely necessary. For finite-system models, like the Rabi, Jaynes-Cummings or the Dicke model, the diamagnetic term is of course important, but these models have a stable ground-state even without the $\bi{A}^2$ term. This is in stark contrast to the 2DEG coupled to the cavity and demonstrates explicitly that finite-system models should be applied to extended condensed matter systems with extra care. Our demonstration strongly suggests that the diamagnetic term has to be included for the correct description of extended systems, like 2D materials, coupled to a cavity. We believe our proof contributes to the ongoing discussion about the proper description of light-matter interactions~\cite{rokaj2017, schaeferquadratic, GalegoCasimir, bernadrdis2018breakdown, DiStefano2019, vukics2014}. 

Lastly, we would like to highlight that our proof can be extended to the case of interacting electrons. This is true because the Coulomb interaction involves only the relative distances between the electrons and satisfies translational invariance. Then, by going to the relative distances and center of mass frame, the relative distances decouple from the quantized vector potential $\bi{A}$ and from the center of mass. However, the center of mass stays coupled to the quantized field $\bi{A}$. Then, one can go along the lines of our proof and show that without the $\bi{A}^2$ term the collective coupling constant has no upper bound and the center of mass can obtain an arbitrarily large momentum which subsequently leads to an arbitrarily negative energy. This of course implies that energy of the system is unbounded from below and that the system consequently has no ground-state.

\chapter{Cavity Modified Response Functions}\label{Cavity Responses}

\begin{displayquote}
\footnotesize{I was still technician, a problem-solver interested only in specific problems, not in buliding the general structure of the subject: I had even certain contempt for general formalism. Kubo saw that there were general possibilities in the correlation function formalism I had pioneered, and he pursued them rather than the specific answers I was after.}
\end{displayquote}
\begin{flushright}
  \footnotesize{Philip Anderson about Ryogo Kubo\\
More and Different~\cite{MoreandDifferent}}
\end{flushright}

Linear response theory, also known as Kubo formalism~\cite{kubo}, is the framework which studies systems under the assumptions that they are originally at rest and then perturbed by an external time-dependent perturbation whose strength is considered to be small. If the latter is the case, then the response of the system to the external perturbation can be expanded into a power series with respect to the strength of the perturbation. To first order in this series, the response of the system is a linear function of the external perturbation. This is the regime of linear response theory and from a physical point of view this is the definition of this particular framework.

Linear response theory finds many applications in the study of atoms, molecules and solid-state systems, because in most cases experimental probes can be regarded as small perturbations to the system. If the experimental probes were not small, they would actually modify the system. This is, for example, the case when materials are driven by strong lasers or when they are studied in the presence of strong magnetic fields like in the quantum Hall effect. Nevertheless, as long as the experimental probes do not modify the system, linear response is applicable and we can describe such experiments with linear response functions which are given by the properties of the original unperturbed system. 

\section{Mathematical Formulation of Linear Response}\label{Formulation Linear Response}

We continue by presenting how linear response theory is usually formulated mathematically. Suppose we have a system described by the Hamiltonian $\hat{H}$ and we apply to the system a time-dependent external perturbation described by the operator $\hat{H}_{\textrm{ext}}(t)$.  The time-dependent Schr\"{o}dinger equation is
\begin{eqnarray}\label{Schrodinger}
    \hat{H}(t)\Psi(t)=\left[\hat{H}+\hat{H}_{\textrm{ext}}(t)\right]\Psi(t)=\textrm{i}\hbar\partial_t \Psi(t).
\end{eqnarray}
In the interaction picture a general state of the system is given by 
\begin{eqnarray}\label{interaction wavefunction}
    \Psi_I(t)=\hat{U}^{\dagger}(t)\Psi(t)=e^{\textrm{i}\hat{H}t/\hbar}\Psi(t)
\end{eqnarray}
where $\Psi(t)$ is the wavefunction in the Schr\"{o}dinger picture. In the interaction picture any operator $\hat{\mathcal{O}}$ is defined as
\begin{eqnarray}
    \hat{\mathcal{O}}_I(t)=\hat{U}^{\dagger}(t)\hat{\mathcal{O}}\hat{U}(t)=e^{\textrm{i}\hat{H}t/\hbar}\hat{\mathcal{O}}e^{-\textrm{i}\hat{H}t/\hbar}.
\end{eqnarray}
Substituting the definition of the wavefunction $\Psi(t)$ (in the Schr\"{o}dinger picture) in terms of the wavefunction $\Psi_I(t)$  (in the interaction picture) into the Schr\"{o}dinger equation in Eq.~(\ref{Schrodinger}) we find for $\Psi_I(t)$
\begin{eqnarray}
    \hat{H}_{\textrm{ext}}(t)e^{-\textrm{i}\hat{H}t/\hbar}\Psi_I(t)=e^{-\textrm{i}\hat{H}t/\hbar}\; \textrm{i}\hbar\partial_t\Psi_I(t).
\end{eqnarray}
We multiply now the equation above with the operator $e^{\textrm{i}\hat{H}t/\hbar}$ and after introducing the operator $\hat{H}_{\textrm{ext},I}(t)=e^{\textrm{i}\hat{H}t/\hbar}\hat{H}_{\textrm{ext}}(t)e^{-\textrm{i}\hat{H}t/\hbar}$ we obtain
\begin{eqnarray}
   \hat{H}_{\textrm{ext},I}(t)\Psi_I(t)=\textrm{i}\hbar\partial_t \Psi_{I}(t). 
\end{eqnarray}
By integrating the equation above we find
\begin{eqnarray}\label{nonclosed equation}
\Psi_I(t)=\Psi(t_0)-\frac{\textrm{i}}{\hbar}\int^t_{t_0} dt^{\prime} \hat{H}_{\textrm{ext},I}(t^{\prime})\Psi_I(t^{\prime}).    
\end{eqnarray}
The equation above on a first glance might look rather simple, but it is actually very complicated because to find the wavefunction $\Psi_I(t)$ at time $t$ one already needs to know the wavefunction $\Psi_I(t^{\prime})$ at all times $t^{\prime}$ prior to $t$. This means that Eq.~(\ref{nonclosed equation}) is not a closed equation. There is a simple but also rather crude solution to this problem. The solution is to approximate the wavefunction in the interaction picture at time $t^{\prime}$ as $\Psi_I(t^{\prime})\simeq \Psi(t_0)$. Under this approximation and after multiplying Eq.~(\ref{nonclosed equation}) with the operator $\hat{U}(t)=e^{-\textrm{i}\hat{H}t/\hbar}$ we obtain a closed equation for the wavefunction $\Psi(t)$ in the Schr\"{o}dinger picture
\begin{eqnarray}\label{Schrodinger pic wavefunction}
    \Psi(t)=\hat{U}(t)\Psi(t_0)-\frac{\textrm{i}}{\hbar}\hat{U}(t)\int^t_{t_0} dt^{\prime} \hat{H}_{\textrm{ext},I}(t^{\prime})\Psi(t_0).
\end{eqnarray}
In linear response theory we are not interested in the time evolution of the wavefunction, but we are interested in observables. Namely, we are interested in how an observable $\hat{\mathcal{O}}$ responds to the external perturbation. The expectation value of the operator $\hat{\mathcal{O}}$ as a function of time  $\langle \hat{\mathcal{O}}(t)\rangle$ in the interaction picture and the Schr\"{o}dinger picture is the same
\begin{eqnarray}
     \langle \hat{\mathcal{O}}(t)\rangle=\langle\Psi_I(t)|\hat{\mathcal{O}}_I(t)|\Psi_I(t)\rangle=\langle\Psi(t)|\hat{\mathcal{O}}|\Psi(t)\rangle.
\end{eqnarray}
For simplicity we denote the ket $|\Psi(t_0)\rangle$ at time $t_0$ as $|\Psi_0\rangle$. It is important to highlight that at time $t_0$ the system is unperturbed and consequently $|\Psi_0\rangle$ is the the ground state of the system before the perturbation, $|\Psi_0\rangle\equiv |\Psi_{gs}\rangle$. Substituting the expression we derived for the wavefunction in the Schr\"{o}dinger picture in Eq.~(\ref{Schrodinger pic wavefunction}) we find
\begin{eqnarray}\label{mean of O}
     \langle \hat{\mathcal{O}}(t)\rangle &=& \langle\hat{\mathcal{O}}\rangle_0-\frac{\textrm{i}}{\hbar}\int^t_{t_0}dt^{\prime}\langle\Psi_0|\hat{U}^{\dagger}(t)\hat{\mathcal{O}}\hat{U}(t)\hat{H}_{\textrm{ext},I}(t^{\prime})|\Psi_0\rangle\\
     &+&\frac{\textrm{i}}{\hbar}\int^t_{t_0}dt^{\prime}\langle\Psi_0|\hat{H}_{\textrm{ext},I}(t^{\prime})\hat{U}^{\dagger}(t)\hat{\mathcal{O}}\hat{U}(t)|\Psi_0\rangle +(2_{\textrm{nd}} \;\textrm{order terms in}\; \hat{H}_{\textrm{ext},I})  \nonumber,
\end{eqnarray}
where $\langle \hat{\mathcal{O}}\rangle_0=\langle \Psi_0|\hat{\mathcal{O}}|\Psi_0\rangle$. Introducing now the definition of the operator $\hat{\mathcal{O}}$ in the interaction picture $\hat{\mathcal{O}}_I(t)=\hat{U}^{\dagger}(t)\hat{\mathcal{O}}\hat{U}(t)$, we can write the expectation value $ \langle \hat{\mathcal{O}}(t)\rangle$ of Eq.~(\ref{mean of O}) in terms of the commutator between $\hat{\mathcal{O}}_I(t)$ and $\hat{H}_{\textrm{ext},I}(t^{\prime})$
\begin{eqnarray}\label{mean of O commutator}
     \langle \hat{\mathcal{O}}(t)\rangle &=& \langle\hat{\mathcal{O}}\rangle_0-\frac{\textrm{i}}{\hbar}\int^t_{t_0}dt^{\prime}\langle\Psi_0|[\hat{\mathcal{O}}_I(t),\hat{H}_{\textrm{ext},I}(t^{\prime})] |\Psi_0\rangle.
\end{eqnarray}
To obtain the above result we neglected the higher order terms of the external perturbation. The change $\delta\langle \hat{\mathcal{O}}(t)\rangle$ in the expectation value of the observable $\hat{\mathcal{O}}$ is the difference between the expectation value $\langle \hat{\mathcal{O}}(t)\rangle$ at time $t$  and the expectation value $\langle\hat{\mathcal{O}}\rangle_0$ at $t_0$ . Thus, we find for $\delta\langle \hat{\mathcal{O}}(t)\rangle$ 
\begin{eqnarray}\label{response Observable}
    \delta\langle \hat{\mathcal{O}}(t)\rangle=-\frac{\textrm{i}}{\hbar}\int^t_{t_0}dt^{\prime}\langle\Psi_0|[\hat{\mathcal{O}}_I(t),\hat{H}_{\textrm{ext},I}(t^{\prime})]|\Psi_0\rangle.
\end{eqnarray}
The equation above gives the response of the observable $\mathcal{O}$, represented by the operator $\hat{\mathcal{O}}$, due to the external perturbation $\hat{H}_{\textrm{ext}}$ as a function of time. In most cases the external perturbation is of the form $\hat{H}_{\textrm{ext}}(t)=f_{\textrm{ext}}(t)\hat{\mathcal{P}}$ where $f_{\textrm{ext}}(t)$ is some external classical force, field, current or potential, which couples to some observable of the system represented by the operator $\hat{\mathcal{P}}$. Substituting now this particular form of the external perturbation into Eq.~(\ref{response Observable}) for the response of $\hat{\mathcal{O}}$ we have
\begin{eqnarray}\label{def response observable}
    \delta\langle \hat{\mathcal{O}}(t)\rangle=-\frac{\textrm{i}}{\hbar}\int^t_{t_0}dt^{\prime}\langle\Psi_0|[\hat{\mathcal{O}}_I(t),\hat{\mathcal{P}}_{I}(t^{\prime})]|\Psi_0\rangle f_{\textrm{ext}}(t^{\prime}).
\end{eqnarray}
By introducing the theta function $\Theta(t-t^{\prime})$ we can re-write the response  $\delta\langle \hat{\mathcal{O}}(t)\rangle$ with the help of a function $\chi^{\mathcal{O}}_{\mathcal{P}}(t-t^{\prime})$ as
\begin{eqnarray}\label{response O time}
    \delta\langle \hat{\mathcal{O}}(t)\rangle=\int^{\infty}_{t_0}dt^{\prime}\chi^{\mathcal{O}}_{\mathcal{P}}(t-t^{\prime})f_{\textrm{ext}}(t^{\prime}),
\end{eqnarray}
with the function $\chi^{\mathcal{O}}_{\mathcal{P}}(t-t^{\prime})$ defined as
\begin{eqnarray}\label{chi def}
    \chi^{\mathcal{O}}_{\mathcal{P}}(t-t^{\prime})=-\frac{\textrm{i}}{\hbar}\Theta(t-t^{\prime})\langle\Psi_0|[\hat{\mathcal{O}}_I(t),\hat{\mathcal{P}}_{I}(t^{\prime})]|\Psi_0\rangle.
\end{eqnarray}
Functions of this form in linear response are known as response functions. Such response functions are of great importance because they give us information about how different observables of the system respond to a (small) external perturbation. We note once more that the wavefunction $\Psi_0$ is the ground state of the original system described by the Hamiltonian $\hat{H}$, before the external perturbation $\hat{H}_{\textrm{ext}}(t)$ was applied. Lastly, by performing a Laplace transform in Eq.(\ref{response O time}), one can also find how the observable $\mathcal{O}$ responds to the perturbation in the frequency domain
\begin{eqnarray}\label{reponse frequency O}
    \delta\langle \hat{\mathcal{O}}(w)\rangle=\chi^{\mathcal{O}}_{\mathcal{P}}(w)f_{\textrm{ext}}(w),
\end{eqnarray}
where $\chi^{\mathcal{O}}_{\mathcal{P}}(w)$ and $f_{\textrm{ext}}(w)$ are the response function and the external perturbation respectively in the frequency domain~\cite{kubo, Vignale, flick2018light}.

\section{Radiation \& Absorption in Linear Response}\label{Photonic Response}

In the previous section we formulated the framework of linear response. Our aim now is to apply the linear response formalism to the free electron gas model coupled to the cavity in order to investigate the radiation and absorption properties of the interacting electron-photon system. To do so we will perform linear response on the photonic sector of the system by computing response functions related to the electromagnetic field. From these response functions we will obtain information about the absorption and the radiation properties of the 2DEG coupled to the cavity field. 

\begin{figure}[H]
 \begin{center}
  \includegraphics[width=0.6\columnwidth]{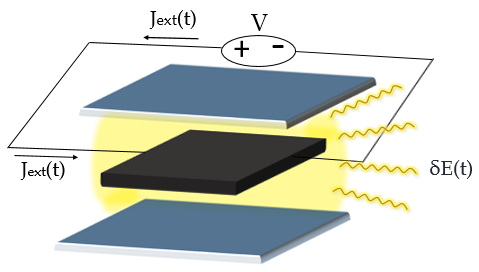}
\caption{\label{Cavity_Induction} 2D material confined inside a cavity, perturbed by an external time-dependent current $\bi{J}_{\textrm{ext}}(t)$. Due to the external current, a time-dependent electric field is induced and the cavity radiates. In an experiment the emitted radiation can be accessed through the openness of the cavity.}   
 \end{center}
\end{figure}

To probe these properties and responses of the electromagnetic field, we will apply an external time dependent current $\bi{J}_{\textrm{ext}}(t)$ as shown in Fig.~\ref{Cavity_Induction}. It is important to mention that in standard quantum mechanics the possibility of perturbing a system with an external current does not exist and only in QED this possibility arises.

To perturb our system by an external current we add the external time dependent term $\hat{H}_{\textrm{ext}}(t)=-\mathbf{J}_{\textrm{ext}}(t)\cdot \hat{\mathbf{A}}$ to the Hamiltonian $\hat{H}$ of Eq.~(\ref{single mode Hamiltonian}), as it is done in QED~\cite{flick2018light, spohn2004, greiner1996}. Here for simplicity we choose the external current to flow only along the $x$-direction $\bi{J}_{\textrm{ext}}(t)=\bi{e}_x|\bi{J}_{\textrm{ext}}(t)|$. With this term included, the full time-dependent Hamiltonian is
\begin{eqnarray}\label{Current Perturbation}
    \hat{H}(t)=\hat{H}-\mathbf{J}_{\textrm{ext}}(t)\cdot \hat{\mathbf{A}}.
\end{eqnarray}
 The external current influences the interacting system inside the cavity, and generates electromagnetic fields, as shown in Fig.~\ref{Cavity_Induction}. The influence of the external current on the observables of the photon field is what we are interested in.

Before we proceed by considering particular response functions we would like first to derive a generic formula for the response of any observable $\mathcal{O}$ due to the external time-dependent current. Following the linear response formalism which we described in section~\ref{Formulation Linear Response}, and using the definition for the response of an observable given by Eq.~(\ref{def response observable}), the response of any observable $\hat{\mathcal{O}}$ due to $\hat{H}_{\textrm{ext}}(t)=-\mathbf{J}_{\textrm{ext}}(t)\cdot \hat{\mathbf{A}}$ is
\begin{eqnarray}
    \delta\langle \hat{\mathcal{O}}(t)\rangle=-\frac{\textrm{i}}{\hbar }\int^{t}_{0}\langle [\hat{\mathcal{O}}_{I}(t),\hat{\bi{A}}_{I}(t^{\prime})]\rangle \left(-\bi{J}_{\textrm{ext}}(t^{\prime})\right),
\end{eqnarray}
where $\hat{\bi{A}}_{I}(t^{\prime})$ is the $\bi{A}$-field in the interaction picture, and the correlator $\langle [\hat{\mathcal{O}}_{I}(t),\hat{\bi{A}}_{I}(t^{\prime})]\rangle$ is defined with respect to the ground state $|\Psi_{gs}\rangle$ of the unperturbed Hamiltonian. Then, the respective response function is
\begin{eqnarray}\label{Response of O}
    \chi^{\mathcal{O}}_{A}(t-t^{\prime})=-\frac{\textrm{i} \Theta(t-t^{\prime})}{\hbar }\langle[\hat{\mathcal{O}}_{I}(t),\hat{\bi{A}}_{I}(t^{\prime})]\rangle.
\end{eqnarray}
In section~\ref{Ground State} we found the ground-state $|\Psi_{gs}\rangle$ of the 2DEG coupled to the cavity (before the current perturbation) in the thermodynamic limit to be given by Eq.~(\ref{Thermodynamic GS}). Having $|\Psi_{gs}\rangle$ we can compute the response function for any observable $\hat{\mathcal{O}}$. To find the response function $\chi^{\mathcal{O}}_{A}(t-t^{\prime})$ of Eq.~(\ref{Response of O}) it is necessary to compute the commutator
\begin{eqnarray}\label{Commutator}
    \langle [\hat{\mathcal{O}}_{I}(t),\hat{\bi{A}}_{I}(t^{\prime})]\rangle&=& \langle \hat{\mathcal{O}}_{I}(t)\hat{\bi{ A}}_{I}(t^{\prime})\rangle-\langle\hat{\mathcal{O}}_{I}(t)\hat{\bi{ A}}_{I}(t^{\prime})\rangle^*.\nonumber\\
\end{eqnarray}
In the equation above we used the hermiticity of the operator $\hat{\mathcal{O}}_{I}(t)\hat{\bi{ A}}_{I}(t^{\prime})$ which implies that $\langle \hat{ \bi{A}}_{I}(t^{\prime}) \hat{\mathcal{O}}_{I}(t)\rangle=\langle\hat{\mathcal{O}}_{I}(t)\hat{\bi{ A}}_{I}(t^{\prime})\rangle^*$. As a consequence we have to compute only the correlator $\langle\hat{\mathcal{O}}_{I}(t)\hat{\bi{ A}}_{I}(t^{\prime})\rangle$. From the fact that $|\Psi_{gs}\rangle$ is the ground state of the Hamiltonian $\hat{H}$ we have
\begin{eqnarray}
e^{-\textrm{i}\hat{H} t^{\prime}/\hbar}|\Psi_{gs}\rangle=e^{-\textrm{i}E_{0,\bi{k}}t^{\prime}/\hbar}|\Psi_{gs}\rangle.
\end{eqnarray}
We use also the definition of the operators in the interaction picture $\hat{\mathcal{O}}_I(t)=e^{\textrm{i}\hat{H}t/\hbar}\hat{\mathcal{O}}e^{-\textrm{i}\hat{H}t/\hbar}$ and we find for the expectation value in the commutator of Eq.~(\ref{Commutator})
\begin{eqnarray}
    \langle\hat{\mathcal{O}}_{I}(t)\hat{\bi{ A}}_{I}(t^{\prime})\rangle=e^{\frac{\textrm{i}E_{0,\bi{k}}(t-t^{\prime})}{\hbar}}\langle\hat{\mathcal{O}}e^{\frac{-\textrm{i}\hat{H}(t-t^{\prime})}{\hbar}} \hat{\bi{A}}\rangle,
\end{eqnarray}
where $E_{0,\bi{k}}=E_{\bi{k}}+\hbar\widetilde{\omega}$ is the ground-state energy given by Eq.~(\ref{eigenspectrum}), with $n_{\lambda}=0$ for both $\lambda=1,2$. To proceed, we have to apply the quantized vector potential $\hat{\bi{A}}$ to the ground state $|\Psi_{gs}\rangle$. For that purpose we need the expression of $\hat{\bi{A}}$ in terms of the bosonic operators $\hat{c}_{\lambda},\hat{c}^{\dagger}_{\lambda}$. From Eqs.~(\ref{Ainb}) and~(\ref{c operators}), and for $\mathbf{K}=0$ (which is true in the ground state) we find  
\begin{eqnarray}\label{AinC}
    \hat{\bi{A}}=\left(\frac{\hbar}{2\epsilon_0\tilde{\omega}V}\right)^{\frac{1}{2}}\left(\hat{c}_1+\hat{c}^{\dagger}_1\right)\bi{e}_x.
\end{eqnarray}
We note that in the expression above for the vector potential, we have only kept the polarization $\bi{e}_x$ which couples to the external current. We apply the vector potential $\hat{\bi{A}}$ to the ground-state $|\Psi_{gs}\rangle$ 
\begin{eqnarray}
    \langle\hat{\mathcal{O}}_{I}(t)\hat{\bi{ A}}_{I}(t^{\prime})\rangle=\left(\frac{\hbar}{2\epsilon_0\tilde{\omega}V}\right)^{\frac{1}{2}}e^{\frac{\textrm{i}E_{0,\bi{k}}(t-t^{\prime})}{\hbar}}\langle \Psi_{gs}|\hat{\mathcal{O}}e^{\frac{-\textrm{i}\hat{H}(t-t^{\prime})}{\hbar}}|\Phi_{0}\rangle\otimes |1,0\rangle_{1}|0,0\rangle_2.
\end{eqnarray}
From the expression above it is clear that the quantized vector potential $\hat{\bi{A}}$ gets the ground-state to the first excited state for $n_1=1$. The state $|\Phi_{0}\rangle\otimes |1,0\rangle_{1}|0,0\rangle_2$ is the first excited state of $\hat{H}$ with eigenenergy $E_{1,\bi{k}}=E_{\bi{k}}+2\hbar\widetilde{\omega}$ and we find
\begin{eqnarray}
    \langle \hat{\mathcal{O}}_{I}(t)\hat{\bi{ A}}_{I}(t^{\prime})\rangle=\left(\frac{\hbar}{2\epsilon_0\tilde{\omega}V}\right)^{\frac{1}{2}}e^{-\textrm{i}\widetilde{\omega}(t-t^{\prime})}\langle \Psi_{gs}|\hat{\mathcal{O}}|\Phi_{0}\rangle\otimes |1,0\rangle_{1}|0,0\rangle_2.
\end{eqnarray}
Finally, using the above result we obtain the expression for the generic commutator of Eq.~(\ref{Commutator}) 
\begin{eqnarray}\label{O commutator}
    \langle [\hat{\mathcal{O}}_{I}(t),\hat{\bi{A}}_{I}(t^{\prime})]\rangle&=&\left(\frac{\hbar}{2\epsilon_0\tilde{\omega}V}\right)^{\frac{1}{2}}\Bigg[e^{-\textrm{i}\widetilde{\omega}(t-t^{\prime})} \langle \Psi_{gs}|\hat{\mathcal{O}}|\Phi_{0}\rangle\otimes |1,0\rangle_{1}|0,0\rangle_2\\
    &-&e^{\textrm{i}\widetilde{\omega}(t-t^{\prime})} \left(\langle \Psi_{gs}|\hat{\mathcal{O}}|\Phi_{0}\rangle\otimes |1,0\rangle_{1}|0,0\rangle_2\right)^*\Bigg].\nonumber
\end{eqnarray}
The formula above is very important because it applies to any observable $\hat{\mathcal{O}}$ and in what follows we will make use of it for the computation of several response functions.

\subsection{$\bi{A}$-Field Response \& Absorption}\label{A field propagator}

\begin{figure}[H]
\begin{center}
  \includegraphics[height=6cm, width=0.6\columnwidth]{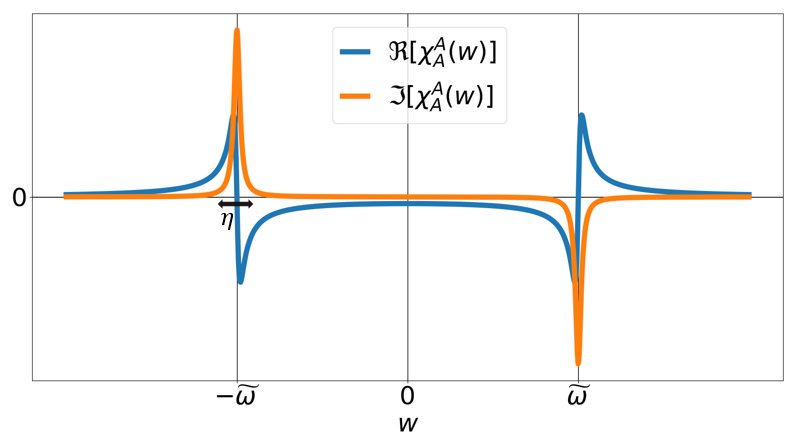}
\caption{\label{A-field Response} Real $\Re[\chi^A_A(w)]$ and imaginary $\Im[\chi^A_A(w)]$ parts of the $\bi{A}$-field response function $\chi^A_A(w)$ in the frequency domain, plotted with a finite value for the artificial broadening parameter $\eta$. The resonances for both parts appear at the plasmon-polariton frequency $w=\pm\widetilde{\omega}$.}  
\end{center}
\end{figure}

The first response that we are interested in, is the response of the $\bi{A}$-field. The response of the vector potential $\delta\langle \hat{\bi{A}}(t)\rangle$ is defined via Eq.~(\ref{response Observable}) and is given by the $\bi{A}$-field response function $\chi^A_A(t-t^{\prime})$. 
\begin{eqnarray}
    \chi^{\mathcal{A}}_{A}(t-t^{\prime})=-\frac{\textrm{i} \Theta(t-t^{\prime})}{\hbar }\langle[\hat{\bi{A}}_{I}(t),\hat{\bi{A}}_{I}(t^{\prime})]\rangle.
\end{eqnarray}
From the generic formula that we derived in Eq.~(\ref{O commutator}), it is clear that to find the response function $\chi^{\mathcal{A}}_{A}(t-t^{\prime})$, all we have to compute is $\langle \Psi_{gs}|\hat{\bi{A}}|\Phi_{0}\rangle\otimes |1,0\rangle_{1}|0,0\rangle_2$. For that we use Eq.~(\ref{AinC}) which gives the vector potential $\hat{\bi{A}}$ in terms of the operators $\hat{c}_1,\hat{c}^{\dagger}_1$ and we have 
\begin{eqnarray}
\langle \Psi_{gs}|\hat{\bi{A}}|\Phi_{0}\rangle\otimes |1,0\rangle_{1}|0,0\rangle_2=\left(\frac{\hbar}{2\epsilon_0\widetilde{\omega}V}\right)^{\frac{1}{2}}.
\end{eqnarray}
Combining the above result together with Eqs.~(\ref{O commutator}) and Eq.~(\ref{Response of O}) we obtain the expression for response function in time $\chi^A_A(t-t^{\prime})$
\begin{eqnarray}\label{A field response}
\chi^A_A(t-t^{\prime})=-\frac{\Theta(t-t^{\prime})\sin(\widetilde{\omega}(t-t^{\prime}))}{\epsilon_0\widetilde{\omega}V} .
\end{eqnarray}
The response function above is also known as the $\bi{A}$-field propagator. Moreover, we use of the integral form of the $\Theta$-function
\begin{eqnarray}
\Theta(\tau)=-\frac{1}{2\pi \textrm{i}}\lim_{\eta \rightarrow 0^{+}}\int^{\infty}_{-\infty}\frac{e^{-\textrm{i}w\tau}}{w+\textrm{i}\eta} dw. \end{eqnarray}
We perform a Fourier transform for the response function $\chi^A_A(t-t^{\prime})$ and we obtain the response of the quantized vector potential $\bi{A}$ in the frequency domain $\chi^A_A(w)$
\begin{eqnarray}\label{A frequencyresponse}
\chi^A_A(w)=\frac{-1}{2\epsilon_0\widetilde{\omega}V}\lim_{\eta \to 0^+}\left[\frac{1}{w+\widetilde{\omega}+\textrm{i}\eta}-\frac{1}{w-\widetilde{\omega}+\textrm{i}\eta}\right].
\end{eqnarray}
Then, from the expression above we can easily deduce the real $\Re[\chi^A_A(w)]$ and the imaginary $\Im[\chi^A_A(w)]$ parts of $\chi^A_A(w)$ 
\begin{eqnarray}\label{Re Im A-field}
    \Re[\chi^A_A(w)]&=&\frac{1}{2\epsilon_0 \widetilde{\omega}V}\left[\frac{w-\widetilde{\omega}}{(w-\widetilde{\omega})^2+\eta^2}-\frac{w+\widetilde{\omega}}{(w+\widetilde{\omega})^2+\eta^2}\right],\nonumber\\
    \Im[\chi^A_A(w)]&=&\frac{\eta}{2\epsilon_0\widetilde{\omega}V}\left[\frac{1}{(w+\widetilde{\omega})^2+\eta^2}-\frac{1}{(w-\widetilde{\omega})^2+\eta^2}\right].
\end{eqnarray}
 From both, the real and the imaginary parts of the response function, we see that the poles are at the frequency $w=\pm\widetilde{\omega}$. This is also depicted in Fig.~\ref{A-field Response}. The frequency $\widetilde{\omega}$ defined in Eq.~(\ref{plasmon polariton}) depends on the cavity frequency $\omega$ and the plasma frequency $\omega_p$ in the cavity. This means that the 2DEG coupled to the cavity field exhibits a plasmon-polariton resonance (or excitation). 
For a self-adjoint operator, like the $\bi{A}$-field, the real and the imaginary part of any response function have to be even and odd respectively~\cite{Vignale}. As we see in Fig.~\ref{A-field Response} this is indeed true. 

Further, before we proceed, we would like to briefly comment on how these response functions should be interpreted. The real part $\Re[\chi^A_A(w)]$ is the component of the response function which is in-phase with the external current and describes a polarization process in which the wavefunction is modified periodically without any energy (on average) being absorbed or released due to the external driving~\cite{Vignale}. On the other side, the imaginary part $\Im[\chi^A_A(w)]$ is the out-of-phase component of $\chi^A_A(w)$ and is responsible for the appearance of energy absorption in the system, with the absorption rate $W$ defined as~\cite{Vignale}
\begin{eqnarray}\label{Absorption Rate}
    W=-w\Im[\chi^A_A(w)]|\bi{J}_{\textrm{ext}}(w)|^2.
\end{eqnarray}

\subsection{Electric Field Response \& Current Induced Radiation}

In addition to the response of the $\bi{A}$-field we would also like to compute the response of the electric field $\bi{E}$ due to the external time-dependent current $\bi{J}_{\textrm{ext}}(t)$. The electric field in dipole approximation and polarized in the $x$-direction is~\cite{rokaj2017}
\begin{equation}\label{Electric field}
	\hat{\mathbf{E}}=\textrm{i}\left(\frac{\hbar\omega }{2\epsilon_0 V}\right)^{\frac{1}{2}}\left(\hat{a}_{1}-\hat{a}^{\dagger}_{1}\right) \bi{e}_x.
\end{equation}
Having the expression for the electric field we can compute the electric field response function $\chi^E_A(t-t^{\prime})$
\begin{eqnarray}
    \chi^{E}_{A}(t-t^{\prime})=-\frac{\textrm{i} \Theta(t-t^{\prime})}{\hbar }\langle[\hat{\bi{E}}_{I}(t),\hat{\bi{A}}_{I}(t^{\prime})]\rangle.
\end{eqnarray}
By writing the electric field in terms of the bosonic operators $\hat{c}_{1},\hat{c}^{\dagger}_{1}$ 
\begin{equation}
	\hat{\mathbf{E}}=\textrm{i}\left(\frac{\hbar\widetilde{\omega} }{2\epsilon_0 V}\right)^{\frac{1}{2}}\left(\hat{c}_{1}-\hat{c}^{\dagger}_{1}\right) \bi{e}_x.
\end{equation}
we can use the generic formula that we derived for any observable in Eq.~(\ref{O commutator}) for the computation of the correlator $\langle[\hat{\bi{E}}_{I}(t),\hat{\bi{A}}_{I}(t^{\prime})]\rangle$, and we obtain the expression for the electric field response function 
\begin{eqnarray}\label{E field response}
    \chi^E_{A}(t-t^{\prime})=\frac{\Theta(t-t^{\prime})\cos(\widetilde{\omega}(t-t^{\prime}))}{\epsilon_0 V}.
\end{eqnarray}
This response function describes the generation of a time-dependent electric field due to the external current $\bi{J}_{\textrm{ext}}(t)$. This means that the external current makes the interacting electron-photon system to radiate. From Eq.~(\ref{E field response}) we see the radiation is at the plasmon-polariton frequency $\widetilde{\omega}$, as the response function in time is a cosine of $\widetilde{\omega}$.

\begin{figure}[H]
\begin{center}
\includegraphics[height=6cm, width=0.6\columnwidth]{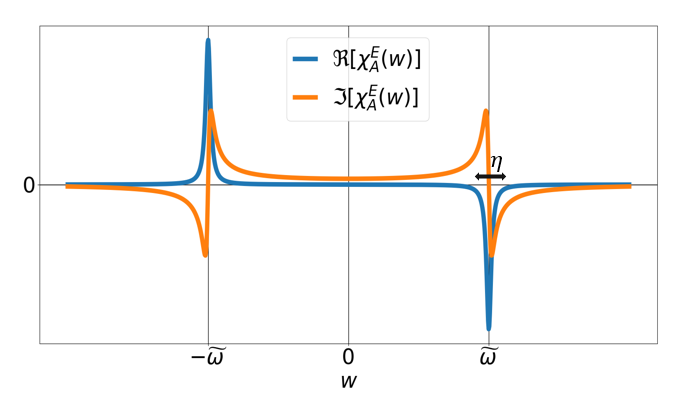}
\caption{\label{E field Responses} Real $\Re[\chi^E_A(w)]$ and imaginary $\Im[\chi^E_A(w)]$ parts of the $\bi{E}$-field response function $\chi^E_A(w)$, plotted with a finite $\eta$. The poles for both parts appear at the frequency $w=\pm\widetilde{\omega}$ and signify the frequency at which a time-dependent electric field is oscillating. }    
\end{center}
\end{figure}

Further, by performing a Fourier transform, we find the response function in the frequency domain
    \begin{eqnarray}\label{E response frequency}
    \chi^{E}_{A}(w)=\frac{\textrm{i}}{2\epsilon_0V}\lim_{\eta \to 0^+}\left[\frac{1}{w+\widetilde{\omega}+\textrm{i}\eta}+\frac{1}{w-\widetilde{\omega}+\textrm{i}\eta}\right],
\end{eqnarray}
from which we deduce the real and the imaginary parts of $ \chi^{E}_{A}(w)$
\begin{eqnarray}
    \Re[\chi^E_A(w)]=\frac{\eta}{2\epsilon_0V}\left[\frac{1}{(w+\widetilde{\omega})^2+\eta^2}-\frac{1}{(w-\widetilde{\omega})^2+\eta^2}\right],\nonumber\\
    \Im[\chi^E_A(w)]=\frac{1}{2\epsilon_0V}\left[\frac{w+\widetilde{\omega}}{(w+\widetilde{\omega})^2+\eta^2}-\frac{w-\widetilde{\omega}}{(w-\widetilde{\omega})^2+\eta^2}\right].
\end{eqnarray}
The fact that the radiation is emitted at the plasmon-polariton resonance $w=\pm\widetilde{\omega}$ can be understood also from the pole structure of the response function in the frequency domain. As depicted in Fig.~\ref{E field Responses} we see that the poles are at $w=\pm\widetilde{\omega}$.

Lastly, we highlight that the response function of the electric field in time $\chi^E_A(t-t^{\prime})$ given by Eq.~(\ref{E field response}), and the response function of the $\bi{A}$-field $\chi^A_A(t-t^{\prime})$ given by Eq.~(\ref{A field response}) satisfy Maxwell's equation~\cite{JacksonEM}. 
\begin{eqnarray}
\chi^{E}_{A}(t-t^{\prime})=-\partial_t \chi^{A}_{A}(t-t^{\prime})
\end{eqnarray}
This is a very nice consistency check of our computations and of the linear response formalism in QED~\cite{flick2018light}, because it demonstrates that linear response theory, even for interacting electron-photon systems, respects the classical Maxwell equations.

\section{Cavity Modified Conductivity}\label{Electronic Response}

In this section we are interested in the conduction properties of the 2DEG inside the cavity. More specifically we want to investigate whether the cavity field modifies the conductive properties of the electron gas. This is a question of current theoretical and experimental interest. Recently, modifications of transport and conduction properties due to cavity confinement have been observed for two-dimensional systems of Landau polaritons~\cite{paravacini2019}, as well as modifications of the critical temperature of superconductors due to strong coupling to the cavity field~\cite{sentef2018, A.Thomas2019}. 

\begin{figure}[h]
 \begin{center}
   \includegraphics[width=0.6\columnwidth]{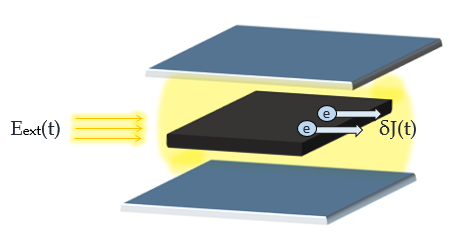}
\caption{\label{Cavity_Conductivity} An external time dependent electric field $\bi{E}_{\textrm{ext}}(t)$ perturbs the combined light-matter system, electrons start to flow, and a current is generated in the material. }  
 \end{center}
\end{figure}

To describe this process we will follow what is commonly done in condensed matter physics, namely perturb the system with a uniform, external, time-dependent electric field $\mathbf{E}_{\textrm{ext}}(t)$, as shown in Fig.~\ref{Cavity_Conductivity}, and then compute how much current is generated and flows through the system due to the external perturbation. Here we choose the electric field to be polarized only along the $x$-direction $\mathbf{E}_{\textrm{ext}}(t)=|\mathbf{E}_{\textrm{ext}}(t)|\bi{e}_x$. The electric field can also be represented as the time derivative of a time-dependent vector potential $\mathbf{E}_{\textrm{ext}}(t)=-\partial_t\bi{A}_{\textrm{ext}}(t)$. We would like to mention that for the external perturbation to be causal, the electric field needs to be zero for all times prior to time $t_0$. This implies that in the frequency domain the electric field and vector potential are related via $\bi{E}_{\textrm{ext}}(w)=\textrm{i}(w+\textrm{i}\eta)\bi{A}(w)$ with $\eta \rightarrow 0^{+}$.
In order to couple the external field we have to add the external vector potential $\bi{A}_{\textrm{ext}}(t)$ in the covariant kinetic energy of the Pauli-Fierz Hamiltonian of Eq.~(\ref{Pauli-Fierz}). With this, the covariant term takes the form $(\textrm{i}\hbar \mathbf{\nabla}_{j}+e \hat{\bi{A}}+e\bi{A}_{\textrm{ext}}(t))^2$~\cite{Landau, TokatlyPRL, spohn2004}. In linear response theory the current is computed to first order in the external perturbation. Then, the conductivity is defined as the function relating the induced current density to the external electric field~\cite{kubo, flick2018light, Vignale}.

The Pauli-Fierz Hamiltonian with the electrons coupled to a single cavity mode, in the dipole approximation, and to first order in the external vector potential $\bi{A}_{\textrm{ext}}(t)$ reads as
\begin{eqnarray}
    \hat{H}(t)=\hat{H}+\sum^{N}_{j=1}\left(\frac{\textrm{i}e\hbar}{m_{\textrm{e}}}\nabla_j+\frac{e^2}{m_{\textrm{e}}}\hat{\mathbf{A}}\right)\cdot \bi{A}_{\textrm{ext}}(t)=\hat{H}-\left(\hat{\bi{J}}_\textrm{p}+\hat{\bi{J}}_{\textrm{d}}\right)\cdot\bi{A}_{\textrm{ext}}(t)
\end{eqnarray}
where $\hat{H}$ is the Hamiltonian of Eq.~(\ref{single mode Hamiltonian}). The external vector potential $\bi{A}_{\textrm{ext}}(t)$ couples to the internal parts of the current operator, namely the paramagnetic part $\hat{\bi{J}}_{\textrm{p}}=(-\textrm{i}e\hbar/m_{\textrm{e}})\sum_j\nabla_j$, and the diamagnetic part $\hat{\bi{J}}_{\textrm{d}}=-e^2N\hat{\bi{A}}/m_{\textrm{e}}$. However, the full physical current includes also the contribution of the external vector potential $\bi{A}_{\textrm{ext}}(t)$ ~\cite{Landau, TokatlyPRL}
\begin{eqnarray}\label{Current Operator}
    \hat{\bi{J}}=\underbrace{-\frac{\textrm{i}e\hbar}{m_{\textrm{e}}}\sum^N_{j=1}\nabla_j}_{\hat{\bi{J}}_{\textrm{p}}}\underbrace{-\frac{e^2N}{m_{\textrm{e}}}\hat{\bi{A}}}_{\hat{\bi{J}}_{\textrm{d}}}-\frac{e^2N}{m_{\textrm{e}}}\bi{A}_{\textrm{ext}}(t).
\end{eqnarray}
Following the linear response formalism, the expectation value for the full physical current is~\cite{Vignale, kubo}
\begin{eqnarray}\label{current Expectation Value}
    \langle \hat{\bi{J}}(t)\rangle=\langle \hat{\bi{J}}\rangle +\delta\langle \hat{\bi{J}}(t)\rangle=\langle \hat{\bi{J}}\rangle-\int^{\infty}_{t_0} dt^{\prime}\chi^J_J(t-t^{\prime})\bi{A}_{\textrm{ext}}(t^{\prime})
\end{eqnarray}
where $\delta\langle \hat{\bi{J}}(t)\rangle$ is the response of the current in time. This can be computed from the current-current response function
 \begin{eqnarray}\label{JJResponse}
     \chi^{J}_{J}(t-t^{\prime})=\frac{-\textrm{i}\Theta(t-t^{\prime})}{\hbar }\langle[\hat{\mathbf{J}}_{I}(t),\hat{\mathbf{J}}_{I}(t^{\prime})]\rangle.
 \end{eqnarray}
To keep the current response $\delta\langle \hat{\bi{J}}\rangle$ to first order in the external field, we neglect all contributions, coming from $\bi{A}_{\textrm{ext}}(t)$. Then, for the commutator of Eq.~(\ref{JJResponse}), we find the following four terms
\begin{eqnarray}
    [\hat{\mathbf{J}}_{I}(t),\hat{\mathbf{J}}_{I}(t^{\prime})]=[\hat{\mathbf{J}}_{\textrm{p}, I}(t),\hat{\mathbf{J}}_{\textrm{p}, I}(t^{\prime})] +[\hat{\mathbf{J}}_{\textrm{d}, I}(t),\hat{\mathbf{J}}_{\textrm{p}, I}(t^{\prime})]+[\hat{\mathbf{J}}_{\textrm{d}, I}(t),\hat{\mathbf{J}}_{\textrm{p}, I}(t^{\prime})] +[\hat{\mathbf{J}}_{\textrm{d}, I}(t),\hat{\mathbf{J}}_{\textrm{d}, I}(t^{\prime})].\nonumber\\
\end{eqnarray}
Using the self-adjointness of the paramagnetic current operator we find 
\begin{eqnarray}
\langle [\hat{\mathbf{J}}_{\textrm{p}, I}(t),\hat{\mathbf{J}}_{\textrm{p}, I}(t^{\prime})]\rangle=\langle\hat{\mathbf{J}}_{\textrm{p}, I}(t)\hat{\mathbf{J}}_{\textrm{p}, I}(t^{\prime})\rangle -\langle\hat{\mathbf{J}}_{\textrm{p}, I}(t)\hat{\mathbf{J}}_{\textrm{p}, I}(t^{\prime})\rangle^*.
\end{eqnarray}
In addition, we use the expression for the paramagnetic current operator in the interaction picture and the fact that the expectation value is computed in the ground-state which has energy $E_{0,\bi{k}}$, and we find 
\begin{eqnarray}
\langle \hat{\mathbf{J}}_{\textrm{p}, I}(t)\hat{\mathbf{J}}_{\textrm{p}, I}(t^{\prime})\rangle=e^{\textrm{i}E_{0,\bi{k}}(t-t^{\prime})/\hbar}\langle \hat{\mathbf{J}}_{p}e^{\textrm{i}\hat{H}(t^{\prime}-t)/\hbar}\hat{\mathbf{J}}_p\rangle.
\end{eqnarray}
Due to translational invariance, the momentum operator commutes with the Hamiltonian $\hat{H}$, and the ground-state $|\Psi_{gs}\rangle=|\Phi_{0}\rangle\otimes|0,0\rangle_1|0,0\rangle_2$ is an eigenstate also of the paramagnetic current operator $\hat{\mathbf{J}}_{\textrm{p}}\sim \sum_j \nabla_j$ as well. Acting with the paramagnetic current operator on the ground-state we obtain the full paramagnetic current $\hat{\mathbf{J}}_{\textrm{p}}|\Psi_{gs}\rangle=\sum_j \mathbf{k}_j|\Psi_{gs}\rangle$. In section~\ref{Ground State} we showed that the ground state in the thermodynamic limit is the Fermi sphere. As a consequence, the total paramagnetic current in the ground state is zero and we have  
\begin{eqnarray}
\hat{\mathbf{J}}_{\textrm{p}}|\Psi_{gs}\rangle=0.
\end{eqnarray}
The above result implies that all expectation values and correlators which involve the paramagnetic current $\hat{\bi{J}}_{\textrm{p}}$ will also be zero. This argument applies of course to the mixed terms $[\hat{\mathbf{J}}_{\textrm{d}, I}(t),\hat{\mathbf{J}}_{\textrm{p}, I}(t^{\prime})]$ and $[\hat{\mathbf{J}}_{\textrm{p}, I}(t),\hat{\mathbf{J}}_{\textrm{d}, I}(t^{\prime})]$ as well. Thus, the response function $\chi^J_{J}(t-t^{\prime})$ in Eq.~(\ref{JJResponse}) is given solely by the diamagnetic terms. Substituting the definition for the diamagnetic current $\hat{\bi{J}}_{\textrm{d}}$ of Eq.~(\ref{Current Operator}) we find the current-current response function $\chi^{J}_{J}(t-t^{\prime})$ to be proportional to the $\bi{A}$-field response function $\chi^A_A(t-t^{\prime})$
 \begin{eqnarray}
     \chi^{J}_{J}(t-t^{\prime})=\left(\frac{e^2 N}{m_{\textrm{e}}}\right)^2 \chi^A_A(t-t^{\prime}).
 \end{eqnarray}
 with $\chi^A_A(t-t^{\prime})$ given by Eq.~(\ref{A field response}). Since $\chi^J_J(t-t^{\prime})$ is proportional to $\chi^A_A(t-t^{\prime})$ the same will also hold in the frequency domain 
 \begin{eqnarray}\label{chiJJ to chiAA}
     \chi^J_J(w)=\left(\frac{e^2N}{m_{\textrm{e}}}\right)^2\chi^A_A(w).
 \end{eqnarray}
 Lastly, we have to compute the expectation value of the current $\langle \hat{\bi{J}}\rangle$ which is
\begin{eqnarray}
    \langle \hat{\bi{J}}\rangle=\langle \hat{\bi{J}}_{\textrm{p}}\rangle+\langle\hat{\bi{J}}_{\textrm{d}}\rangle-\frac{e^2N}{m_{\textrm{e}}}\langle\bi{A}_{\textrm{ext}}(t)\rangle.
\end{eqnarray}
As we already explained, the contribution of the paramagnetic $\hat{\bi{J}}_{\textrm{p}}$ current is zero in the ground state $|\Psi_{gs}\rangle$. The diamagnetic current $\hat{\bi{J}}_{\textrm{d}}$ is proportional to the vector potential $\hat{\bi{J}}_{\textrm{d}}\sim \hat{\bi{A}}$, which is the sum of an annihilation and a creation operator. The expectation values of these operators in the ground-state is zero. Consequently only the external field contributes to the expectation value of the current
\begin{eqnarray}
    \langle \hat{\bi{J}}\rangle=-\frac{e^2N}{m_{\textrm{e}}}\bi{A}_{\textrm{ext}}(t).
\end{eqnarray}
This contribution comes from the the full background charge of the $N$ electrons in our system. Moreover, from Eq.~(\ref{current Expectation Value}) by performing a Fourier transformation we can derive the relation between the current and the external $\bi{A}_{\textrm{ext}}(w)$-field, in the frequency domain 
\begin{eqnarray}\label{J n A }
    \langle \hat{\mathbf{J}}(w)\rangle=\left(-\frac{e^2N}{m_{\textrm{e}}}-\chi^{J}_{J}(w)\right)\mathbf{A}_{\textrm{ext}}(w).
\end{eqnarray}
The electric field and vector potential in the frequency domain are related via the relation $\bi{A}_{\textrm{ext}}(w)=\bi{E}_{\textrm{ext}}(w)/\textrm{i}(w+\textrm{i}\eta)$. Using the latter relation and dividing Eq.~(\ref{J n A }) by the volume $V$ in order to introduce the current density $\langle\hat{\bi{j}}(w)\rangle=\langle\hat{\mathbf{J}}(w)\rangle/V$ we can define the frequency dependent conductivity $\sigma(w)$ as the ratio between the external electric field $\bi{E}_{\textrm{ext}}(w)$ and the current density $\langle\hat{\bi{j}}(w)\rangle$~\cite{Mermin,Vignale} 
\begin{eqnarray}
 \langle\hat{\mathbf{j}}(w)\rangle=\left(-\frac{e^2n_{\textrm{e}}}{m_{\textrm{e}}}-\frac{\chi^J_J(w)}{V}\right)\frac{\mathbf{E}_{\textrm{ext}}(w)}{\textrm{i}(w+\textrm{i}\eta)}=\sigma(w)\mathbf{E}_{\textrm{ext}}(w).
\end{eqnarray}
The frequency dependent conductivity $\sigma(w)$ is also known as the optical conductivity and the previous equation is known as the Kubo formula for the electrical conductivity~\cite{kubo, Vignale}. Using the result for the current-current response function $\chi^J_J(w)$ given by Eqs.~(\ref{chiJJ to chiAA}) and (\ref{A frequencyresponse}), and introducing $\omega_p^2=e^2n_{\textrm{e}}/m_{\textrm{e}}\epsilon_0$ which is the plasma frequency in the cavity, we obtain the expression for the optical conductivity $\sigma(w)$
\begin{eqnarray}\label{Conductivity}
    \sigma(w)=\frac{\textrm{i}\epsilon_0\omega^2_p}{w+\textrm{i}\eta}- \frac{\textrm{i}\epsilon_0\omega^4_p}{(w+\textrm{i}\eta)2\widetilde{\omega}}\left[\frac{1}{w+\widetilde{\omega}+\textrm{i}\eta}-\frac{1}{w-\widetilde{\omega}+\textrm{i}\eta}\right] \;\;\textrm{with}\;\; \eta \rightarrow 0^{+}.
    \end{eqnarray}
The real $\Re[\sigma(w)]$ and imaginary $\Im[\sigma(w)]$ parts of the optical conductivity are 
\begin{eqnarray}\label{Real and Imaginary sigma}
\Re[\sigma(w)]&=&\frac{\epsilon_0\eta\omega^2_p}{w^2+\eta^2}-\frac{\eta\epsilon_0\omega^4_p}{2\widetilde{\omega}(w^2+\eta^2)}\left[\frac{2w+\widetilde{\omega}}{(w+\widetilde{\omega})^2+\eta^2}-\frac{2w-\widetilde{\omega}}{(w-\widetilde{\omega})^2+\eta^2}\right],\\
\Im [\sigma(w)]&=&\frac{\epsilon_0w\omega^2_p}{w^2+\eta^2}-\frac{\epsilon_0\omega^4_p}{2\widetilde{\omega}(w^2+\eta^2)}\left[\frac{w^2-\eta^2+w\widetilde{\omega}}{(w+\widetilde{\omega})^2+\eta^2}-\frac{w^2-\eta^2-w\widetilde{\omega}}{(w-\widetilde{\omega})^2+\eta^2}\right].\nonumber
\end{eqnarray}

\begin{figure}[h]
\begin{center}
  \includegraphics[height=6.5cm,width=0.65\columnwidth]{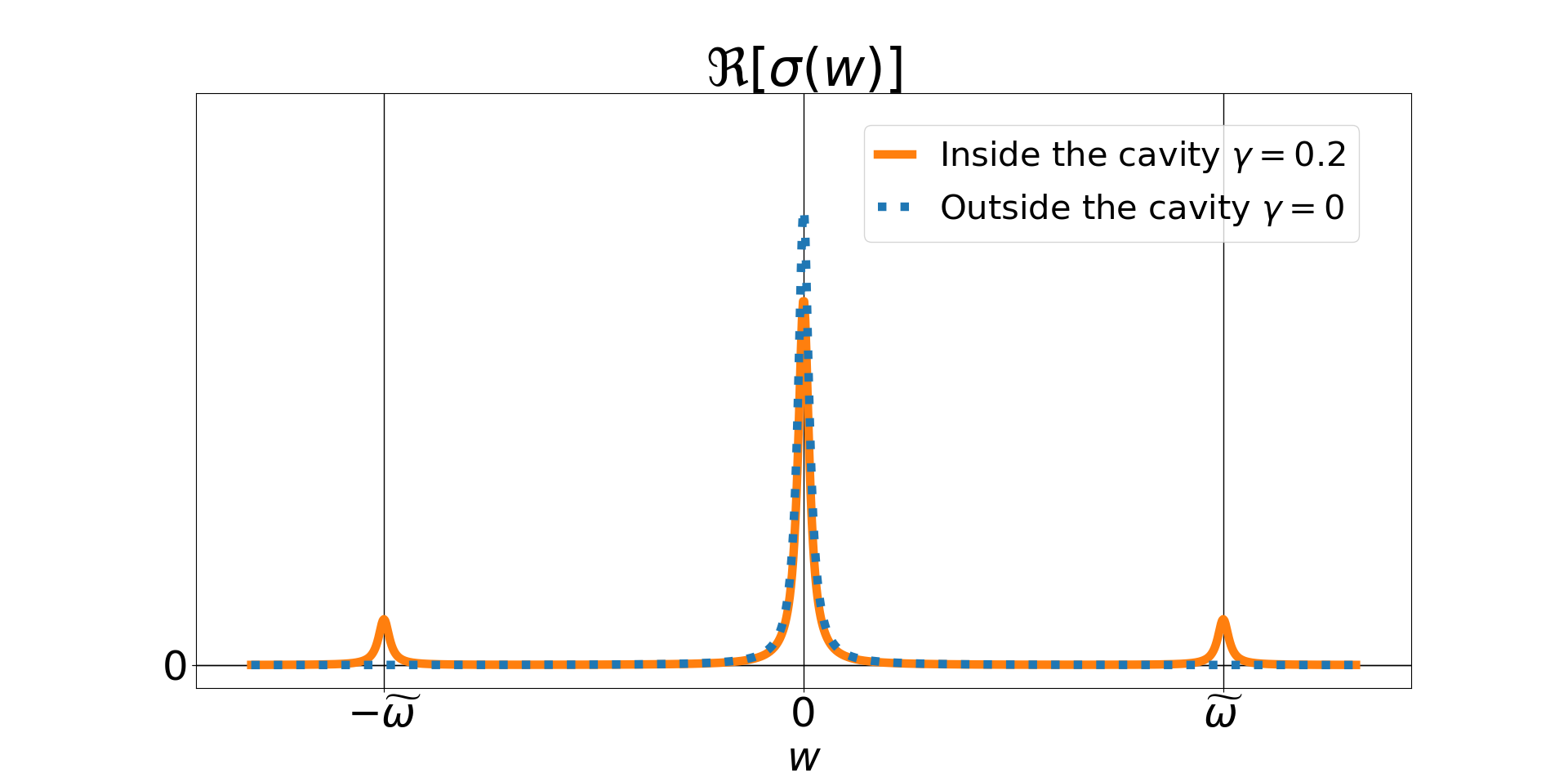}
\caption{\label{Real_Conductivity}Real part $\Re[\sigma(w)]$ of the conductivity $\sigma(w)$ of the 2DEG outside (blue dashed line) and inside (orange solid line) the cavity for coupling $\gamma=0.2$. Inside the cavity the real part of the conductivity of the 2DEG exhibits poles at the plasmon-polariton frequency $w=\pm\widetilde{\omega}$. At frequency $w=0$ the Drude peak of the 2DEG gets suppressed due to the cavity field.  }  
\end{center}
\end{figure}

\begin{figure}[h]
\begin{center}
  \includegraphics[height=6.5cm,width=0.65\columnwidth]{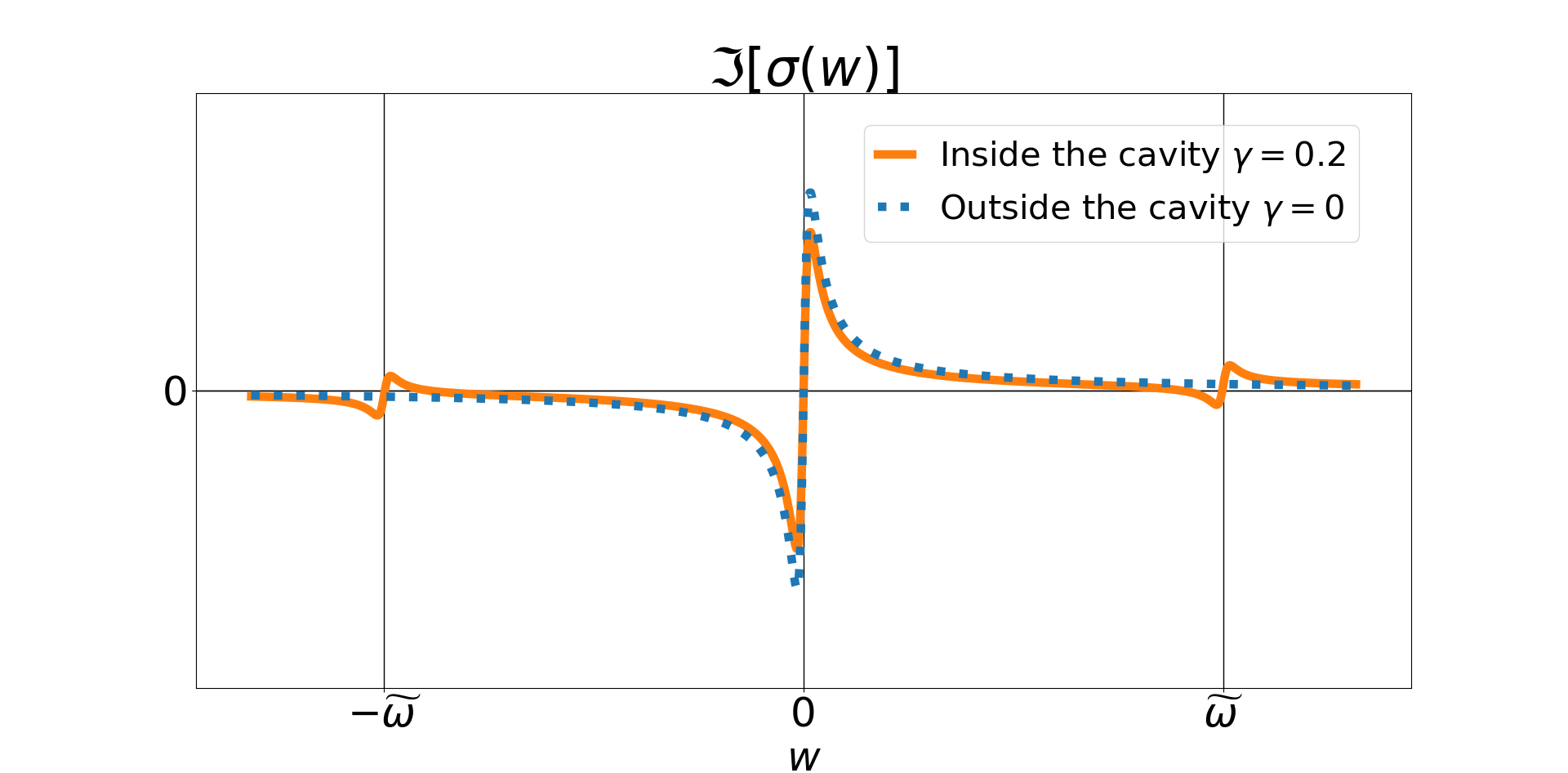}
\caption{\label{Imaginary_Conductivity}Imaginary part $\Im[\sigma(w)]$ of the conductivity $\sigma(w)$ of the 2DEG outside (blue dashed line) and inside (orange solid line) the cavity for coupling $\gamma=0.2$. Inside the cavity the imaginary part of the conductivity of the 2DEG has poles at the plasmon-polariton frequency $w=\pm\widetilde{\omega}$. The peak at $w=0$ gets suppressed by the cavity field. }  
\end{center}
\end{figure}
In the optical conductivity $\sigma(w)$ given by Eq.~(\ref{Conductivity}) there are two contributions. The first one is due to the full electron density $n_{\textrm{e}}$ and comes from the plasma frequency $\omega^2_p=n_{\textrm{e}}e^2/m_{\textrm{e}}\epsilon_0$ and is of second order in $\omega_p$. This is the standard Drude term of the free electron gas~\cite{Vignale}. The second contribution comes from the current-current response function $\chi^J_J(w)$. This contribution is purely due to the cavity field because $\chi^J_J(w)$ is proportional to the $\bi{A}$-field response function $\chi^A_A(w)$. The current-current response $\chi^J_J$ is of fourth order in $\omega_p$ and is a diamagnetic modification to the standard free electron gas conductivity. 

More specifically, both the real and the imaginary part of the optical conductivity, depicted in Figs.~\ref{Real_Conductivity} and~\ref{Imaginary_Conductivity} respectively, exhibit resonances at the plasmon-polariton frequency $w=\pm\widetilde{\omega}$, which modify the optical conductivity of the 2DEG. Most importantly, we see that in the real part of the conductivity the Drude peak~\cite{Basov, BasovHighTc} of the 2DEG at $w=0$ is suppressed by the cavity field due to the higher-order diamagnetic contributions. Since the Drude peak is of great importance for condensed matter systems and materials let us have a closer look at it.

\subsection{Cavity Suppression of the Drude Peak}

The Drude peak is defined as the $w\rightarrow 0$ limit of the real part of the optical conductivity and gives the DC electrical conductivity of a particular system $\sigma_{\textrm{dc}}=\lim_{w\rightarrow 0}\Re[\sigma(w)]$~\cite{Mermin, Basov, Vignale}. In the standard free electron gas (without a cavity) the DC conductivity is $\sigma^0_{\textrm{dc}}=\epsilon_0\omega^2_p/\eta$, which is the first term in Eq.~(\ref{Real and Imaginary sigma}) for $w\rightarrow 0$.

However, in our case we consider the 2DEG coupled to the cavity and we have the extra diamagnetic contributions and we find that the DC conductivity of the electron gas in the cavity depends also on the the collective coupling constant $\gamma$ (defined in Eq.~(\ref{collective coupling}))
\begin{eqnarray}
\sigma_{\textrm{dc}}(\gamma)=\sigma^0_{\textrm{dc}}\left(1-\frac{\gamma}{1+\eta^2/\widetilde{\omega}^2}\right) \;\; \textrm{with}\;\; \eta \rightarrow 0^+.
\end{eqnarray}
Upon neglecting the infinitesimal parameter $\eta$ we find that the DC conductivity in the cavity, i.e., the Drude peak, decreases linearly as function of the coupling $\gamma$
\begin{eqnarray}\label{DC Conductivity}
\sigma_{\textrm{dc}}(\gamma)=\sigma^0_{\textrm{dc}}\left(1-\gamma\right). 
\end{eqnarray}
This is an important result because it shows that by confining 2D materials inside a cavity, the cavity field does not only modify the optical properties of the material, like the optical conductivity, but the cavity can also influence the static electrical conductivity of the material.

The fact that the cavity decreases the conduction of electrons means that the cavity field acts like a viscous medium which slows down the motion of the charged particles. Within this picture the suppression of the Drude peak can also be interpreted as an increase in the effective mass of the electrons due to the cavity photons. From Eq.~(\ref{DC Conductivity}) for the DC conductivity we find that the effective (or renormalized) electron mass is  
\begin{eqnarray}
m_{\textrm{e}}(\gamma)=\frac{m_{\textrm{e}}}{(1-\gamma)}.
\end{eqnarray}
Such an increase of the effective electron mass we will encounter again in section~\ref{Effective QFT} when we will couple the 2DEG to the full continuum of electromagnetic modes.

Finally, it is important to mention that due to the fact that the coupling constant has an upper bound $\gamma <1$ (see Eq.~\ref{collective coupling}) the Drude peak is always positive and the 2DEG is a conductor. However, if $\gamma$ could reach the critical value $1$ (which is forbidden) then the DC conductivity would be zero. This would mean that the cavity can turn the 2DEG from a conductor to an insulator. For $\gamma >1$ the Drude peak becomes negative which means that the system is no longer stable. This explains from the linear-response point of view why the collective coupling $\gamma$ must not exceed the value of 1.

\section{Mixed Responses: Matter-Photon \& Photon-Matter}

What we have done so far is to perform linear response on the photonic and on the electronic sectors of our system separately. In the photonic sector we applied an external current and we computed response functions related to the photon field. This is what is commonly done in the fields of quantum optics and photonics~\cite{cohen1997photons, JacksonEM}. In the electronic sector we perturbed the system  with an external electric field and computed the current response of the system, as it is done in condensed matter physics~\cite{Vignale, Mermin}.

Quantum electrodynamics combines both perspectives under a common unified framework. In addition to the matter-matter and photon-photon responses QED allows to access also cross-correlated response functions, like matter-photon and photon-matter~\cite{ruggenthaler2017b, flick2018light}. The aim of this section is exactly to compute these cross-correlated responses and to investigate how these responses relate to the standard matter-matter and photon-photon response functions.

\subsection{Matter-Photon Response}

First we would like to compute the response of the current $\delta\langle \hat{\bi{J}}(t)\rangle$ due to the external time-dependent current $\bi{J}_{\textrm{ext}}(t)$. The response of the current $\delta\langle \hat{\bi{J}}(t)\rangle$ is defined via Eq.~(\ref{response Observable}) and can be computed from the mixed response function $\chi^J_A(t-t^{\prime})$ 
\begin{eqnarray}
\chi^J_A(t-t^{\prime})=\frac{-\textrm{i}\Theta(t-t^{\prime})}{\hbar }\langle[\hat{\mathbf{J}}_{I}(t),\hat{\mathbf{A}}_{I}(t^{\prime})]\rangle.
\end{eqnarray}
As we explained in the previous section, the contributions of the paramagnetic part of the current is zero. As a consequence only the diamagnetic component contributes. Substituting the expression for the diamagnetic current $\hat{\bi{J}}_{\textrm{d}}$ we find that the mixed response $\chi^{J}_{A}(t-t^{\prime})$ is proportional to the $\bi{A}$-field response function
\begin{eqnarray}
\chi^{J}_{A}(t-t^{\prime})=-\frac{e^2N}{m_{\textrm{e}}}\chi^A_A(t-t^{\prime})
\end{eqnarray}
where $\chi^A_A(t-t^{\prime})$ is given by Eq.~(\ref{A field response}). The previous relation between the two response functions will hold also in the frequency domain 
\begin{eqnarray}\label{ChiJA and ChiAA}
  \chi^{J}_{A}(w)=\left(\frac{-e^2N}{m_{\textrm{e}}}\right)\chi^A_A(w).  
\end{eqnarray}
Finally, we note that the mixed response function $\chi^J_A(w)$ is dimensionless and describes the ratio between the induced current $\delta\langle \hat{\bi{J}}(w)\rangle$ and the external current $\bi{J}_{\textrm{ext}}(w)$ 
\begin{eqnarray}
\delta\langle \hat{\bi{J}}(w)\rangle=\chi^J_A(w)\bi{J}_{\textrm{ext}}(w).
\end{eqnarray}

\subsection{Photon-Matter Response}

Having computed the matter-photon response function $\chi^J_A$ we would also like to compute the photon-matter response function $\chi^A_J$ which corresponds to the inverse physical process. For this, we need to look into the response of the vector potential $\delta\langle \hat{\bi{A}}(t)\rangle$. The response of the vector potential is given by the photon-matter response function $\chi^A_J(t-t^{\prime})$
\begin{eqnarray}
\chi^A_J(t-t^{\prime})=\frac{-\textrm{i}\Theta(t-t^{\prime})}{\hbar }\langle[\hat{\mathbf{A}}_{I}(t),\hat{\mathbf{J}}_{I}(t^{\prime})]\rangle.
\end{eqnarray}
To remain within linear response, we neglect the contribution of $\bi{A}_{\textrm{ext}}(t)$ which would result into higher order corrections. As we already stated the paramagnetic contribution is zero. Substituting the definition for the diamagnetic current $\hat{\bi{J}}_{\textrm{d}}$ we find that the mixed response function $\chi^{A}_{J}(t-t^{\prime})$ is 
\begin{eqnarray}
\chi^{A}_{J}(t-t^{\prime})=-\frac{e^2N}{m_{\textrm{e}}}\chi^{A}_{A}(t-t^{\prime})
\end{eqnarray}
The above relation between the two responses will also hold in the frequency domain
\begin{eqnarray}\label{chiAJ to chiAA}
     \chi^{A}_{J}(w)=\left(\frac{-e^2N}{m_{\textrm{e}}}\right)\chi^{A}_{A}(w).
 \end{eqnarray}
From the equation above we see that the photon-matter response function $\chi^A_J(w)$ is equal to the matter-photon response $\chi^J_A(w)$. Finally, we would like to emphasize that the photon-matter response $\chi^A_J(w)$ describes the dimensionless ratio between the induced $\hat{\bi{A}}$-field and the external field $\bi{A}_{\textrm{ext}}$
\begin{eqnarray}
\delta\langle \hat{\bi{A}}(w)\rangle=\chi^A_J(w)\bi{A}_{\textrm{ext}}(w).
\end{eqnarray}
The above relation implies that the external field gets screened by the internal dynamical system, and the response function $\chi^{A}_{J}(w)$ describes precisely this screening process~\cite{flick2018light}. 

\section{Equivalence Between the Electronic and the Photonic Sector}\label{Duality}

Let us now finally compare the four fundamental response sectors which we introduced and computed throughout this chapter, the photon-photon, matter-matter, photon-matter and matter-photon. From all the responses in the different sectors we can can construct the following response matrix
\begin{eqnarray}\label{Response Table}
    \left(\begin{tabular}{c}
		$\delta\langle \hat{\bi{J}}(w)\rangle$ \\
	    $\delta\langle\hat{\bi{A}}(w)\rangle$  
	\end{tabular}\right)=\left(\begin{tabular}{ c c }
		$\chi^J_{J}(w)$ &$ \chi^J_{A}(w)$ \\
	    $\chi^A_{J}(w)$ &$ \chi^A_{A}(w)$ 
	\end{tabular}\right) \left(\begin{tabular}{c}
		$\bi{A}_{\textrm{ext}}(w)$ \\
	    $\bi{J}_{\textrm{ext}}(w)$  
	\end{tabular}\right)\nonumber\\
\end{eqnarray}
which summarizes all the fundamental responses of the system. Using now the Eqs.~(\ref{chiJJ to chiAA}), (\ref{chiAJ to chiAA}) and~(\ref{ChiJA and ChiAA}) which give the response functions $\chi^J_J(w)$, $\chi^J_A(w)$ and $\chi^A_{J}(w)$ respectively, we see that all responses are proportional to the $\bi{A}$-field response function $\chi^A_A(w)$. Thus, all elements of the response table can be written in terms of $\chi^A_A(w)$
\begin{eqnarray}\label{Response Table chiAA}
    \left(\begin{tabular}{c}
		$\delta\langle \hat{\bi{J}}(w)\rangle$ \\
	    $\delta\langle\hat{\bi{A}}(w)\rangle$  
	\end{tabular}\right)=\chi^A_{A}(w)\left(\begin{tabular}{ c c }
		$\left(e^2N/m_{\textrm{e}}\right)^2$ &\; $ -e^2N/m_{\textrm{e}}$ \\\\
	    $-e^2N/m_{\textrm{e}}$ &$1$ 
	\end{tabular}\right) \left(\begin{tabular}{c}
		$\bi{A}_{\textrm{ext}}(w)$ \\
	    $\bi{J}_{\textrm{ext}}(w)$  
	\end{tabular}\right).
\end{eqnarray}
The fact that all response functions are proportional to $\chi^A_A(w)$ means that all response functions have precisely the same pole structure. This demonstrates a fundamental relation between the two sectors of the theory, namely that the photonic and the electronic sectors have exactly the same excitations and resonances. From an experimental point of view this means, that in an experiment perturbing an interacting light-matter system with an external time-dependent current (which couples to the photon field) and perturbing with an external electric field (which couples to the current) would result in exactly the same information about the excitations and the resonances of the interacting system.

Moreover, from the response table in Eq.~(\ref{Response Table chiAA}) we see that the current-current response scales quadratically with the number of electrons $\chi^J_J(w)\sim N^2\chi^A_A(w)$, while the mixed responses linearly $\chi^J_A(w)=\chi^A_J(w)\sim N \chi^A_A(w)$. The photon-photon response function $\chi^A_A(w)$ given by Eq.~(\ref{Re Im A-field}) depends also on the area of the 2DEG as $1/S$. As a consequence in the large $N,S$ limit only the responses involving matter ($\chi^J_J, \chi^J_A, \chi^A_J$) are finite, due to the dependence on $N$, while $\chi^A_A$ strictly speaking goes to zero. This is the same behavior that shows up also for the energy densities of the two sectors as mentioned in section~\ref{Ground State}. Again, this hints towards the fact that, to have a finite photon-photon response, we need to include an infinite amount of modes by treating the photon field in the continuum.

 Lastly, we would like to emphasize that all response functions we computed depend on the arbitrarily small auxiliary parameter $\eta$ which needs to be introduced in linear response to have a well-defined Fourier transform~\cite{Vignale, flick2018light}. However, in the limit $\eta\rightarrow0^+$ the response functions go to zero (see for example Eq.~(\ref{Re Im A-field})) except of the frequencies $w=\pm\widetilde{\omega}$ where they diverge. This means that the broadening parameter $\eta$ works like a regulator which spreads the resonance over a finite range of frequencies and describes the coupling of the system to an artificial environment and the dissipation of energy to this environment~\cite{Vignale}. To get rid of the artificial broadening parameter $\eta$, one has to treat matter and photons on an equal footing and perform the continuum-limit also for the electromagnetic field. This is a direction which we explore in the next chapter, and as we will see allows for the description of absorption and dissipation without the need of any artificial broadening. 

\chapter{Effective Quantum Field Theory in the Continuum}\label{Effective QFT}
\begin{displayquote}
\footnotesize{In practice, quantum field theory is marvelously good for calculating answers to many physics questions. The answers involve approximations. These approximations seem to work very well: that is, they [produce] answers that match experiments. Unfortunately we do not fully understand, in a mathematically rigorous way, what these approximations are supposed to be approximating. }
\end{displayquote}
\begin{flushright}
  \footnotesize{John C.~Baez\\
Struggles with the Continuum~\cite{BaezContinuum}}
\end{flushright}

So far we have studied in generality the behavior of the 2DEG coupled to the cavity, in the large $N$ limit for the electronic sector, with the cavity photon-field being treated in the single-mode approximation. In quantum optics and cavity QED~\cite{faisal1987, cohen1997photons} the single-mode approximation has been proven very insightful and extremely successful for the description of a wide range of interacting light-matter systems~\cite{kockum2019ultrastrong}. 

However, the single-mode approximation is far from providing a complete description of the interaction between matter and the photon field, which contains an infinite number of degrees of freedom (photon-modes). Moreover, as it is known since the early times of the quantum theory of radiation, and the seminal work of Einstein~\cite{Einstein:1917zz}, to describe even one of the most fundamental processes of light-matter interaction like spontaneous emission, the full continuum of modes of the electromagnetic field needs to be considered. In addition, we should always bear in mind that in a cavity of course a particular set of modes of the photon field are selected (due to the cavity-environment), but it is never the case that only a single mode of the photon field contributes to the light-matter coupling. The single-mode models like the Rabi, the Jaynes-Cummings and the Dicke model, describe effectively (using an effective coupling) the exchange of energy between matter and the photon field as if there were only a single mode coupled to matter~\cite{harochekleppner}.  

 For our system this problem becomes even more severe because we consider a macroscopic system like the 2DEG, where the propagation of the in-plane modes becomes important. This implies that it is of utmost importance to include the 2D continuum of modes into the description of the electromagnetic field. To make the argument even clearer we would like to give some further justifications on why such a theory in the continuum is necessary in order to describe and capture, particular observables, physical processes, and effects of interest for our system.
 
 From the point of view of observables and physical processes the main reasons are:
 
 \begin{itemize}
    \item In section~\ref{Ground State} we highlighted that the contribution of the single-mode photon field to the ground-state energy density $E_{p}/S$ in the thermodynamic limit ($N,S\rightarrow\infty$) becomes arbitrary small and tends to zero. This means that in the single-mode case the contribution of the cavity field to the ground-state energy of the system is negligible, due the discrepancy between the amount of the electrons and the amount of photon-modes.
    
    \item Since the contribution of the cavity-field to the ground-state energy density is zero (in the large $N,S$ limit) no real contribution to the renormalized or effective mass of the electron can occur. This of course is true because we consider a single-mode photon field, and as it known from QED, mass renormalization effects show up when electrons couple to the full continuum  of electromagnetic modes~\cite{Weinberg, Srednicki, Frohlich2010, CHEN20082555, Mandl}.
     
     \item As we emphasized in the end of subsection~\ref{Duality}, absorption processes and dissipation can only be described consistently when a continuum of modes is taken into account~\cite{Vignale}.

     \item In the single-mode approximation, no macroscopic forces can appear between the cavity mirrors, like the Casimir-Polder forces~\cite{casimir1948influence}. As it is well known from the literature, such forces show up only when the electromagnetic field is treated in the full continuum~\cite{buhmann2013dispersionI, buhmann2013dispersionII}. 
 \end{itemize}

    For all these reasons we proceed with the construction of an effective quantum field theory in order to treat the photon field in the continuum.

\section{Effective Field Theory, Coupling and Cutoff}\label{Cutoff}

In order to promote the single-mode theory to a quantum field theory, we need to perform the continuum limit for the photon field and sum over all the in-plane photon-modes. For the free electron gas this procedure can be performed for an arbitrary amount of photon-modes in full generality, as we show in appendix~\ref{Mode-Mode Interactions}. However, this treatment would make the theory analytically non-solvable, especially in the thermodynamic (continuum) limit. 

For this reason, we will follow a different approach and we will perform the summation over the photon-modes in an effective way. What is meant by this is, that we will neglect the mode-mode interactions and we will integrate the single mode spectrum of Eq.~(\ref{eigenspectrum}) over all the in-plane modes. This will allow us to construct an analytically solvable effective quantum field theory, in the continuum, for both light and matter. We note that the validity of the approximation to neglect the mode-mode interactions depends on how large the diamagnetic shift $\omega_p$~\cite{faisal1987} is. 

For the construction of our effective quantum field theory, first we introduce back the dependence on the photonic momenta $\bm{\kappa}=(2\pi n_x/L,2\pi n_y/L,\pi n_z/L_z)$ of all the parameters in the theory. The bare frequencies $\omega$ of the quantized electromagnetic field in terms of the momenta $\bm{\kappa}$ are $\omega(\bm{\kappa})=c|\bm{\kappa}|$. This applies also to the dressed frequency $\widetilde{\omega}=\sqrt{\omega^2+\omega^2_p}$  which gets replaced by $\widetilde{\omega}(\bm{\kappa})=\sqrt{\omega^2(\bm{\kappa})+\omega^2_p}$. Subsequently, the single-mode coupling constant $\gamma=\omega^2_p/\widetilde{\omega}^2$ becomes $\bm{\kappa}$-dependent $\gamma(\bm{\kappa})=\omega^2_p/\widetilde{\omega}^2(\bm{\kappa})$.

With these substitutions and summing the single-mode energy spectrum of Eq.~(\ref{eigenspectrum}) over all the in-plane photon-momenta $(\kappa_x,\kappa_y)$, we obtain the analytic expression for the ground state energy (where $n_{\lambda}=0$ for both $\lambda=1,2$) of the effective quantum field theory 
\begin{eqnarray}\label{effective energy}
E_{\mathbf{k}}(\Lambda)=\frac{\hbar^2}{2m_{\textrm{e}}} \left[\sum\limits^{N}_{j=1}\mathbf{k}^2_j-\left(\sum^{\Lambda}_{\kappa_x,\kappa_y}\gamma(\bm{\kappa})\right)\frac{1}{N}\sum^2_{\lambda=1}\left(\bm{\varepsilon}_{\lambda}\cdot \mathbf{K}\right)^2\right]+\sum^{\Lambda}_{\kappa_{x},\kappa_{y}}\hbar\widetilde{\omega}(\bm{\kappa}).
\end{eqnarray}
\begin{figure}[h]
\begin{center}
\includegraphics[height=6.5cm,width=7cm]{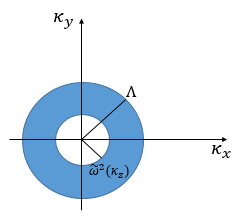}
\caption{\label{Photon Continuum}Graphic representation of the frequency range on which the photon field is defined in the effective field theory. The natural lower cutoff of the theory is $\widetilde{\omega}^2(\kappa_z)$, while the highest allowed frequency is $\Lambda$. }
\end{center}
\end{figure}
In the energy expression of the effective theory we introduced the cutoff $\Lambda$ which defines the highest frequency that we allow for the photon field, as it is also shown in Fig.~\ref{Photon Continuum}. Such a cutoff is necessary for effective field theories and it is commonly introduced also in QED~\cite{greiner1996, spohn2004}. The sum over the in-plane single-mode coupling constants $\gamma(\bm{\kappa})$ defines the effective coupling constant $g(\Lambda)$ in the effective field theory
\begin{eqnarray}
    g(\Lambda)=\sum^{\Lambda}_{\kappa_x,\kappa_y}\gamma(\bm{\kappa})=\frac{e^2N}{\epsilon_0 m_{\textrm{e}}L_z}\frac{1}{S}\sum^{\Lambda}_{\kappa_x,\kappa_y}\frac{1}{\omega^2(\bm{\kappa})+\omega^2_p}.\nonumber\\
\end{eqnarray}
In the limit where the area of the cavity mirrors becomes very large, $S \rightarrow \infty$, the momenta $(\kappa_x,\kappa_y)$ of the photon field become continuous variables and the sum turns into an integral
\begin{eqnarray}\label{effectivecoupling}
    g(\Lambda)=\frac{e^2N}{\epsilon_0 m_{\textrm{e}}L_z}\frac{1}{4\pi^2}\iint\limits^{\Lambda}_{0}\frac{d \kappa_xd\kappa_y}{c^2\bm{\kappa}^2+\omega^2_p}=N\alpha\ln\left(\frac{\Lambda}{\widetilde{\omega}^2(\kappa_z)}\right)\nonumber\\
\end{eqnarray}
where we introduced the parameters
\begin{eqnarray}\label{alpha parameter}
    \alpha=\frac{e^2}{4\pi c^2\epsilon_0 m_{\textrm{e}}L_z} \;\; \textrm{and}\;\; \widetilde{\omega}^2(\kappa_z)=c^2\kappa^2_z+\omega^2_p,
\end{eqnarray}
and the momentum $\kappa_z=\pi/L_z$ (for $n_z=1$) depends on the distance between the cavity mirrors $L_z$ (see Fig.~\ref{HEG_Cavity}). We note that the parameter $\alpha$ in Eq.~(\ref{alpha parameter}) is dimensionless. After substituting the value for all the fundamental constants we find for the dimensionless parameter $\alpha=2.81\times 10^{-15}\textrm{m}/L_z$.

Here comes a crucial point, the effective coupling $g(\Lambda)$ in Eq.~(\ref{effectivecoupling}) exhibits a linear dependence on the number of electrons $N$. It is important to emphasize that this explicit dependence of the effective coupling on the number of electrons shows up due to dipolar coupling, i.e., because in the effective theory we couple all modes to all particles in the same way. However, beyond the dipole approximation, in QED we know that each mode has a spatial profile which means that each mode couples to the local charge density and not to the full amount of charges. This is a second point in which the effectiveness of our field theory becomes manifest. This has important implications because in the thermodynamic limit $N\rightarrow \infty$ the effective coupling $g(\Lambda)$ becomes arbitrarily large. Despite this fact, for the effective coupling $g(\Lambda)$ rigorous conditions can be derived under which the effective coupling constant remains finite, and the effective theory is stable and well defined.  

In section~\ref{Ground State} we found the ground state of the 2DEG coupled to the cavity, in the thermodynamic limit, for all values of the single-mode coupling $\gamma$. More precisely we proved that for $\gamma<1$ the system has a stable ground state, while for $\gamma>1$ (which is in principle forbidden see Eq.~(\ref{collective coupling}) for $\gamma$) the system becomes unstable and has no ground state. Having promoted the single mode theory into an effective field theory, we still need to guarantee the stability of the electron-photon system by forbidding the effective coupling to exceed 1, $ 0 \leq g(\Lambda) \leq 1$. Given this condition and the definition of the effective coupling $g(\Lambda)$ in Eq.~(\ref{effectivecoupling}) we obtain the allowed values for the upper cutoff $\Lambda$ in the effective theory 
\begin{eqnarray}\label{Cutoffrange}
    \widetilde{\omega}^2(\kappa_z)\leq \Lambda \leq \widetilde{\omega}^2(\kappa_z)e^{1/N\alpha}.
\end{eqnarray}
From the above inequality we see that the highest allowed momentum for the photon field is $\widetilde{\omega}^2(\kappa_z)e^{1/N\alpha}$. Beyond this value the effective coupling $g(\Lambda)$ becomes larger than 1 and the electron-photon system becomes unstable. In relativistic QED the finite energy scale (or momentum) for which the theory diverges is known as the Landau pole~\cite{Srednicki}. Due to this historic reason we will also refer to the highest allowed momentum in our effective theory as the Landau pole 
\begin{eqnarray}\label{Landaupole}
    \Lambda_{\textrm{pole}}=\widetilde{\omega}^2(\kappa_z)e^{1/N\alpha}.
\end{eqnarray}
Further, from Eq.~(\ref{Cutoffrange}) we see that the upper cutoff $\Lambda$ is a multiple of the lower natural cutoff $\widetilde{\omega}^2(\kappa_z)$. This implies that we can define $\Lambda$ with the use of a dimensionless parameter $\Lambda_0$ 
\begin{eqnarray}\label{Lambda0}
    \Lambda=\widetilde{\omega}^2(\kappa_z)\Lambda_0\;\; \textrm{with}\;\; 1\leq \Lambda_0\leq e^{1/N\alpha}.
\end{eqnarray}
The range chosen above for $\Lambda_0$ guarantees that the effective coupling stays in the desired regime $0\leq g(\Lambda)\leq 1$ and our effective theory remains stable. To gain a more thorough understanding of our effective theory, we would also like to investigate the infrared (IR) and the ultraviolet (UV) behavior of the effective coupling constant $g(\Lambda)$. From the expression of $g(\Lambda)$ in Eq.~(\ref{effectivecoupling}) we see that $g(\Lambda)$ diverges if we allow the cutoff to go to infinity. This means that our theory is UV divergent. This is the logarithmic divergence of QED which is known to exist for both relativistic and non-relativistic QED~\cite{Weinberg, Srednicki, greiner1996, spohn2004, HiroshimaSpohn}. However, the effective coupling $g(\Lambda)$ has no IR divergence because for arbitrarily small momenta $\kappa_z=\pi/L_z$ the coupling goes to zero due to the parameter $\alpha$. Our theory is IR divergent-free is due to the diamagnetic shift $\omega_p$ in Eq.~(\ref{effectivecoupling}) which defines the natural lower cutoff of the photon field~\cite{rokaj2019}. As we have already seen, the diamagnetic shift originates from the $\bi{A}^2$ term in the Pauli-Fierz Hamiltonian. Thus, we conclude that the diamagnetic term is of major importance, because it makes non-relativistic QED IR divergent-free, while relativistic QED suffers from both UV and IR divergences.

\section{Renormalized \& Effective Mass}
\begin{displayquote}
\footnotesize{He often asked, ``How do you know that the mass in the Dirac equation is the same as the experimental mass?'' }
\end{displayquote}
\begin{flushright}
  \footnotesize{M. Dresden about H.~A.~Kramers\\
H.~A.~Kramers Between Tradition and Revolution~\cite{Dresden}}
\end{flushright}

From relativistic QED it is known that when electrons interact with the full electromagnetic vacuum (continuum of modes) the mass and the charge of the electron need to be renormalized. These renormalization effects lead to observable radiative corrections like the vacuum polarization, the anomalous magnetic moment and the Lamb shift~\cite{Weinberg, Mandl}. On the other hand, in non-relativistic QED there is no need to renormalize the charge of the electrons due to the elimination of the positrons from the theory~\cite{HiroshimaSpohn}. However, mass renormalization effects show up, and this is precisely what we are interested in here. Namely, the renormalization of the electron mass due the interaction with the continuum of modes of the cavity.
\begin{figure}[h]
\begin{center}
  \includegraphics[height=5cm,width=6cm]{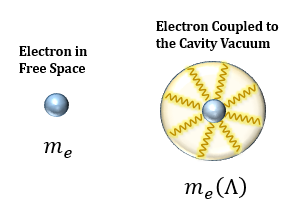}
\caption{\label{Penormalized Mass}Schematic depiction of an electron in free space with mass $m_{\textrm{e}}$ and an electron coupled to the full electromagnetic vacuum. The virtual photons of the cavity ``dress'' the electron and renormalize the electron mass $m_{\textrm{e}}(\Lambda)$. }  
\end{center}
\end{figure}

In general, to compute the renormalized mass of the electron is a rather laborious task. In most cases it is performed perturbatively with methods ranging from dimensional regularization~\cite{Mandl}, renormalization group techniques~\cite{Srednicki, Wilson, Weinberg} or causal perturbation theory~\cite{EpsteinGlaser}. In the non-relativistic theory the renormalized (or effective) mass in the case of free particles, is defined by the curvature of the energy dispersion around $\bi{k}=0$ and more precisely by the formula~\cite{Frohlich2010, CHEN20082555}
\begin{eqnarray}\label{Def Renormalized Mass}
    m_{\textrm{e}}(\Lambda)=\left(\frac{1}{\hbar^2}\frac{\partial^2 E_{\bi{k}}(\Lambda)}{\partial \bi{k}^2_i}\right)^{-1},
\end{eqnarray}
 where $E_{\bi{k}}(\Lambda)$ is the energy dispersion of the system, which depends on the momenta of the electrons and the cutoff of the theory.

\begin{figure}[H]
\begin{center}
 \includegraphics[width=0.5\linewidth]{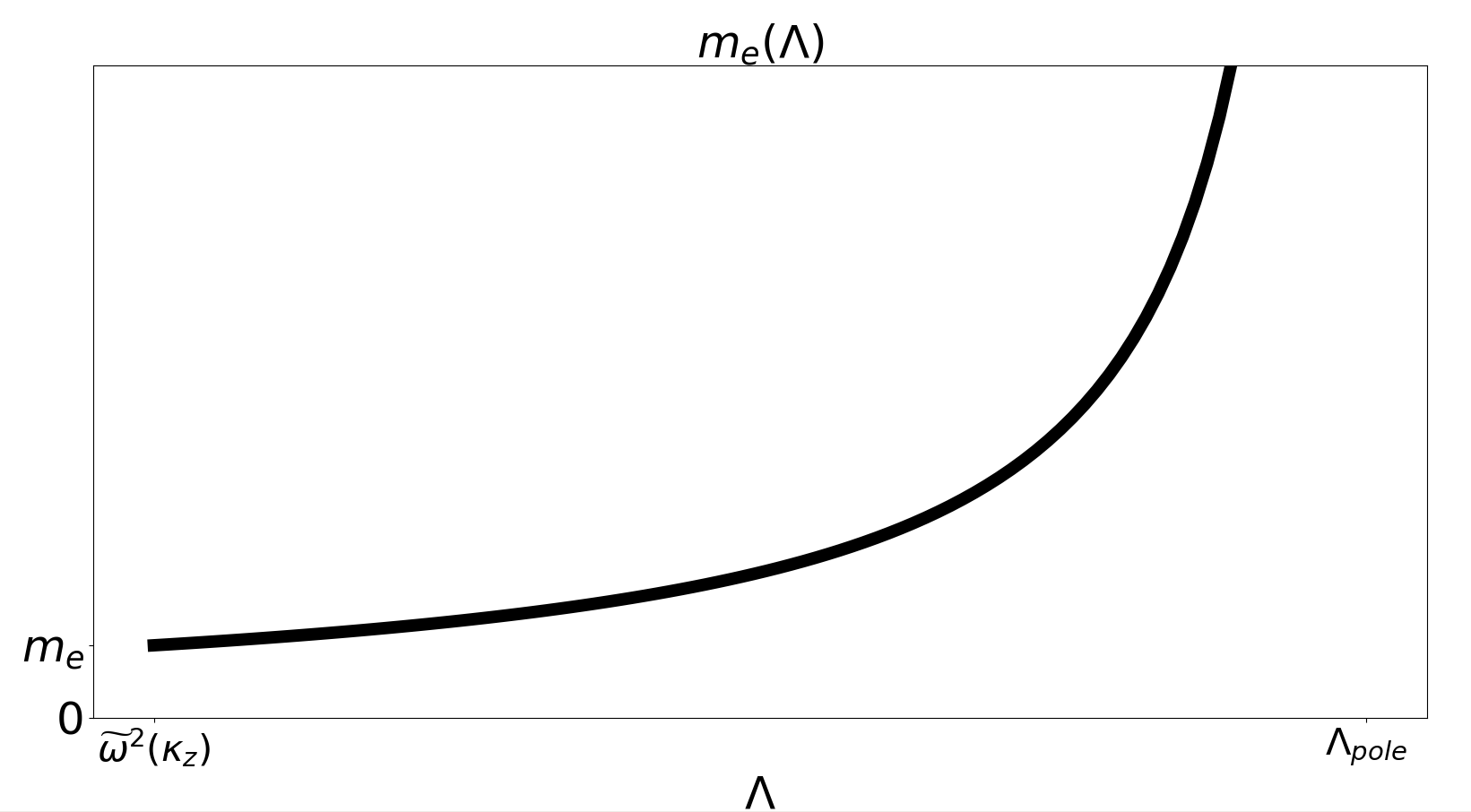}
\caption{\label{mass_running} Plot of the renormalized mass $m_{\textrm{e}}(\Lambda)$ as a function of the upper cutoff $\Lambda$. If upper cutoff $\Lambda$ is equal to the lower cutoff $m_{\textrm{e}}(\Lambda)$ is equal to the free-space mass $m_{\textrm{e}}$. As $\Lambda$ increases the renormalized mass $m_{\textrm{e}}(\Lambda)$ increases and eventually goes to infinity for $\Lambda=\Lambda_{\textrm{pole}}$. }   
\end{center}
\end{figure}

However in our case, because we have an analytic expression for the energy spectrum $E_{\bi{k}}(\Lambda)$ of the effective theory given by Eq.~(\ref{effective energy}), we do not have to use any of the (mainly) perturbative techniques that we mentioned before, but we can straightforwardly use the definition for $m_{\textrm{e}}(\Lambda)$ given in Eq.~(\ref{Def Renormalized Mass}). Thus, we find the following analytic expression for the renormalized electron mass 
\begin{eqnarray}\label{Renormalized Mass}
    m_{\textrm{e}}(\Lambda)=m_{\textrm{e}}\left(1-\alpha\ln\left(\frac{\Lambda}{\widetilde{\omega}^2(\kappa_z)}\right)\right)^{-1}.
\end{eqnarray}
From the above result we conclude that the renormalized electron mass $m_{\textrm{e}}(\Lambda)$ is larger than the electron mass in free space $m_{\textrm{e}}$ and increases as a function of the upper cutoff $\Lambda$. This behavior is in agreement with results coming from both relativistic and non-relativistic QED~\cite{Srednicki, Weinberg, Frohlich2010, CHEN20082555, HiroshimaSpohn}. In the range given by Eq.~(\ref{Cutoffrange}) the renormalized mass is positive and the effective theory is well-defined (see Fig.~\ref{mass_running}). If the cutoff goes beyond the pole $\Lambda_{\textrm{pole}}$ the renormalized mass becomes negative and this signifies that the theory becomes unstable. This is in analogy to the single-mode case, when the collective coupling constant $\gamma$ becomes larger than 1. This explains from another point of view why the upper cutoff $\Lambda$ must not exceed the pole $\Lambda_{\textrm{pole}}$. In the limit where the upper cutoff $\Lambda$ is equal to the lower cutoff $\widetilde{\omega}^2(\kappa_z)$ the renormalized mass $m_{\textrm{e}}(\Lambda)$ is equal to the bare electron mass $m_{\textrm{e}}$ (see Fig.~\ref{mass_running}).

In addition, from Eq.~(\ref{Renormalized Mass}) we see that $m_{\textrm{e}}(\Lambda)$ depends also on the electron density inside the cavity $n_{\textrm{e}}$ via the dressed frequency $\widetilde{\omega}(\kappa_z)$ given by Eq.~(\ref{alpha parameter}). This means that there is a many-body effect in the renormalized mass $m_{\textrm{e}}(\Lambda)$. This many-body effect shows up because we are considering a many-body system consisting of $N$ free electrons coupled to the photon field and our treatment is non-perturbative. We note that such a many-body effect does not show up for the usual single-particle mass renormalization~\cite{Mandl, BetheRenorm} and is potentially very small for any finite system, but clearly not for extended systems like a 2DEG. To the best of our knowledge such a many-body effect for the renormalized electron mass has not been reported before. 
  \begin{figure}[H]
\begin{center}
 \includegraphics[width=0.7\linewidth]{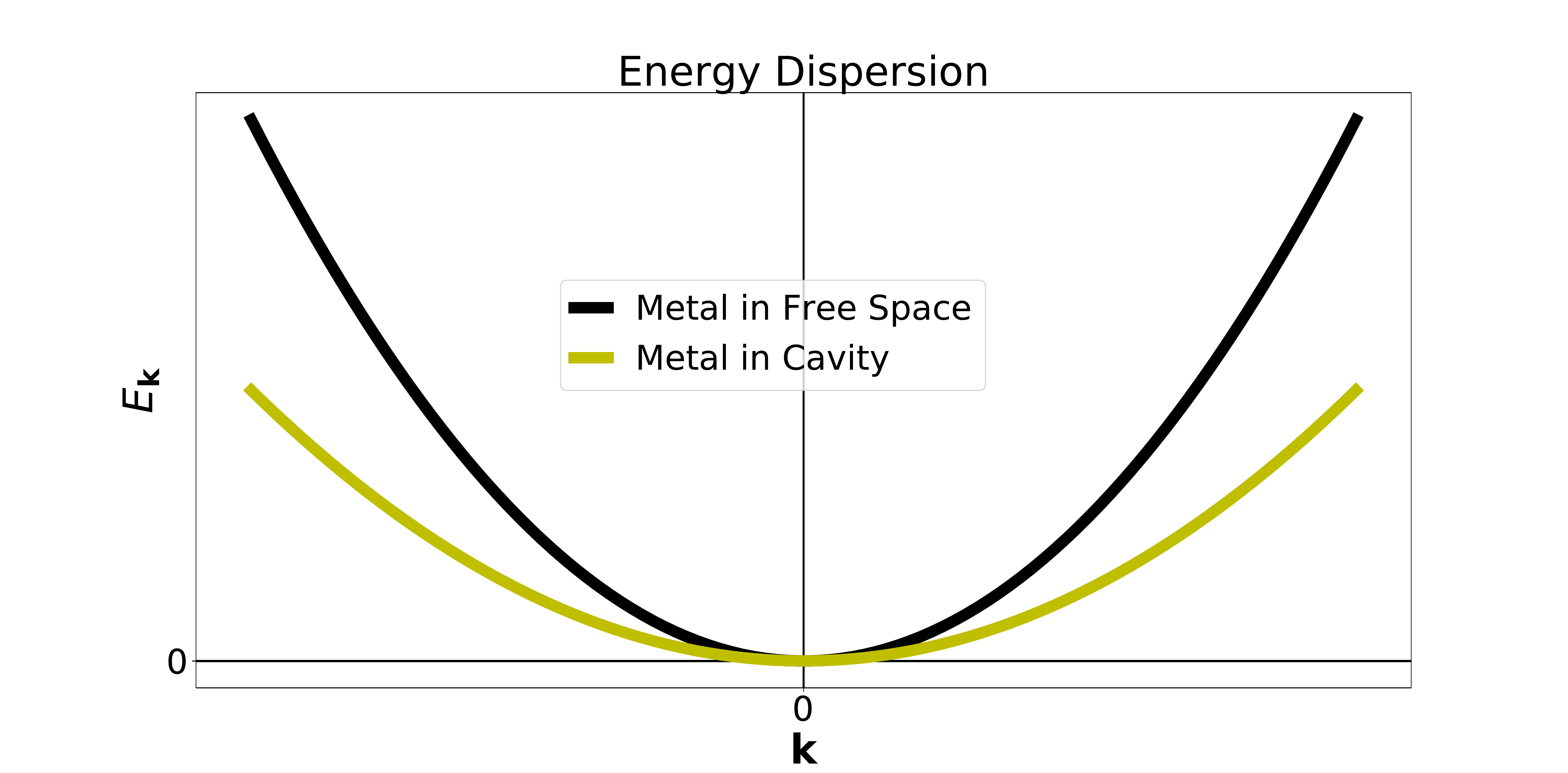}
\caption{\label{Parabola Renormalization} Energy dispersion for electrons in a metal outside (black parabola) and inside a cavity (yellow parabola). From the curvature of the parabolas the effective electron mass in the respective environment can be obtained. The dispersion of the electrons inside the cavity is less steep, because the electron mass is larger than in free space, due the cavity photons. }   
\end{center}
\end{figure}

Finally, we would like to emphasize that the renormalization of the electron mass has experimental implications and can be measured experimentally. This can be done by measuring the dispersion of the electrons in a metallic material (which has a parabolic band dispersion) with angle-resolved photoemission spectroscopy (ARPES)~\cite{Damascelli_2004} inside and outside the cavity. From the curvature of the parabolas around $\bi{k}=0$ one can obtain the effective mass inside and outside the cavity and the ratio between them (see also Fig.~\ref{Parabola Renormalization}). Having the ratio $m_{\textrm{e}}(\Lambda)/m_{\textrm{e}}$ and the analytic formula given by Eq.~(\ref{Renormalized Mass}) one can deduce directly what the highest momentum (the cutoff) $\Lambda$ is to which the electrons couple to. Finally, using Eq.~(\ref{effectivecoupling}), we can also find what the value for the electron-photon coupling $g(\Lambda)$ is in the effective quantum field theory. This provides a novel way to measure the electron-photon coupling for extended systems in cavity QED, which goes beyond standard quantum optics models, via the effective mass. 

\section{Modification of the Fermi Liquid Quasi-particle Excitations}

We would like to proceed now by demonstrating some further implications of our effective quantum field theory. In section~\ref{Ground State} we showed that the ground-state distribution of the electrons in $\bi{k}$-space is the standard 2D Fermi sphere and the electrons in the ground-state occupy all single particle states with momenta less than the Fermi momentum $|\bi{p}_{\textrm{F}}|=|\hbar \bi{k}_{\textrm{F}}|$. This means that our system is a Fermi liquid~\cite{LandauFermiLiquid, Nozieres, Baym}. 

What we are interested in here is to investigate how the photon field can alter the behavior of the fermionic quasiparticle excitations of the Fermi liquid. The fundamental fermionic quasi-particle excitations are generated by adding electrons with momentum greater than the Fermi momentum $|\bi{p}_{\textrm{F}}|$~\cite{Nozieres, Baym}. The energy of the quasi-particle at the Fermi surface is
\begin{eqnarray}
    \mu= E_{\bi{k}}(\Lambda, N+1)-E_{\bi{k}}(\Lambda, N),
\end{eqnarray}
where $E_{\bi{k}}(\Lambda, N)$ is the ground-state energy of the system for $N$ electrons distributed on the 2D Fermi sphere with their wavevectors in the region $0\leq\bi{k}<\bi{k}_{\textrm{F}}$ and $E_{\bi{k}}(\Lambda, N+1)$ is the energy of the system containing one more electron with $\bi{k}=\bi{k}_{\textrm{F}}$. In the ground-state, because the electrons are distributed on the Fermi sphere, the collective momentum is zero, $\bi{K}=0$. As a consequence in the energy $E_{\bi{k}}(\Lambda, N)$ the negative term which depends on the collective momentum  does not contribute. However, in the $N+1$ electron-state this is not the case, because the last electron added on the Fermi surface with $\bi{k}=\bi{k}_{\textrm{F}}$ introduces a non-zero collective momentum which now gives a non-trivial contribution to the energy spectrum. Thus, we find that the quasi-particle excitation energy at the Fermi surface is
\begin{eqnarray}
    \mu= \frac{\hbar^2}{2m_{\textrm{e}}}\left[\bi{k}^2_{\textrm{F}}-\alpha\ln\left(\frac{\Lambda}{\widetilde{\omega}^2(\kappa_z)}\right)\sum^2_{\lambda=1}\left(\bm{\varepsilon}_{\lambda}\cdot \bi{k}_{\textrm{F}}\right)^2\right].
\end{eqnarray} 
We note that the quasi-particle excitation at the Fermi surface $\mu$ is also known as the chemical potential. To obtain the above result we made use of the fact that the effective coupling per particle is $g(\Lambda)/(N+1)=\alpha\ln\left(\Lambda/\widetilde{\omega}^2(\kappa_z)\right)$, as given by Eq.~(\ref{effectivecoupling}), and that the polarization vectors are orthogonal. In addition, by introducing the renormalized mass $m_{\textrm{e}}(\Lambda)$ the chemical potential takes the form
\begin{eqnarray}\label{chemical potential}
    \mu=\frac{\hbar^2\bi{k}^2_{\textrm{F}}}{2m_{\textrm{e}}(\Lambda)}.
\end{eqnarray}
From this result we see that the chemical potential is a function of the upper cutoff of the photon field $\Lambda$, and particularly depends on the renormalized electron mass $m_{\textrm{e}}(\Lambda)$ given by Eq.~(\ref{Renormalized Mass}). This means that the photon field in the continuum modifies the chemical potential. Further, in Fermi liquid theory, the quasi-particle excitations in the neighborhood of the Fermi surface depend on the chemical potential and are given by the expression~\cite{Baym}
\begin{eqnarray}
    \epsilon^x_{\bi{k}}=\mu+\hbar v_{\textrm{F}}(\bi{k}-\bi{k}_{\textrm{F}})=\frac{\hbar^2\bi{k}^2_{\textrm{F}}}{2m_{\textrm{e}}(\Lambda)}+\hbar v_{\textrm{F}}(\bi{k}-\bi{k}_{\textrm{F}})\nonumber\\
\end{eqnarray}
where $v_{\textrm{F}}$ is the Fermi momentum at the Fermi surface. From the fact that the quasiparticle excitations of the Fermi liquid depend on the chemical potential we see that they also get modified and depend on the renormalized electron mass $m_{\textrm{e}}(\Lambda)$. This shows that our effective quantum field theory has direct implications for Fermi liquid theory. Finally, we would like to mention that in the limit where the upper cutoff goes to the lower cutoff, $\Lambda \rightarrow \widetilde{\omega}^2(\kappa_z)$, the renormalized mass becomes equal to the bare electron mass, $m_{\textrm{e}}(\Lambda) \rightarrow m_{\textrm{e}}$, and in this case the quasi-particle excitations do not get modified. This explains from another viewpoint why in the single-mode theory there is no fundamental modification of the properties of the Fermi liquid and why it is important to include the full continuum of electromagnetic modes for the description of the the photon field.

\section{Zero-Point Energy \& Casimir Force}\label{Repulsive Casimir Forces}
  
 In the single-mode case the zero-point energy of the photon field was $E_p=\hbar\widetilde{\omega}$, and as it was explained in section~\ref{Ground State} the corresponding energy density $E_p/S$ in the thermodynamic limit is negligible. The question that naturally arises now is: what is the zero-point energy of the photon field in the continuum, particularly for the effective quantum field theory that we constructed?
 
The zero-point energy of the photon field (in the continuum) is known to be responsible for the emergence of forces like the interatomic van der Waals forces, the Casimir-Polder forces between an atom and a macroscopic body~\cite{casimir1948influence, buhmann2013dispersionI}, and the Casimir force between two parallel conducting plates~\cite{Casimir:1948dh}. Here, we are considering a 2D material inside a cavity and consequently we fall in the third category, which means that the macroscopic forces in our system will be Casimir forces. To find the Casimir force between the mirrors of the cavity we have to compute the zero-point energy of the photon field $E_p $ per area $S$. From the expression for energy spectrum of the effective theory in Eq.~(\ref{effective energy}) we find that the ground-state energy ($n_{\lambda}=0$) per area is
  \begin{eqnarray}
     \frac{E_p}{S}=\frac{1}{S}\sum_{\kappa_x,\kappa_y}\hbar\widetilde{\omega}(\bm{\kappa})=\frac{\hbar}{4\pi^2}\iint\limits^{\Lambda}_{0}d\kappa_xd\kappa_y\sqrt{c^2\bm{\kappa}^2+\omega^2_p}.
 \end{eqnarray}
In the expression above we also took the limit $S\rightarrow\infty$ in which the sum gets promoted into an integral. To perform the integration above we go to polar coordinates and we find the photon energy per area
 \begin{eqnarray}
     \frac{E_p}{S}=\frac{\hbar(\Lambda^{3/2}_0-1)}{6\pi c^2}\widetilde{\omega}^3(\kappa_z),
 \end{eqnarray}
where to obtain the above result we used Eq.~(\ref{Lambda0}) which gives $\Lambda$ in terms of $\Lambda_0$. Further, we use the expression for $\widetilde{\omega}(\kappa_z)$ given by Eq.~(\ref{alpha parameter}) and we compute the derivative of the photon energy density $E_p/S$ with respect to the distance of the cavity mirrors $L_z$ and we obtain the force per area (the pressure) 
 \begin{equation}\label{Casimir force}
     \frac{F_{\textrm{c}}}{S}=-\frac{\partial(E_p/S)}{\partial L_z}=\frac{\hbar(\Lambda^{3/2}_0-1)}{4\pi c^2}\left(\frac{2\pi^2 c^2}{L^3_z}+\frac{e^2n_{\textrm{2D}}}{m_{\textrm{e}}\epsilon_0 L^2_z}\right)\sqrt{\frac{\pi^2 c^2}{L^2_z}+\frac{e^2n_{\textrm{2D}}}{m_{\textrm{e}}\epsilon_0 L_z}}.
 \end{equation}
We would like to mention that to obtain the result above the dependence of $\omega_p$ on the distance between the cavity mirrors $L_z$, as given by Eq.~(\ref{plasma frequency}), was taken into account. The force (or pressure) above describes the force that the mirrors of the cavity exhibit due to the zero-point energy of the electromagnetic field, which is modified by the interaction between the 2DEG and the cavity field. The Casimir force given by Eq.~(\ref{Casimir force}) has a positive sign because $\Lambda^{3/2}_0\geq 1$, which implies that the force is repulsive. The possibility of repulsive Casimir forces has been studied theoretically in several publications and in many different settings~\cite{HoyeBrevik, Boyer1974, Milonni, Butcher_2012, Henkel} and has also been experimentally observed for interacting materials immersed in a fluid~\cite{Munday2009}. In the system that we consider here we do not have a fluid, but a 2DEG which interacts with the cavity field.

\section{Absorption and Dissipation in the Effective Field Theory}\label{Linear Response EFT}
 
In chapter~\ref{Cavity Responses} we performed linear response for the 2DEG inside the cavity, in the single mode case. Our aim now in this section is to perform linear response in the continuum by employing the effective quantum field theory that we constructed. In section~\ref{Duality} we showed that both the photonic and the electronic sector share their resonances and excitations and as a consequence it is adequate to simply focus on one sector of the system. 
 
Here we will focus on the photonic sector and for that we will perturb the system by applying an external time-dependent current $\bi{J}_{\textrm{ext}}(t)$ which couples to the photon field, as shown in Fig.~\ref{Cavity_Induction}. Thus, the external perturbation is $\hat{H}_{\textrm{ext}}(t)=-\mathbf{J}_{\textrm{ext}}(t)\cdot\hat{\mathbf{A}}$. The external current is chosen to be in the $x$-direction $\bi{J}_{\textrm{ext}}(t)=\bi{e}_xJ_{\textrm{ext}}(t)$. 

In the effective field theory the vector potential is given by the sum over all the in-plane photon-modes 
\begin{eqnarray}\label{manymodeA}
	\hat{\mathbf{A}}=\left(\frac{\hbar}{\epsilon_0 V}\right)^{\frac{1}{2}}\sum_{\kappa_x,\kappa_y}\frac{\bi{e}_x}{\sqrt{2\omega(\bm{\kappa}})}\left(\hat{a}_{\bm{\kappa}}+\hat{a}^{\dagger}_{\bm{\kappa}}\right).
\end{eqnarray}
For the $\bi{A}$-field we included only the polarization in the $x$-direction, because it is the only one that couples to the external current. To perform linear response first we have to introduce and define the Hamiltonian of the effective theory $\hat{H}_{\textrm{eff}}$. To define $\hat{H}_{\textrm{eff}}$ it is not necessary to give an explicit expression in terms of electronic and photonic operators. We can define $\hat{H}_{\textrm{eff}}$ in a simpler fashion by giving a definition of the ground-state of $\hat{H}_{\textrm{eff}}$ and its excited states. We define the ground-state of $\hat{H}_{\textrm{eff}}$ as
\begin{eqnarray}
    |\Psi_{gs}\rangle=|\Phi_{0}\rangle \otimes \prod_{\kappa_x,\kappa_y}|0,0\rangle_{\kappa_x,\kappa_y}
\end{eqnarray}
where $|\Phi_0\rangle$ is the electronic ground state given by the Slater determinant in Eq.~(\ref{Slater determinant}), with the electrons distributed on the 2D Fermi sphere. Further, the bosonic states $|0,0\rangle_{\kappa_x,\kappa_y}$ get annihilated by the operator $\hat{c}_{\bm{\kappa}}$, $\hat{c}_{\bm{\kappa}} |0,0\rangle_{\kappa_x,\kappa_y} =0, \;\; \forall\;\; \bm{\kappa}$. The excited states can now be defined by applying the creation operators $\hat{c}^{\dagger}_{\bm{\kappa}}$ on the ground-state, and as a consequence the excited states of $\hat{H}_{\textrm{eff}}$ satisfy the equation 
\begin{eqnarray}
    \hat{H}_{\textrm{eff}}\frac{(\hat{c}^{\dagger}_{\bm{\kappa}})^m}{\sqrt{m!}}|\Psi_{gs}\rangle=\left(E_{\mathbf{k}}+\hbar\widetilde{\omega}(\bm{\kappa})\left(m+\frac{1}{2}\right)\right)\frac{(\hat{c}^{\dagger}_{\bm{\kappa}})^m}{\sqrt{m!}}|\Psi_{gs}\rangle
\end{eqnarray}
where $E_{\bi{k}}=\sum_j\hbar^2\bi{k}^2_j/2m_{\textrm{e}}$ is the kinetic energy of the electrons\footnote{We note that the electronic excitations are not taken into account for the definition of $\hat{H}_{\textrm{eff}}$, because the perturbation we consider here couples only to the photonic states}. We would like to mention that the operators $\hat{c}_{\bm{\kappa}},\hat{c}^{\dagger}_{\bm{\kappa}^{\prime}}$ satisfy the bosonic algebra $[\hat{c}_{\bm{\kappa}},\hat{c}^{\dagger}_{\bm{\kappa}^{\prime}}]=\delta_{\bm{\kappa}\bm{\kappa}^{\prime}}$ $\forall \;\; \bm{\kappa},\bm{\kappa}^{\prime}$.

Having defined the effective Hamiltonian, the full time dependent Hamiltonian with the external perturbation included is $\hat{H}(t)=\hat{H}_{\textrm{eff}}-\bi{J}_{\textrm{ext}}(t)\cdot\hat{\bi{A}}$. The vector potential in terms of the renormalized annihilation and creation operators of Eq.~(\ref{c operators}) is
\begin{eqnarray}\label{manymodeAinC}
	\hat{\mathbf{A}}=\left(\frac{\hbar}{\epsilon_0 V}\right)^{\frac{1}{2}}\sum_{\kappa_x,\kappa_y}\frac{\bi{e}_x}{\sqrt{2\widetilde{\omega}(\bm{\kappa}})}\left(\hat{c}_{\bm{\kappa}}+\hat{c}^{\dagger}_{\bm{\kappa}}\right).
\end{eqnarray}
With the use of the effective Hamiltonian we can define also the operators in the interaction picture  $\hat{\mathcal{O}}_{I}(t)=e^{\textrm{i}t\hat{H}_{\textrm{eff}}/\hbar}\hat{\mathcal{O}}e^{-\textrm{i}t\hat{H}_{\textrm{eff}}/\hbar}$ and the wavefunctions respectively as $\Psi_{I}(t)=e^{\textrm{i}t\hat{H}_{\textrm{eff}}/\hbar}\Psi(t)$. Having defined everything we needed we can continue now with the linear response.

Here we aim to compute the $\bi{A}$-field response function $\chi^A_A(t-t^{\prime})$ which is defined via Eq.~(\ref{chi def}). We substitute the expression for the $\bi{A}$-field given by Eq.~(\ref{manymodeAinC}) and using the fact that $\hat{H}_{\textrm{eff}}$ is a sum of non-interacting modes and that the ground state $|\Psi_{gs}\rangle$ is a tensor product of the photonic states of all the modes, we find that in the effective theory $\chi^A_A$ is the sum of all the single-mode response functions given by Eq.~(\ref{A field response})
\begin{eqnarray}
    \chi^{A}_{A}(t-t^{\prime})=-\sum_{\kappa_x,\kappa_y}\frac{\Theta(t-t^{\prime})\sin(\widetilde{\omega}(\bm{\kappa})(t-t^{\prime}))}{\epsilon_0V\widetilde{\omega}(\bm{\kappa})}.
\end{eqnarray}
Subsequently, the same relation will also hold in the frequency domain
\begin{eqnarray}\label{EFT response frequency}
    \chi^A_A(w)=\sum_{\kappa_x,\kappa_y}\frac{-1}{2\epsilon_0\widetilde{\omega}(\bm{\kappa})V}\lim_{\eta \to 0^+}\left[\frac{1}{w+\widetilde{\omega}(\bm{\kappa})+\textrm{i}\eta}-\frac{1}{w-\widetilde{\omega}(\bm{\kappa})+\textrm{i}\eta}\right].
\end{eqnarray}
To obtain the above expression we used Eq.~(\ref{A frequencyresponse}) which gives the $\bi{A}$-field response function in the frequency domain in the single mode case. In the continuum limit the sum turns into an integral and the real and the imaginary part of the response function $\chi^A_A(w)$ are 
\begin{eqnarray}\label{Re and Im Integrals}
    &&\Re\left[\chi^A_A(w)\right]=\frac{1}{8\pi^2\epsilon_0L_z}\iint\limits^{\Lambda}_0\left[\frac{w-\widetilde{\omega}(\bm{\kappa})}{\widetilde{\omega}(\bm{\kappa})[(w-\widetilde{\omega}(\bm{\kappa}))^2+\eta^2]}-\frac{w+\widetilde{\omega}(\bm{\kappa})}{\widetilde{\omega}(\bm{\kappa})[(w+\widetilde{\omega}(\bm{\kappa}))^2+\eta^2]}\right]d\kappa_xd\kappa_y,\nonumber\\
    &&\Im\left[\chi^A_A(w)\right]=\frac{\eta}{8\pi^2\epsilon_0 L_z}\iint\limits^{\Lambda}_0\left[\frac{1}{\widetilde{\omega}(\bm{\kappa})[(w+\widetilde{\omega}(\bm{\kappa}))^2+\eta^2]}-\frac{1}{\widetilde{\omega}(\bm{\kappa})[(w-\widetilde{\omega}(\bm{\kappa}))^2+\eta^2]}\right]d\kappa_xd\kappa_y.\nonumber\\
\end{eqnarray}
After performing the above integrals we find the final analytic expressions for the real and the imaginary parts of the $\bi{A}$-field response function in the effective quantum field theory
\begin{eqnarray}\label{Re and Im of EFT Response}
    \Re[\chi^A_A(w)]&=&\frac{1}{8\pi c^2\epsilon_0L_z}\left[\ln\left(\frac{(w-\widetilde{\omega}(\kappa_z))^2+\eta^2}{(w-\sqrt{\Lambda})^2+\eta^2}\right)+\ln\left(\frac{(w+\widetilde{\omega}(\kappa_z))^2+\eta^2}{(w+\sqrt{\Lambda})^2+\eta^2}\right)\right]\;\;\nonumber \textrm{and}\\
    \Im[\chi^A_A(w)]&=&\frac{1}{4\pi c^2\epsilon_0L_z}\Bigg[\tan^{-1}\left(\frac{\sqrt{\Lambda}+w}{\eta}\right)-\tan^{-1}\left(\frac{\widetilde{\omega}(\kappa_z)+w}{\eta}\right)\nonumber\\
    &+&\tan^{-1}\left(\frac{\widetilde{\omega}(\kappa_z)-w}{\eta}\right)-\tan^{-1}\left(\frac{\sqrt{\Lambda}-w}{\eta}\right)\Bigg].
\end{eqnarray}

\begin{figure}[H]
\begin{center}
   \includegraphics[height=6.cm, width=0.65\linewidth]{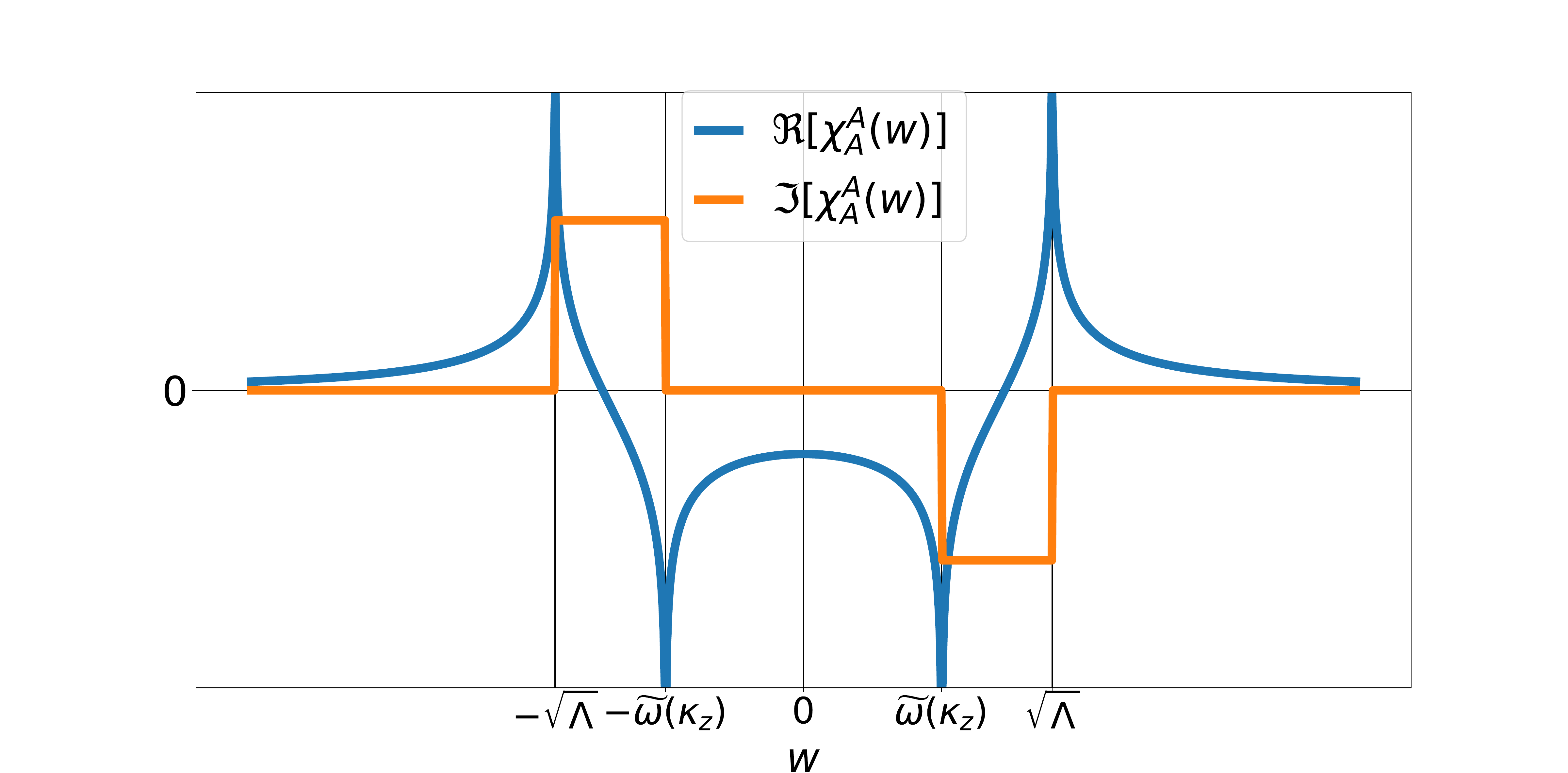}
\caption{\label{EFT Response} Real $\Re[\chi^A_A(w)]$ and imaginary $\Im[\chi^A_A(w)]$ parts of the $\bi{A}$-field response function $\chi^A_A(w)$ in the effective quantum field theory for $\eta=0$. The imaginary part has a finite value in the frequency window $\widetilde{\omega}(\kappa_z)<|w|<\sqrt{\Lambda}$ which means that the system can absorb energy in this frequency range. The real part diverges at the lower $w=\pm \widetilde{\omega}(\kappa_z)$ and the upper cutoffs $w=\pm \sqrt{\Lambda}$ and shows that there are two particular scales in the effective field theory.} 
\end{center}
\end{figure}

Taking now the limit $\eta \rightarrow 0^+$ for the artificial broadening $\eta$ the imaginary part of the response function takes the form
\begin{eqnarray}\label{Im EFT}
    \Im[\chi^A_A(w)]&=&
	\begin{cases}
		\dfrac{1}{4 c^2\epsilon_0L_z}\;,\;\; \textrm{for} \;\; -\sqrt{\Lambda}<w<-\widetilde{\omega}(\kappa_z)  \\\\
		-\dfrac{1}{4 c^2\epsilon_0L_z}\;,\;\; \textrm{for} \;\; \widetilde{\omega}(\kappa_z)<w<\sqrt{\Lambda} \\\\
		0\;\;, \;\;\;\;\;\;\; \textrm{elsewhere}. 
	\end{cases}
\end{eqnarray}
From the above result we see that $\Im[\chi^A_A(w)]$ is well-defined in the limit $\eta \rightarrow 0^+$ for all $w$ without the appearance of any divergences. This is in contrast to single-mode case in which the imaginary part of the response function $\chi^A_A$ given by Eq.~(\ref{Re Im A-field}) for $\eta \rightarrow 0$ was diverging at the frequency $w=\pm \widetilde{\omega}$. The fact that the imaginary part of the response function in the continuum is well-defined and does not diverge, means that absorption described consistently in the effective quantum field theory and the absorption rate $W$ in Eq.~(\ref{Absorption Rate}) is also well-defined and can be computed without any ambiguity. This proves our claim in the beginning of the chapter, that by constructing a field theory in the continuum, we can describe absorption processes and dissipation from first-principles, without having to introduce some kind of environment for our system and without the need of the artificial broadening parameter $\eta$. Moreover, as it is depicted in Fig.~\ref{EFT Response}, the imaginary part given by Eq.~(\ref{Im EFT}) takes a constant value in the region $\widetilde{\omega}(\kappa_z)<|w|<\sqrt{\Lambda}$ and is zero elsewhere. This means that the electron-photon system in the cavity absorbs energy continuously with the same strength in the frequency window $\widetilde{\omega}(\kappa_z)<|w|<\sqrt{\Lambda}$. This is true because this is the frequency range in which the effective quantum field theory was defined, see Fig.~\ref{Photon Continuum}, and all modes of the photon field are excited by the external current with exactly the same strength.
On the other hand, the real part $\Re[\chi^A_A(w)]$ of the response function for $\eta\rightarrow 0^+$ diverges at the frequencies $w=\pm\widetilde{\omega}(\kappa_z)$ and $w=\pm\sqrt{\Lambda}$, and gives us information about the resonances of the system. In the single-mode case, in section~\ref{Photonic Response}, there was only one resonance appearing at frequency $w=\pm \widetilde{\omega}$. Now, in the effective field theory we have two resonances at the frequency of the lower intrinsic cutoff $\widetilde{\omega}(\kappa_z)$ and the cutoff $\sqrt{\Lambda}$. This indicates that in the effective theory there are two energy scales.

Finally, we would like to mention that in the thermodynamic limit the imaginary and the real parts of the response function $\chi^A_A(w)$ have a well-defined value and do not vanish. This is in contrast to the single-mode response function given by Eq.~(\ref{Re Im A-field}). This shows again that by going to the continuum for the description of the photon field, the photonic observables become well-defined and have a non-zero contribution for the macroscopic 2DEG coupled to the cavity.

\part{Quantum Hall Systems in Cavity QED}

\chapter{Landau Levels \& Quantum Hall Effect}\label{Landau Levels QHE}

The integer quantum Hall effect discovered by von Klitzing, Dorda and Pepper in 1980~\cite{Klitzing} is one of the most interesting and fundamental phenomena in condensed matter physics. It has led to a tremendous amount of developments and to the re-definition of the international system of units~\cite{vonKlitzingUnits}. What was found in 1980 is that for two-dimensional materials at low temperatures the total macroscopic Hall conductance $\sigma_{xy}$ exhibits quantized plateaus whose value depends solely on the Planck constant $h$ and the charge of the electron $e$
\begin{eqnarray}
\sigma_{xy}=\frac{e^2}{h}\nu \;\;\; \textrm{with}\;\; \; \nu \in \mathbb{N}.
\end{eqnarray}
Soon after the discovery of the quantized Hall effect, Laughlin~\cite{LaughlinPRB} showed that this effect can be nicely understood in terms of non-interacting electrons in Landau levels and that it is a consequence of gauge invariance. The integer $\nu$ in the simple picture of non-interacting electrons corresponds to the integer filling factor of the Landau levels~\cite{Landau}.

In the four decades after the discovery of the integer effect, a great number of related phenomena have been observed, like the fractional quantum Hall effect~\cite{TsuifractionalQHE, Laughlingfractional}, the quantum spin Hall effect~\cite{QuantumSpinHall}, the quantum anomalous Hall effect~\cite{anomalousHalleffect} and more recently the light-induced anomalous Hall effect~\cite{LightHalleffect}. All these exciting developments have beautifully been reviewed in the 40 year anniversary article of the quantum Hall effect~\cite{40QHE}.

In this chapter however, we will not try to present all of these developments and phenomena, but rather we will just focus on the integer effect which we will describe within the picture of non-interacting electrons in Landau levels, as it was done originally by Laughlin~\cite{LaughlinPRB}.

\section{Free Electrons in a Homogeneous Magnetic Field}

Free electrons in a material in the presence of a classical homogeneous magnetic field along the $z$ direction $\bi{B}_{\textrm{ext}}=B\bi{e}_z$ of strength $B$ are described by the minimally-coupled Schr\"{o}dinger Hamiltonian~\cite{Landau}
\begin{eqnarray}\label{Hamiltonian LLs}
\hat{H}=\frac{1}{2m_{\textrm{e}}}\left(\mathrm{i}\hbar \mathbf{\nabla}+e \mathbf{A}_{\textrm{ext}}(\bi{r})\right)^2,
\end{eqnarray}
where in the Landau gauge the external vector potential which gives rise to the magnetic field is $\mathbf{A}_{\textrm{ext}}(\mathbf{r})=-\mathbf{e}_x B y$~\cite{Landau}. The Landau gauge is very convenient because it preserves translational invariance in two out of the three spatial dimensions, specifically in our case in the $z$ and $x$ directions. This implies that the Hamiltonian of Eq.~(\ref{Hamiltonian LLs}) commutes with the translation operator for the $x$ and $z$ directions and consequently the eigenfunctions of $\hat{H}$ in $x$ and $z$ will be plane waves
\begin{eqnarray}
\phi_{k_x,k_z}(x,z)=e^{\textrm{i}k_x x} e^{\textrm{i}k_z z} \;\;\; \textrm{with}\;\;\; k_x, k_z \in \mathbb{R}.
\end{eqnarray}
Applying $\hat{H}$ on the plane waves above we have
\begin{eqnarray}\label{HLL on plane waves}
\hat{H}\phi_{k_x,k_z}=\left[\frac{\hbar^2k^2_z}{2m_{\textrm{e}}}-\frac{\hbar^2}{2m_{\textrm{e}}}\frac{\partial^2}{\partial y^2}+ \frac{m_{\textrm{e}}\omega^2_c}{2}\left(y+\frac{\hbar k_x}{eB}\right)^2 \right]\phi_{k_x,k_z},
\end{eqnarray}
where we introduced also the cyclotron frequency $\omega_c$
\begin{eqnarray}\label{cyclotron frequency}
\omega_c=\frac{eB}{m_{\textrm{e}}}.
\end{eqnarray}
In the equation~(\ref{HLL on plane waves}) the part depending on the variable $y$ remains to be treated. The part of $\hat{H}$ depending on $y$ is a shifted harmonic oscillator
\begin{eqnarray}
\hat{H}_y=-\frac{\hbar^2}{2m_{\textrm{e}}}\partial^2_y+ \frac{m_{\textrm{e}}\omega^2_c}{2}\left(y+\frac{\hbar k_x}{eB}\right)^2
\end{eqnarray}
and the eigenfunctions of the operator above are Hermite functions of the variable $y+\hbar k_x/eB$
\begin{eqnarray}
\psi_n\left(y+\frac{\hbar k_x}{eB}\right)=\frac{1}{\sqrt{n! 2^n}}\left(\frac{m_{\textrm{e}}\omega_c}{\pi \hbar}\right)^{1/4}e^{-\frac{m_{\textrm{e}}\omega_c}{2\hbar}\left(y+\frac{\hbar k_x}{eB}\right)^2}H_n\left(\sqrt{\frac{m_e\omega_c}{\hbar}}\left(y+\frac{\hbar k_x}{eB}\right)\right)\nonumber\\
\end{eqnarray}
with eigenvalues $\hbar\omega_c(n+1/2)$
\begin{eqnarray}
\hat{H}_y\psi_n\left(y+\frac{\hbar k_x}{eB}\right)= \hbar\omega_c\left(n+\frac{1}{2}\right)\psi_n\left(y+\frac{\hbar k_x}{eB}\right)  \;\;\; \textrm{with}\;\;\; n\in \mathbb{N}.
\end{eqnarray}
Thus, applying now $\hat{H}\phi_{k_x,k_z}$ on the shifted Hermite functions $\psi_n\left(y+\hbar k_x/eB\right)$ we obtain 
\begin{eqnarray}
\hat{H}\phi_{k_x,k_z}\psi_n=\left[\frac{\hbar^2k^2_z}{2m_{\textrm{e}}}+\hbar\omega_c\left(n+\frac{1}{2}\right)\right]\phi_{k_x,k_z}\psi_n.
\end{eqnarray}
From the expression above we deduce that the full set of eigenfuctions for an electron in a classical homogeneous magnetic field is
\begin{eqnarray}
\Psi_{k_x,k_z,n}(\bi{r})=\phi_{k_x,k_z}(x,z)\psi_n\left(y+\frac{\hbar k_x}{eB}\right),
\end{eqnarray}
with eigenenergies
\begin{eqnarray}\label{3D Landau levels}
E_{n,k_z,k_x}=\frac{\hbar^2k^2_z}{2m_{\textrm{e}}}+\hbar\omega_c\left(n+\frac{1}{2}\right) \;\; \textrm{with}\;\; k_x, k_z\in\mathbb{R}, \; n\in\mathbb{N}.
\end{eqnarray}
This analytic solution for the a free electron in a homogeneous magnetic field was derived by Landau~\cite{Landau} and the associated energy levels of this system are known as Landau levels. The most interesting property of the Landau levels is that they are completely degenerate with respect to the momentum $k_x$ in the $x$ direction. This massive degeneracy is what makes this particular system so special and we will see that it is also responsible for the quantization of the Hall conductance in the case of the 2D electron gas. 

In what follows we restrict our considerations in the case of a 2D electron gas in a homogeneous magnetic field.

\section{Landau-Level Filling \& Density of States}

Before getting to the the Hall conductance we need first to understand how to actually fill the Landau levels, because the system that we are interested in, it does not consist of a single particle but it is actually a gas of many non-interacting electrons.

In principle, with free boundary conditions we would be able to accommodate all electrons of the material in the lowest Landau level, because in this case the degeneracy is infinite. But here we are considering a two-dimensional material of area $S$, which for simplicity is chosen to be a rectangle with sides of length $L_x$ and $L_y$. In the $y$ direction the eigenfunctions are shifted Hermite functions $\psi_n\left(y+\hbar k_x/m_{\textrm{e}}\omega_c\right)$. The $y$ coordinate is confined in the region $0\leq y \leq L_y$ and requiring also the argument $y+\hbar k_x/m_{\textrm{e}}\omega_c$ of the Hermite functions to be in this region we find that the allowed values for the momentum $k_x$ are~\cite{Peierls, Ziman}
\begin{eqnarray}\label{k region}
0\leq k_x\leq eBL_y/\hbar.
\end{eqnarray}
The above equation is very important because it provides a way to determine how many electrons can occupy a Landau level. The number of electrons per Landau level $N_{LL}$ is 
\begin{eqnarray}
N_{LL}=\frac{L_x}{2\pi}\int^{\frac{eBL_y}{\hbar}}_{0}dk_x=\frac{BL_xL_y}{\Phi_0}=\frac{\Phi_{\textrm{s}}}{\Phi_0},
\end{eqnarray}
where as $\Phi_{\textrm{s}}=BL_xL_y$ is the flux that goes through the whole sample of area $S=L_xL_y$ and $\Phi_0=h/e$ is the fundamental magnetic flux quantum. Knowing the number of electrons per Landau level, the number of occupied Landau levels $\nu$, also known as Landau level filling factor, can be determined. The filling factor is then the ratio between the number of electrons in the 2D material $N$ divided by the number of electrons per Landau level $N_{LL}$~\cite{Vignale}
\begin{eqnarray}
\nu=\frac{N}{N_{LL}}=\frac{n_{\textrm{2D}}\Phi_0}{B},
\end{eqnarray}
where $n_{\textrm{2D}}$ is the electron density of the 2D material.

\subsection{Oscillatory Density of States}

A further consequence of this particular way in which Landau levels get filled is the fact that the density of states of the electrons exhibits an oscillatory behavior. These oscillations of the density of states are experimentally accessible and they are related to the oscillations of the magnetization in metals, as a function of the magnetic field strength, known as the de Haas-van Alphen effect, and the oscillations of the longitudinal resistivity known as the Shubnikov-de Hass effect~\cite{Ziman, Mermin}. Here however, the aim of the section is not to give a description of these two important phenomena, but rather only to describe the oscillations of the density of states which is responsible for these effects~\cite{Ziman}.  

\begin{figure}[H]
\begin{center}
  \includegraphics[height=6cm, width=0.6\columnwidth]{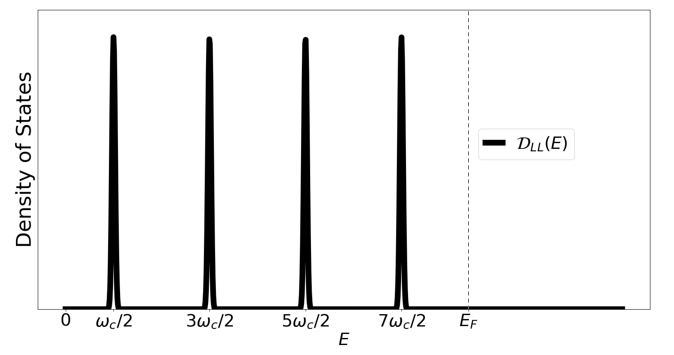}
\caption{\label{DOS_LL}Density of states of the two-dimensional Landau levels as function of the energy. The density of states has an oscillatory behavior with peaks showing up with periodicity $\omega_c=eB/m_{\textrm{e}}$ all the way to the Fermi energy $E_{\textrm{F}}$.}  
\end{center}
\end{figure}

The energy spectrum of the Landau levels in the case of a two-dimensional electron gas is
\begin{eqnarray}
E_{n,k_x}=\hbar\omega_c\left(n+\frac{1}{2}\right)\;\; \textrm{with}\;\; k_x\in\mathbb{R}, \; n\in\mathbb{N}.
\end{eqnarray}
Then, the two-dimensional density of states of the Landau levels is defined as 
\begin{eqnarray}
\mathcal{D}_{\textrm{LL}}(E)=\frac{1}{2\pi L_y}\sum_n\int dk_x \delta\left(E-E_{n,k_x}\right).
\end{eqnarray}
The range of the momenta is given by Eq.~(\ref{k region}). Because the eigenenergies are independent of $k_x$ the integration over $k_x$ is trivial and we find 
\begin{eqnarray}
\mathcal{D}_{\textrm{LL}}(E)=\frac{eB}{2\pi \hbar}\sum_n \delta\left(E-\hbar\omega_c\left(n+\frac{1}{2}\right)\right) \;\;\; \textrm{for}\;\;\; E<E_{\textrm{F}}.
\end{eqnarray}
From the expression above it is clear that the density of states for the 2D Landau levels exhibits strong van Hove singularities ($\delta$-function peaks) at the energies of the Landau levels $E_n=\hbar\omega_c(n+1/2)$~\cite{Ziman}. Further, in Fig.~\ref{DOS_LL} where we plot the density of states for the 2D Landau levels, we see that the density of states exhibits an oscillatory behavior with $\delta$-peaks showing periodically all the way to the Fermi energy $E_{\textrm{F}}$. The periodicity of the peaks is the cyclotron frequency $\omega_c$.

\section{Quantization of the Hall Conductance}

To describe the conduction properties of the 2D electron gas in the presence of the uniform magnetic field, we also need to include an external electric field $\bi{E}$ which perturbs the electrons and forces them to flow and produce a current. For that purpose we consider an electric field along the $y$ direction $\bi{E}=E\bi{e}_y$. The potential energy due to the electric field is
\begin{eqnarray}\label{E field potential}
V_{E}(\bi{r})=e\bi{E}\cdot \bi{r}=eEy.
\end{eqnarray}
adding the potential energy of the electric field in the Hamiltonian of Eq.~(\ref{Hamiltonian LLs}) the full Hamiltonian of the system is
\begin{eqnarray}
\hat{H}_E=\frac{1}{2m_{\textrm{e}}}\left(\mathrm{i}\hbar \mathbf{\nabla}+e \mathbf{A}_{\textrm{ext}}(\bi{r})\right)^2+eEy.
\end{eqnarray}
We note that this Hamiltonian was employed also by Laughlin in~\cite{LaughlinPRB} for the description of the integer quantum Hall effect. The Hamiltonian above preserves translational invariance in the $x$ coordinate. Thus, the eigenfunctions of $\hat{H}_E$ are plane waves of the form $\phi_{k_x}(x)=e^{\textrm{i}k_xx}$ and applying $\hat{H}_E$ on $\phi_{k_x}(x)$ one finds
\begin{eqnarray}
\hat{H}_E\phi_{k_x}(x)=\left[-\frac{\hbar^2}{2m_{\textrm{e}}}\frac{\partial^2}{\partial y^2}+ \frac{m_{\textrm{e}}\omega^2_c}{2}\left(y+\frac{\hbar k_x}{eB}\right)^2 +eEy\right]\phi_{k_x}(x).
\end{eqnarray}
Performing a square completion, the part of the Hamiltonian depending on $y$, can be brought into the form of a shifted harmonic oscillator
\begin{eqnarray}
\hat{H}_E\phi_{k_x}(x)=\left[-\frac{\hbar^2}{2m_{\textrm{e}}}\frac{\partial^2}{\partial y^2}+ \frac{m_{\textrm{e}}\omega^2_c}{2}\left(y+\frac{\hbar k_x}{eB}+\frac{eE}{m_{\textrm{e}}\omega^2_c}\right)^2 -\frac{e^2E^2}{2m_{\textrm{e}}\omega^2_c}-\frac{\hbar k_xE}{B}\right]\phi_{k_x}(x).
\end{eqnarray}
As in the previous section, the eigenfunctions with respect to the operator depending on $y$ are Hermite functions 
\begin{eqnarray}
\psi_n\left(s\right)=\frac{1}{\sqrt{n! 2^n}}\left(\frac{m_{\textrm{e}}\omega_c}{\pi \hbar}\right)^{1/4}\exp\left(-\frac{m_{\textrm{e}}\omega_c}{2\hbar}s^2\right)H_n\left(\sqrt{\frac{m_e\omega_c}{\hbar}}s\right)
\end{eqnarray}
which depend on the variable $s$
\begin{eqnarray}
s=y+\frac{\hbar k_x}{eB}+\frac{eE}{m_{\textrm{e}}\omega^2_c}
\end{eqnarray}
Applying the operator $\hat{H}_E\phi_{k_x}(x)$ on the these Hermite functions one obtains
\begin{eqnarray}
\hat{H}_E\phi_{k_x}(x)\psi_n\left(s\right) = \left[\hbar\omega_c\left(n+\frac{1}{2}\right)-\frac{e^2E^2}{2m_{\textrm{e}}\omega^2_c}-\frac{\hbar k_xE}{B}\right] \phi_{k_x}(x)\psi_n\left(s\right).\nonumber\\
\end{eqnarray}
From the expression above one deduces that the energy spectrum is 
\begin{eqnarray}
E_{n,k_x}=\hbar\omega_c\left(n+\frac{1}{2}\right)-\frac{e^2E^2}{2m_{\textrm{e}}\omega^2_c}-\frac{\hbar k_xE}{B}.
\end{eqnarray}
It is clear that degeneracy with respect to $k_x$ is now lifted due to the electric field. But here the strength of the electric field $E$ is considered to be much smaller than the strength of the magnetic field $B$, which implies that $E/B\approx 0$. As a consequence the degeneracy and the filling of the Landau levels remain practically the same. 

Now the question that arises is: How much current flows due to the external electric field? To figure this out, we need to compute the expectation value of the current operator. The current operator is~\cite{Landau}
\begin{eqnarray}
\hat{\bi{J}}=\sum^N_{j=1}\left(-\frac{\textrm{i}e\hbar}{m_{\textrm{e}}}\nabla_j-\frac{e^2}{m_{\textrm{e}}}\bi{A}_{\textrm{ext}}(\bi{r}) \right)=\sum^N_{j=1}\left(-\frac{\textrm{i}e\hbar}{m_{\textrm{e}}}\nabla_j+\frac{e^2B y }{m_{\textrm{e}}}\mathbf{e}_x\right).
\end{eqnarray}
Let us start with the $y$ component of the current operator. With $\nu$ Landau levels filled the expectation value of the $x$ component of the current operator is
\begin{eqnarray}
\langle \hat{J}_y \rangle= -\frac{e}{m_{\textrm{e}}}\sum^{\nu}_{n=1} \sum_{k_x} \langle \Psi_{n,k_x}| -\textrm{i}\hbar \partial_y|\Psi_{n,k_x}\rangle= -\frac{e}{m_{\textrm{e}}}\sum^{\nu}_{n=1} \sum_{k_x} \langle \psi_n(s)|-\textrm{i}\hbar \partial_y |\psi_n(s)\rangle.
\end{eqnarray}
The Hermite functions $\psi_n(s)$ are functions of the coordinate $s$. The momentum operator $-\textrm{i}\hbar \partial_y$ transforms trivially with respect to the shifted coordinate $s$: $-\textrm{i}\hbar \partial_y \rightarrow -\textrm{i}\hbar \partial_s$. Then, the expectation value of the momentum operator on the Hermite functions is zero $\langle \psi_n(s)|\textrm{i}\hbar \partial_s |\psi_n(s)\rangle=0$ and as a consequence the current in the $y$ direction is zero as well
\begin{eqnarray}
\langle \hat{J}_y\rangle=0.
\end{eqnarray}
 In the same fashion the $x$ component of the current operator is
\begin{eqnarray}
\langle \hat{J}_x\rangle &=&-\frac{e}{m_{\textrm{e}}}\sum^{\nu}_{n=1} \sum_{k_x} \langle \Psi_{n,k_x}| -\textrm{i}\hbar \partial_x+ eBy |\Psi_{n,k_x}\rangle\nonumber\\
&=&-\frac{e}{m_{\textrm{e}}}\sum^{\nu}_{n=1}\sum_{k_x} \langle \Psi_{n,k_x}| \hbar k_x+ eBy |\Psi_{n,k_x}\rangle.
\end{eqnarray}
Introducing the variable $s$ we find for $\langle \hat{J}_x\rangle $
\begin{eqnarray}
\langle \hat{J}_x\rangle = -\frac{e^2B}{m_{\textrm{e}}}\sum^{\nu}_{n=1}\sum_{k_x} \langle \Psi_{n,k_x}| s-\frac{eE}{m_{\textrm{e}}\omega^2_c} |\Psi_{n,k_x}\rangle.
\end{eqnarray}
The expectation value of $s$ on the Hermite functions $\psi_n(s)$ is zero, $\langle \psi_n(s)|s|\psi_n(s)\rangle=0$. Promoting also the sum over the momenta $k_x$ into an integral as it was done for the computation of the Landau level filling factor we get
\begin{eqnarray}
\langle \hat{J}_x\rangle = \frac{e^2B}{m_{\textrm{e}}}\frac{eE}{m_{\textrm{e}}\omega^2_c}\sum^{\nu}_{n=1}\frac{L_x}{2\pi}\int^{eB L_y/\hbar}_0 dk_x= \frac{e^2E S}{h} \nu.
\end{eqnarray}
In the last step the cyclotron frequency $\omega_c=eB/m_{\textrm{e}}$ and the area of the material $S=L_xL_y$ where introduced, and the sum was also computed. Dividing the expectation value of the current operator by the area $S$ in order to introduce the current density $\langle \hat{j}_x\rangle= \langle \hat{J}_x\rangle/S $ the final result for the current density is obtained
\begin{eqnarray}
\langle \hat{j}_x\rangle= \nu \frac{e^2}{h} E.
\end{eqnarray}
Combining the results for both components of the current density 
\begin{eqnarray}\label{current density}
\langle \;\hat{\bi{j}}\;\rangle= \left( \frac{e^2\nu }{h} ,0\right) E.
\end{eqnarray}
The conductivity tensor $\sigma$ 
\begin{eqnarray}
\sigma= \left(\begin{tabular}{c c}
    $\sigma_{xx}$ & $\sigma_{xy}$ \\
     $\sigma_{yx}$ & $\sigma_{yy}$
\end{tabular} \right)
\end{eqnarray}
is defined as the ratio between the external electric field $\bi{E}=E\bi{e}_y$ and the induced current density~\cite{Mermin, Vignale}
\begin{eqnarray}
\langle\; \hat{\bi{j}}\;\rangle= \sigma \bi{E}. 
\end{eqnarray}
Comparing the definition of the conductivity with the result obtained for the current density $\langle \;\hat{\bi{j}}\;\rangle$ it is clear that the longitudinal conductivity $\sigma_{yy}=0$ is zero while the Hall conductivity $\sigma_{xy}$ is
\begin{eqnarray}\label{Hall conductance}
\sigma_{xy}=\frac{e^2}{h} \nu \;\;\; \textrm{with}\;\; \nu \in \mathbb{N}.
\end{eqnarray}
From the result above one concludes that the Hall conductance is quantized since it is a multiple of the filling factor $\nu$ which is an integer for fully occupied Landau levels. The Hall conductance depends only on two fundamental constants of Nature, the electron charge $e$ and Planck's constant $h$. This is the famous quantization of the macroscopic Hall conductance discovered by von Klitzing, Dorda and Pepper~\cite{Klitzing}. As a last comment we would like to mention that for a deeper understanding of the quantization of the Hall conductance one also needs to consider the topological description of the quantum Hall effect coming from the TKNN (Thouless, Kohmoto, Nightingale, and den Nijs) formula~\cite{Thouless}. Further, for the accurate description of the longitudinal resistance and conductance of a 2D material in the presence of a magnetic field it is important also the scattering by impurities to be taken into account~\cite{ButtikerQHE}.

\chapter{Bloch's Theorem \& the Magnetic Translation Group}\label{Bloch MTG}
\begin{displayquote}
\footnotesize{Different geometrical figures have qualitative differences, although, being all alike merely spatial shapes, they have no material peculiarites, only formal ones. Building on this new foundation, Pythagoras suggested that the qualitative differences in nature were based on differences of geometrical structure.}
\end{displayquote}
\begin{flushright}
  \footnotesize{R.~G.~Collingwood\\
The Idea of Nature~\cite{Collingwood}}
\end{flushright}

\section{Bloch's Theorem in One Dimension}

The aim of this section is to present a proof of Bloch's theorem and to present how the Bloch theory of periodic solids is constructed. Bloch's theorem is one of the cornerstones of solid-state physics and materials science as it allows for the description of periodic materials like metals, semiconductors and insulators~\cite{Mermin, Callaway}. 

Here, for simplicity, we focus on the case of a one-dimensional periodic crystal. Although this one-dimensional setting might seem a bit restricted, it actually captures the essence of Bloch's theorem. For the general case of a three-dimensional crystalline solid we advise the reader to look into standard textbooks~\cite{Mermin, Callaway}. The Hamiltonian for an electron in a one-dimensional periodic solid is 
\begin{eqnarray}\label{periodic H}
\hat{H}=-\frac{\hbar^2}{2m_{\textrm{e}}}\frac{\partial^2}{\partial x^2}+v_{\textrm{ext}}(x).
\end{eqnarray}
In Bloch theory the external potential $v_{\textrm{ext}}(x)$ is assumed to be infinitely periodic. This implies that the external potential is invariant under the lattice translations $x\rightarrow x+R_n$, $v_{\textrm{ext}}(x+R_n)=v_{\textrm{ext}}(x)$, where $R_n=an$ are Bravais lattice vectors, $a$ is the lattice constant of the 1D crystal, and $n \in \mathbb{Z}$.

With respect to the Bravais lattice we can define the corresponding set of translation operators $\mathcal{G}=\{\hat{T}_n\}$ which act on a wavefunction and translate it from $x$ to $x+R_n$
\begin{eqnarray}
\hat{T}_n\psi(x)=\psi(x+R_n) \;\;\; n\in \mathbb{Z}.
\end{eqnarray}
The translation operators $\hat{T}_n$ are given by the expression
\begin{eqnarray}
\hat{T}_n=e^{\frac{\textrm{i}}{\hbar}R_n \hat{P}_x}=e^{R_n\partial_x} \;\;\; \textrm{where}\;\;\; \partial_x\equiv \frac{\partial}{\partial x},
\end{eqnarray}
An important property of the set of translation operators $\mathcal{G}=\{\hat{T}_n\}$ is that they form a group, because they satisfy the four group axioms: 

\textit{Closure}.---For all translation operators $\hat{T}_n, \hat{T}_m$ with $n,m\in \mathbb{Z}$ the product of them $\hat{T}_n\hat{T}_m$ is also a translation operator.
\begin{eqnarray}\label{closure}
\hat{T}_n\hat{T}_m=e^{R_n\partial_x}e^{R_m\partial_x}=e^{R_{n+m}\partial_x}=\hat{T}_{n+m} \;\;
\end{eqnarray}
where we used the fact that the sum of two Bravais lattice vectors $R_n$ and $R_m$ is also a Bravais lattice vector $R_n+R_m=R_{n+m}$.
    
\textit{Associativity}.---For all translation operators $\hat{T}_n, \hat{T}_m$ and $\hat{T}_l$ with $n,m,l\in \mathbb{Z}$ it holds that 
\begin{eqnarray}\label{associativity}
\left(\hat{T}_n\hat{T}_m\right)\hat{T}_l=\hat{T}_n\left(\hat{T}_m\hat{T}_l\right).
\end{eqnarray}
To show this we use the fact that $\hat{T}_n\hat{T_m}=\hat{T}_{n+m}$ and $\hat{T}_m\hat{T}_l=\hat{T}_{m+l}$. Then, we substitute these two relations into Eq.~(\ref{associativity}) and we have
\begin{eqnarray}
\hat{T}_{n+m}\hat{T}_l=\hat{T}_n\hat{T}_{m+l} \Longrightarrow \hat{T}_{n+m+l}=\hat{T}_{n+m+l}.
\end{eqnarray}
The result above shows that indeed the translation operators satisfy the associativity.
    
\textit{Existence of Identity}.---The identity element of the set of translation operators is $\hat{T}_0=1$. The product of the identity operator $\hat{T}_0$ with any other translation operator $\hat{T}_n$ gives the translation $\hat{T}_n$ 
\begin{eqnarray}
\hat{T}_0\hat{T}_n=\hat{T}_n \hat{T}_0=\hat{T}_n.
\end{eqnarray}
    
\textit{Existence of Inverse}.---For every translation operator $\hat{T}_n$ with $n\in \mathbb{Z}$, there exists the element $\hat{T}_{-n}$ such that the product of these two operators gives us the identity $\hat{T}_0$
\begin{eqnarray}
\hat{T}_n\hat{T}_{-n}=\hat{T}_{n-n}=\hat{T}_0.
\end{eqnarray}
To obtain the result above we used the fact that $R_{-n}=-R_{n}$.

Before we continue with the proof of Bloch's theorem we would like to mention two further, important properties of the translation operators, which we will need in order to establish the proof of Bloch's theorem. 

\textit{Abelianity}.---The first one has do to do with the group-classification of the translation group. The set of translation operators $\mathcal{G}=\{\hat{T}_n\}$ does not only form a group but actually forms an Abelian group because all translation operators $\hat{T}_n$
and $\hat{T}_m$ commute with each other
\begin{eqnarray}
\left[\hat{T}_n,\hat{T}_m\right]=0\;\; \forall\;\; n,m\in \mathbb{Z}.
\end{eqnarray}
This fact is true because the translation operators are all exponentials of the differential operator $\partial_x$ which of course commutes with itself. 

\textit{Unitarity}.---The last important property of the translation operators is that they are unitary operators
\begin{eqnarray}
\hat{T}^{\dagger}_n\hat{T}_n=\hat{T}_n\hat{T}^{\dagger}_n=\hat{T}_0=1.
\end{eqnarray}
This holds because the adjoint of the differential operator $\partial_x$ is the operators itself with a minus $(\partial_x)^{\dagger}=-\partial_x$. This implies that the adjoint of $\hat{T}_n$ is actually its inverse $\hat{T}_{-n}$
\begin{eqnarray}
\hat{T}^{\dagger}_n=\hat{T}_{-n}.
\end{eqnarray}
Having laid the necessary mathematical basis, we proceed with establishing an important lemma for the proof of Bloch's theorem.

\begin{lemma}
The Hamiltonian $\hat{H}$ of Eq.~(\ref{periodic H}) commutes with all the translation operators $\hat{T}_n$. 
\end{lemma}
\textit{Proof}. To prove this statement we consider a test wavefunction\footnote{For the properties of the standard test wavefunctions usually considered in quantum mechanics, the reader may look in mathematical-physics textbooks~\cite{Blanchard, Teschl}. The discussion of these properties is important if one is interested in establishing the proofs that we present in this section, to a higher level of mathematical rigor.} $\psi(x)$ on which first we apply the Hamiltonian $\hat{H}$ of Eq.~(\ref{periodic H}), $\hat{H}\psi(x)$. Then, subsequently, we apply the translation operator $\hat{T}_n$ and we have
\begin{eqnarray}\label{lemma eq}
\hat{T}_n\hat{H}\psi(x)=\hat{T}_n\left(-\frac{\hbar^2}{2m_{\textrm{e}}}\frac{\partial^2}{\partial x^2}+v_{\textrm{ext}}(x)\right)\psi(x).
\end{eqnarray}
The translation operator obviously commutes with the kinetic energy operator and we have
\begin{eqnarray}
\hat{T}_n\hat{H}\psi(x)=-\frac{\hbar^2}{2m_{\textrm{e}}}\frac{\partial^2}{\partial x^2}\hat{T}_n\psi(x)+ \hat{T}_n\left(v_{\textrm{ext}}(x)\psi(x)\right)
\end{eqnarray}
For $\hat{T}_n\left(v_{\textrm{ext}}(x)\psi(x)\right)$ we have that $\hat{T}_n\left(v_{\textrm{ext}}(x)\psi(x)\right)=v_{\textrm{ext}}(x+R_n)\psi(x+R_n)$. The external potential is periodic $v_{\textrm{ext}}(x+R_n)=v_{\textrm{ext}}(x)$. Further, the translated function $\psi(x+R_n)$ can be written with the use of the translation operator $\hat{T}_n$ as $\psi(x+R_n)=\hat{T}_n\psi(x)$, and thus we obtain
\begin{eqnarray}
\hat{T}_n\hat{H}\psi(x)=-\frac{\hbar^2}{2m_{\textrm{e}}}\frac{\partial^2}{\partial x^2}\hat{T}_n\psi(x)+ v_{\textrm{ext}}(x)\hat{T}_n\psi(x)=\hat{H}\hat{T}_n\psi(x).
\end{eqnarray}
The equation above implies that $\hat{H}$ and $\hat{T}_n$ commute
\begin{eqnarray}
[\hat{H},\hat{T}_n]=0.
\end{eqnarray}
\hfill$\blacksquare$

The fact that the Hamiltonian $\hat{H}$ commutes with the translation operators $\hat{T}_n$ implies that the group of translation operators $\mathcal{G}=\{\hat{T}_n\}$  and the Hamiltonian $\hat{H}$ form a set of commuting operators. From a a fundamental theorem of quantum mechanics~\cite{Mermin, GriffithsQM} it follows that the eigenfunctions of $\hat{H}$ can be chosen to be simultaneous eigenfunctions of all the translation operators $\hat{T}_n$
\begin{eqnarray}
\hat{H}\Psi(x)=E\Psi(x)\;\;\;\textrm{and}\;\;\; \hat{T}_n\Psi(x)=\lambda_n\Psi(x),
\end{eqnarray}
where $E$ and $\lambda_n$ are the eigenvalues of $\hat{H}$ and $\hat{T}_n$ respectively. This result is very important and is the basic building block on which we are going to build the proof of Bloch's theorem.

\begin{theorem}[Bloch's Theorem]
A complete basis of eigenfunctions of the one-electron Hamiltonian $\hat{H}=-\hbar^2\partial^2_x/2m_{\textrm{e}}+v_{\textrm{ext}}(x)$, where $v_{\textrm{ext}}(x+R_n)=v_{\textrm{ext}}(x)$ for all $R_n$ in a Bravais lattice, can be chosen to have the form $\Psi_k(x)=e^{\textrm{i}kx}U^k(x)$ where $U^k(x)$ is a periodic function respecting the periodicity of the lattice $U^k(x+R_n)=U^k(x)$.
\end{theorem}
\textit{Proof}. Since we showed that the $\hat{H}$ and $\hat{T}_n$ commute, in order to find the form of the eigenfunctions of $\hat{H}$ we can investigate what kind of form the eigenfunctions of the translation operators have. Suppose that $\Psi(x)$ is an eigenfunction of all the operators $\hat{T}_n$ with eigenvalue $\lambda_n$
\begin{eqnarray}
\hat{T}_n\Psi(x)=\lambda_n\Psi(x).
\end{eqnarray}
Conjugating the above equation and taking the inner product between the above and its conjugate we obtain
\begin{eqnarray}
\langle \Psi|\hat{T}^{\dagger}_n\hat{T}_n|\Psi\rangle=\lambda_n\lambda^*_n\langle\Psi|\Psi\rangle.
\end{eqnarray}
Using the fact that the translation operators are unitary $\hat{T}^{\dagger}_n\hat{T}_n=1$ we find that the norm of the eigenvalues $\lambda_n$ must be equal to one
\begin{eqnarray}
|\lambda_n|=1.
\end{eqnarray}
This implies that we can write the eigenvalues $\lambda_n$ in the form
\begin{eqnarray}\label{form of lambdas}
\lambda_n=e^{\textrm{i}\theta_n}.
\end{eqnarray}
We now apply now the translation operators $\hat{T}_n$ and $\hat{T}_m$ on $\Psi(x)$. Because $\Psi(x)$ is an eigenfunction of both operators we have
\begin{eqnarray}
\hat{T}_n\hat{T}_m\Psi(x)=\lambda_n\lambda_m\Psi(x).
\end{eqnarray}
In Eq.~(\ref{closure}) we showed that the translation operators satisfy closure, which implies that $\hat{T}_n\hat{T}_m=\hat{T}_{n+m}$. Using this relation in the equation above we obtain
\begin{eqnarray}
\hat{T}_{n+m}\Psi(x)=\lambda_n\lambda_m\Psi(x).
\end{eqnarray}
The wavefunction $\Psi(x)$ is an eigenfunction of the operator $\hat{T}_{n+m}$ as well, because it is an element of the translation group $\mathcal{G}$. The eigenvalue of $\hat{T}_{n+m}$ is $\lambda_{n+m}$, and thus we have
\begin{eqnarray}
\lambda_{n+m}=\lambda_n\lambda_m.
\end{eqnarray}
The eigenvalues $\lambda_n$ are given by the form in Eq.~(\ref{form of lambdas}) and we find that
\begin{eqnarray}
e^{\textrm{i}\theta_{n+m}}=e^{\textrm{i}\theta_n}e^{\textrm{i}\theta_m}.
\end{eqnarray}
The above equation means that $\theta_{n+m}$ is the sum of $\theta_n$ and $\theta_m$
\begin{eqnarray}
\theta_{n+m}=\theta_n+\theta_m.
\end{eqnarray}
This implies that $\theta_n$ is a linear function of $n$. Thus, we can write $\theta_n$, and the eigenvalues $\lambda_n$, as a function of the Bravais lattice vectors $R_n=an$ (which is linear in $n$) and an arbitrary (quantum) number $k\in \mathbb{R}$
\begin{eqnarray}\label{theta and lambda}
\theta_n=kR_n\;\;\; \textrm{and}\;\;\; \lambda_n=e^{\textrm{i}kR_n}.
\end{eqnarray}
The quantum number $k$ is very important because it allows us to label (and also classify) the eigenfunctions of the translation operators $\hat{T}_n$ as $\Psi_k(x)$. The quantum number $k$ in solid-state physics is known as the crystal momentum. Since the eigenvalues $\lambda_n$ are given by Eq.~(\ref{theta and lambda}) we find that the eigenfunctions $\Psi_k(x)$ have to satisfy the equation
\begin{eqnarray}\label{eigenvalue phase}
\hat{T}_n\Psi_k(x)=e^{\textrm{i}kR_n}\Psi_k(x).
\end{eqnarray}
The operator $\hat{T}_n$ by definition translates the wavefunction from $x$ to $x+R_n$ and we have
\begin{eqnarray}
\Psi_k(x+R_n)=e^{\textrm{i}kR_n}\Psi_k(x).
\end{eqnarray}
The equation above means that all eigenfunctions $\Psi_k(x)$ of the translation operators $\hat{T}_n$ and of the Hamiltonian $\hat{H}$ are invariant under the lattice translations $x\rightarrow x+R_n$, up to the phase $e^{\textrm{i}kR_n}$. This particular behavior of the eigenfunctions $\Psi_k(x)$ is satisfied by the following ansatz
\begin{eqnarray}\label{Bloch Ansatz}
\Psi_k(x)=e^{\textrm{i}kx}U^k(x) \;\; \textrm{with}\;\; U^k(x+R_n)=U^k(x)\;\; \forall\; R_n.\nonumber\\
\end{eqnarray}
The ansatz above is known as the Bloch ansatz and generally wavefunctions of this form are known as Bloch waves. The Bloch ansatz given by Eq.~(\ref{Bloch Ansatz}) is the product of a plane wave $e^{\textrm{i}kx}$ and the function $U^k(x)$ which is periodic under the lattice translations.\hfill$\blacksquare$

Having proven Bloch's theorem and having constructed Bloch's ansatz does not mean that only with these two ingredients we can describe electrons in periodic solids. This is because we have still not specified what is the domain for the quantum number $k$. The Bloch ansatz by itself does not provide this information. The answer to this point is given by the next corollary to Bloch's theorem. 

\begin{corollary}\label{Corollary}The Bloch waves $\Psi_k(x)$ and $\Psi_{k+G_q}(x)$ differing by a reciprocal lattice vector $G_q=2\pi q/a$, with $q\in \mathbb{Z}$, are degenerate with respect to the group of translation operators $\mathcal{G}=\{\hat{T}_n\}$. 
\end{corollary}
\textit{Proof}. To prove the above statement, we consider the Bloch waves $\Psi_k(x)$ and $\Psi_{k+G_q}(x)$ with $G_q=2\pi q/a$. Applying the translation operator $\hat{T}_n$ to the Bloch wave $\Psi_k(x)$ yields the eigenvalue $e^{\textrm{i}kR_n}$
\begin{eqnarray}
\hat{T}_n\Psi_k(x)=e^{\textrm{i}kR_n}\Psi_k(x)
\end{eqnarray}
as we showed in Eq.~(\ref{eigenvalue phase}). Let us check now what is the eigenvalue corresponding to the Bloch wave $\Psi_{k+G_q}(x)$. We apply the operator $\hat{T}_n$ and we have
\begin{eqnarray}
\hat{T}_n\Psi_{k+G_q}(x)=e^{\textrm{i}(k+G_q)R_n}\Psi_{k+G_q}(x).
\end{eqnarray}
Substituting now the definition for the reciprocal lattice vector $G_q=2\pi q/a$ and the Bravais lattice vector $R_n=an$ we find
\begin{eqnarray}
\hat{T}_n\Psi_{k+G_q}(x)=e^{\textrm{i}2\pi qn}e^{\textrm{i}kR_n}\Psi_{k+G_q}(x)=e^{\textrm{i}kR_n}\Psi_{k+G_q}(x).\nonumber\\
\end{eqnarray}
Thus, the Bloch wave $\Psi_{k+G_q}(x)$ yields the same eigenvalue as the Bloch wave $\Psi_{k}(x)$.
\hfill$\blacksquare$

Corollary~\ref{Corollary} is of major importance because it implies that for the description of an electron in a periodic potential we can consider only the Bloch waves $\Psi_k(x)$ with crystal momentum in the range $k\in [-\pi/a,\pi/a]$. This particular domain of $k$-space is called the first Brillouin zone. In addition, if we assume that the translation operators have no degenerate states, then from Corollary~\ref{Corollary} we conclude that the Bloch waves $\Psi_k(x)$ and $\Psi_{k+G_q}(x)$ are the same state
\begin{eqnarray}\label{k-periodic gauge}
\Psi_k(x)=\Psi_{k+G_q}(x)\;\; \forall\;\; G_q.
\end{eqnarray}
As explained in~\cite{Vanderbilt} the above relation is a ``periodic gauge condition'', but in general one may adopt different boundary conditions in $k$-space. The periodic gauge condition of Eq.~(\ref{k-periodic gauge}) has as a consequence that also the energies $E_n(k)$ of the Hamiltonian $\hat{H}$ are periodic with respect to the crystal momentum $k$
\begin{eqnarray}
E_n(k)=E_n(k+G_q) \;\; \forall\;\; G_q,
\end{eqnarray}
where the index $n$ signifies the fact that for each $k$ we have multiple energy levels. This is how standard Bloch theory is typically constructed. To conclude, the importance of Bloch theory (and Bloch's theorem) is that it provides a ansatz with which we can compute the energy levels of an electron in a periodic solid~\cite{Vanderbilt, Mermin, Callaway}. The eigenfunctions of the electrons are fully classified with respect to the continuous quantum number $k$ which is usually restricted to the first Brillouin zone. Due to the fact that $k$ is a continuous number we have the formation of a continuous manifold of energy levels which are known as energy bands. Knowing the energy bands and the number of bands filled by the electrons in the crystal, one can classify whether the material under consideration is a metal, a semiconductor or an insulator.

\section{The Problem with Solids in a Uniform Magnetic Field}

In what follows we will take a close look into another problem which, from a physical point of view, seems to be translationally invariant as well. Namely, the problem of a periodic solid under the influence of an external classical homogeneous magnetic field. For the sake of simplicity, but without loss of generality, we will restrict our considerations in the case of a two-dimensional periodic solid in which a perpendicular uniform magnetic field is applied.

\begin{figure}[H]
\begin{center}
  \includegraphics[height=7cm, width=0.7\columnwidth]{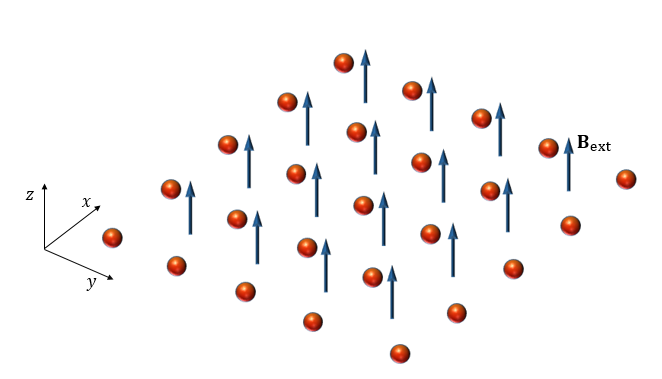}
\caption{\label{Solid B-field}Schematic depiction of a 2D periodic solid in the $(x,y)$ plane in the presence of an external perpendicular homogeneous magnetic field $\bi{B}_{\textrm{ext}}$ pointing in the $z$ direction. The external magnetic field is the same from unit cell of the crystal to the other and such a system from a physical point of view is translationally invariant.}  
\end{center}
\end{figure}

As it becomes clear also from Fig.~\ref{Solid B-field} from a physical point of view this system is translationally invariant. Because from one cell of the solid to the other, nothing changes. The external scalar potential is the same, because it is periodic, and the external magnetic field is also same because it is homogeneous. Thus, one would expect that the Hamiltonian describing such a system would be translationally invariant, and Bloch's theorem should be applicable, as in the case of electrons in a periodic solid which we described in great detail in the previous section.

However, if we consider the minimal-coupling Schr\"{o}dinger Hamiltonian for an electron in a crystal, in the presence of a classical homogeneous magnetic field~\cite{Landau} 
\begin{eqnarray}\label{Hext} 
\hat{H}&=&\frac{1}{2m_{\textrm{e}}}\left(\mathrm{i}\hbar \mathbf{\nabla}+e \mathbf{A}_{\textrm{ext}}(\bi{r})\right)^2 +v_{\textrm{ext}}(\mathbf{r}),
\end{eqnarray}
we see that the electrons do not couple directly to the magnetic field $\bi{B}_{\textrm{ext}}$, which is homogeneous, but rather to the vector potential $\bi{A}_{\textrm{ext}}(\bi{r})$. The magnetic field is equal to the curl of the vector potential, $\bi{B}_{\textrm{ext}}=\nabla \times \bi{A}_{\textrm{ext}}(\bi{r})$, and as a consequence the vector potential must have some dependence on $\bi{r}$ to produce a uniform magnetic field. Here we choose the vector potential in the Landau gauge $\mathbf{A}_{\textrm{ext}}(\mathbf{r})=-\mathbf{e}_x B y$~\cite{Landau} in which the vector potential is linear in the electronic coordinate $y$. Due to this dependence on the electronic coordinate $y$ the classical vector potential breaks translational invariance for this system. This is also depicted schematically in the figure below.
\begin{figure}[H]
\begin{center}
  \includegraphics[height=6.5cm, width=0.6\columnwidth]{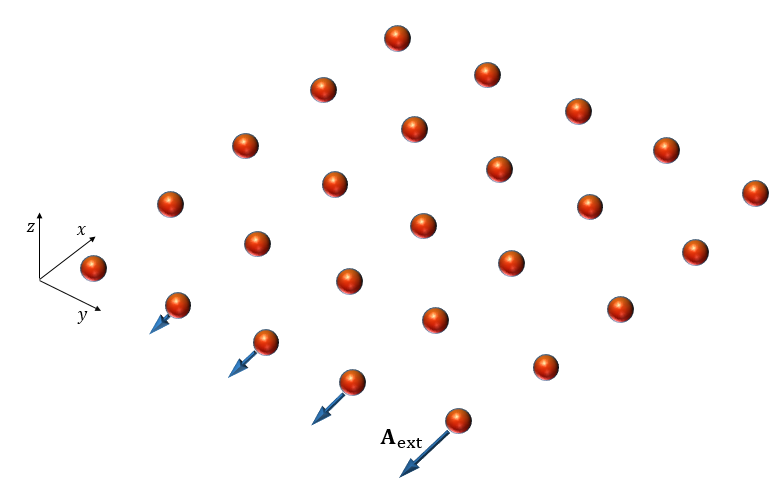}
\caption{\label{Solid A-field}Schematic depiction of a 2D periodic solid in the $(x,y)$ plane coupled to the classical vector potential $\bi{A}_{\textrm{ext}}(\bi{r})$ which produces a perpendicular homogeneous magnetic field in the $z$ direction. The vector potential is chosen in the Landau gauge $\mathbf{A}_{\textrm{ext}}(\mathbf{r})=-\mathbf{e}_x B y$. The vector potential is polarized in the $x$ direction while it increases linearly as a function of $y$, and consequently breaks translational symmetry along the $y$ direction.}  
\end{center}
\end{figure}
The fact that translational symmetry is broken implies that also Bloch's theorem cannot be applied and a periodic solid in a homogeneous magnetic field cannot be treated within the framework of Bloch theory.

Although the Hamiltonian in Eq.~(\ref{Hext}) is not translationally invariant and the translation operators do not commute with it, there is another set of operators which actually commute with $\hat{H}$. These operators are the magnetic translation operators, which were introduced by Brown~\cite{BrownMTG} and Zak~\cite{MTG_I, MTG_II} independently. But before we come in the topic of the magnetic translations and the magnetic translation group we would like to specify the geometry of the 2D crystal, its Bravais lattice and its reciprocal lattice.

\textit{Setting the Geometry}.---To describe a periodic solid the external potential is chosen to be periodic $v_{\textrm{ext}}(\mathbf{r})=v_{\textrm{ext}}(\mathbf{r}+\mathbf{R}_{\mathbf{n}})$ where $\mathbf{R}_{\mathbf{n}}$ is a 2D Bravais lattice vector with $\mathbf{n}=(n,m)\in \mathbb{Z}^2$. The Bravais lattice vector in general is $\mathbf{R}_{\mathbf{n}}=n\mathbf{a}_1+m\mathbf{a}_2$ where $\mathbf{a}_1$ and $\mathbf{a}_2$ are primitive vectors which lie in different directions and span the lattice. Without loss of generality we can choose the vector $\mathbf{a}_1$ to be in the $x$-direction $\mathbf{a}_1=a_1\mathbf{e}_x$. The second primitive vector in this case is $\mathbf{a}_2=a_2\cos\theta\mathbf{e}_x+a_2\sin\theta\mathbf{e}_y$ where $\theta$ is the angle between the vectors $\mathbf{a}_2$ and $\mathbf{a}_1$. Thus, the Bravais lattice vectors are 
\begin{eqnarray}\label{bravaisvectors}
\mathbf{R}_{\mathbf{n}}=\left(na_1+ma_2\cos\theta\right)\mathbf{e}_x+ma_2\sin\theta\mathbf{e}_y.
\end{eqnarray}
Further, the reciprocal lattice vectors are $\mathbf{G}_{\mathbf{n}^{\prime}}=n^{\prime}\mathbf{b}_1+m^{\prime}\mathbf{b}_2$ with $\mathbf{n}^{\prime}=(n^{\prime},m^{\prime})\in \mathbb{Z}^2$. The defining relation for the vectors $\mathbf{b}_{1}$ and $\mathbf{b}_{2}$ is~\cite{Mermin, Callaway}
\begin{eqnarray}\label{reciprocal}
\mathbf{b}_{i}\cdot \mathbf{a}_j=2\pi \delta_{ij}, \;\;\; i,j=1,2.
\end{eqnarray}
With the choice we made for the primitive vectors the reciprocal primitive vectors satisfying Eq.~(\ref{reciprocal}) are
\begin{eqnarray}
\mathbf{b}_1=\frac{2\pi}{a_1}\mathbf{e}_x-\frac{2\pi\cos\theta}{a_1\sin\theta}\mathbf{e}_y\;\; \textrm{and}\;\; \mathbf{b}_2=\frac{2\pi}{a_2\sin\theta}\mathbf{e}_y.
\end{eqnarray}
Thus, the reciprocal lattice vectors are
\begin{eqnarray}
\mathbf{G}_{\mathbf{n}^{\prime}}=\frac{2\pi n^{\prime}}{a_1}\mathbf{e}_x +\left[\frac{2\pi m^{\prime}}{a_2\sin\theta}-\frac{2\pi n^{\prime}\cos\theta}{a_1\sin\theta}\right] \mathbf{e}_y
\end{eqnarray}
which for convenience we will write as
\begin{eqnarray}\label{reciprocallattice}
&&\mathbf{G}_{\mathbf{n}^{\prime}}=\left(G^x_{n^{\prime}}, G_{m^{\prime},n^{\prime}}\right)\;\; \textrm{where}\;\;G_{m^{\prime},n^{\prime}}=\frac{G^y_{m^{\prime}}}{\sin\theta}-\frac{G^x_{n^{\prime}}\cos\theta}{\sin\theta}\nonumber\\
 &&\textrm{and}\;\;  G^x_{n^{\prime}}=\frac{2\pi n^{\prime}}{a_1 },\;\;\;G^y_{m^{\prime}}=\frac{2\pi m^{\prime}}{a_2}.
\end{eqnarray}
With these choices we have now defined our setting.

\section{Magnetic Translations}\label{Magnetic Translations}

 We proceed now by investigating more systematically the translational properties of the Hamiltonian of Eq.~(\ref{Hext}) which describes our system. Here, we are considering a 2D periodic potential which means that we have two sets of Bravais lattice vectors. One set is for $m=0$ where $\mathbf{R}_{(n,0)}=na_1\mathbf{e}_x$ and one set for $n=0$ where $\mathbf{R}_{(0,m)}=m\left(a_2\cos\theta\mathbf{e}_x+a_2\sin\theta\mathbf{e}_y\right)$. Comparing the Bravais vectors $\mathbf{R}_{(n,0)}$ to $\mathbf{R}_{(0,m)}$, we see that there is an important difference. The vectors $\mathbf{R}_{(n,0)}$ are parallel to $\mathbf{e}_x$, while the vectors $\mathbf{R}_{(0,m)}$ have non-zero projections both on $\mathbf{e}_x$ and $\mathbf{e}_y$. Although the Hamiltonian does not respect translational invariance in $y$, it is invariant under the Bravais translations $\mathbf{R}_{(n,0)}$ in the $x$ direction. This means that the Hamiltonian commutes with the respective set of translation operators $\hat{T}_{(n,0)}=e^{\mathbf{R}_{(n,0)}\cdot\nabla}=e^{na_1\partial_x}$
\begin{eqnarray}
\left[\hat{H},\hat{T}_{(n,0)}\right]=0.
\end{eqnarray}
On the contrary under the translations $\mathbf{R}_{(0,m)}$ the Hamiltonian is not invariant, because they involve the $y$ direction. As a consequence the Hamiltonian does not commute with the respective translation operators $\hat{T}_{(0,m)}=e^{\mathbf{R}_{(0,m)}\cdot\nabla}$.

Although we do not have full translational symmetry over all the possible lattice translations, we have translational symmetry with respect to the Bravais lattice vectors $\mathbf{R}_{(n,0)}=na_1\mathbf{e}_x$ in the $x$ direction. This means that we can make use of Bloch's theorem in the $x$ coordinate and write the wavefunction of our system in terms of Bloch waves in the $x$-coordinate~\cite{Mermin, Callaway} 
\begin{eqnarray}\label{ansatz}
\Psi_{k}(\mathbf{r})=e^{\textrm{i}k x} U^{k}(x,y),
\end{eqnarray}
where $k$ is the crystal momentum in the $x$ direction. The function $U^{k}(x,y)$ due to Bloch's theorem is periodic in the $x$ coordinate, with periodicity $a_1$, $U^{k}(x+na_1,y)=U^{k}(x,y)$~\cite{Mermin, Callaway}. It is important to highlight again that in the $y$ coordinate Bloch's theorem does not hold. As a consequence the function $U^{k}(x,y)$ is not periodic in $y$. To keep our analysis general, the form of $U^{k}(x,y)$ with respect to $y$ is left unspecified and generic.

We would like now to derive what is in many cases called the Bloch Hamiltonian~\cite{Kane}. To do so we project the Hamiltonian of Eq.~(\ref{Hext}) on the Bloch wavefunction of Eq.~(\ref{ansatz}) 
\begin{eqnarray}
\hat{H}\Psi_{k}(\mathbf{r})=e^{\textrm{i}kx}\left[\frac{1}{2m_{\textrm{e}}}\left(\mathrm{i}\hbar \mathbf{\nabla}-\hbar \mathbf{k}+e \mathbf{A}_{\textrm{ext}}(\bi{r})\right)^2 +v_{\textrm{ext}}(\mathbf{r})\right]U^{k}(x,y)=e^{\textrm{i}kx}E_{k}U^{k}(x,y).\nonumber\\
\end{eqnarray}
where $\mathbf{k}=\mathbf{e}_xk$. By dividing the Schr\"{o}dinger equation above by the plane wave $e^{\textrm{i}kx}$ we can define the Bloch Hamiltonian $\hat{H}\left(\mathbf{k},\mathbf{r}\right)$ as
\begin{eqnarray}\label{Hk}
\hat{H}(\mathbf{k},\mathbf{r})=\frac{1}{2m_{\textrm{e}}}\left(\mathrm{i}\hbar \mathbf{\nabla}-\hbar \mathbf{k}+e \mathbf{A}_{\textrm{ext}}(\bi{r})\right)^2 +v_{\textrm{ext}}(\mathbf{r}).\nonumber\\
\end{eqnarray}
The Bloch Hamiltonian $\hat{H}(\mathbf{k},\mathbf{r})$ and the Schr\"{o}dinger Hamiltonian $\hat{H}$ are related via the transformation
\begin{eqnarray}\label{HnqH}
\hat{H}=e^{\textrm{i}\mathbf{k}\cdot\mathbf{r}}\hat{H}(\mathbf{k},\mathbf{r})e^{-\textrm{i}\mathbf{k}\cdot\mathbf{r}}.
\end{eqnarray}
We note that the Bloch Hamiltonian $\hat{H}(\mathbf{k},\mathbf{r})$ does not act on the Bloch wavefunction given by Eq.~(\ref{ansatz}), but on the quotient between $\Psi_{k}$ and $e^{\textrm{i}k x}$ \begin{eqnarray}
\Psi_{k}(\mathbf{r})/e^{\textrm{i}k x}=U^{k}(x,y).
\end{eqnarray}

Let us check now the translational properties of the Bloch Hamiltonian. To do so we apply the translation operator $\hat{T}_{\bi{n}}$ on $\hat{H}(\mathbf{k},\mathbf{r})$
\begin{eqnarray}\label{TnHk}
\hat{T}_{\bi{n}}\hat{H}(\mathbf{k},\mathbf{r})=\hat{H}(\mathbf{k},\mathbf{r}+\bi{R}_{\bi{n}})\hat{T}_{\bi{n}}.
\end{eqnarray}
From the definition of the of the Bloch Hamiltonian in Eq.~(\ref{Hk}) and because the external potential respects the Bravais lattice symmetry, we find that 
\begin{eqnarray}\label{HkBm}
\hat{H}(\mathbf{k},\mathbf{r}+\bi{R}_{\bi{n}})=\hat{H}(\mathbf{k}+\bi{B}_m,\mathbf{r}),
\end{eqnarray}
where 
\begin{eqnarray}\label{Bm}
\mathbf{B}_{m}=\frac{\mathbf{e}_x eBma_2\sin\theta}{\hbar}.
\end{eqnarray}
We substitute Eq.~(\ref{HkBm}) into Eq.~(\ref{TnHk}) and we have
\begin{eqnarray}
\hat{T}_{\bi{n}}\hat{H}(\mathbf{k},\mathbf{r})=\hat{H}(\mathbf{k}+\bi{B}_m,\mathbf{r})\hat{T}_{\bi{n}}.
\end{eqnarray}
Further, we use the relation between the Bloch Hamiltonian and the Schr\"{o}dinger Hamiltonian given by Eq.~(\ref{HnqH}) and we have
\begin{eqnarray}
\hat{T}_{\bi{n}}e^{-\textrm{i}\bi{k} \cdot \bi{r}}\hat{H}e^{\textrm{i}\bi{k}\cdot\bi{r}}=e^{-\textrm{i}\left(\bi{k} +\bi{B}_m\right)\cdot \bi{r}}\hat{H}e^{\textrm{i}\left(\bi{k}+\bi{B}_m\right)\cdot\bi{r}}\hat{T}_{\bi{n}}
\end{eqnarray}
we multiply the equation above from the left with $e^{\textrm{i}\left(\bi{k} +\bi{B}_m\right)\cdot \bi{r}}$ and from the right with $e^{-\textrm{i}\bi{k}\cdot\bi{r}}$
\begin{eqnarray}
e^{\textrm{i}\left(\bi{k} +\bi{B}_m\right)\cdot \bi{r}}\hat{T}_{\bi{n}}e^{-\textrm{i}\bi{k} \cdot \bi{r}}\hat{H}=\hat{H}e^{\textrm{i}\left(\bi{k}+\bi{B}_m\right)\cdot\bi{r}}\hat{T}_{\bi{n}}e^{-\textrm{i}\bi{k}\cdot\bi{r}}.
\end{eqnarray}
For $\hat{T}_{\bi{n}}$ on the plane wave $e^{\textrm{i}\bi{k}\cdot\bi{r}}$ it holds that $\hat{T}_{\bi{n}}e^{-\textrm{i}\bi{k}\cdot\bi{r}}=e^{-\textrm{i}\bi{k}\cdot\bi{r}}e^{-\textrm{i}\bi{k}\cdot\bi{R}_{\bi{n}}}\hat{T}_{\bi{n}}$ and using this property we obtain
\begin{eqnarray}
e^{\textrm{i}\bi{B}_m\cdot \bi{r}}\hat{T}_{\bi{n}}\hat{H}=\hat{H}e^{\textrm{i}\bi{B}_m\cdot\bi{r}}\hat{T}_{\bi{n}},
\end{eqnarray}
where we eliminated the constant phase $e^{-\textrm{i}\bi{k}\cdot\bi{R}_{\bi{n}}}$ which showed up on both sides of the equation. From the last equation we conclude that the operators $\mathcal{M}_{\bi{n}}=e^{\textrm{i}\bi{B}_m\cdot \bi{r}}\hat{T}_{\bi{n}}$ commute with the Schr\"{o}dinger Hamiltonian $\hat{H}$
\begin{eqnarray}
[\mathcal{M}_{\bi{n}},\hat{H}]=0.
\end{eqnarray}

The set of operators $\mathcal{M}_{\bi{n}}$ are known as the magnetic translations and were first introduced by Brown~\cite{BrownMTG} and soon after independently by Zak~\cite{MTG_I, MTG_II}. 
\begin{eqnarray}\label{magnetic translations Bm}
\textrm{Magnetic Translations}:\; \Bigg\{\mathcal{M}_{\bi{n}}=e^{\textrm{i}\mathbf{B}_m\cdot\mathbf{r}}\hat{T}_{{\mathbf{n}}}\;\; \textrm{with}\;\; \bi{n}\in \mathbb{Z}^2 \;\; \textrm{and}\;\; \mathbf{B}_{m}=\frac{\mathbf{e}_x eBma_2\sin\theta}{\hbar} \Bigg \}.\nonumber\\
\end{eqnarray}
However, there is more information that we can extract from the Bloch Hamiltonian $\hat{H}(\bi{k},\bi{r})$. As we showed in corollary~\ref{Corollary} we know that two crystal momenta,  $k$ and $k^{\prime}$, differing by a reciprocal lattice vector $G^x_q=2\pi a/a_1 $ 
\begin{eqnarray}\label{equivalent momenta}
k^{\prime}_x=k+G^x_q 
\end{eqnarray}
are equivalent because the respective Bloch waves $\Psi_{k}$ and $\Psi_{k^{\prime}}$ yield exactly the same eigenvalue with respect to the translation operators\cite{Mermin, Callaway}. The fact that the crystal momenta related by Eq.~(\ref{equivalent momenta}) are equivalent means that the $\bi{k}$-space in Bloch theory has the structure of a torus.
Moreoever, in Eq.~(\ref{HkBm}) we found that for the Bloch Hamiltonian $\hat{H}(\bi{k},\bi{r})$ a translation in real-space  is equal to a translation in $\bi{k}$-space. Enforcing the condition~(\ref{equivalent momenta}) (the torus structure) on Eq.~(\ref{HkBm}), we find that for the Bloch Hamiltonian under a lattice translation $\bi{r} \rightarrow \bi{r}+\bi{R}_{\bi{n}}$ must hold 
\begin{eqnarray}
\hat{H}(\bi{k}+\bi{B}_m,\bi{r})=\hat{H}(\bi{k}+\bi{G}^x_q,\bi{r}).
\end{eqnarray}
From the above equation we find that $B_m$ has to be equal to a reciprocal lattice vector $G^x_q$ for all $m$,
\begin{eqnarray}
B_m=G^x_q \;\; \forall\; m \; \;\Longrightarrow \; \frac{ eBa_2 a_1\sin\theta}{\hbar 2\pi}=q.
\end{eqnarray}
Upon introducing the magnetic flux through the area of the fundamental unit cell $\Phi= B|\bi{a}_1\times \bi{a}_2|=B a_1a_2\sin\theta$ and the magnetic flux quantum $\Phi_0=h/e$ we find the following conditions for the relative magnetic flux through the fundamental unit cell
\begin{eqnarray}\label{flux conditions torus}
 \frac{\Phi}{\Phi_0}=q \;\;\; \textrm{with}\;\;\; q \in \mathbb{Z}.
\end{eqnarray}
The above conditions are the well-known magnetic flux conditions of the magnetic translation group~\cite{MTG_I}. These conditions were originally derived by enforcing an abelian group structure on the magnetic translation operators~\cite{MTG_I, MTG_II}. Here however, we followed a different approach and derived the flux conditions by enforcing the torus structure on the Bloch Hamiltonian. To the best of our knowledge this alternative derivation has not been demonstrated before in the literature. The fact that the flux conditions can be derived from both ways, means that enforcing the torus structure and imposing the abelian group structure result in the same conditions for the magnetic flux, and consequently are equivalent. 

Having found a set of operators which commute with the Schr\"{o}dinger Hamiltonian, the following questions arise: What is the structure of the magnetic translation operators? Do they form a group? If yes, is this group abelian like the translation group in Bloch theory? Is it possible to construct an ansatz wavefunction for solids in magnetic fields out of the eigenfunctions of the magnetic translations, like it was done in Bloch theory using the eigenfunctions of the translation group?

Most of the questions about the structure of the magnetic translations were answered in the seminal papers of Brown~\cite{BrownMTG} and Zak~\cite{MTG_I, MTG_II}. But whether an ansatz construction out of the eigenfunctions of the magnetic translations is possible remains still open because it is not clear what kind of form the eigenfunctions of the magnetic translation operators should have.   

In what follows we present what is the current state of the art of the magnetic translations and the magnetic translation group (in two dimensions), and we discuss some of the properties of the magnetic unit cell~\cite{Kohmoto}. In this thesis we will not make use of the magnetic translation group for the description of 2D materials in strong magnetic fields. It is important however, to study in detail what can be accomplished with this formalism, in order to compare to the results that we will obtain from our quantum electrodynamical Bloch theory~\cite{rokaj2019} in the next chapter.

\section{Magnetic Translation Group}\label{}

Our aim now is to check under which conditions the magnetic translations form a group and what kind of group structure they obey. For the magnetic translation operators to form a group, they need to satisfy the four basic axioms of a group:

\textit{Existence of Identity}.---The identity element of the set of magnetic translation operators is $\mathcal{M}_0=1$. The product the identity $\mathcal{M}_0$ with any other magnetic translation $\mathcal{M}_{\bi{n}}$ gives the magnetic translation $\mathcal{M}_{\bi{n}}$ 
    \begin{eqnarray}
    \mathcal{M}_0\mathcal{M}_{\bi{n}}=\mathcal{M}_{\bi{n}} \mathcal{M}_0=\mathcal{M}_{\bi{n}}.
    \end{eqnarray}

\textit{Closure}.---The product of any two magnetic translations must also be a magnetic translation. To check closure we consider two magnetic translations $\mathcal{M}_{{\mathbf{n}}}$ and $\mathcal{M}_{{\mathbf{n}^{\prime}}}$
\begin{eqnarray}\label{groupstructure}
\mathcal{M}_{{\mathbf{n}}} \mathcal{M}_{{\mathbf{n}^{\prime}}}&=&e^{\textrm{i}\mathbf{B}_m\cdot\mathbf{r}}\hat{T}_{{\mathbf{n}}}e^{\textrm{i}\mathbf{B}_{m^{\prime}}\cdot\mathbf{r}}\hat{T}_{{\mathbf{n}^{\prime}}}=e^{\textrm{i}\mathbf{B}_m\cdot\mathbf{r}}e^{\textrm{i}\mathbf{B}_{m^{\prime}}\cdot\left(\mathbf{r}+\mathbf{R}_{\mathbf{n}}\right)}\hat{T}_{{\mathbf{n}}}\hat{T}_{{\mathbf{n}^{\prime}}}\nonumber\\
&=&e^{\textrm{i}\mathbf{B}_{m^{\prime}}\cdot\mathbf{R}_{\mathbf{n}}}e^{\textrm{i}\mathbf{B}_{m+m^{\prime}}\cdot\mathbf{r}}\hat{T}_{{\mathbf{n}+\mathbf{n}^{\prime}}}=e^{\textrm{i}\mathbf{B}_{m^{\prime}}\cdot\mathbf{R}_{\mathbf{n}}}\mathcal{M}_{{\mathbf{n}+\mathbf{n}^{\prime}}}.\nonumber\\
\end{eqnarray}
From the above result it is clear that in general the magnetic translation operators do not form a group, since the product of two magnetic translation operators is not a magnetic translation operator, i.e., $\mathcal{M}_{{\mathbf{n}}} \mathcal{M}_{{\mathbf{n}^{\prime}}} \neq \mathcal{M}_{\bi{n}+\bi{n}^{\prime}}$ . 

At this point two possibilities arise for the magnetic translations to form a closed set: 

(i) We can add to the set of magnetic translations $\{\mathcal{M}_{\bi{n}}\}$ the infinite amount of operators $e^{\textrm{i}\mathbf{B}_{m^{\prime}}\cdot\mathbf{R}_{\mathbf{n}}}\mathcal{M}_{{\mathbf{n}+\mathbf{n}^{\prime}}}$. These operators also commute with the Hamiltonian of Eq.~(\ref{Hext}), because they are magnetic translation operators multiplied by a constant exponential prefactor. Then, what we have is an extended set of magnetic translations
\begin{eqnarray}
\textrm{Extended Magnetic Translations}: \Big\{ \mathcal{M}_{\bi{n}}\cup e^{\textrm{i}\mathbf{B}_{m^{\prime}}\cdot\mathbf{R}_{\mathbf{n}}}\mathcal{M}_{{\mathbf{n}+\mathbf{n}^{\prime}}}  \Big\}.
\end{eqnarray}
More or less, this is the path that was taken by Zak in his seminal paper~\cite{MTG_I}. Such a construction then leads to the non-abelian magnetic translation group~\cite{MTG_I, MTG_II}. This is a beautiful mathematical construction which  also introduces a path-dependence on the allowed magnetic translations. However, this is a rather difficult and complicated construction which due to the fact that it is inherently non-abelian is opposite to the standard translation operators which form an abelian group and allow to establish Bloch's theorem. For all these reasons we will follow an alternative path.

(ii) The second possibility for the magnetic translations to satisfy closure is the exponential $e^{\textrm{i}\mathbf{B}_{m^{\prime}}\cdot\mathbf{R}_{\mathbf{n}}}$ to be equal to one, $e^{\textrm{i}\mathbf{B}_{m^{\prime}}\cdot\mathbf{R}_{\mathbf{n}}}=1$. Then, the product of two magnetic translations is again a magnetic translation. The exponential $e^{\textrm{i}\mathbf{B}_{m^{\prime}}\cdot\mathbf{R}_{\mathbf{n}}}$ upon substituting the expressions for $\bi{B}_{m^{\prime}}$ Eq.~(\ref{magnetic translations Bm}) and for the Bravais lattice vector is  
\begin{eqnarray}
e^{\textrm{i}\mathbf{B}_{m^{\prime}}\cdot\mathbf{R}_{\mathbf{n}}}=e^{\textrm{i}\frac{eBm^{\prime}a_2\sin\theta}{\hbar}na_1}e^{\textrm{i}\frac{eBm^{\prime}a_2\sin\theta}{\hbar}ma_2\cos\theta},
\end{eqnarray}
we also introduce the magnetic flux quantum $\Phi_0=h/e$ and the magnetic flux through the primitive unit cell $\Phi=B|\mathbf{a}_1\times\mathbf{a}_2|=Ba_1a_2\sin\theta$ and we have
\begin{eqnarray}
e^{\textrm{i}\mathbf{B}_{m^{\prime}}\cdot\mathbf{R}_{\mathbf{n}}}=e^{\textrm{i}\frac{\Phi }{\Phi_{0}}2\pi m^{\prime}n}e^{\textrm{i}\frac{\Phi }{\Phi_0}\frac{a_2\cos\theta}{a_1}2\pi m^{\prime}m}.
\end{eqnarray}
Requiring the above exponential to be equal to 1 for all integers $n,m,n^{\prime},m^{\prime}$ we obtain the following conditions for the relative magnetic flux $\Phi/\Phi_0$ and the lattice constants 
\begin{eqnarray}\label{Closure conditions}
\frac{\Phi}{\Phi_0}= q \;\;\; \textrm{and}\;\;\; a_2\cos\theta=la_1 \;\;\textrm{where}\;\; q,l\in \mathbb{Z}.\nonumber\\
\end{eqnarray}
Under the above conditions the magnetic translations satisfy the property of closure
\begin{eqnarray}
\mathcal{M}_{{\mathbf{n}}} \mathcal{M}_{{\mathbf{n}^{\prime}}}=\mathcal{M}_{\bi{n}+\bi{n}^{\prime}}.
\end{eqnarray}
We would like to highlight that the first part of the closure conditions in Eq.~(\ref{Closure conditions}) which have to do with the magnetic flux are the same conditions that emerged in Eq.~(\ref{flux conditions torus}) when we imposed the torus structure on the Bloch Hamiltonian $\hat{H}(\bi{k},\bi{r})$ of Eq.~(\ref{Hk}). As we will see the flux conditions guarantee that the magnetic translations are an abelian group.

The other two properties which we need to check for the magnetic translations to form a group, are the existence of an inverse and associativity.

\textit{Existence of Inverse}.---For every magnetic translation $\mathcal{M}_{\bi{n}}$ with $\bi{n}\in \mathbb{Z}^2$, there exists the element $\mathcal{M}_{-\bi{n}}$ such that the product of these two operators gives the identity $\mathcal{M}_0$. To show this, we multiply $\mathcal{M}_{\bi{n}}$ with $\mathcal{M}_{-\bi{n}}$ and using Eq.~(\ref{groupstructure}) we obtain
    \begin{eqnarray}
    \mathcal{M}_{\bi{n}}\mathcal{M}_{-\bi{n}}=e^{\textrm{i}\mathbf{B}_{-m}\cdot\mathbf{R}_{\mathbf{n}}}\mathcal{M}_{{\mathbf{n}-\mathbf{n}}}.
    \end{eqnarray}
Under the conditions for closure in Eq.~(\ref{Closure conditions}) the exponential $e^{\textrm{i}\mathbf{B}_{m^{\prime}}\cdot\mathbf{R}_{\mathbf{n}}}$ is equal to one, and using the fact that $\bi{B}_{-m}=-\bi{B}_{m}$ and $\bi{R}_{-\bi{n}}=-\bi{R}_{\bi{n}}$ one finds that indeed $\mathcal{M}_{-\bi{n}}$ is the inverse of $\mathcal{M}_{\bi{n}}$,
    \begin{eqnarray}
    \mathcal{M}_{\bi{n}}\mathcal{M}_{-\bi{n}}=\mathcal{M}_0=1.
    \end{eqnarray}

\textit{Associativity}.---For all translation operators $\mathcal{M}_{\bi{n}_1}, \mathcal{M}_{\bi{n}_2}$ and $\mathcal{M}_{\bi{n}_3}$ with $\bi{n}_1,\bi{n}_2,\bi{n}_3 \in \mathbb{Z}^2$ it holds that 
\begin{eqnarray}\label{associativityMTG}
\left(\mathcal{M}_{\bi{n}_1} \mathcal{M}_{\bi{n}_2}\right)\mathcal{M}_{\bi{n}_3}=\mathcal{M}_{\bi{n}_1}\left(\mathcal{M}_{\bi{n}_2}\mathcal{M}_{\bi{n}_3}\right) .
\end{eqnarray}
Under the closure conditions of Eq.~(\ref{Closure conditions}) it holds that $\mathcal{M}_{\bi{n}_1} \mathcal{M}_{\bi{n}_2}=\mathcal{M}_{\bi{n}_1+\bi{n}_2}$ and $\mathcal{M}_{\bi{n}_2}\mathcal{M}_{\bi{n}_3}=\mathcal{M}_{\bi{n}_2+\bi{n}_3}$ and using these relations we have
\begin{eqnarray}
\mathcal{M}_{\bi{n}_1+\bi{n}_2}\mathcal{M}_{\bi{n}_3}=\mathcal{M}_{\bi{n}_1}\mathcal{M}_{\bi{n}_2+\bi{n}_3} \;\;
\Longrightarrow \;\; \mathcal{M}_{\bi{n}_1+\bi{n}_2+\bi{n}_3}=\mathcal{M}_{\bi{n}_1+\bi{n}_2+\bi{n}_3}.
\end{eqnarray}
Thus, the magnetic translations under the conditions of Eq.~(\ref{Closure conditions}) satisfy associativity as well. Thus, we conclude that the magnetic translation operators under the conditions of Eq.~(\ref{Closure conditions}) satisfy all four axioms of a group.

\textit{Abelianity}.---The great advantage of standard lattice translations in Bloch theory is that the translation operators not only commute with the Hamiltonian but they also form an abelian group. This means that the eigenstates of the Hamiltonian $\hat{H}$ can be chosen to be simultaneous eigenfunctions of all the lattice translation operators $\hat{T}_{{\mathbf{n}}}$~\cite{Mermin}. From this viewpoint it is important also to check whether the magnetic translation group is commutative.

To check this property we consider the magnetic translations $\mathcal{M}_{{\mathbf{n}}}$ and $\mathcal{M}_{{\mathbf{n}^{\prime}}}$ and compute their commutator
\begin{eqnarray}
[\mathcal{M}_{{\mathbf{n}}},\mathcal{M}_{{\mathbf{n}^{\prime}}}]=\mathcal{M}_{{\mathbf{n}}}\cdot\mathcal{M}_{{\mathbf{n}^{\prime}}}-\mathcal{M}_{{\mathbf{n}^{\prime}}}\cdot\mathcal{M}_{{\mathbf{n}}}
\end{eqnarray}
Using Eq.~(\ref{groupstructure}) we have
\begin{eqnarray}
[\mathcal{M}_{{\mathbf{n}}},\mathcal{M}_{{\mathbf{n}^{\prime}}}]&=&\left(e^{\textrm{i}\mathbf{B}_{m^{\prime}}\cdot\mathbf{R}_{\mathbf{n}}}-e^{\textrm{i}\mathbf{B}_{m}\cdot\mathbf{R}_{\mathbf{n}^{\prime}}}\right)\mathcal{M}_{{\mathbf{n}+\mathbf{n}^{\prime}}}\\
&=&\left(e^{\textrm{i}\frac{\Phi}{\Phi_0}2\pi m^{\prime}n}-e^{\textrm{i}\frac{\Phi}{\Phi_0}2\pi mn^{\prime}}\right)e^{\textrm{i}\frac{\Phi }{\Phi_0}\frac{a_2\cos\theta}{a_1}2\pi mm^{\prime}}\mathcal{M}_{{\mathbf{n}+\mathbf{n}^{\prime}}}.\nonumber
\end{eqnarray}
Under the magnetic flux conditions $\Phi/\Phi_0=q$ given in Eq.~(\ref{Closure conditions}) the magnetic translations actually commute 
\begin{eqnarray}
[\mathcal{M}_{{\mathbf{n}}},\mathcal{M}_{{\mathbf{n}^{\prime}}}]=0
\end{eqnarray}
which means that they form an abelian group~\footnote{We would like to point out that the magnetic translation operators commute also if $m^{\prime}n=mn^{\prime}$ irrespective of the group structure conditions in Eq.~(\ref{Closure conditions}). The condition $m^{\prime}n=mn^{\prime}$ though mean that whether two magnetic translations commute depends on the path that they follow. This is the path depedence that we mentioned previously, for the non-abelian case}. From the above result we see that the magnetic flux conditions $\Phi/\Phi_0$ guarantee that the magnetic translations commute. The magnetic flux conditions also showed up in the previous section in Eq.~(\ref{flux conditions torus}) when we imposed the torus structure on the Bloch Hamiltonian $\hat{H}(\bi{k},\bi{r})$. This relation hints towards a novel connection, namely that the torus structure of the Bloch Hamiltonian (which is a topological property) forces commutativity on the magnetic translation operators (which is a purely algebraic property). This connection shows that these two properties are actually equivalent.
\begin{figure}[H]
\begin{center}
  \includegraphics[height=3.5cm, width=0.7\columnwidth]{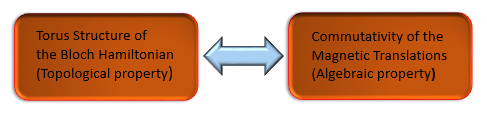}
\caption{\label{EquivalenceMTG}Schematic depiction of the equivalence between the torus structure of the Bloch Hamiltonian and the commutativity of the magnetic translations.}  
\end{center}
\end{figure}
The equivalence between the torus structure of the Bloch Hamiltonian and the commutativity of the magnetic translations to the best of our knowledge has not been reported before. To conclude, the abelian magnetic translation group (MTG) is defined as
\begin{eqnarray}
\textrm{ Abelian MTG}:\; \Big\{\mathcal{M}_{\bi{n}}=e^{\textrm{i}\mathbf{B}_m\cdot\mathbf{r}}\hat{T}_{{\mathbf{n}}}\;\; \big|\;\; \Phi=q \Phi_0\;\;  \textrm{and}\;\; a_2\cos\theta=la_1, \;\; q,l \in \mathbb{Z}, \;\bi{n} \in \mathbb{Z}^2 \Big\}.\nonumber\\
\end{eqnarray}

\subsection{Magnetic Unit Cell}

Taking now a closer look on the conditions of Eq.~(\ref{Closure conditions}), which guarantee commutativity and group structure, one realizes that they are rather restrictive, because only integer multiples of the flux quantum are permitted, and in addition not all geometries are allowed. From a physical point of view this is quite unsatisfactory. Because for standard materials with a lattice constant of the order of $1$ \AA, the allowed strength of magnetic field would be $B\sim 10^5$ T. Such field-strengths are completely out of reach and in principle would make the whole MTG construction useless. 

\begin{figure}
\centering
\begin{subfigure}{.5\textwidth}
  \centering
  \includegraphics[height=1.8in ,width=0.7\linewidth]{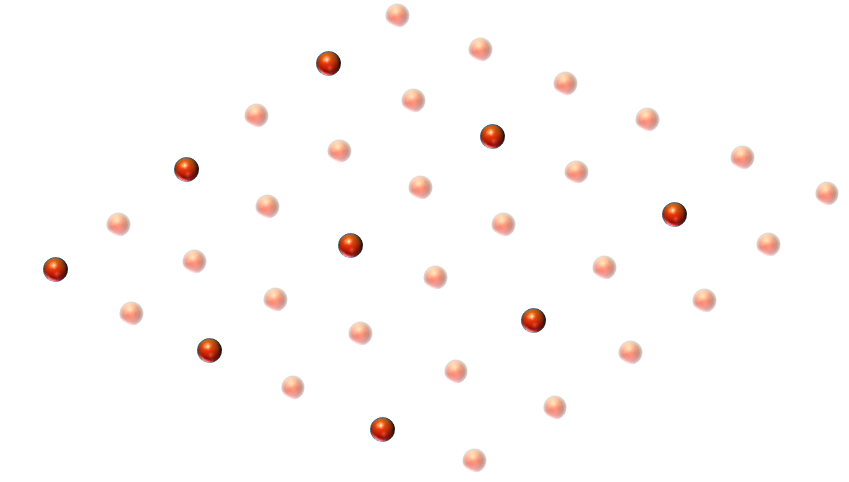}
  \caption{Magnetic unit cell for $\Phi/\Phi_0=1/2$.}
\end{subfigure}%
\begin{subfigure}{.5\textwidth}
  \centering
  \includegraphics[height=1.8in ,width=0.7\linewidth]{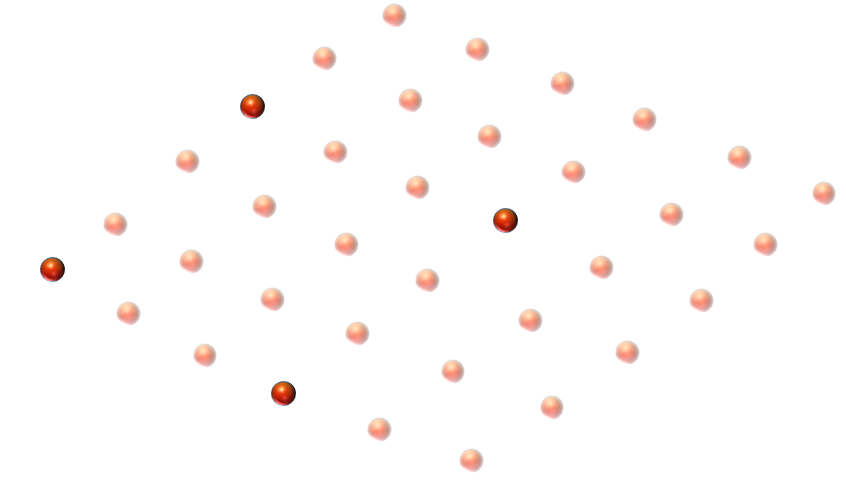}
  \caption{Magnetic unit cell for $\Phi/\Phi_0=1/3$.}
\end{subfigure}
\label{fig:test}
\caption{Graphic representation of two different magnetic unit cells. The bright spheres represent the enlarged magnetic cells. Each magnetic unit cell contains a number of fundamental unit cells which is equal to the square of the inverse relative flux $(\Phi_0/\Phi)^2$.}\label{magnetic unit cell}
\end{figure}

A solution to this problem is to introduce an enlarged unit cell~\cite{Kohmoto} by multiplying one of the lattices constants by an integer $p$
\begin{eqnarray}\label{enlargedlattice}
a_1 \longrightarrow pa_1 \;\; \textrm{where}\;\; p\in \mathbb{Z}.
\end{eqnarray}
With this one can define a new set of Bravais vectors 
\begin{eqnarray}\label{Magneticbravaisvectors}
\mathbf{R}^p_{\mathbf{n}}=\left(n (p a_1)+ma_2\cos\theta\right)\mathbf{e}_x+ma_2\sin\theta\mathbf{e}_y.
\end{eqnarray}
The fact that $p$ is an integer guarantees that the new set of Bravais vectors $\mathbf{R}^p_{\mathbf{n}}$ are a subset of the original set of Bravais vectors $\mathbf{R}_{\mathbf{n}}$, \{$\mathbf{R}^p_{\mathbf{n}}\} \subseteq \{\mathbf{R}_{\mathbf{n}}\}$. Substituting now the new enlarged lattice constant $pa_1$ into Eq.~(\ref{Closure conditions}) for the flux conditions, one finds that the new conditions for the allowed magnetic flux are
\begin{equation}\label{rationalFluxes}
    \frac{\Phi}{\Phi_0}=\frac{q}{p} \;\;\; \textrm{where} \;\;\; p,q \in \mathbb{Z}.
\end{equation}
From the equation above we see that by introducing the enlarged cell we can also treat rational multiples of the flux quantum. Due to the fact that for each value of the magnetic field we need to construct a different Bravais lattice, as depicted in Fig.~\ref{magnetic unit cell}, this construction is called the magnetic unit cell~\cite{Kohmoto}. 
Although this is an insightful construction which increases substantially the applicability of the magnetic translation group, and allows to treat rational fluxes, still from a practical point of view it has certain limitations. The main problem with the magnetic unit cell is that for smaller and smaller magnetic fluxes the magnetic unit cell needs to become larger and larger. For example if one is interested in magnetic fields of the order of 1 Tesla, which can be achieved experimentally, then the magnetic flux ratio should be roughly $1/1000$ which means that the magnetic unit cell should be 1000 times larger than the fundamental unit cell. This means that in such a computation a thousand unit cells of the solid need to be included. Obviously, such a computation becomes rather cumbersome and expensive from a numerical point of view.

In conclusion, the magnetic translation group is the fundamental structure that governs the behavior of Bloch electrons in the presence of a homogeneous magnetic field, and has led to several very important discoveries. Like the topological description of the quantization of the Hall conductance~\cite{Kohmoto, Thouless}. Further, it has allowed for the fundamental understanding of the band splitting that emerges in the presence of a magnetic field, which leads to the formation of the fractal spectrum of the Hofstadter butterfly~\cite{Hofstadter}, and has found many applications in tight-binding models with the Peierls substitution~\cite{Hofstadter, ClaroWannier, Wannier, PedersenGrapheneButterfly, ButterflyOpticallattice}. However, the question of whether the magnetic translations can allow for the description of solids in magnetic fields and provide an ansatz analogous to Bloch's ansatz remains open. So far, such an ansatz solution based on eigenfunctions of the magnetic translation group, which would allow for the computation of energy bands directly from the Schr\"{o}dinger equation, has not been constructed. As a consequence a complete description of periodic solids in magnetic fields is still missing.

\chapter{Quantum Electrodynamical Bloch Theory}\label{QED Bloch theory}
\begin{displayquote}
\footnotesize{A common mistake of beginners is the desire to understand everything completely right away.}
\end{displayquote}
\begin{flushright}
  \footnotesize{A.~B.~Migdal\\
Qualitative Methods in Quantum Theory~\cite{MigdalQualitativeMethods}}
\end{flushright}
In the previous chapter we saw how the Schr\"{o}dinger equation for electrons in a periodic potential coupled to a classical, homogeneous magnetic field, fails to capture the fundamental symmetry of this system, namely translational invariance. The question that we aim to answer here is: Can translational symmetry for this system be restored by embedding the problem into QED?
\begin{figure}[H]
\begin{center}
  \includegraphics[width=0.5\columnwidth]{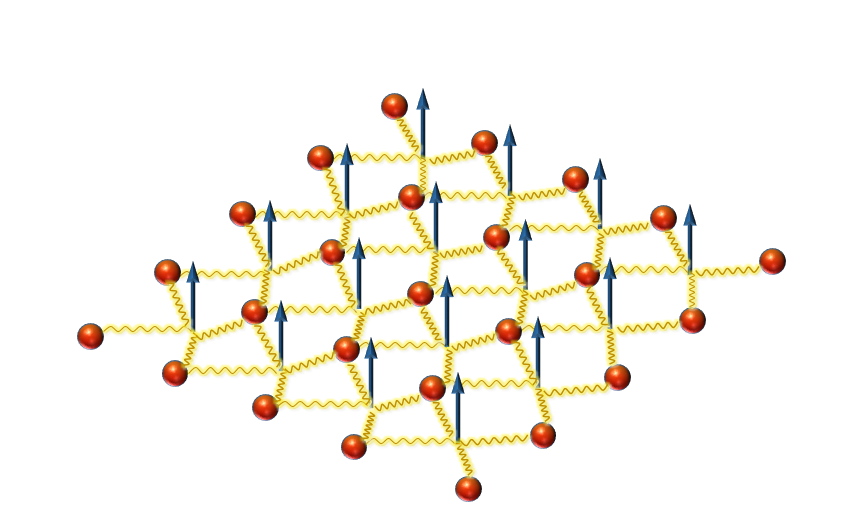}
\caption{\label{QED Bloch Cartoon}Cartoon depiction of a 2D periodic material in a uniform perpendicular magnetic field embedded in the QED vacuum.}  
\end{center}
\end{figure}
If translational symmetry can be restored, then the tools of Bloch theory, which we described in great detail in the previous section, would be applicable. This would imply that we can use Bloch's theorem in order to construct a generalized Bloch ansatz for electron-photon systems in non-relativistic QED. 

Further, the question about translational inavariance it is not just an academic or theoretical question. It is actually motivated by the progress in 2D Moire materials and by the fact that such 2D periodic systems in strong magnetic fields, or large fractions of the flux quantum, can now be probed experimentally with high accuracy~\cite{DeanButterfly, WangButterfly, BarrierButterfly, ForsytheButterfly}. Consequently, answering the question about translational invariance, could potentially lead to novel observable effects and applications in the field of cavity QED, as it was proposed in~\cite{rokaj2019}.

\section{Translational Symmetry with Homogeneous Magnetic Fields}

To address the question we posed, we will employ the non-relativistic version of QED, described by the Pauli-Fierz Hamiltonian of Eq.~(\ref{Pauli Fierz Hamiltonian}). More specifically, our starting point is the Pauli-Fierz Hamiltonian for $N$ interacting electrons, in a periodic potential in the presence of a classical, homogeneous, magnetic field and coupled to a single quantized mode of the photon field~\cite{cohen1997photons, spohn2004, rokaj2017} 
\begin{eqnarray}\label{velgauge} 
\hat{H}&=&\sum\limits^{N}_{j=1}\left[\frac{1}{2m_{\textrm{e}}}\left(\mathrm{i}\hbar \mathbf{\nabla}_{j}+e\hat{\mathbf{A}}(\bi{r}_j)+e\mathbf{A}_{\textrm{ext}}(\bi{r}_j)\right)^2 +v_{\textrm{ext}}(\mathbf{r}_{j})\right]\nonumber\\
&+&\frac{1}{4\pi\epsilon_0}\sum\limits^{N}_{j< k}\frac{e^2}{|\mathbf{r}_j-\mathbf{r}_k|}+\hbar\omega\left(\hat{a}^{\dagger}\hat{a}+\frac{1}{2}\right).
\end{eqnarray}
Here $\mathbf{A}_{\textrm{ext}}(\mathbf{r})$ is the external vector potential which gives rise to a homogeneous magnetic field $\mathbf{B}_{\textrm{ext}}=\nabla \times \mathbf{A}_{\textrm{ext}}(\mathbf{r})=\mathbf{e}_z B$ in the $z$-direction, and is given by the expression $\mathbf{A}_{\textrm{ext}}(\mathbf{r})=-\mathbf{e}_x B y$~\cite{Landau}. Moreover, $\hat{\mathbf{A}}(\bi{r})$ is the quantized vector potential of the photon field, beyond the dipole approximation,~\cite{spohn2004}
\begin{equation}\label{eq2.4b}
\hat{\bi{A}}(\bi{r})=\left(\frac{\hbar}{\epsilon_0 V}\right)^{\frac{1}{2}}\frac{\bm{\varepsilon}}{\sqrt{2\omega}}\left( \hat{a}e^{\mathrm{i}\bm{\kappa}\cdot\bi{r}}+\hat{a}^{\dagger}e^{-\mathrm{i}\bm{\kappa}\cdot\bi{r}}\right),
\end{equation}
where $\bm{\kappa}$ is the wave vector of the photon mode, $\omega=c|\bm{\kappa}|$ is the frequency, and $\bm{\varepsilon}$ is the polarization vector of the field~\cite{cohen1997photons, spohn2004}. We would also like to remind the reader that the annihilation and creation operators in terms of the displacement coordinates $q$ and their conjugate momenta $\partial_q=\partial/\partial q$, are  $\hat{a}=\left(q+\partial_q\right)/\sqrt{2}$ and $\hat{a}^{\dagger}=\left(q-\partial_q\right)/\sqrt{2}$, as defined in Eq.~(\ref{q coordinate and momenta}). 

The quantized photon field in QED captures the back-reaction of matter to the electromagnetic field. For that purpose the quantized field is chosen to have the same polarization with the external field, $\bm{\epsilon}=\mathbf{e}_x$, because otherwise the classical field would not be able to to influence the quantized field, and vise versa. These back-reaction effects are very important in solid-state physics, e.g., in the semi-classical microscopic-macroscopic connection that determines the induced fields inside a material~\cite{Mochan, Maki, Ehrenreich}. 

Here we are interested in periodic materials and for that purpose the external potential is taken to be periodic, $v_{\textrm{ext}}(\mathbf{r})=v_{\textrm{ext}}(\mathbf{r}+\mathbf{R}_{\mathbf{n}})$, where $\mathbf{R}_{\mathbf{n}}$ is a Bravais lattice vector~\cite{Mermin}. To simplify the analysis we choose the lattice vectors $\mathbf{R}_{\mathbf{n}}=na_x\mathbf{e}_x+ma_y\mathbf{e}_y+la_z\mathbf{e}_z$. Although, the scalar potential is periodic, it is clear that the classical external vector potential $\mathbf{A}_{\textrm{ext}}(\mathbf{r})$ breaks translational symmetry, because it is linear in $y$, and that the quantized vector potential~(\ref{eq2.4b}) breaks translational symmetry as well. As a consequence the generic Pauli-Fierz Hamiltonian~(\ref{velgauge}) for a classical magnetic field and the photon field, beyond the dipole approximation, is not periodic and Bloch theory is not applicable.

However, recently it was proposed that the problem of the broken translational symmetry can be resolved in the long-wavelength or optical limit~\cite{rokaj2019}. As we explained in chapter~\ref{Length Gauge QED}, in the optical limit the quantized vector potential is assumed to be spatially uniform  
\begin{eqnarray}
\hat{\bi{A}}=\sqrt{\frac{\hbar}{\epsilon_0V\omega}}\mathbf{e}_x q.
\end{eqnarray}
But this approximation raises another question: what is exactly the meaning of the optical limit for a solid?  

The optical limit (or dipole approximation) is usually employed when the size of the electronic system is much smaller than the wavelength of the electromagnetic field. But solids are macroscopic systems, and compared to the size of an atom are infinitely large, especially in the framework of Bloch theory where infinite periodicity is assumed. This means that for the optical limit to be applied in the case of a solid, the wavelength of the field should be infinite and respectively the frequency should become arbitrarily small and tend to zero. Naively, taking the limit $\omega\rightarrow 0$ for the quantized vector potential $\hat{\bi{A}}$ seems to lead to divergencies in~(\ref{eq2.4b}). However, there is a way to perform this limit in a consistent fashion without encountering any divergences, by taking into account the back-reaction of matter due to the diamagnetic $\hat{\bi{A}}^2$ term~\cite{rokaj2019}. 

\begin{figure}[H]
\begin{center}
  \includegraphics[width=0.6\columnwidth]{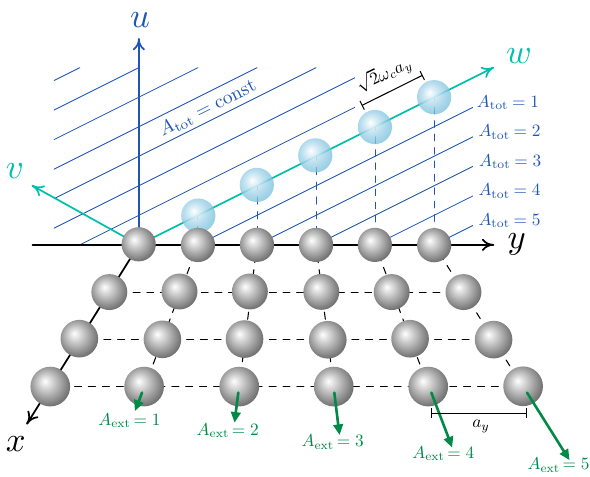}
\caption{\label{QED translation}Schematic depiction of a 2D periodic material in the presence of a homogeneous magnetic field, coupled also to a quantized photon-mode. The classical external field $\mathbf{A}_{\textrm{ext}}$ breaks periodicity along $y$. By including the quantized field $\hat{\mathbf{A}}$, proportional to the photonic coordinate $u$, translational symmetry gets restored in the polaritonic direction $w$ which is a linear combination of $y$ and $u$. The lattice periodicity along $w$ is $\sqrt{2} \omega_c a_{y}$, where $\omega_c=eB/m_{\textrm{e}}$. Thus, when embedding the $(x,y)$ plane into the higher-dimensional space involving the coordinate $u$, periodicity gets restored.}  
\end{center}
\end{figure}

To do so, we isolate the purely photonic part of the Hamiltonian $\hat{H}$ which includes the energy of the bare photon mode $\omega$ plus the square of the vector potential
\begin{eqnarray}
\hat{H}_p=\hbar\omega\left(\hat{a}^{\dagger}\hat{a}+\frac{1}{2}\right)+\frac{Ne^2}{2m_{\textrm{e}}}\hat{\mathbf{A}}^2.
\end{eqnarray}
The photonic part given in terms of $q$ and $\partial/\partial q$ takes the form
\begin{eqnarray}
\hat{H}_p=\frac{\hbar\omega}{2}\left(-\frac{\partial^2}{\partial q^2}+q^2\right)+q^2\frac{\hbar Ne^2}{2m_{\textrm{e}}\omega\epsilon_0 V},
\end{eqnarray}
and after introducing the dressed frequency $\widetilde{\omega}^2=\omega^2+\omega^2_p$ and the coordinate $u=q\sqrt{\widetilde{\omega}/\omega}$, the purely photonic operator $\hat{H}_p$ takes the form a harmonic oscillator
\begin{eqnarray}
\hat{H}_p=\frac{\hbar\widetilde{\omega}}{2}\left(-\frac{\partial^2}{\partial u^2}+u^2\right)
\end{eqnarray}
where the frequency $\omega_p$ is the diamagnetic shift frequency that we also encountered in section~\ref{2DEG in QED} in Eq.~(\ref{plasma frequency}), which depends on the electron density $n_{\mathrm{e}}$ and is given by $\omega_p=\sqrt{n_{\mathrm{e}} e^2/m_{\textrm{e}}\epsilon_0}$. The quantized vector potential in terms of the scaled coordinate $u$ is 
\begin{eqnarray}
\hat{\bi{A}}=u\mathbf{e}_x\sqrt{\frac{\hbar}{\epsilon_0 V\widetilde{\omega}}}.
\end{eqnarray}
Performing now the optical limit the dressed frequency $\widetilde{\omega}$ goes to the plasma frequency $\omega_p$, without any divergence showing up for the quantized vector ptoential, as it was promised. Substituting then, the expressions for the purely photonic part $\hat{H}_p$ and the vector potential $\hat{\mathbf{A}}$ back into~(\ref{velgauge}) we obtain the Pauli-Fierz Hamiltonian in the (strict) optical limit
\begin{eqnarray}\label{optical}
\hat{H}&=&\sum^{N}_{j=1}\left[-\frac{\hbar^2}{2m_{\textrm{e}}}\nabla^2_j+\frac{\textrm{i}\hbar e}{m_{\textrm{e}}}\left(\hat{\mathbf{A}}+\mathbf{A}_{\textrm{ext}}(\mathbf{r}_j)\right)\cdot\nabla_j +v_{\textrm{ext}}(\mathbf{r}_j)\right]\\
&+&\frac{1}{4\pi\epsilon_0}\sum\limits^{N}_{j< k}\frac{e^2}{|\mathbf{r}_j-\mathbf{r}_k|}+\frac{e^2}{2m_{\textrm{e}}}\sum^{N}_{j=1}\left(\hat{\mathbf{A}}+\mathbf{A}_{\textrm{ext}}(\textbf{r}_j)\right)^2-\frac{\hbar\omega_p}{2}\frac{\partial^2}{\partial u^2}.\nonumber
\end{eqnarray}
We note that in the optical limit the quantized vector potential is  
\begin{eqnarray}
\hat{\mathbf{A}}=\mathbf{e}_xu\sqrt{\frac{\hbar}{\epsilon_0V\omega_p}}.
\end{eqnarray}
Let us check now the translational properties of the Hamiltonian in the optical limit. For a periodic potential $\hat{H}$ is still not periodic in the electronic coordinates, because $\mathbf{A}_{\textrm{ext}}(\mathbf{r})$ is linear in $y$. But the optical Hamiltonian $\hat{H}$ is periodic under the generalized translation in the full electronic plus photonic configuration space
\begin{eqnarray}\label{symmetry}
(\mathbf{r}_j,u)\longrightarrow\left(\mathbf{r}_j+\mathbf{R}_{\mathbf{n}},u+Bma_y\sqrt{\epsilon_0 V\omega_p/\hbar}\right).
\end{eqnarray}
This proves our claim that in the optical limit the broken translational symmetry, due to a homogeneous magnetic field, gets restored when the problem is embedded in QED. This fact is also depicted geometrically in Fig.~\ref{QED translation}.

\newpage

\section{Effective Hamiltonian \& QED-Bloch Expansion }
\begin{displayquote}
\footnotesize{It is very common for a physicist to start from a very crude model, without knowing whether his assumptions are valid. You have no need to worry about your result.}
\end{displayquote}
\begin{flushright}
  \footnotesize{Heisenberg to Tomonaga\\
QED and the Men Who Made It~\cite{SchweberQEDHistory}}
\end{flushright}

Having restored translational symmetry, by embedding the purely electronic problem into QED, our aim now is to go one step further and construct a Bloch-type ansatz in the polaritonic (electronic plus photonic) space and to derive a Bloch-type central equation for the description of solids in a classical, uniform magnetic field, coupled also to the quantized electromagnetic field. 

To make the problem tractable, instead of expressing the unfeasible many-body interacting Hamiltonian of Eq.~(\ref{optical}), we will employ the independent electron approximation which is similar to the standard approach of density-functional theory (DFT). We note that this independent-electron approach is consistent with Bloch theory, which is not a theory of a single electron in a periodic potential, but of many non-interacting electrons. 

To incorporate the fact that the charged particles couple collectively to the photon field, we will use an effective electron density, which will allow us to capture the back-reaction of matter to the photon field. For the inclusion of any further effects, like exchange and correlation effects, one would need the inclusion of effective fields as introduced in quantum-electrodynamical DFT~\cite{ruggenthaler2014, ruggenthaler2015, TokatlyPRL}. Introducing the cyclotron frequency we obtain the effective Hamiltonian in the independent electron approximation which was proposed in~\cite{rokaj2019}
\begin{eqnarray}\label{Approptical}
\hat{H}_{eff}&=&-\frac{\hbar^2}{2m_{\textrm{e}}}\nabla^2+\mathrm{i}\hbar\mathbf{e}_x\left(u\sqrt{\hbar\omega_p/m_{\textrm{e}}}-y\omega_c\right)\cdot\nabla\\
&+&v_{\textrm{ext}}(\mathbf{r})+\frac{m_{\textrm{e}}}{2}\left(u\sqrt{\hbar\omega_p/m_{\textrm{e}}}-y\omega_c\right)^2-\frac{\hbar\omega_p}{2}\frac{\partial^2}{\partial u^2}.\nonumber
\end{eqnarray}
We note that the effective Hamiltonian is invariant under the following translation
\begin{eqnarray}\label{1psymmetry}
(\mathbf{r},u)\longrightarrow \left(\mathbf{r}+\mathbf{R}_{\mathbf{n}},u+ma_y\omega_c \sqrt{m_{\textrm{e}}/\hbar\omega_p}\right)
\end{eqnarray}
that acts on both, the electronic and the photonic coordinates. To describe properly this symmetry in the polaritonic space we will go to a new set of coordinates. First, we introduce the scaled coordinates $\widetilde{u}$ and $\widetilde{y}$ 
\begin{eqnarray}
\widetilde{u}=u\sqrt{\frac{\hbar\omega_p}{m_{\textrm{e}}}} \;\;\; \textrm{and}\;\;\; \widetilde{y}= \omega_c y.
\end{eqnarray}
and the mass parameters $m_{p}$ and $m_{c}$
\begin{eqnarray}\label{scaled masses}
m_p=\frac{m_{\textrm{e}}}{\omega^2_p}\;\;\; \textrm{and}\;\;\; m_{c}=\frac{m_{\textrm{e}}}{\omega^2_c},
\end{eqnarray}
and the effective Hamiltonian takes the form
\begin{eqnarray}
\hat{H}_{eff}&=&-\frac{\hbar^2}{2m_{\textrm{e}}}\left(\frac{\partial^2}{\partial x^2}+\frac{\partial^2}{\partial z^2}\right) -\frac{\hbar^2}{2m_{c}}\frac{\partial^2}{\partial\widetilde{y}^2}+v_{\textrm{ext}}(\bi{r})\\
&+&\textrm{i}\hbar\left(\widetilde{u}-\widetilde{y}\right)\frac{\partial}{\partial x} +\frac{m_{\textrm{e}}}{2}\left(\widetilde{u}-\widetilde{y}\right)^2-\frac{\hbar^2}{2m_p}\frac{\partial^2}{\partial\nonumber \widetilde{u}^2}.\nonumber
\end{eqnarray}
In addition, we introduce the relative distance and center of mass (like) coordinates between $\widetilde{u}$ and $\widetilde{y}$
\begin{eqnarray}\label{w and v coordinates}
w=\frac{m_p\widetilde{u}+m_c\widetilde{y}}{\sqrt{2}M}\;\;\; \textrm{and}\;\;\; v=\frac{\widetilde{u}-\widetilde{y}}{\sqrt{2}},
\end{eqnarray}
and the Hamiltonian $\hat{H}_{eff}$ simplifies to
\begin{eqnarray}
\hat{H}_{eff}=-\frac{\hbar^2}{2m_{\textrm{e}}}\left(\frac{\partial^2}{\partial x^2}+\frac{\partial^2}{\partial z^2}\right) -\frac{\hbar^2}{2M}\frac{\partial^2}{\partial w^2}+v_{\textrm{ext}}(\bi{r})+ \textrm{i}\hbar\sqrt{2}v\frac{\partial}{\partial x} +m_{\textrm{e}}v^2-\frac{\hbar^2}{2\mu}\frac{\partial^2}{\partial v^2}.\nonumber\\
\end{eqnarray}
where the mass parameters $M$ and $\mu$ are
\begin{eqnarray}\label{mass polaritonic parameters}
M=\frac{m_p+m_c}{2}\;\;\; \textrm{and}\;\;\; \mu=\frac{m_pm_c}{M}.
\end{eqnarray}
Furthermore, the effective Hamiltonian by performing a square completion can be written in the compact form
\begin{eqnarray}\label{Heff compact}
\hat{H}_{eff}=-\frac{\hbar^2}{2m_{\textrm{e}}}\frac{\partial^2}{\partial z^2}-\frac{\hbar^2}{2M}\frac{\partial^2}{\partial w^2}+ v_{\textrm{ext}}(\bi{r})-\frac{\hbar^2}{2\mu}\frac{\partial^2}{\partial v^2 }+ \frac{\mu\Omega^2}{2}\left(v+\frac{\textrm{i}\hbar}{\sqrt{2}m_{\textrm{e}}}\frac{\partial}{\partial x}\right)^2.
\end{eqnarray}
where the dressed frequency $\Omega^2$ is 
\begin{eqnarray}\label{upper polariton frequency}
\Omega^2=\frac{2m_{\textrm{e}}}{\mu}=\omega^2_p+\omega^2_c.
\end{eqnarray}
Further, the original electronic vector $\bi{r}=(x,y,z)$ in the new polaritonic coordinate system is
\begin{eqnarray}
\bi{r}=(x,y,z)=\left(x,\frac{w}{\sqrt{2}\omega_c}-\frac{m_p v}{\sqrt{2}M\omega_c},z\right).
\end{eqnarray}
It is important to note that the coordinates $v$ and $w$ are independent because their respective coordinates and momenta commute.

\textit{Setting the Geometry.}---The external potential in a solid is of course periodic $v_{\textrm{ext}}(\mathbf{r})=v_{\textrm{ext}}(\mathbf{r}+\mathbf{R}_{\mathbf{n}})$ where $\mathbf{R}_{\mathbf{n}}$ is a Bravais lattice vector with $\mathbf{n}=(n,m,l)\in \mathbb{Z}^2$. The Bravais lattice vector in general is $\mathbf{R}_{\mathbf{n}}=n\mathbf{a}_1+m\mathbf{a}_2+l\bi{a}_3$ where $\mathbf{a}_1$, $\bi{a}_2$ and $\mathbf{a}_3$ are the primitive vectors which lie in different directions and span the lattice. Without loss of generality we can choose the vector $\mathbf{a}_1$ to be in the $x$-direction $\mathbf{a}_1=a_1\mathbf{e}_x$. The second primitive vector in this case is $\mathbf{a}_2=a_2\cos\theta\mathbf{e}_x+a_2\sin\theta\mathbf{e}_y$ where $\theta$ is the angle between the vectors $\mathbf{a}_2$ and $\mathbf{a}_1$. Here, we are mostly interested in 2D materials and the $z$-direction is not of much relevance for us. Due to this reason we choose the third primitive vector to be $\bi{a}_3=a_3\bi{e}_z$, which however limits the possible 3D Bravais lattices that we can treat. 

Thus, the Bravais lattice vectors are 
\begin{eqnarray}\label{3DBravaisvectors}
\mathbf{R}_{\mathbf{n}}=\left(na_1+ma_2\cos\theta\right)\mathbf{e}_x+ma_2\sin\theta\mathbf{e}_y+la_3\bi{e}_z.
\end{eqnarray}
Then, the reciprocal lattice vectors are $\mathbf{G}_{\mathbf{n}^{\prime}}=n^{\prime}\mathbf{b}_1+m^{\prime}\mathbf{b}_2+l^{\prime}\bi{b}_3$ with $\mathbf{n}^{\prime}=(n^{\prime},m^{\prime})\in \mathbb{Z}^2$. The defining relation for the vectors $\mathbf{b}_{1}, \bi{b}_2$ and $\mathbf{b}_{3}$ is~\cite{Mermin, Callaway}
\begin{eqnarray}\label{3Dreciprocal}
\mathbf{b}_{i}\cdot \mathbf{a}_j=2\pi \delta_{ij}, \;\;\; i,j=1,2,3.
\end{eqnarray}
With the choice we made for the primitive vectors, the reciprocal primitive vectors satisfying Eq.~(\ref{3Dreciprocal}) are
\begin{eqnarray}
\mathbf{b}_1=\frac{2\pi}{a_1}\mathbf{e}_x-\frac{2\pi\cos\theta}{a_1\sin\theta}\mathbf{e}_y\;\;\;\; \mathbf{b}_2=\frac{2\pi}{a_2\sin\theta}\mathbf{e}_y\;\;\;\textrm{and}\;\; \bi{b}_3=\frac{2\pi}{a_3}\bi{e}_z.
\end{eqnarray}
Thus, the reciprocal lattice vectors are
\begin{eqnarray}
\mathbf{G}_{\mathbf{n}^{\prime}}=\frac{2\pi n^{\prime}}{a_1}\mathbf{e}_x +\left[\frac{2\pi m^{\prime}}{a_2\sin\theta}-\frac{2\pi n^{\prime}\cos\theta}{a_1\sin\theta}\right] \mathbf{e}_y+\frac{2\pi l^{\prime}}{a_3}\bi{e}_z.
\end{eqnarray}
which for convenience we will write as
\begin{eqnarray}\label{3Dreciprocallattice}
&&\mathbf{G}_{\mathbf{n}^{\prime}}=\left(G^x_{n^{\prime}}, G_{m^{\prime},n^{\prime}},G^z_{l^{\prime}}\right)\;\; \textrm{where}\;\;G_{m^{\prime},n^{\prime}}=\frac{G^y_{m^{\prime}}}{\sin\theta}-\frac{G^x_{n^{\prime}}\cos\theta}{\sin\theta}\nonumber\\
 &&\textrm{and}\;\;  G^x_{n^{\prime}}=\frac{2\pi n^{\prime}}{a_1 },\;\;\;G^y_{m^{\prime}}=\frac{2\pi m^{\prime}}{a_2}\;\;\textrm{and}\;\; G^z_{l^{\prime}}=\frac{2\pi l^{\prime}}{a_3}.
\end{eqnarray}
With these choices we have defined our setting and the lattices that we aim to describe. Then, the external vector potential can be written in terms of a Fourier series as follows
\begin{eqnarray}
v_{\textrm{ext}}(\bi{r})=\sum_{\bi{n}^{\prime}}V_{\bi{n}^{\prime}} e^{\textrm{i}\bi{G}_{\bi{n}^{\prime}}\cdot \bi{r}},
\end{eqnarray}
which in terms of the polaritonic coordinates $w$ and $u$ is
\begin{eqnarray}\label{potential Fourier}
&&v_{\textrm{ext}}(\bi{r})=\sum_{\bi{n}^{\prime}}V_{\bi{n}^{\prime}}e^{\textrm{i}\bi{G}^w_{\bi{n}^{\prime}}\cdot \bi{r}_w}e^{-\textrm{i}G^v_{m^{\prime},n^{\prime}}v} \;\;\; \textrm{where}\;\; G^v_{m^{\prime},n^{\prime}}=\frac{m_pG_{m^{\prime},n^{\prime}}}{\sqrt{2}M\omega_c}\\
&&\bi{G}^w_{\bi{n}^{\prime}}=(G^x_{n^{\prime}},G^w_{m^{\prime},n^{\prime}},G^z_{l^{\prime}})=(G^x_{n^{\prime}},G_{m^{\prime},n^{\prime}}/\sqrt{2}\omega_c,G^z_{l^{\prime}})  .
\end{eqnarray}

\textit{QED-Bloch Ansatz.}---The Hamiltonian $\hat{H}_{eff}$ of Eq.~(\ref{Heff compact}) is invariant under the translations in the polaritonic configuration space 
\begin{eqnarray}\label{polariton Bravais}
(x,w,z)\longrightarrow (x+na_1+ma_2\cos\theta,w+ma_2\sin\theta,z+l a_3).
\end{eqnarray}
This implies that we can use Bloch's theorem in $(x,w,z)$. Consequently, the eigenfunctions of $\hat{H}_{eff}$ can be written with the ansatz 
\begin{eqnarray}\label{BlochAnsatz}
\Psi_{\mathbf{k}}(\mathbf{r}_{w},v)=e^{\mathrm{i}\mathbf{k}\cdot\mathbf{r}_w}U^{\mathbf{k}}(\mathbf{r}_w,v)
\end{eqnarray}
where $\mathbf{r}_w=(x,w,z)$. Here the function $U^{\mathbf{k}}(\mathbf{r}_w,v)$ is periodic under the translations in the polaritonic space defined in Eq~(\ref{polariton Bravais}). The crystal momentum $\mathbf{k}=(k_x,k_w,k_z)$ corresponds to $\mathbf{r}_w$ and $k_w$ is a polaritonic quantum number. The polaritonic unit cell in the $w$-direction scales linearly with the strength of the magnetic field (see Fig.~\ref{QED translation}). The same feature appears also for the magnetic unit cell. But in the case of the magnetic unit cell only field strengths which generate a rational magnetic flux through a unit cell, are allowed~\cite{Kohmoto}. On the contrary, the polaritonic unit cell puts no restrictions on the strength of the magnetic field.

Since the function $U^{\mathbf{k}}(\mathbf{r}_w,v)$ is periodic in $\mathbf{r}_w$ we expand it in a Fourier series in $\mathbf{r}_w$, while for the $v$-dependent part we consider a generic wavefunction $\phi^{\bi{k}}_{\bi{n}}(v)$  
\begin{eqnarray}
\Psi_{\mathbf{k}}(\mathbf{r}_{w},v)=e^{\mathrm{i}\mathbf{k}\cdot\mathbf{r}_w}\sum_{\mathbf{n}}U^{\mathbf{k}}_{\mathbf{n}}e^{\mathrm{i}\mathbf{G}^w_{\mathbf{n}}\cdot\mathbf{r}_w}\phi^{\bi{k}}_{\bi{n}}(v),
\end{eqnarray}
where $\mathbf{G}^w_{\mathbf{n}}=(G^x_n,G^w_{m,n},G^z_l)$ is the reciprocal lattice vector in the $(x,w,z)$-space. We substitute the above ansatz wavefunction into $\hat{H}_{eff}$ and we have
\begin{eqnarray}
&&\sum_{\bi{n}}U^{\bi{k}}_{\bi{n}}e^{\textrm{i}(\bi{k}+\bi{G}^w_{\bi{n}})\cdot\bi{r}_w} \Bigg[\frac{\hbar^2(k_z+G^z_l)^2}{2m_{\textrm{e}}}+\frac{\hbar^2(k_w+G^w_{m,n})^2}{2M}+v_{\textrm{ext}}(\bi{r})\nonumber\\
&&-\frac{\hbar^2}{2\mu}\frac{\partial^2}{\partial v^2}+\frac{\mu \Omega}{2}\left(v-\frac{\hbar(k_x+G^x_n)}{\sqrt{2}m_{\textrm{e}}}\right)^2-E_{\bi{k}}\Bigg]\phi^{\bi{k}}_{\bi{n}}(v)=0
\end{eqnarray}
The Hamiltonian contains a harmonic oscillator which is shifted along the ``scaled" momentum in the $x$-direction $\hbar(k_x+G^x_n)/\sqrt{2}m_{\textrm{e}}$ 
\begin{eqnarray}
\hat{H}_v=-\frac{\hbar^2}{2\mu}\frac{\partial^2}{\partial v^2}+\frac{\mu \Omega^2}{2}\left(v-\frac{\hbar(k_x+G^x_n)}{\sqrt{2}m_{\textrm{e}}}\right)^2
\end{eqnarray}
and the eigenfunctions of this operator are Hermite functions $\phi_j$ of the coordinate $v-A^{k_x}_n$
\begin{eqnarray}\label{Landaupolaritons}
\phi_{j} \left(v-A^{k_x}_n\right)\;\; \textrm{where}\;\; A^{k_x}_n=\frac{\hbar(k_x+G^x_n)}{\sqrt{2}m_{\textrm{e}}}
\end{eqnarray}
with eigenenergies 
\begin{eqnarray}
\mathcal{E}_j=\hbar\Omega\left(j+\frac{1}{2}\right)=\hbar\sqrt{\omega^2_p+\omega^2_c}\left(j+\frac{1}{2}\right)
\end{eqnarray}
which are degenerate with respect to the momentum in $x$-direction. These eigenstates $\phi_{j} \left(v-A^{k_x}_n\right)$ as it was shown in~\cite{rokaj2019}
 correspond to Landau polaritons~\cite{Keller2020} and have many structural similarities to the well-known Landau levels~\cite{Landau}. We will now make use of these eigenfunctions and we will expand the wavefunction $\phi^{\bi{k}}_{\bi{n}}(v)$ in terms of this basis. Then, the polaritonic Bloch ansatz takes the form 
\begin{eqnarray}\label{BlochAnsatzHermite}
\Psi_{\mathbf{k}}(\mathbf{r}_{w},v)=e^{\mathrm{i}\mathbf{k}\cdot\mathbf{r}_w}\sum_{\mathbf{n},j}U^{\mathbf{k}}_{\mathbf{n},j}e^{\mathrm{i}\mathbf{G}^w_{\mathbf{n}}\cdot\mathbf{r}_w}\phi_j(v-A^{k_x}_n).
\end{eqnarray} 
We note that due to our choice to expand the generic wavefunction $\phi^{\bi{k}}_{\bi{n}}$ on the basis $\{\phi_j(v-A^{k_x}_n)\}$, the $v$-dependent part of the polaritonic Bloch ansatz now depends only on the Fourier index $n$ and the crystal momentum $k_x$. Substituting the above ansatz into our Schr\"{o}dinger equation we have
\begin{eqnarray}
&&\sum_{\bi{n},j}U^{\bi{k}}_{\bi{n},j}e^{\textrm{i}\bi{G}^w_{\bi{n}}\cdot\bi{r}_w} \phi_j(v-A^{k_x}_n) \Bigg[\frac{\hbar^2(k_z+G^z_l)^2}{2m_{\textrm{e}}}+\frac{\hbar^2(k_w+G^w_{m,n})^2}{2M}+v_{\textrm{ext}}(\bi{r})+\mathcal{E}_j-E_{\bi{k}}\Bigg]=0.\nonumber\\
\end{eqnarray}
Now we also use the Fourier expansion of the external potential given in Eq.~(\ref{potential Fourier}) and we have
\begin{eqnarray}
&&\sum_{\bi{n},j}U^{\bi{k}}_{\bi{n},j}e^{\textrm{i}\bi{G}^w_{\bi{n}}\cdot\bi{r}_w} \phi_j\left(v-A^{k_x}_n\right) \Bigg[\frac{\hbar^2(k_z+G^z_l)^2}{2m_{\textrm{e}}}+\frac{\hbar^2(k_w+G^w_{m,n})^2}{2M}+\mathcal{E}_j-E_{\bi{k}}\Bigg]\nonumber\\
&&+\sum_{\bi{n},\bi{n}^{\prime},j}V_{\bi{n}^{\prime}} U^{\mathbf{k}}_{\mathbf{n},j} e^{\textrm{i}\bi{G}^w_{\bi{n}+\bi{n}^{\prime}}\cdot \bi{r}_w}e^{-\textrm{i}G^v_{m^{\prime},n^{\prime}}v}\phi_j(v-A^{k_x}_n)=0
\end{eqnarray}
To eliminate the plane waves depending on $\bi{r}_w$ we multiply the above expression by $e^{\textrm{i}\bi{G}^w_{\bi{q}}\cdot \bi{r}_w}$ and we integrate over $\bi{r}_w$ and we have
\begin{eqnarray}
&&\sum_{j}U^{\bi{k}}_{\bi{n},j} \phi_j\left(v-A^{k_x}_n\right) \Bigg[\frac{\hbar^2(k_z+G^z_l)^2}{2m_{\textrm{e}}}+\frac{\hbar^2(k_w+G^w_{m,n})^2}{2M}+\mathcal{E}_j-E_{\bi{k}}\Bigg]\nonumber\\
&&+\sum_{\bi{n}^{\prime},j}V_{\bi{n}-\bi{n}^{\prime}} U^{\mathbf{k}}_{\mathbf{n}^{\prime},j} e^{-\textrm{i}G^v_{m-m^{\prime},n-n^{\prime}}v}\phi_j(v-A^{k_x}_{n^{\prime}})=0,
\end{eqnarray}
we note that after the integration over $\bi{r}_w$ we exchanged the index $\bi{q}$ with $\bi{n}$. Further, we apply from the left the bra\footnote{We note, that the standard bra and ket notation does not depend on the chosen coordinate. Here however, we keep the coordinate dependence as it will be useful to perform some shift transformations on these states and to obtain the matrix representation of the displacement operators on these states.}~$\langle \phi_i(v-A^{k_x}_n)|$ 
\begin{eqnarray}\label{beforeMatrixeq}
&&U^{\bi{k}}_{\bi{n},i} \Bigg[\frac{\hbar^2(k_z+G^z_l)^2}{2m_{\textrm{e}}}+\frac{\hbar^2(k_w+G^w_{m,n})^2}{2M}+\mathcal{E}_i-E_{\bi{k}}\Bigg]\\
&&+\sum_{\bi{n}^{\prime},j}V_{\bi{n}-\bi{n}^{\prime}} U^{\mathbf{k}}_{\mathbf{n}^{\prime},j}\; \langle \phi_i(v-A^{k_x}_n)|e^{-\textrm{i}G^v_{m-m^{\prime},n-n^{\prime}}v}|\phi_j(v-A^{k_x}_{n^{\prime}})\rangle=0.\nonumber
\end{eqnarray}
The only thing left to be computed in order to derive our QED-Bloch central equation is the matrix elements
\begin{eqnarray}
\langle \phi_i(v-A^{k_x}_n)|e^{-\textrm{i}G^v_{m-m^{\prime},n-n^{\prime}}v}|\phi_j(v-A^{k_x}_{n^{\prime}})\rangle.
\end{eqnarray}
To calculate this matrix we will perform first a change of coordinates which will give us an overall phase, independent of the integration
\begin{eqnarray}
s=v-A^{k_x}_n
\end{eqnarray}
and we have for the matrix elements
\begin{eqnarray}
e^{-\textrm{i}A^k_nG^v_{m-m^{\prime},n-n^{\prime}}}\langle \phi_i\left(s\right)|e^{-\textrm{i}G^v_{m-m^{\prime},n-n^{\prime}}s}|\phi_j\left(s+A^0_{n-n^{\prime}}\right)\rangle.
\end{eqnarray}
In order to compute the matrix elements above we will use the algebra of displacement operators~\cite{Glauber}. The plane wave $\exp(-\textrm{i}G^v_{m-m^{\prime},n-n^{\prime}}s)$ can be written as a displacement operator by using the expression for the coordinate $s$ in terms of the annihilation and creation operators $\hat{b},\hat{b}^{\dagger}$~\cite{GriffithsQM}
\begin{eqnarray}
s=\sqrt{\frac{\hbar}{2\mu\Omega}}\left(\hat{b}+\hat{b}^{\dagger}\right).
\end{eqnarray}
Using the latter we have for the plane wave $\exp(-\textrm{i}G^v_{m-m^{\prime},n-n^{\prime}}s)$
\begin{eqnarray}
e^{-\textrm{i}G^v_{m-m^{\prime},n-n^{\prime}}s}=\hat{D}\left(-\textrm{i}\sqrt{\frac{\hbar}{2\mu\Omega}} G^v_{m-m^{\prime},n-n^{\prime}}\right)
\end{eqnarray}
In addition, the wavefunction 
\begin{eqnarray}
\phi_j\left(s+A^0_{n-n^{\prime}}\right)
\end{eqnarray}
can be written using the translation operator
\begin{eqnarray}
\phi_j\left(s+A^0_{n-n^{\prime}}\right)=\hat{T}\left(A^0_{n-n^{\prime}}\right)\phi_j(s).
\end{eqnarray}
The translation operator is given by the expression~\cite{Mermin}
\begin{eqnarray}
\hat{T}\left(A^0_{n-n^{\prime}}\right)=\exp\left(A^0_{n-n^{\prime}}\partial_s\right).
\end{eqnarray}
The differential operator $\partial_s$ in terms of annihilation and creation operators is 
\begin{eqnarray}
\partial_s\equiv\frac{\partial}{\partial s}=\sqrt{\frac{\mu\Omega}{2\hbar}}\left(\hat{b}-\hat{b}^{\dagger}\right).
\end{eqnarray}
This implies that the translation operator can also be written as a displacement operator~\cite{Glauber}
\begin{eqnarray}
\hat{T}\left(A^0_{n-n^{\prime}}\right)=\hat{D}\left(-\sqrt{\frac{\mu\Omega}{2\hbar}}A^0_{n-n^{\prime}}\right).
\end{eqnarray}
Using the expressions we derived in terms of the displacement operators we obtain the following expression for the matrix elements
\begin{eqnarray}\label{matrixeq}
\exp(-\textrm{i}A^{k_x}_n G^v_{m-m^{\prime},n-n^{\prime}}) \langle \phi_i\left(s\right)|\hat{D}\left(-\frac{\textrm{i}\sqrt{\hbar}G^v_{m-m^{\prime},n-n^{\prime}}}{\sqrt{2\mu\Omega}}\right)\hat{D}\left(-\sqrt{\frac{\mu\Omega}{2\hbar}}A^0_{n-n^{\prime}}\right)|\phi_j\left(s\right)\rangle\nonumber.\\
\end{eqnarray}
We use now the Baker-Hausdorf formula~\cite{Glauber}
\begin{eqnarray}
\hat{D}(\alpha)\hat{D}(\beta)=\hat{D}(\alpha+\beta)\exp((\alpha\beta^*-\alpha^*\beta)/2)
\end{eqnarray}
and we obtain the following result for the product of displacement operators
\begin{eqnarray}
&&\hat{D}\left(-\textrm{i}\sqrt{\frac{\hbar}{2\mu\Omega}} G^v_{m-m^{\prime},n-n^{\prime}}\right)\hat{D}\left(-\sqrt{\frac{\mu\Omega}{2\hbar}}A^0_{n-n^{\prime}}\right)=\\
&&=\hat{D}\left(-\textrm{i}\sqrt{\frac{\hbar}{2\mu\Omega}} G^v_{m-m^{\prime},n-n^{\prime}}-\sqrt{\frac{\mu\Omega}{2\hbar}}A^0_{n-n^{\prime}}\right)\exp\left(\frac{\textrm{i}}{2}G^v_{m-m^{\prime},n-n^{\prime}}A^0_{n-n^{\prime}}\right)\nonumber
\end{eqnarray}
We substitute the expression above into Eq.~(\ref{matrixeq}) and we have
\begin{eqnarray}\label{Matrixeq}
&&\exp(-\textrm{i}G^v_{m-m^{\prime},n-n^{\prime}}A^{k_x}_{(n+n^{\prime})/2})\times\langle\phi_i|\hat{D}\left(-\frac{\textrm{i}\sqrt{\hbar}G^v_{m-m^{\prime},n-n^{\prime}}}{\sqrt{2\mu\Omega}}-\sqrt{\frac{\mu\Omega}{2\hbar}}A^0_{n-n^{\prime}}\right)|\phi_j\rangle.\nonumber\\
\end{eqnarray}
The matrix representation of this displacement operator in the basis $\{\phi_i(s)\}$ is given by~\cite{Glauber}
\begin{eqnarray}\label{displacementeq}
\langle \phi_i|\hat{D}(\alpha_{n-n^{\prime},m-m^{\prime}})|\phi_j\rangle=\sqrt{\frac{j!}{i!}}\alpha^{i-j}_{n-n^{\prime},m-m^{\prime}}e^{-\frac{|\alpha_{n-n^{\prime},m-m^{\prime}}|^2}{2}}L^{(i-j)}_j(|\alpha_{n-n^{\prime},m-m^{\prime}}|^2),\nonumber
\end{eqnarray}
where $i\geq j$ and $L^{(i-j)}_j(|\alpha_{n-n^{\prime},m^{\prime}}|^2)$ are the associated Laguerre polynomials. We note that for $j>i$ one needs to take 
\begin{eqnarray}
\langle \phi_i|\hat{D}(\alpha_{n-n^{\prime},m-m^{\prime}})|\phi_j\rangle=(-1)^{j-i}\langle \phi_j|\hat{D}(\alpha_{n-n^{\prime},m-m^{\prime}})|\phi_i\rangle^{*}
\end{eqnarray}
because $\hat{D}^{\dagger}(\alpha)=\hat{D}(-\alpha)$~\cite{Glauber}. Moreover, the matrix elements $\alpha_{n-n^{\prime},m-m^{\prime}}$ are
\begin{eqnarray}\label{alphamatrix}
\alpha_{n-n^{\prime},m-m^{\prime}}&=&-\sqrt{\frac{\mu\Omega}{2\hbar}}A^0_{n-n^{\prime}}-\textrm{i}\sqrt{\frac{\hbar}{2\mu\Omega}}G^v_{m-m^{\prime},n-n^{\prime}}.
\end{eqnarray}
Substituting Eq.~(\ref{Matrixeq}) and~(\ref{displacementeq}) for the matrix representation of the displacement operator into Eq.~(\ref{beforeMatrixeq}), we obtain the QED-Bloch central equation
\begin{eqnarray}\label{QED-Bloch Central}
&&U^{\bi{k}}_{\bi{n},i} \Bigg[\frac{\hbar^2(k_z+G^z_l)^2}{2m_{\textrm{e}}}+\frac{\hbar^2(k_w+G^w_{m,n})^2}{2M}+\mathcal{E}_i-E_{\bi{k}}\Bigg]+\\
&&\sum_{\bi{n}^{\prime},j}V_{\bi{n}-\bi{n}^{\prime}} U^{\mathbf{k}}_{\mathbf{n}^{\prime},j}\; \exp\left(-\textrm{i}G^v_{m-m^{\prime},n-n^{\prime}}A^{k_x}_{(n+n^{\prime})/2}\right)\langle \phi_i|\hat{D}(\alpha_{n-n^{\prime},m-m^{\prime}})|\phi_j\rangle=0.\nonumber
\end{eqnarray}
The equation above is the main result of this chapter and of QED-Bloch theory. The QED-Bloch central equation provides a unified framework for the description of periodic materials in the presence of homogeneous magnetic fields, coupled also to the quantized electromagnetic field~\cite{rokaj2019}. 

\section{Landau Polaritons}

\begin{figure}[h]
\begin{center}
  \includegraphics[width=0.5\columnwidth]{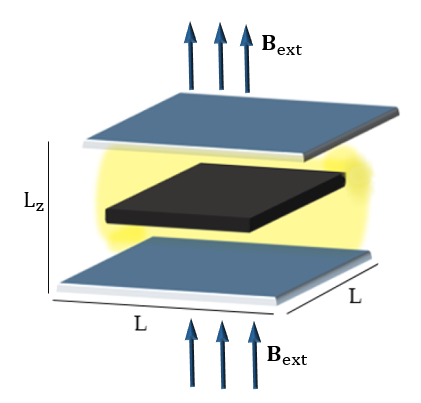}
\caption{\label{Landau Levels Cavity}Cartoon depiction of a 2D electron gas (material in black) confined inside a cavity. The whole system is placed perpendicular to a classical homogeneous magnetic field $\bi{B}_{\textrm{ext}}$. }  
\end{center}
\end{figure}

As a first application of the QED-Bloch framework we would like to consider the case where we have a two-dimensional free electron gas in the presence of a homogeneous magnetic field coupled to a quantized field originating from a cavity (see Fig.~\ref{Landau Levels Cavity}). So the system that we are interested in is: 2D Landau levels inside a cavity. Such Landau levels systems confined inside a cavity have been studied theoretically~\cite{Hagenmuller2010cyclotron, rokaj2019} and experimentally~\cite{Keller2020, ScalariScience, paravacini2019}. 
In the independent electron approximation, as we explained in the previous section, such a system is described by the effective Hamiltonian introduced in Eq.~(\ref{Approptical}). The effective Hamiltonian for no periodic potential $v_{\textrm{ext}}(\bi{r})=0$ reads as
\begin{eqnarray}\label{effective}
\hat{H}_{eff}=-\frac{\hbar^2}{2m_{\textrm{e}}}\nabla^2+\mathrm{i}\hbar\mathbf{e}_x\left(u\sqrt{\frac{\hbar\omega_p}{m_{\textrm{e}}}}-y\omega_c\right)\cdot\nabla+\frac{m_{\textrm{e}}}{2}\left(u\sqrt{\frac{\hbar\omega_p}{m_{\textrm{e}}}}-y\omega_c\right)^2-\frac{\hbar\omega_p}{2}\frac{\partial^2}{\partial u^2}.\nonumber\\
\end{eqnarray}
The energy spectrum of this system can be directly obtained from the QED-Bloch central equation~(\ref{QED-Bloch Central}) by simply taking the limit of no external potential which implies that the Fourier components of the potential are zero, $V_{\bi{n}-\bi{n}^{\prime}}=0$. Further, since there is no external potential the reciprocal lattice vectors have no role and have to be taken to be equal to zero. This is done by simply taking $n=m=l=0$.
\begin{eqnarray}
U^{\bi{k}}_{i} \left(\frac{\hbar^2k^2_z}{2m_{\textrm{e}}}+\frac{\hbar^2k^2_w}{2M}+\mathcal{E}_i-E_{\bi{k}}\right)=0.\nonumber
\end{eqnarray}
Then, from the above equation it is clear that the eigenspectrum of the 2D Landau levels coupled to the cavity is 
\begin{eqnarray}\label{Landau polaritons}
E_{\bi{k},i}=\frac{\hbar^2k^2_z}{2m_{\textrm{e}}}+\frac{\hbar^2k^2_w}{2M}+\hbar\Omega\left(i+\frac{1}{2}\right).
\end{eqnarray}
Further, the components of the $U^{\bi{k}}_{i}$ of the QED-Bloch ansatz defined in Eq.~(\ref{BlochAnsatzHermite}) are trivial and we find that the full set of eigenfuctions corresponding to the 2D Landau levels in the cavity are  
\begin{eqnarray}\label{Landau polariton States}
\Psi_{\mathbf{k},i}(\mathbf{r}_{w},v)=e^{\mathrm{i}\mathbf{k}\cdot\mathbf{r}_w}\phi_i\left(v-\frac{\hbar k_x}{\sqrt{2}m_{\textrm{e}}}\right).
\end{eqnarray}
This exact analytic solution for the 2D Landau levels in the cavity was found in~\cite{rokaj2019}. We note that here we have kept the momentum $k_z$ just for generality, but for the pure 2D case the momentum in the $z$ direction is taken equal to zero $k_z=0$. The eigenfunctions above are plane waves in the directions $x,z$ and $w$ because in these directions we have translational invariance, as it is also shown in Fig.~\ref{QED translation}. The eigenfunctions in Eq.~(\ref{Landau polariton States}) are functions of the combined polaritonic coordinates $w$ and $v$ defined in Eq.~(\ref{w and v coordinates}) and describe quasi-particles formed between the Landau levels and the photons. Consequently, they are interpreted as Landau polaritons. Such Landau polariton states have been studied theoretically~\cite{Hagenmuller2010cyclotron, rokaj2019} and have been observed experimentally~\cite{ScalariScience, Keller2020, li2018}.

\begin{figure}[H]
\begin{center}
    \includegraphics[width=0.6\columnwidth]{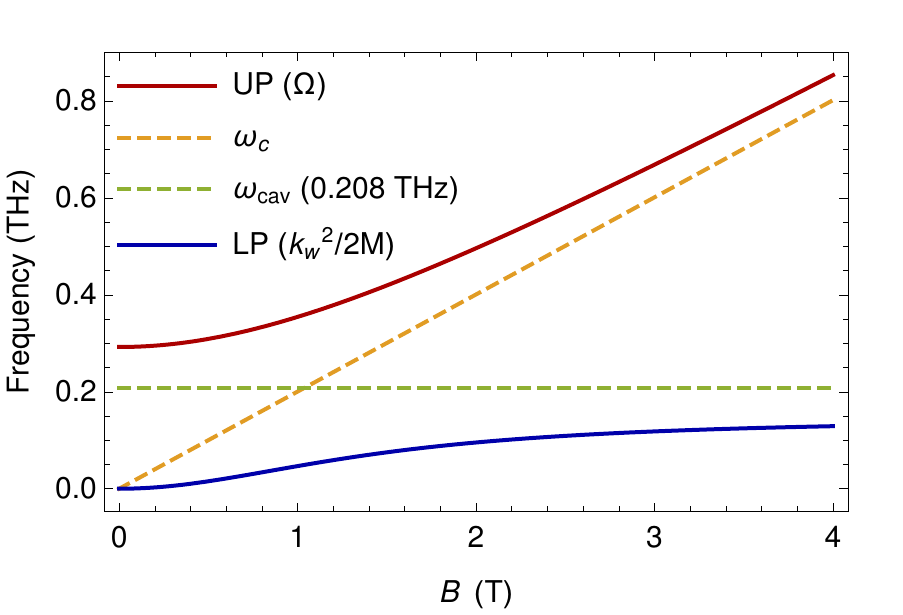}
\caption{\label{Landau_pol_fig} Upper (red line) and lower (blue line) Landau polariton excitations given by the energy spectrum~(\ref{Landau polaritons}) as a function of the strength of the magnetic field $B$ (T). The upper polariton (UP) asymptotically reaches the dispersion of the cyclotron transition $\omega_c=eB/m^*$ (orange dashed line). The lower polariton (LP) does not reach the empty cavity frequency $\omega_{\textrm{cav}}$ and a polariton gap emerges in accordance to the experimental findings in~\cite{Keller2020}.}
\end{center}
\end{figure}

To be more specific in~\cite{Keller2020} Landau polariton quasi-particle excitations were observed for a 2D hole gas in strained Germanium with 2D density $n^{\textrm{2D}}=1.3\times10^{12}\;\textrm{cm}^{-2}$ confined in a cavity with fundamental frequency $\omega_{\textrm{cav}}=0.208\;\textrm{THz}$. In this setting the diamagnetic frequency $\omega_p$ can be defined in terms of the 2D density and the cavity frequency $\omega_{\textrm{cav}}=2\pi c/L_z$ as
\begin{eqnarray}
\omega_p=\sqrt{\frac{e^2n_{\mathrm{2D}}\omega_{\textrm{cav}}}{2\pi c m^*\epsilon_0}}.
\end{eqnarray}
Using the parameters reported in~\cite{Keller2020} and the effective mass $m^*=0.336\;m_{\textrm{e}}$ the frequency $\omega_p$ takes the value $\omega_p=0.292$ THz and reproduces the gap for $B=0$ in~\cite{Keller2020}. Having obtained the value for $\omega_p$ we can compute the Landau polariton excitations given by Eq.~(\ref{Landau polaritons}). Figure~\ref{Landau_pol_fig} shows the upper and lower Landau polariton excitations as a function of the magnetic field. Analyzing the asymptotic behavior of the lower polariton $k^2_w/2M$ as a function of the magnetic field  we find its highest allowed value to be $\omega_p/2=0.146$ THz. In this case the lower polariton  does not reach the empty cavity frequency $\omega_{\textrm{cav}}=0.208$ THz as depicted in Fig.~\ref{Landau_pol_fig}. Our model reproduces the data reported in~\cite{Keller2020}, in contrast to the Hopfield model~\cite{Hagenmuller2010cyclotron} which as discussed in~\cite{Keller2020}, fails to account for the behavior of the lower polariton. Finally, in the case of no cavity the Landau polariton spectrum of Eq.~(\ref{Landau polaritons}) goes to the energy spectrum of the usual Landau levels because $\Omega \rightarrow \omega_c$ and $M\rightarrow \infty$.

\subsection{Landau Polaritons As a Screening Effect}\label{Landau polaritons Screening}

The hybridization of the Landau levels due to the interaction with the cavity photons can also be understood from more conventional condensed matter perspective, as a screening of the external magnetic field. So let us see how this can be done.

The spectrum of the Landau polaritons given by Eq.~(\ref{Landau polaritons}) is similar to the spectrum of the Landau levels~\cite{Landau}, but with the difference that it describes the combined electronic plus photonic energy levels. Due to this crucial difference, in order to extract the information about the electronic part of the energy, we need to subtract the photonic contribution. The purely photonic part of the Hamiltonian is
\begin{eqnarray}
\hat{H}_p=-\frac{\hbar \omega_p}{2}\frac{\partial^2}{\partial u^2}+\frac{\hbar\omega_p}{2}u^2.
\end{eqnarray}
To find the photonic contribution on the discrete part of the spectrum, we need to compute the expectation value on the eigenstate $\phi_j(v)$ which is related to discrete part of the spectrum. We note as we are interested in the purely discrete part, the momentum $k_x$ is taken equal to zero, $k_x=0$.

To do so, we need to write $\hat{H}_p$ in terms of the polaritonic coordinates. From the definition of $v$ and $w$ in Eq.~(\ref{w and v coordinates}) we find the expression for $u$ and $\partial/\partial u$
\begin{eqnarray}
\frac{\partial}{\partial u}=\sqrt{\frac{\hbar\omega_p}{2m_{\textrm{e}}}}\left(\frac{m_p}{M}\frac{\partial}{\partial w}+\frac{\partial}{\partial v}\right) \;\;\; \& \;\;\; u=\sqrt{\frac{m_{\textrm{e}}}{2\hbar\omega_p}}\left(w-\frac{m_c}{M}v\right).
\end{eqnarray}
Then, the part of $\hat{H}_p$ that depends only on the coordinate $v$, which is the relevant for the states $\phi_j(v)$, is
\begin{eqnarray}
\hat{H}_p=-\frac{\hbar^2\omega^2_p}{4m_{\textrm{e}}}\frac{\partial^2}{\partial v^2}+m_{\textrm{e}}\left(\frac{\omega_p}{\Omega}\right)^4v^2 + \mathcal{O}(w,\partial_w),
\end{eqnarray}
where in the last step we used that $m_c/M=2\omega^2_p/\Omega^2$. Having the above expression for $\hat{H}_p$ we can now compute the expectation value $\mathcal{E}^p_j\equiv\langle \phi_j|\hat{H}_p|\phi_j\rangle$ which consistutes the photonic contribution to the discrete part of the Landau-polariton energy spectrum. After some tedious algebra we find
\begin{eqnarray}
\mathcal{E}^p_j=\hbar\Omega\left(j+\frac{1}{2}\right)\left[\frac{1}{2}\left(\frac{\omega_p}{\Omega}\right)^2+\frac{1}{2}\left(\frac{\omega_p}{\Omega}\right)^4\right].
\end{eqnarray}
Subtracting from the full spectrum the photonic part we obtain the purely electronic part $\mathcal{E}^{\textrm{e}}_j$ of the energy
\begin{eqnarray}
\mathcal{E}^{\textrm{e}}_j=\hbar\Omega\left(j+\frac{1}{2}\right)\left[1-\frac{1}{2}\left(\frac{\omega_p}{\Omega}\right)^2-\frac{1}{2}\left(\frac{\omega_p}{\Omega}\right)^4\right].
\end{eqnarray}
By introducing the parameter $g=\omega_p/\omega_c$, which describes the light-matter coupling in this setting, the electronic energies take the form of Landau levels but with a screened magnetic field
\begin{eqnarray}
\mathcal{E}^{\textrm{e}}_j= \frac{\hbar e B(g)}{m_{\textrm{e}}}\left(j+\frac{1}{2}\right)=\frac{\hbar e B\chi(g)}{m_{\textrm{e}}}\left(j+\frac{1}{2}\right)\;\;\; \textrm{with}\;\; \chi(g)=\frac{2+3g^2}{2\left(1+g^2\right)^{3/2}},
\end{eqnarray}
where $B(g)=B\chi(g)$ is the screened magnetic field depending on the light-matter coupling $g$, and $\chi(g)$ is the response function describing this screening process. 

\begin{figure}[H]
\begin{center}
    \includegraphics[width=0.6\columnwidth]{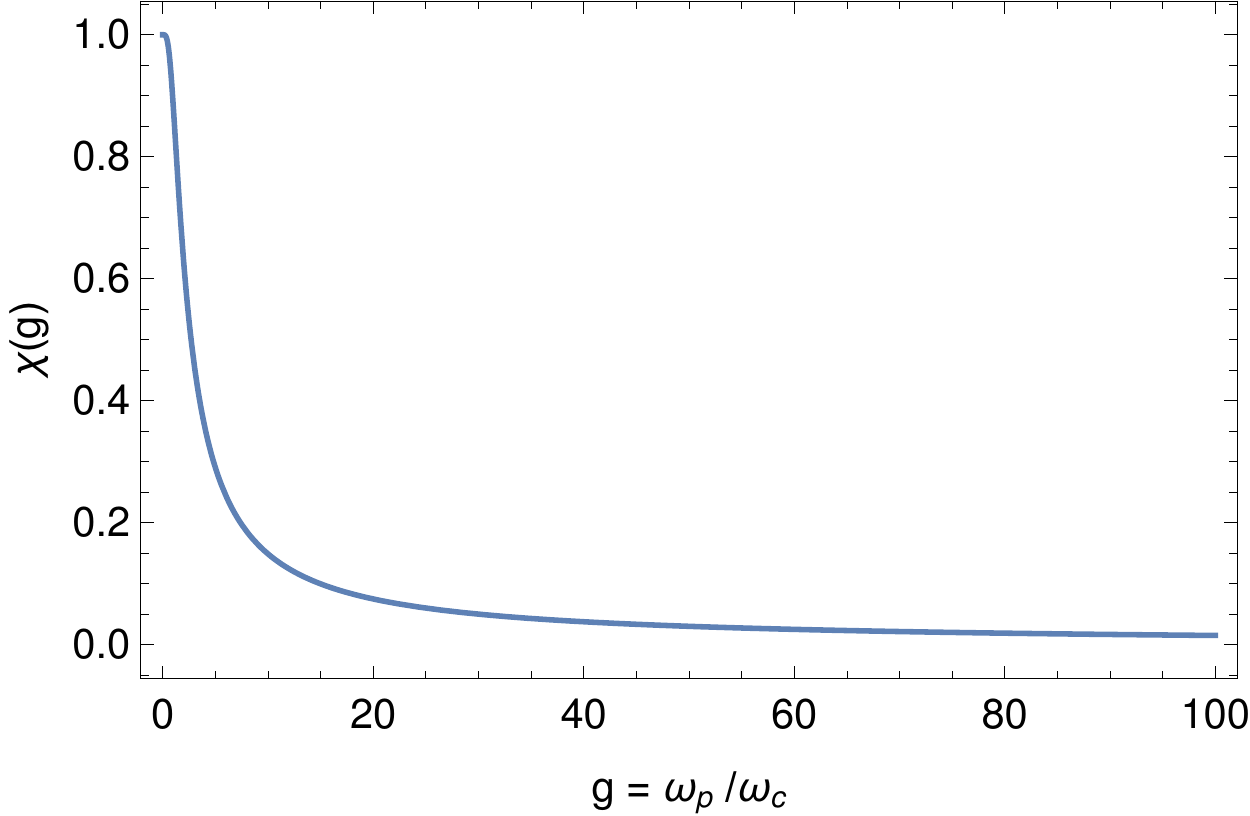}
\caption{\label{B_field_Screening}Depiction of the response function $\chi(g)$ as a function of the light-matter coupling $g$. As we see $\chi(g)$ decreases monotonically as a function of $g$. The response function describes the screening of the external magnetic field due to the light-matter coupling.  }
\end{center}
\end{figure}
Finally, we would like to mention that the modification of the Landau levels can also be understood as a renormalization of the electron mass due to the light matter coupling $m_{\textrm{e}}(g)=m_{\textrm{e}}/\chi(g)$. This interpretation is analogous to the one in the case of the modification of the Drude peak in section~\ref{Electronic Response} and to the computation for the mass renormalization we performed in chapter~\ref{Effective QFT}. In accordance with our previous results, the renormalized mass also in this case increases as function of the the light-matter coupling, $m_{\textrm{e}}(g)\geq m_{\textrm{e}}$.

\newpage

\section{2D Materials in Homogeneous Magnetic Fields\\ (No Quantized Field Limit) }

The aim of this section is to apply the framework of QED-Bloch theory for the description of two-dimensional materials in the presence only of a homogeneous magnetic field, without any quantized field. This is the standard purely electronic problem and constitutes the standard quantum Hall setting in which for example the famous TKNN (Thouless, Kohmoto, Nightingale, and den Nijs) formula~\cite{Thouless, Kohmoto} for the quantization of the Hall conductance was derived, as well as the setting in which the fractal spectrum of the Hofstadter butterfly was obtained~\cite{Hofstadter}.

In order to describe purely two-dimensional materials we need to project the QED-Bloch central equation in~(\ref{QED-Bloch Central}) from three dimensions to two. For that purpose, we simply eliminate the degrees of freedom associated with the $z$-direction in our QED-Bloch ansatz and the external potential. This implies that the crystal momentum in the $z$-direction and the Fourier index $l$, have both to be taken equal to zero, $k_z=l=0$. Then for $U^{\bi{k}}_{\bi{n}}$ and $V_{\bi{n}-\bi{n}^{\prime}}$ we have
\begin{eqnarray}
U^{\bi{k}}_{\bi{n}} \longrightarrow U^{k_x,k_w}_{n,m} \;\;\; \textrm{and}\;\;\;  V_{\bi{n}-\bi{n}^{\prime}} \longrightarrow V_{n-n^{\prime},m-m^{\prime}}.
\end{eqnarray}
Then, the central equation in (\ref{QED-Bloch Central}) takes the form
\begin{eqnarray}\label{QED-Bloch Central2D}
&&U^{k_x,k_w}_{n,m,i} \Bigg[\frac{\hbar^2(k_w+G^w_{m,n})^2}{2M}+\mathcal{E}_i-E_{k_x,k_w}\Bigg]+\\
&&\sum_{n^{\prime},m^{\prime},j}V_{n-n^{\prime},m-m^{\prime}} U^{k_x,k_w}_{n^{\prime},m^{\prime},j}\; \exp\left(-\textrm{i}G^v_{m-m^{\prime},n-n^{\prime}}A^{k_x}_{(n+n^{\prime})/2}\right)\langle \phi_i|\hat{D}(\alpha_{n-n^{\prime},m-m^{\prime}})|\phi_j\rangle=0.\nonumber
\end{eqnarray}
Now comes the most crucial part for the description of the purely electronic problem, namely to take the limit of no quantized field. The quantized vector potential $\hat{\bi{A}}$ depends on the diamagnetic shift frequency $\omega_p$ defined in Eq.~(\ref{plasma frequency}). This is the frequency in which the quantized field oscillates and in order to take the limit of no quantized field we have to take the limit $\omega_p \rightarrow 0$ for all the parameters involved in our central equation.

First, we consider the limit $\omega_p \rightarrow 0$ for the mass parameter $M$ defined in Eq.~(\ref{mass polaritonic parameters}) and we find that 
\begin{eqnarray}
\lim_{\omega\rightarrow 0}M = \infty,
\end{eqnarray}
this implies that the kinetic term depending on $k_w$ in the central equation vanishes and the Fourier components of our polaritonic Bloch wave no longer depends on $k_w$. Due to the vanishing of the $w$ degree of freedom the index $m$ in the Bloch wave becomes redundant and no longer plays any role
\begin{eqnarray}
U^{k_x,k_w}_{n,m,j} \longrightarrow U^{k_x}_{n,j}.
\end{eqnarray}
Consequently, the central equation reduces to
\begin{eqnarray}
U^{k_x}_{n,i} \left[\mathcal{E}_i-E_{k_x}\right]+\sum_{n^{\prime},m^{\prime},j}V_{n-n^{\prime},m^{\prime}} U^{k_x}_{n^{\prime},j}\; \exp\left(-\textrm{i}G^v_{m^{\prime},n-n^{\prime}}A^{k_x}_{(n+n^{\prime})/2}\right)\langle \phi_i|\hat{D}(\alpha_{n-n^{\prime},m^{\prime}})|\phi_j\rangle=0.\nonumber\\
\end{eqnarray}
To obtain the result above we also relabelled the index $-m^{\prime} \rightarrow m^{\prime}$. Now what is left to be done is to perform the $\omega_p\rightarrow 0$ limit for the rest of the parameters in the central equation which depend on $\omega_p$. By doing so we find
\begin{eqnarray}
&&  \lim_{\omega_p\rightarrow 0}\Omega =\omega_c\;\;\; \textrm{and}\;\;\; \lim_{\omega_p\rightarrow 0 }\mu\Omega=\frac{2m_{\textrm{e}}}{\omega_c}\\
&&\lim_{\omega_p \rightarrow 0} G^v_{m^{\prime},n-n^{\prime}}=\frac{\sqrt{2}}{\omega_c}G_{m^{\prime},n-n^{\prime}}\;\;\; \textrm{and}\\
&&\lim_{\omega_p\rightarrow 0}\alpha_{n-n^{\prime},m^{\prime}}=\sqrt{\frac{\hbar}{2m_{\textrm{e}}\omega_c}}\left(- G^x_{n-n^{\prime}}-\textrm{i}G_{m^{\prime},n-n^{\prime}}\right)\equiv \beta_{n-n^{\prime},m^{\prime}}.\label{beta matrix}
\end{eqnarray}
Substituting all the above results and the definition for $A^{k_x}_{(n+n^{\prime})/2}$ given by Eq.~(\ref{Landaupolaritons}) and we have
\begin{eqnarray}\label{LLB Central}
&&U^{k_x}_{n,i}\left[\hbar\omega_c\left(i+\frac{1}{2}\right)-E_{k_x}\right]+\\
&&\sum_{n^{\prime},m^{\prime},j}V_{n-n^{\prime},m^{\prime}} U^{k_x}_{n^{\prime},j} \exp\left(\frac{-\textrm{i}\hbar(k_x+\frac{1}{2}G^x_{n+n^{\prime}})G_{m^{\prime},n-n^{\prime}}}{m_{\textrm{e}}\omega_c}\right)\langle \phi_i|\hat{D}(\beta_{n-n^{\prime},m^{\prime}})|\phi_j\rangle=0,\nonumber
\end{eqnarray}
where the matrix $\beta_{n-n^{\prime},m^{\prime}}$ is defined in Eq.~(\ref{beta matrix}). The central equation derived above depends solely on electronic parameters like the electronic crystal momentum $k_x$, the mass of the electron $m_{\textrm{e}}$ and the cyclotron frequency $\omega_c=eB/m_{\textrm{e}}$ which is characteristic for electrons in a uniform magnetic field~\cite{Landau}. As a consequence the above central equation describes consistently the physics of two-dimensional periodic systems in the presence of a perpendicular homogeneous magnetic field. From this equation we can compute the energy bands for all values of the magnetic field because our approach is non-perturbative.   

For completeness, we would also like to give the expression of the polaritonic Bloch ansatz defined in Eq.~(\ref{BlochAnsatzHermite}) in the limit of no quantized field. The QED-Bloch ansatz depends on the polaritonic coordinates $w$ and $v$ defined in Eq.~(\ref{w and v coordinates}). Taking $\omega_p\rightarrow 0$ the coordinate $w$ vanishes while the coordinate $v$ becomes $v=-\omega_c y$. Thus, we find that the polaritonic QED-Bloch ansatz in the limit of no quantized field is 
\begin{eqnarray}\label{LLB ansatz}
\Psi_{k_x}(x,y)=e^{\textrm{i}k_x}\sum_{n,j}U^{k_x}_{n,j}e^{\textrm{i}G^x_n}\phi_j\left(-\frac{\omega_cy}{\sqrt{2}}-A^{k_x}_n\right).
\end{eqnarray}
The above wavefunction corresponds to a correlated expansion between Bloch waves in the $x$ coordinate and Landau levels $\phi_j\left(-\frac{\omega_cy}{\sqrt{2}}-A^{k_x}_n\right)$ in the $y$ coordinate. Such an expansion has been used for the description of 2D materials in homogeneous magnetic fields in several publications~\cite{PfannkucheButterfly, Langbein1969, Rauh1975, GeiselButterflyChaos} and central equations analogous to Eq.~(\ref{LLB Central}) have been derived.

\subsection{Harper Equation \& Hofstadter Butterfly }

Our aim now is to connect the central equation we derived in Eq.~(\ref{LLB Central}) which describes 2D periodic solids in the presence of a classical homogeneous magnetic field, to the well-known Harper equation~\cite{Harper_1955} and the fractal spectrum of the Hofstadter butterfly~\cite{Hofstadter}. 

\begin{figure}[h]
\begin{center}
    \includegraphics[height=9cm,width=0.8\columnwidth]{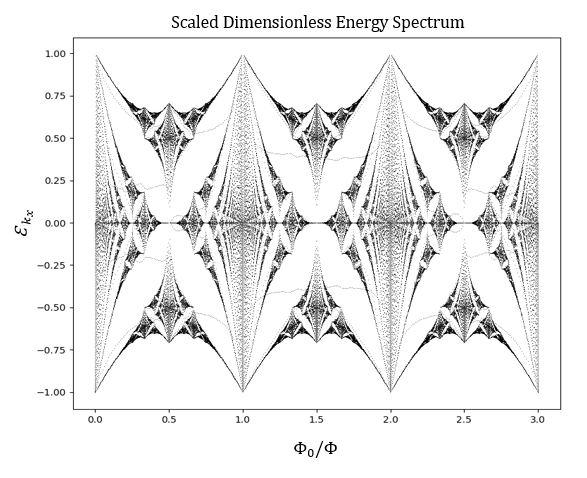}
\caption{\label{Reciprocal Butterfly}Scaled dimensionless energies $\mathcal{E}_{k_x}$ of the system, given by the Harper equation~\ref{Harper equation}, as a function of the reciprocal magnetix flux $\Phi_0/\Phi$. }
\end{center}
\end{figure}

In the seminal papers by Harper~\cite{Harper_1955} and Hofstadter~\cite{Hofstadter} the Harper equation and the butterfly spectrum were derived for an orthogonal square lattice. For that purpose we choose $\theta=\pi/2$ and $a_1=a_2=a$. Moreover, to connect Eq.~(\ref{LLB Central}) to the the Harper equation we will reduce our central equation (\ref{LLB Central}) to the case where all electrons lie in the lowest Landau level $i=0$. In the lowest Landau level and for an orthogonal square lattice Eq.~(\ref{LLB Central}) simplifies to
\begin{eqnarray}
U^{k_x}_{n}\left(\frac{\hbar\omega_c}{2}-E_{k_x}\right)+\sum_{n^{\prime},m^{\prime}}V_{n-n^{\prime},m^{\prime}} U^{k_x}_{n^{\prime}} \exp\left(\frac{-\textrm{i}\hbar(k_x+\frac{1}{2}G^x_{n+n^{\prime}})G^y_{m^{\prime}}}{m_{\textrm{e}}\omega_c}\right)e^{-\frac{|\beta_{n-n^{\prime},m^{\prime}}|^2}{2}}=0.\nonumber\\
\end{eqnarray}
To obtain the above result we used that 
\begin{eqnarray}
\langle \phi_0|\hat{D}(\beta_{n-n^{\prime},m^{\prime}})|\phi_0\rangle=e^{-\frac{|\beta_{n-n^{\prime},m^{\prime}}|^2}{2}}
\end{eqnarray}
and that for $\theta=\pi/2$ the reciprocal lattice vector $G_{m^{\prime},n-n^{\prime}}$ is
\begin{eqnarray}
G_{m^{\prime},n-n^{\prime}}=G^y_{m^{\prime}}.
\end{eqnarray}
The Harper equation was derived for a tight-binding model with the Peierls substitution and next-neighbor hopping. Due to this we choose a cosine potential for the external potential. Such a potential introduces only next-neighbor hopping in Fourier space, and the Fourier components of the potential are $V_{\pm 1,0}=V_{0,\pm 1}=V_0$ and all the other Fourier components are zero. We note that $V_0$ is the strength of periodic potential. Then, for this simple periodic potential we have
\begin{eqnarray}
&&U^{k_x}_{n}\left(\frac{\hbar\omega_c}{2}-E_{k_x}\right) + V_0 U^{k_x}_{n-1}e^{-\frac{|\beta_{1,0}|^2}{2}}+V_0 U^{k_x}_{n+1}e^{-\frac{|\beta_{-1,0}|^2}{2}}+\\
&+&V_0U^{k_x}_n \exp\left(\frac{-\textrm{i}\hbar(k_x+G^x_n)G^y_1}{m_{\textrm{e}\omega_c}}\right) e^{-\frac{|\beta_{0,1}|^2}{2}}+V_0U^{k_x}_n \exp\left(\frac{-\textrm{i}\hbar(k_x+G^x_n)G^y_{-1}}{m_{\textrm{e}\omega_c}}\right) e^{-\frac{|\beta_{0,-1}|^2}{2}}\nonumber.
\end{eqnarray}
Further, we use the fact for a square lattice $a_1=a_2=a$, and the norm of the four lowest components of the $\beta$-matrix, defined in Eq.~(\ref{beta matrix}), are all equal
\begin{eqnarray}
|\beta_{1,0}|^2=|\beta_{-1,0}|^2=|\beta_{0,1}|^2=|\beta_{0,-1}|^2=\frac{\hbar (2\pi)^2}{2m_{\textrm{e}}\omega_c a^2}=\frac{\pi\Phi_0}{\Phi}.
\end{eqnarray}
Also we make use of the expressions for the reciprocal lattice vectors $G^y_{1}=-G^y_{-1}=2\pi/a$ and $G^x_n=2\pi n/a$, and we obtain
\begin{eqnarray}\label{Unscaled Harper}
U^{k_x}_{n}\left(\frac{\hbar\omega_c}{2}-E_{k_x}\right) + V_0 e^{\frac{-\pi \Phi_0 }{2\Phi}}\left[U^{k_x}_{n-1}+U^{k_x}_{n+1}+2U^{k_x}_n \cos\left(\frac{2\pi \Phi_0}{\Phi} \left(\frac{ak_x}{2\pi}+n\right)\right)\right]=0.\nonumber\\
\end{eqnarray}
In the last step we also introduced the magnetic flux quantum $\Phi_0=h/e$ and the magnetic flux through the unit cell $\Phi=Ba^2$. Moreover, in the work of Hofstadter~\cite{Hofstadter} the fractal spectrum appears not for the energy itself, $E$, but for the unitless scaled energy, $\mathcal{E}=E/t$, divided by the constant hopping parameter $t$. Of course this does not make a difference within the tight-binding model because the hopping parameter $t$ is a constant. On the contrary for the minimally coupled Schr\"{o}dinger equation (\ref{Hext}), the magnetic field is part of the covariant (physical) momentum of the electron. Thus, the kinetic energy of the electrons naturally depends on the magnetic field and as a consequence the hopping (which represents the kinetic energy in the tight-binding approach) should be a function of the magnetic field. In our setting we define the flux-dependent hopping parameter $t(\Phi)$ as
\begin{eqnarray}
t(\Phi)=V_0 e^{\frac{-\pi \Phi_0 }{2\Phi}}
\end{eqnarray}
and the unitless scaled energies as
\begin{eqnarray}\label{scaled energies}
\mathcal{E}_{k_x}=\frac{1}{t(\Phi)}\left(E_{k_x}-\frac{\hbar\omega_c}{2}\right)=\frac{e^{\frac{\pi\Phi_0}{\Phi}}(E_{k_x}-\hbar \omega_c/2)}{V_0},
\end{eqnarray}
and we find the following equation for the scaled dimensionless energies of the system
\begin{eqnarray}\label{Harper equation}
\mathcal{E}_{k_x}U^{k_x}_{n}= U^{k_x}_{n-1}+U^{k_x}_{n+1}+2U^{k_x}_n \cos\left(\frac{2\pi \Phi_0}{\Phi} \left(\frac{ak_x}{2\pi}+n\right)\right).
\end{eqnarray}
The equation above is known as the Harper equation~\cite{Harper_1955} and plotting the eigenenergies of this equation we obtain the fractal spectrum of the Hofstadter butterfly~\cite{Hofstadter}, which is depicted in Fig~\ref{Reciprocal Butterfly}. However, there is one important difference. In the original Harper equation the energy spectrum is a function of the relative magnetic flux $\Phi/\Phi_0$~\cite{Harper_1955} and the corresponding butterfly spectrum also appears as a function of the relative flux $\Phi/\Phi_0$~\cite{Hofstadter}. In our case the energies are a function of the reciprocal relative flux $\Phi_0/\Phi$ and the fractal spectrum appears with respect to the reciprocal flux $\Phi_0/\Phi$. This fact, that starting from the minimal-coupling Schr\"{o}dinger Hamiltonian shows the Hofstadter butterfly as function of the reciprocal flux $\Phi_0/\Phi$, and this fundamental difference to tight-binding models with the Peierls phase, has been shown and discussed in several publications~\cite{PfannkucheButterfly, Langbein1969, Rauh1975, GeiselButterflyChaos}. In these works, to obtain the butterfly spectrum in the reciprocal flux, the correlated expansion on Landau levels and Bloch waves given by Eq.~(\ref{LLB ansatz}) was performed. It is worth to mention the precise words of Gregory Wannier on this important issue~\cite{Wannier} 
\begin{displayquote}
\footnotesize{In the framework of the Peierls-Onsager method~[tight-binding models with the Peierls phase] it
shows, within an assumed Bloch band, the allowed energies as function of the field.
In the Regensburg interpretation~[minimal-coupling Hamiltonian with Landau-level Bloch wave expansion] it shows, for an assumed periodic potential, the fine structure of a free electron Landau level as a function of the reciprocal
field. The first interpretation is easier, but only in the second is the basic function
space clearly defined. The model has the full complexity of the general problem, but involves only one parameter which will be called $\Phi$~[the magnetic flux]. We have now ample numerical results for the model. The figure appears as an infinite strip periodic in the parameter $\Phi$. This is physically surprising because it implies that certain finite magnetic fields are equivalent to either zero or infinite field. This periodicity feature should not be taken too seriously in either interpretation. In the case of the Onsager interpretation it has been shown that the method is correct to all powers of the field only if, in addition, the band structure is allowed to vary parametrically as function of the field. By the time $\Phi$ has reached 1 (an impossibly large field by today’s laboratory
standards) such changes are no doubt very drastic. In the Regensburg interpretation there could be some truth in a periodic repetition of the fine structure with $1/H$~[$H$ being the strength of the magnetic field]. But it is not very likely that such a pattern is quantitatively the same as that for
strong fields. The method suppresses the inter-level matrix elements which are responsible for the transformation of the free electron Landau levels into the Bloch band
Landau levels. This must involve a thorough scrambling of the levels at intermediate
fields. }
\end{displayquote}
\begin{flushright}
  \footnotesize{G.~Wannier~\cite{Wannier}}
\end{flushright}
So what Gregory Wannier is pointing towards in this paragraph is the fact that it is highly surprising for the energy spectrum to have a periodicity either as a function of $\Phi/\Phi_0$ or as a function of $\Phi_0/\Phi$. Because in the Schr\"{o}dinger equation the magnetic field enters via the minimal coupling and the Hamiltonian has a linear and a quadratic dependence with respect to the magnetic field. Both periodic behaviors are an artefact. In the case of the tight-binding models with the Peierls phase, this periodic behavior is an artefact introduced by the way the magnetic field is coupled to the Bloch electrons, i.e., the Peierls phase. In the case of the minimal-coupling Hamiltonian the periodicity is introduced due to the redefinition (the scaling) of the energies in Eq.~(\ref{scaled energies}). This redefinition of the energy spectrum cuts out the linear dependence of the energy of the lowest Landau level $\omega_c=eB/m_{\textrm{e}}$ and the exponential increase due to the flux dependent hopping parameter $t(\Phi)$. To understand what is the actual dependence of the energy spectrum we plot below the un-scaled dimensionful energies of our system given by Eq.~(\ref{Unscaled Harper}).

\begin{figure}[H]
\begin{center}
    \includegraphics[height=9cm,width=0.7\columnwidth]{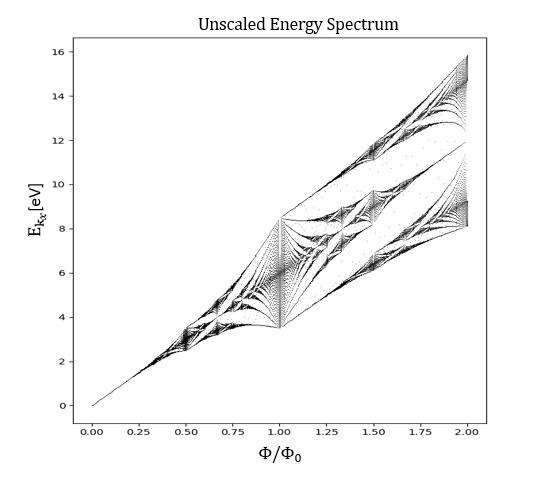}
\caption{\label{Unscaled Butterfly}Energy spectrum for an electron in a square lattice potential as function of the relative magnetic flux $\Phi/\Phi_0$ as it is given by Eq.~(\ref{Unscaled Harper}). The potential strength of the potential is $V_0=3$eV while the lattice constant of the lattice $a=2$\AA.}
\end{center}
\end{figure}

In Fig.~\ref{Unscaled Butterfly} we see that for small fluxes we have the linear dispersion due to the energy of the lowest Landau level. As the magnetic field increases the Landau levels starts to split and gaps show up. Then, for $\Phi/\Phi_0>1/2$ the fractal nature of the spectrum shows up and the Hofstadter butterfly becomes clearly visible. The Hofstadter butterfly however is not periodic, but it actually spreads out due to the flux-dependent hopping $t(\Phi)$ parameter.  Figure~\ref{Unscaled Butterfly} reconciles the two fundamental properties of the minimal-coupling Hamiltonian: (i) the energy has to increase as a function of the magnetic field and (ii) due to the magnetic translation group a splitting of the energy bands for every fractional value of the relative magnetic flux $\Phi/\Phi_0=p/q$ needs to occur, which subsequently leads to the formation of the fractal~\cite{BrownMTG, MTG_I, MTG_II}.

\textit{Dual Descriptions.}---The question that finally arises is: why the minimal-coupling Hamiltonian differs in such a fundamental way from the tight-binding model with the Peierls phase? 

In many cases these two descriptions for electrons in periodic structures coupled to electromagnetic fields are equivalent descriptions and match at least to some certain degree of approximation. Typically this is true for slowly varying electromagnetic fields within the unit cell of the solid. However, the problem with our particular system is that although the magnetic field is constant the vector potential that actually couples to the electrons is linear in space $\bi{A}_{\textrm{ext}}=-\bi{e}_xBy$. This already makes the Peierls substitution questionable. 

This fact was pointed out by Luttinger~\cite{LuttingerTBPeierls} in one of the first papers deriving the tight-binding model with the Peierls phase, starting from the minimal-coupling Schr\"{o}dinger Hamiltonian. To derive this model, Luttinger had to explicitly drop a term from the Hamiltonian~\cite{LuttingerTBPeierls}. Eliminating this term breaks the actual correspondence/relation between the minimal-coupling Hamiltonian and the tight-binding model with the Peierls phase. 

\begin{figure}[H]
\begin{center}
    \includegraphics[height=7cm,width=\columnwidth]{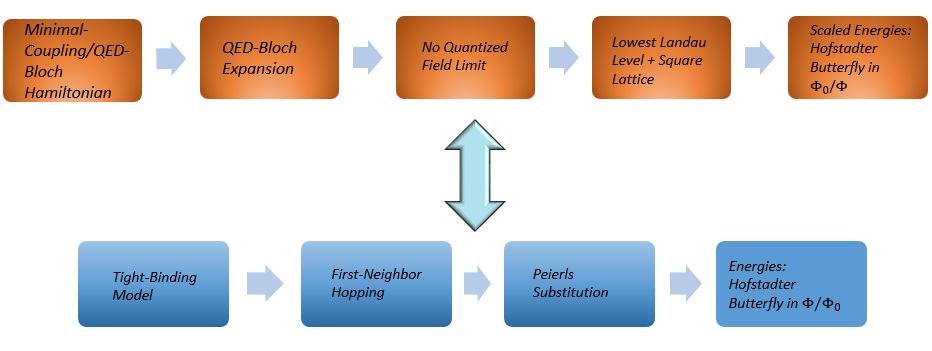}
\caption{\label{Butterfly Duality}Schematic illustration of the duality between the minimal-coupling Hamiltonian and the tight-binding model with the Peierls substitution. In both models the Hofstadter butterfly emerges but in a dual fashion. Namely, in the one case as a function of the reciprocal flux while in the other as a function of the flux.}
\end{center}
\end{figure}

However, this does not mean that the two descriptions are completely disconnected. They both result into the Harper equation and the Hofstadter butterfly, with the difference that in the one case it shows up as a function of the magnetic flux $\Phi/\Phi_0$ while in the other as a function of the reciprocal flux $\Phi_0/\Phi$. This means that the minimal-coupling Hamiltonian and the tight-binding model with the Peierls phase are not equivalent but they are actually dual. As we will see also in the next section this duality holds only in the lowest Landau level. The duality between the two approaches and the steps to obtain the respective Hofstadter butterflies is summarized in Fig.~\ref{Butterfly Duality}. 

\subsection{Butterfly Spectra for All 2D Bravais Geometries}

\begin{figure}[H]
\begin{subfigure}{.5\textwidth}
  \centering
  \includegraphics[height=5cm, width=0.8\linewidth]{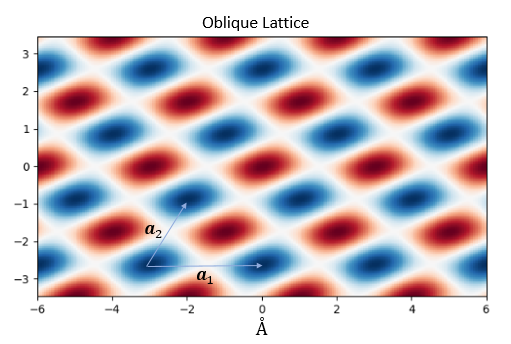}
  \caption{Oblique potential for $a_1=3$\AA,\\
  $a_2=2$\AA, and angle $\theta=\pi/3$. }
  \label{Oblique}
\end{subfigure}%
\begin{subfigure}{.5\textwidth}
  \centering
  \includegraphics[height=5cm, width=0.8\linewidth]{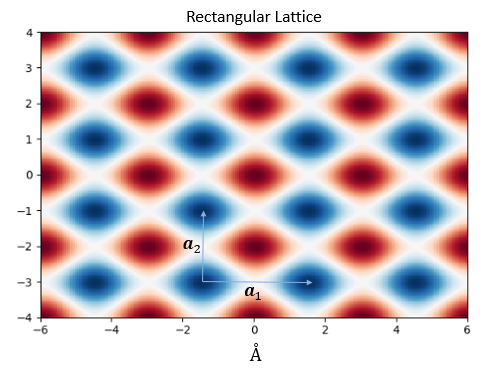}
  \caption{Rectangular potential for $a_1=3$\AA\\and $a_2=2$\AA. }
  \label{Rectangular}
\end{subfigure}
\begin{subfigure}{.5\textwidth}
  \centering
  \includegraphics[height=5cm, width=0.8\linewidth]{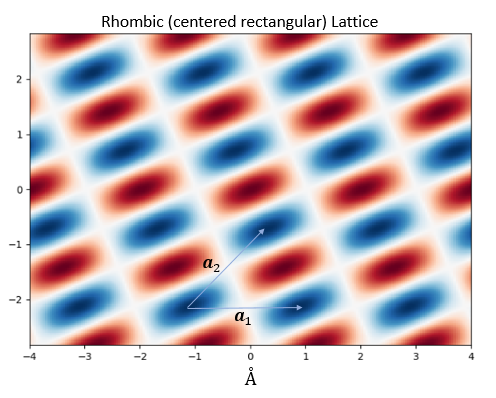}
  \caption{Centered rectangular potential with\\ $a_1=a_2=2$\AA.}
  \label{Rhombic}
\end{subfigure}
\begin{subfigure}{.5\textwidth}
  \centering
  \includegraphics[height=5cm, width=0.8\linewidth]{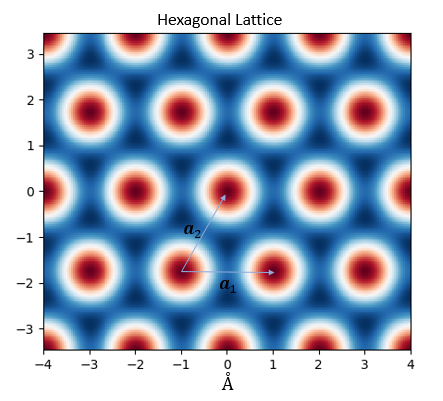}
  \caption{Hexagonal potential with $a_1=a_2=2$\AA.}
  \label{Hexagonal}
\end{subfigure}
\begin{subfigure}{.5\textwidth}
  \centering
  \includegraphics[height=5cm,width=0.95\linewidth]{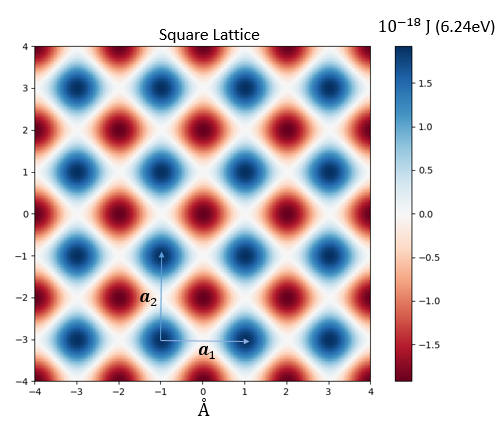}
  \subcaption{Square potential with $a_1=a_2=2$\AA.}
  \label{Square}
\end{subfigure}
\caption{Depiction of all distinct 2D Bravais lattices. The potential strength for all potentials is chosen to be $V_0=3$eV and all potentials are constructed using the cosine function.}
\label{Bravais Lattices}
\end{figure}

So far we have applied our QED-Bloch formalism to the case of a periodic solid with a square lattice potential, with the electrons in the lowest Landau level. Our aim now is to consider more complicated periodic structures and to investigate what happens beyond the lowest Landau level. To do so, we will employ the central equation~(\ref{LLB Central}). In the 2D plane there are only five distinct Bravais geometries~\cite{Mermin}: (i) the oblique lattice, (ii) the rectangular lattice, (iii) the centered rectangular lattice, (iv) the hexagonal lattice, and (v) the square lattice. For the construction of the five fundamental Bravais lattices, we use the simple cosine function and for all potentials we use the same potential strength $V_0$. The five Bravais lattices are shown in the Fig.~\ref{Bravais Lattices}.

\subsubsection{Oblique Lattice Butterfly Spectra}
\begin{figure}[H]
\begin{subfigure}{.5\textwidth}
  \centering
  \includegraphics[height=6cm, width=0.9\linewidth]{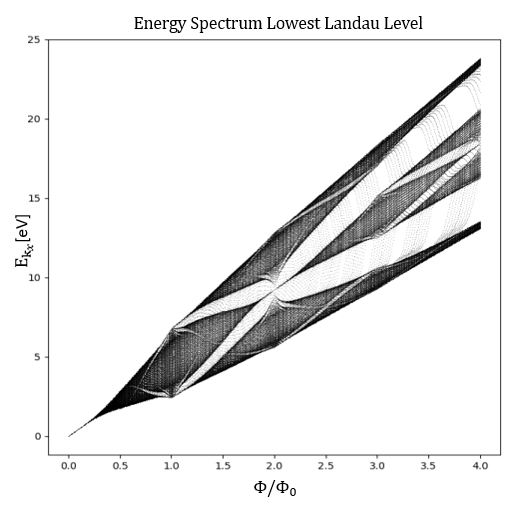}
  \caption{Energy spectrum in the \\lowest Landau level.}
  \label{Oblique 0LL}
\end{subfigure}%
\begin{subfigure}{.5\textwidth}
  \includegraphics[height=6cm, width=0.9\linewidth]{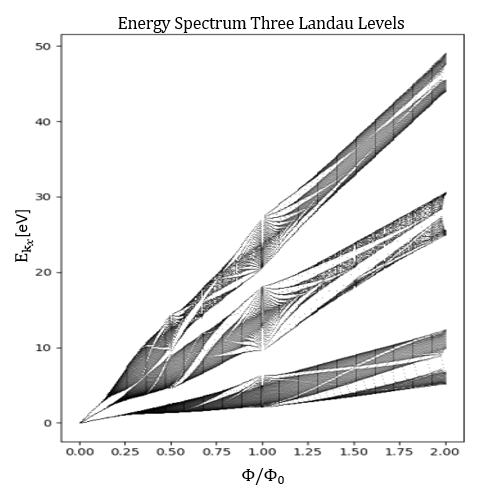}
  \caption{Energy spectrum with \\three Landau levels included. }
  \label{Oblique 3LL}
\end{subfigure}
\caption{Energy bands as a function of the relative magnetic flux $\Phi/\Phi_0$ for the oblique lattice.}
\end{figure}

Let us start first with the oblique lattice shown in Fig.~\ref{Oblique}, which has the least symmetry. First we compute the energy spectrum as a function of the magnetic flux in the case where all the electrons in the 2D material lie in the lowest Landau level, which is shown below. The Landau level spreads as a function of the flux and around one flux quantum the gaps are clearly visible. Above $\Phi/\Phi_0>1$ we see the typical Butterfly pattern emerge. In Fig.\ref{Oblique 3LL} we include three Landau levels. In this case we see that in the higher Landau levels the fractal spectrum shows up for smaller values of the magnetic flux. 

\subsubsection{Rectangular Lattice Butterfly Spectra}

Using again the central equation~(\ref{LLB Central}) we compute the energy bands for the rectangular cosine potential shown in Fig.~\ref{Rectangular}, in the lowest Landau level and for the case where we have three Landau levels occupied. In the lowest Landau level the spectrum is similar to the one of the oblique lattice, while for the three Landau levels we see that the gaps between the Landau levels are smaller.
\begin{figure}[H]
\begin{subfigure}{.5\textwidth}
  \centering
  \includegraphics[height=6cm, width=0.9\linewidth]{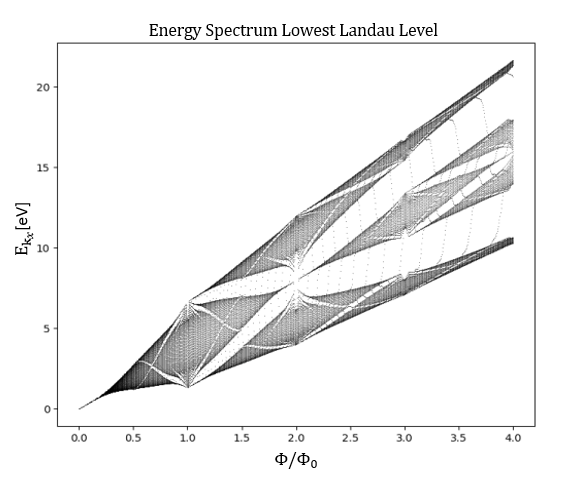}
  \caption{Energy spectrum in \\the lowest Landau level. }
  \label{Rectangular 0LL}
\end{subfigure}%
\begin{subfigure}{.5\textwidth}
  \includegraphics[height=6cm, width=0.9\linewidth]{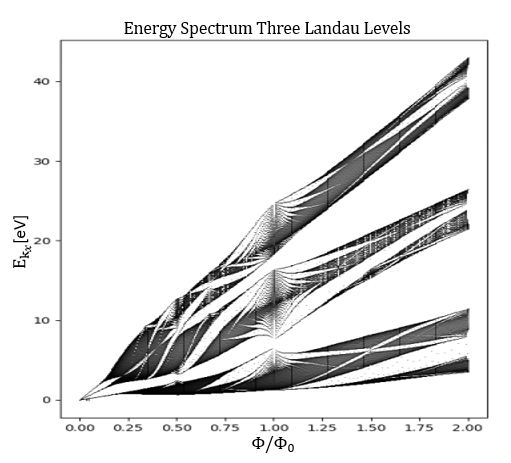}
  \caption{Energy spectrum with\\ three Landau levels included.}
  \label{Rectangular 3LL}
\end{subfigure}
\caption{Energy bands as a function of the relative magnetic flux $\Phi/\Phi_0$ for the rectangular lattice.}
\end{figure}

\subsubsection{Centered Rectangular Lattice Butterfly Spectra}
We also compute the energy bands for the cosine rhombic (centered rectangular) potential depicted in Fig.~\ref{Rhombic}. In the lowest Landau level the energy bands look similar to the one for the square lattice shown in Fig.~\ref{Unscaled Butterfly} but with the butterfly pattern to be narrower than the one in Fig.~\ref{Unscaled Butterfly}. We also consider the case of three Landau levels. The three Landau levels here are very well separated, unlike the rectangular and the oblique lattices. 
\begin{figure}[H]
\begin{subfigure}{.5\textwidth}
  \centering
  \includegraphics[height=6cm, width=0.9\linewidth]{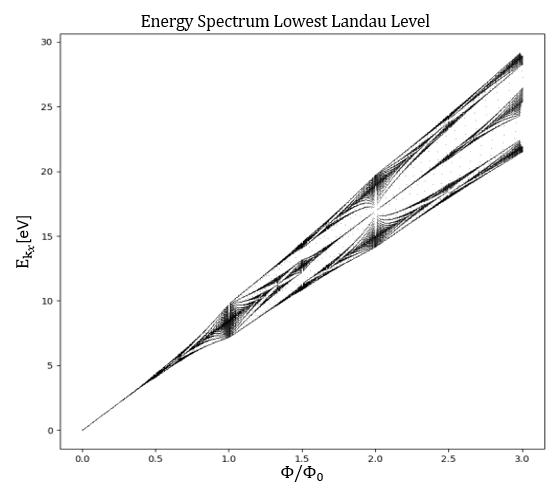}
  \caption{Energy Spectrum in\\ the lowest Landau level. }
  \label{RHO 0LL}
\end{subfigure}%
\begin{subfigure}{.5\textwidth}
  \includegraphics[height=6cm, width=0.9\linewidth]{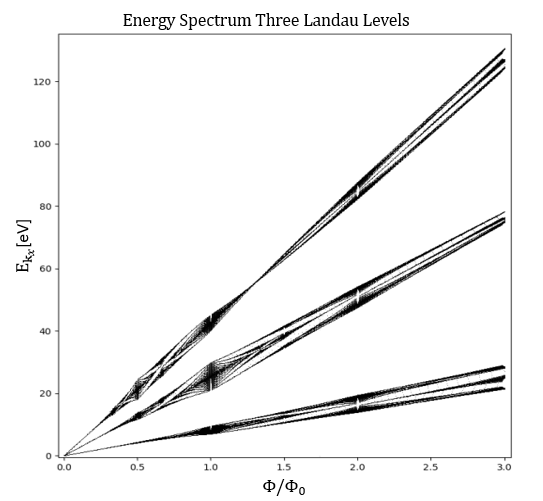}
  \caption{Energy spectrum with\\ three Landau levels included.}
  \label{Rhombic 3LL}
\end{subfigure}
\caption{Energy bands as a function of the relative magnetic flux $\Phi/\Phi_0$ for the cosine centered rectangular lattice.}
\end{figure}

\subsubsection{Hexagonal Lattice Butterfly Spectra}
\begin{figure}[H]
\begin{subfigure}{.5\textwidth}
  \centering
  \includegraphics[height=6cm, width=0.9\linewidth]{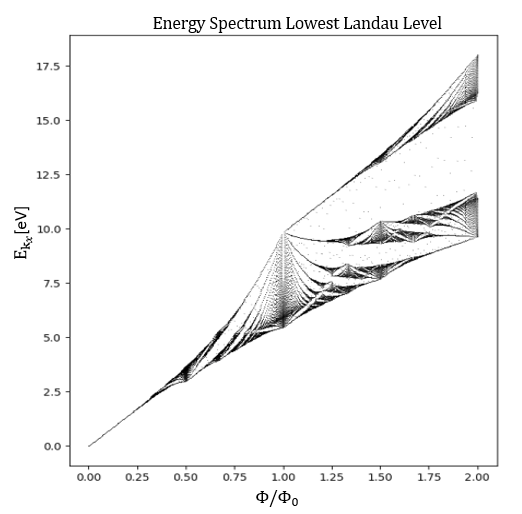}
  \caption{Energy spectrum in\\ the lowest Landau level. }
  \label{Hexagonal 0LL}
\end{subfigure}%
\begin{subfigure}{.5\textwidth}
  \centering
  \includegraphics[height=6cm, width=0.9\linewidth]{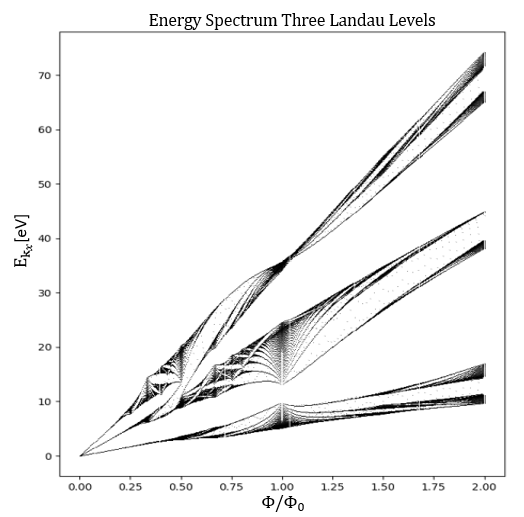}
  \caption{Energy specturm with \\ three Landau levels.}
  \label{Hexagonal 3LL}
\end{subfigure}
\caption{Energy bands as a function of the relative magnetic flux $\Phi/\Phi_0$ for the hexagonal cosine lattice.}
\end{figure}

Further, we compute the energy bands for the cosine hexagonal potential of Fig.~\ref{Hexagonal}. In the lowest Landau levels the energy spectrum exhibits a beautiful asymmetric self-similar (fractal) pattern which is clearly distinct from the standard Hofstadter butterfly shown in Fig.~\ref{Unscaled Butterfly}. Beyond the lowest Landau level we see in Fig.~\ref{Hexagonal 3LL} that the second and the third Landau level touch and influence each other.
\subsubsection{Square Lattice Butterfly Spectrum}
Finally, we compute the energy spectrum beyond the lowest Landau level for the cosine square lattice potential depicted in Fig.~\ref{Square}. Also for the square lattice the second and the third Landau level influence each other. We would like to emphasize that beyond the lowest Landau level there is no clear connection with respect to the tight-binding model with the Peierls phase.
\begin{figure}[H]
\begin{center}
    \includegraphics[height=6cm,width=0.5\columnwidth]{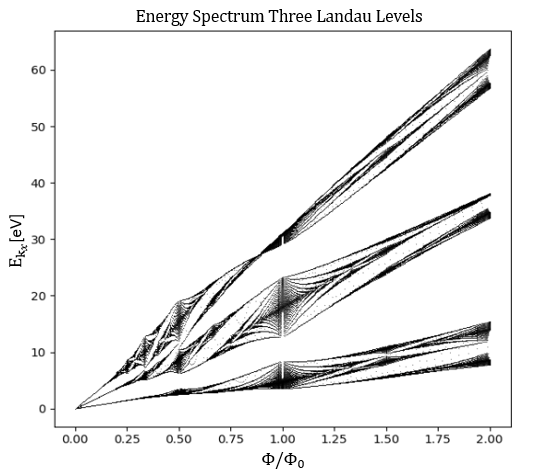}
    \caption{\label{Square 3LL}Energy spectrum as a function of the relative magnetic flux for the square cosine potential. }
\end{center}
\end{figure}

\textit{Concluding Remarks.}---Throughout this section we focused mainly in showing the butterfly spectra for Bravais lattices using the cosine function for the construction of the potentials. However, we would like to emphasize that this does not mean that other type of potentials, like Coulombic potentials, cannot be treated with the presented formalism. Because the presented theory is constructed in real space every type of potential can be treated without any problem by simply expanding the potential in a Fourier series. Furthermore, non-Bravais lattice structures like the one of graphene can be treated easily again by expanding the honeycomb potential of graphene in a Fourier series. Finally, we note that Moire superlattices, on which current experimental tests of the Hofstadter spectrum are performed~\cite{DeanButterfly, WangButterfly, ForsytheButterfly, BarrierButterfly}, can also be treated consistently. These topics and the modeling of these experiments are still under investigation and will be presented in a forthcoming publication.

\section{Fractal Polaritons: Polaritonic Hofstadter Butterfly}

The aim of this last section is to further explore the framework of QED-Bloch theory, by considering the effect of the quantized photon field on the fractal spectrum of the Hofstadter butterfly. To do so, we will focus on a 2D periodic material confined inside a cavity, placed perpendicular to a homogeneous magnetic field. In the case of a 2D material the QED-Bloch central equation~(\ref{QED-Bloch Central}) takes the form
\begin{eqnarray}
&&U^{k_x,k_w}_{n,m,i} \Bigg[\frac{\hbar^2(k_w+G^w_{m,n})^2}{2M}+\mathcal{E}_i-E_{k_x,k_w}\Bigg]+\\
&&\sum_{n^{\prime},m^{\prime},j}V_{n-n^{\prime},m-m^{\prime}} U^{k_x,k_w}_{n^{\prime},m^{\prime},j}\; \exp\left(-\textrm{i}G^v_{m-m^{\prime},n-n^{\prime}}A^{k_x}_{(n+n^{\prime})/2}\right)\langle \phi_i|\hat{D}(\alpha_{n-n^{\prime},m-m^{\prime}})|\phi_j\rangle=0.\nonumber
\end{eqnarray}
To obtain the above equation we eliminated the momentum in the $z$-direction $k_z$ and the index $l$ in the QED-Bloch ansatz~(\ref{BlochAnsatzHermite}) which is associated with the Fourier expansion in the $z$ direction. For simplicity we will consider a 2D square cosine potential as we did also for the derivation of the Harper equation and we will consider the case where only the lowest Landau polariton $\phi_0(v-A^{k_x}_n)$ is occupied.

In the case of the square cosine potential (see also Fig.~\ref{Square}) the angle between the lattice vectors is $\theta=\pi/2$, the two lattice constants are equal $a_1=a_2=a$ and the non-zero Fourier components of the potential are $V_{\pm1,0}=V_{0,\pm1}=V_0$. Further, we note that for the square cosine potential holds $G^v_{m,n}=m_pG^y_m/\sqrt{2}M\omega_c$ and $G^w_{m,n}=G^y_m/\sqrt{2}\omega_c$, see for this Eqs.~(\ref{3Dreciprocallattice}) and (\ref{potential Fourier}). With these assumptions for the periodic potential, the QED-Bloch central equation becomes
\begin{eqnarray}
&&U^{k_x,k_w}_{n,m} \Bigg[\frac{\hbar^2(k_w+\frac{G^y_{m}}{\sqrt{2}\omega_c})^2}{2M}+\mathcal{E}_0-E_{k_x,k_w}\Bigg]+V_0U^{k_x,k_w}_{n-1,m}e^{-\frac{|\alpha_{1,0}|^2}{2}}+V_0U^{k_x,k_w}_{n+1,m}e^{-\frac{|\alpha_{-1,0}|^2}{2}} +\nonumber\\
&&V_0U^{k_x,k_w}_{n,m-1}\exp\left(\frac{-\textrm{i}m_pG^y_1A^{k_x}_n}{\sqrt{2}M\omega_c}\right)e^{-\frac{|\alpha_{0,1}|^2}{2}}+V_0U^{k_x,k_w}_{n,m+1}\exp\left(\frac{-\textrm{i}m_pG^y_{-1}A^{k_x}_n}{\sqrt{2}M\omega_c}\right)e^{-\frac{|\alpha_{0,-1}|^2}{2}}=0\nonumber\\
\end{eqnarray}
Further, for $\theta=\pi/2$ and $a_1=a_2=a$ the $\alpha$-matrix defined in Eq.~(\ref{alphamatrix}) is
\begin{eqnarray}
\alpha_{n,m}=(-1)\sqrt{\frac{\hbar}{2m_{\textrm{e}}\Omega}}\left(G^x_n+\textrm{i}\frac{\omega_c}{\Omega}G^y_m\right)=\frac{-2\pi}{a}\sqrt{\frac{\hbar}{2m_{\textrm{e}}\Omega}}\left(n+\textrm{i}\frac{\omega_c}{\Omega}m\right).
\end{eqnarray}
Using the above expression we find for the four components of the $\alpha$ matrix entering our equation
\begin{eqnarray}
&&|\alpha_{1,0}|^2=|\alpha_{-1,0}|^2=\frac{4\pi^2}{a^2}\frac{\hbar}{2m_{\textrm{e}}\Omega}=\frac{\pi\Phi_0}{\Phi \left(1+\omega^2_p/\omega^2_c\right)^{1/2}} \;\;\; \textrm{and}\\
&&|\alpha_{0,1}|^2=|\alpha_{0,-1}|^2=\frac{\pi \Phi_0}{\Phi\left(1+\omega^2_p/\omega^2_c\right)^{3/2}}.
\end{eqnarray}
To obtain the above results we used the definition for $\mu$ and $\Omega$ given in Eqs.~(\ref{mass polaritonic parameters}) and (\ref{upper polariton frequency}). Moreover, we use the definitions for $m_p$, $M$, $\Omega$, and $A^{k_x}_n$ given respectively in Eqs.~(\ref{scaled masses}),~(\ref{mass polaritonic parameters}),~(\ref{upper polariton frequency}) and~(\ref{Landaupolaritons}) and we find for the quantity
\begin{eqnarray}
\frac{m_p}{\sqrt{2}M\Omega}G^y_1A^{k_x}_n=\frac{1}{1+\omega^2_p/\omega^2_c}\frac{2\pi\Phi_0}{\Phi}\left(\frac{ak_x}{2\pi}+n\right)
\end{eqnarray}
Since everything is now expressed in terms of the dimensionless ratio $\omega_p/\omega_c$ between the two fundamental scales in the electron-photon system, we introduce the parameter $g$
\begin{eqnarray}\label{fractal coupling}
g=\frac{\omega_p}{\omega_c}.
\end{eqnarray}
As we already saw in section~\ref{Landau polaritons Screening}, $g$ defines the light-matter coupling between the Landau levels and the cavity mode. After all these manipulations and having introduced the light-matter coupling $g$ we obtain
\begin{eqnarray}
 &0&=\left[\frac{\hbar^2\left(\sqrt{2}\omega_c k_w+G^y_{m}\right)^2}{2m_{\textrm{e}}(1+g^{-2})}+\mathcal{E}_0-E_{k_x,k_w}\right]U^{k_x,k_w}_{n,m}+ t_1(\Phi,g)\left(U^{k_x,k_w}_{n-1,m}+U^{k_x,k_w}_{n+1,m}\right)\\
&+&t_2(\Phi,g)\left[U^{k_x,k_w}_{n,m-1}\exp\left(\frac{-\textrm{i}2\pi\Phi_0}{\Phi(1+g^2)}\left(\frac{ak_x}{2\pi}+n\right)\right)+U^{k_x,k_w}_{n,m+1}\exp\left(\frac{\textrm{i}2\pi\Phi_0}{\Phi(1+g^2)}\left(\frac{ak_x}{2\pi}+n\right)\right)\right]\nonumber
\end{eqnarray}
where we defined the hopping-like functions
\begin{eqnarray}
t_{1}(\Phi,g)=V_0e^{-\frac{\pi\Phi_0}{2\Phi(1+g^2)^{1/2}}} \;\;\; \textrm{and}\;\;\; t_2(\Phi,g)=V_0e^{-\frac{\pi\Phi_0}{2\Phi(1+g^2)^{3/2}}}
\end{eqnarray}
which depend on the relative magnetic flux and the light-matter coupling. Finally, in analogy to the scaling of the energies defined in Eq.~(\ref{scaled energies}) for the Harper equation, we divide the previous equation by $S(\Phi,g)$
\begin{eqnarray}\label{polariton scaling}
S(\Phi,g)=t_1(\Phi,g)+t_2(\Phi,g)
\end{eqnarray}
and we obtain 
\begin{eqnarray}\label{Polariton Harper}
 &0&=\left[\frac{\hbar^2\left(\sqrt{2}\omega_c k_w+G^y_{m}\right)^2}{2m_{\textrm{e}}(1+g^{-2})}+\mathcal{E}_0-E_{k_x,k_w}\right]\frac{U^{k_x,k_w}_{n,m}}{S(\Phi,g)}+ \frac{t_1(\Phi,g)}{S(\Phi,g)}\left(U^{k_x,k_w}_{n-1,m}+U^{k_x,k_w}_{n+1,m}\right)\\
&+&\frac{t_2(\Phi,g)}{S(\Phi,g)}\left[U^{k_x,k_w}_{n,m-1}\exp\left(\frac{-\textrm{i}2\pi\Phi_0}{\Phi(1+g^2)}\left(\frac{ak_x}{2\pi}+n\right)\right)+U^{k_x,k_w}_{n,m+1}\exp\left(\frac{\textrm{i}2\pi\Phi_0}{\Phi(1+g^2)}\left(\frac{ak_x}{2\pi}+n\right)\right)\right]\nonumber
\end{eqnarray}
Equation~(\ref{Polariton Harper}) is the polaritonic analogue of the Harper equation. In the limit of the light-matter coupling to zero, $g\rightarrow 0$, the polariton Harper equation~(\ref{Polariton Harper}) boils down to the Harper equation (\ref{Harper equation}). This is true because the kinetic term depending on the polaritonic momentum $k_w$ goes to zero and the equation becomes completely independent of $k_w$, and as a consequence the Fourier index $m$ can be dropped. This is a nice consistency check of the polaritonic Harper equation.

However, there are several important differences to the standard Harper equation. First of all, Eq.~(\ref{Polariton Harper}) does not describe Landau levels in a periodic potential but actually Landau polaritons on a lattice. Further, with respect to the standard Harper equation there is an additional degree of freedom $k_w$ corresponding to the polaritonic Bloch wave in the $w$ direction. Most importantly, the energy spectrum of the polaritonic Harper-like equation is not just a function of the relative magnetic flux $\Phi/\Phi_0$, but also a function of the light-matter coupling constant $g=\omega_p/\omega_c$. 

This opens the possibility of not having a fractal/self-similar spectrum as a function of the relative flux but also a fractal as a function of the light-matter coupling constant $g$. To test the existence of this polaritonic fractal we plot the energy spectrum of our system, coming from the polaritonic Harper equation~(\ref{Polariton Harper}), as function of the light-matter coupling $g$ for different values of the relative magnetic flux $\Phi/\Phi_0$.

First, we start with computing the spectrum for a small magnetic flux, $\Phi/\Phi_0=5\times 10^{-3}$. 
\begin{figure}[h]
\centering
  \includegraphics[height=7cm, width=0.7\linewidth]{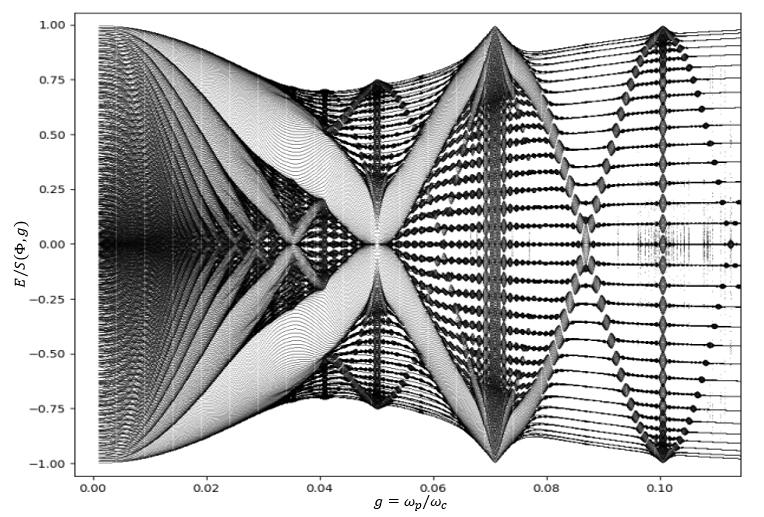}
 \caption{\label{Fractal_pol_1}Dimensionless scaled energy spectrum as a function of the light matter coupling $g=\omega_p/\omega_c$ for magnetic flux ratio $\Phi/\Phi_0=5\times10^{-3}$ of a 2D square cosine potential with lattice constant $a=2$\AA~and potential strength $V_0=3$eV.}
\end{figure}
In figure~\ref{Fractal_pol_1} we clearly see on the left hand side of the figure a beautiful self-similar butterfly pattern showing up as a function of the light-matter coupling $g$. The butterfly pattern appears in the range of light-matter coupling $0<g<0.07$. Then, the butterfly desolves into discrete energy levels with internal oscillatory behavior. This is because the light-matter interaction becomes strong  and the cavity induced quantum fluctuations of the electromagnetic field dominate and ruin the self-similar pattern.

\begin{figure}[H]
\centering
	\includegraphics[height=8cm, width=0.7\linewidth]{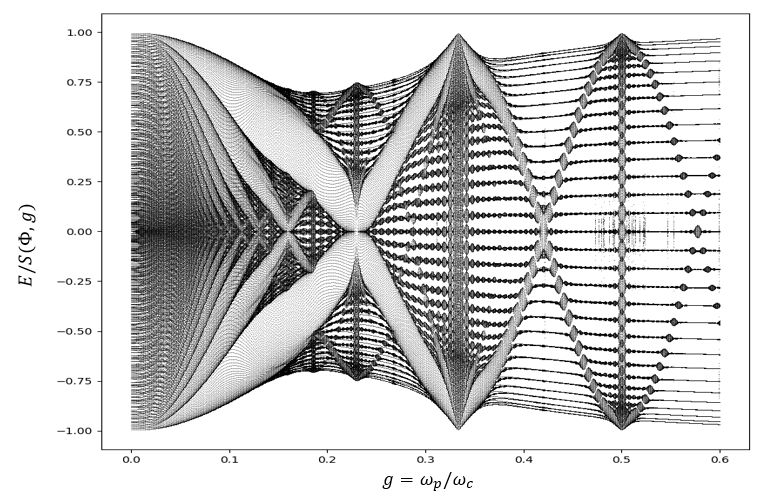}
	\caption{\label{Fractal_pol_2}Dimensionless energy spectrum as a function of the light-matter coupling $g=\omega_p/\omega_c$ with magnetic flux $\Phi/\Phi_0=0.1$ for a 2D square cosine potential with lattice constant $a=2$\AA~and potential strength $V_0=3$eV.}
\end{figure}

We proceed by considering a much larger value for the magnetic flux, $\Phi/\Phi_0=0.1$, twenty times larger than the one considered previously, and we compute the respective spectrum which is depicted in Fig.~(\ref{Fractal_pol_2}).
As we see even for a twenty times larger magnetic flux the pattern remains exactly the same with the one obtained in Fig~(\ref{Fractal_pol_1}) one, with the only change to be the range within which the butterfly pattern shows up. In this case the self-similar pattern appears in the range $0<g<0.35$. This indicates some kind of approximate scaling symmetry in the polariton Harper equation.  

Lastly, we consider the case where the magnetic flux is equal to one flux quantum, $\Phi/\Phi_0=1$.
\begin{figure}[H]
    \centering
    \includegraphics[height= 8cm, width=0.8 \linewidth]{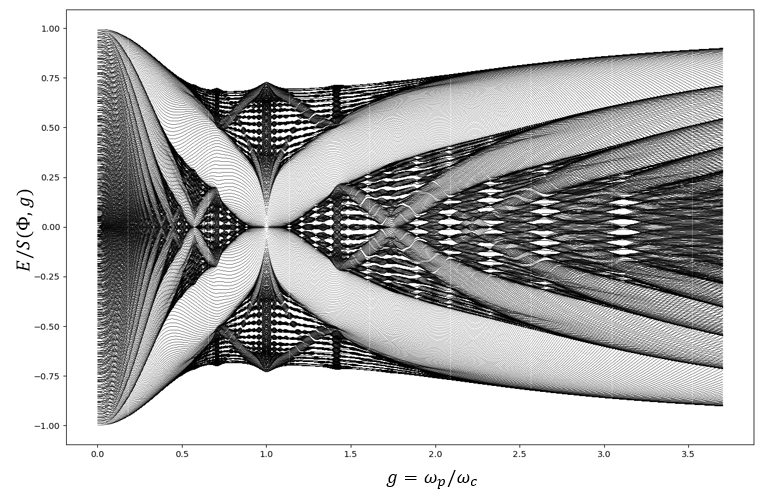}
    \caption{ \label{Fractal Pol 3}Dimensionless energy spectrum as a function of the light-matter coupling $g=\omega_p/\omega_c$ with magnetic flux $\Phi/\Phi_0=1$, for a a 2D square cosine potential with lattice constant $a=2$\AA~and potential strength $V_0=3$eV.}
   \end{figure}
For such a magnetic flux we see that the fractal pattern appears for a very large range of light-matter coupling. With respect to $g=1$ we see that left and right there is self-similarity but clearly the pattern is different on the two sides of the plot. 

From these computations of the energy spectrum for different magnetic fluxes and over different regimes of light-matter interaction, we conclude that for 2D periodic materials strongly coupled to the quantized cavity field and placed perpendicular to a uniform magnetic field there is not only a fractal spectrum emerging as a function of the magnetic flux but there is also a novel self-similar spectrum showing up as a function of the light-matter coupling $g$. This implies that fractal structures do not only appear due to the magnetic field but also due to the quantized cavity field and the interaction of the Landau polariton states with the periodic potential of the material. Thus, what we have presented here introduces the novel concept of fractal polaritons~\cite{RokajButterfly2021}. To the best of our knowledge such a phenomenon has not been reported before and we believe that it can be measured by transport measurements on 2D Moir\'{e} materials under cavity confinement and in the presence of a strong magnetic field. For the observation of the fractal polaritons Moir\'{e} materials are necessary in order to achieve a substantial fraction of the flux quantum~\cite{DeanButterfly, WangButterfly}. How exactly the polariton fractal can be observed for such systems will be the topic of an upcoming publication. Finally, we would like to comment that the polaritonic Hofstadter butterfly it is not an exact fractal, but only an approximate one. This can be easily seen from Figs.~\ref{Fractal_pol_1} and~\ref{Fractal_pol_2} where the self-similar pattern shows up only in the left hand side of the figures, for small values of $g$. From a mathematical point of view this can be understood from the fact that the QED-Bloch Hamiltonian is periodic under translations in the full polaritonic configuration space. This is unlike the Harper equation which is quasi-periodic. The quasi-periodicity of the Harper equation is related to the fact that the Hofstadter butterfly is a Cantor set, and consequently a fractal, as it was proven by Avila and Jitomirskaya~\cite{Avila2006}.


\chapter{Epilogue}

\begin{displayquote}
\footnotesize{We live on an island surrounded by a sea of ignorance. As our island of knowledge grows, so does the shore of our ignorance.\footnote{I would like to thank my friend Perseas Christodoulidis for bringing to my attention this quote by John A. Wheeler.}}
\end{displayquote}
\begin{flushright}
  \footnotesize{John A. Wheeler}
\end{flushright}

\section{Summary}

Quantum electrodynamics and condensed matter physics have been long considered as distinct disciplines. However, this situation is rapidly changing with the progress in materials science and the control of quantum matter with the use of electromagnetic fields. The need of an extensive interchange between these two fields has become now crucial for the explanation of recent experimental developments.

It is a great privilege to enter the field of cavity QED materials at the time where such a tremendous progress is taking place and the need for the development of novel theoretical approaches for the description of many-body condensed matter systems strongly coupled to the photon field is more pressing than ever. Typically, it is in such times when paradigm shifts occur and well established concepts and methods need to be re-invented and go through a radical change~\cite{KuhnScienceRevol}.

In this thesis we tried, to the extent possible, to go into this direction and develop new methods and construct new paradigmatic models for extended condensed matter systems coupled to the photon field. The main principle that was explored and studied throughout the thesis was translational invariance in the context of condensed matter systems strongly coupled to the photon field. Translational invariance is one of the defining and most basic properties of solid-state and condensed matter systems. Due to its fundamental nature and importance it was necessary the fundamentals of QED to be understood and studied thoroughly as well.

\textit{Fundamentals of Quantum Electrodynamics}.---In the first part of the thesis and particularly in chapter~\ref{Quantum Electrodynamics} we showed how the Pauli-Fierz theory~\cite{spohn2004} can be obtained as the non-relativistic limit of QED. Then, in chapter~\ref{Length Gauge QED} we presented how the Pauli-Fierz theory looks in the so called long-wavelength limit or dipole approximation, which is in most cases employed in cavity QED. The main benefit of the dipole approximation is that the photon field becomes spatially homogeneous. This simplifies the light-matter interaction and preserves translational invariance in the electronic configuration space, which makes it ideal for the description of solid-state systems coupled to the light field. In the dipole approximation there is a unitarily equivalent form of the Pauli-Fierz Hamiltonian which is known as the length gauge~\cite{rokaj2017, schaeferquadratic, power1982quantum, power1959coulomb, woolley1980gauge}. In the length gauge the light-matter Hamiltonian takes a different form and a new interaction term arises known as the dipole self-energy. The light-matter interaction and the dipole self-energy both break translational invariance in the electronic configuration space. However, as translational invariance is a physical property, it is preserved also in the length gauge, but now manifests itself in the full electronic plus photonic (polaritonic) configuration space~\cite{rokaj2017}. It is important to note that translational invariance is not preserved if the dipole self-energy is omitted~\cite{rokaj2017}. The dipole self-energy in the recent years has been the source of an ongoing debate in the field of cavity QED, as it has been claimed that this term can be safely neglected from the length gauge Hamiltonian~\cite{GalegoCasimir, Triana, faisal1987}. Despite these claims, the dipole self-energy as we showed in chapter~\ref{Length Gauge QED}, is absolutely necessary for the stability of the Pauli-Fierz Hamiltonian and without this term the Hamiltonian has no ground state as it was proven in~\cite{rokaj2017}. In addition, without the dipole self-energy, gauge invariance is broken, photonic observables are not described correctly and the Maxwell's equations in matter are not satisfied~\cite{rokaj2017, schaeferquadratic}. For all these reasons the dipole self-energy cannot be discarded from the length gauge Hamiltonian.

\textit{The Free Electron Gas in Cavity QED}.---In the second part of the thesis we focused on the Sommerfeld model~\cite{Sommerfeld1928} of the free electron gas coupled to the quantized photon field originating from a cavity. The Sommerfeld model is paradigmatic for condensed matter physics and has been used for the development of several many-body theories and models, like the jellium model~\cite{Mermin, Vignale}, the local density approximation in density functional theory~\cite{kohn1965} and Landau's Fermi liquid theory~\cite{LandauFermiLiquid}. In chapter~\ref{Free Electron Gas} we gave a brief overview of the free electron gas and some of its basic properties, and in chapter~\ref{Free Electron Gas in cavity QED} we revisited the Sommerfeld model in cavity QED and we provided the exact analytic solution of this many-electron system for a finite amount of modes (see also appendix~\ref{Mode-Mode Interactions}) in the long-wavelength limit~\cite{rokaj2020}. The main ingredient to accomplish this analytic solution was the use of translational invariance and the fact that momentum is a good conserved quantum number. Then, performing a Bogolyubov transformation and a coherent shift on the photonic operators the exact solution was achieved. To the best of our knowledge, this is the first exact analytic solution of such an extended macroscopic system and we hope that it will serve as a new paradigm for the emerging field of condensed matter QED, and for the exploration of collective and superradiant phenomena beyond the Dicke model~\cite{dicke1954}. Moreover, in chapter~\ref{Free Electron Gas in cavity QED} we showed that the combined electron-photon ground state in the thermodynamic limit is a Fermi liquid dressed with virtual photons~\cite{rokaj2020}. Further, we investigated the stability of this system and proved that without the diamagnetic $\bi{A}^2$ term the system becomes unstable because the light-matter coupling has no upper bound. 

To make a connection to experimentally accessible properties of the free electron gas in the cavity, in chapter~\ref{Cavity Responses} we performed linear response for the coupled electron-photon system and we showed that the cavity field modifies the conductive properties of the electron gas by introducing new resonances and suppressing the DC conductivity and the Drude peak of the electron gas~\cite{rokaj2020}. Finally, to go beyond the finite-mode approximation for the photon field, in chapter~\ref{Effective QFT} we constructed an effective field theory in the continuum. In this effective field theory we found that the electron gets a many-body mass renormalization. Further, we computed the zero-point energy in this effective field theory and we showed that it leads to a macroscopic repulsive Casimir force due to the strong light-matter interaction. Also, in this effective field theory due to the continuum of modes dissipation can be treated from first principles without the need of an artificial broadening parameter~\cite{rokaj2020}.

\textit{Quantum Hall Systems in Cavity QED}.---In the third part of the thesis we focused on another fundamental phenomenon of condensed matter physics, in which translational symmetry is explicitly broken due to an external magnetic field, namely the quantum Hall effect~\cite{Klitzing}. In chapter~\ref{Landau Levels QHE} we reviewed how the quantization of the Hall conductance can be described in terms of non-interacting electrons in fully occupied Landau levels, as it was done originally by Laughlin~\cite{LaughlinPRB}. Then, in chapter~\ref{Bloch MTG} we gave a rigorous proof of Bloch's theorem for periodic materials~\cite{Mermin} and a thorough presentation of the fundamental problems that arise for periodic materials in a homogeneous magnetic field due to the breaking of translational symmetry. Although, Bloch's theorem cannot be applied in this setting a new symmetry group arises, the magnetic translation group~\cite{BrownMTG, MTG_I, MTG_II}. Unfortunately, the magnetic translation group does not provide a complete solution for the description of periodic materials in homogeneous magnetic fields. Because it puts strict conditions on the strength of the magnetic field but most importantly because no ansatz, analogous to the Bloch ansatz, has been constructed so far from eigenfunctions of the magnetic translation operators.

In chapter~\ref{QED Bloch theory}, motivated by the observation that from a physical point of view, a homogeneous magnetic field should not break translational symmetry, because the field is the same from one point of space to another, we decided to revisit this problem in the framework of non-relativistic QED. We showed that translational symmetry can be restored in the enlarged electronic plus photonic configuration space, by including the quantized photon field~\cite{rokaj2019}. In this framework we can make use of Bloch's theorem in the enlarged space of electrons and photons, and this framework was named quantum electrodynamical Bloch (QED-Bloch) theory. We further constructed a polaritonic Bloch ansatz that allows for the description of periodic materials in homogeneous magnetic fields but also in the presence of their quantum fluctuations. As a first application of QED-Bloch theory we considered Landau levels coupled to the cavity field. This system we solved analytically and we demonstrated that it describes hybrid quasi-particle excitations between the Landau levels and the photon field, known as Landau polaritons~\cite{rokaj2019}, which have also been observed experimentally~\cite{ScalariScience, Keller2020}. As a further application of QED-Bloch theory we considered the case where we have a 2D periodic material confined in a cavity and perpendicular to a homogeneous magnetic field. In this case we showed that fractal polaritonic spectra emerge, as a function of the light-matter coupling~\cite{RokajButterfly2021}. This is a truely novel phenomenon that, to the best of our knowledge, has not been reported before and we believe can be measured by coupling such 2D systems in cavities. Finally, we showed that in the limit of no quantized field, QED-Bloch theory recovers the Harper's equation~\cite{Harper_1955} and the fractal spectrum of the Hofstadter butterfly and thus can be applied also to the purely electronic problem, of periodic materials in strong magnetic fields.

\section{Future Directions}

With every piece of research performed a set of questions is settled to some degree of satisfaction and certainty, but always an equal (if not larger) amount of questions arise and new lines of research come within reach. Thus, we would like to conclude with some open questions and potential next steps to be followed as a natural extension of this thesis.

\subsection*{Fermi Liquid Theory in QED}

Landau's Fermi liquid theory~\cite{Landau} is paradigmatic for many-body and condensed matter physics and has been applied for the description of metals and liquid $^3\textrm{H}_{\textrm{e}}$. Its further extensions have enabled the understanding of superfluid $^4\textrm{H}_{\textrm{e}}$ and superconductivity~\cite{Nozieres}. In Fermi liquid theory the Coulomb interaction and electron correlations are treated with the use of field theoretic Green's function methods. These methods are powerful and allow for the description of correlated systems. 

In the second part of the thesis, we showed that the electron-photon ground state of the free electron gas in the cavity, is a Fermi liquid dressed with photons~\cite{rokaj2020}. Further, it was demonstrated that the cavity modifies the effective mass and the excitations of the fermionic quasi-particles of the Fermi liquid~\cite{rokaj2020}. These results pave the way for the generalization of Fermi liquid theory in the framework of QED. This development would be an important leap towards formulating a first-principles theory for correlated electron systems strongly coupled to the quantized electromagnetic field. This would allow to describe real materials in cavity QED, like metals, superconductors and Bose-Einstein condensates. Lastly, we believe that such a theory will set a new paradigm for experimental efforts towards unraveling novel correlated phases between quantum matter and light. 

\subsection*{Superradiant Phases}
    
In 1973 Hepp and Lieb~\cite{Lieb} demonstrated that when a system of many two-level atoms couples to the electromagnetic field it undergoes a so-called superradiant phase transition. Since then there exists an ongoing debate and several no-go theorems have questioned the existence of this phase transition~\cite{Birula}. More recently also other kind of superradiant phases have been suggested like ferroelectric~\cite{Demlerferro} or magnonic~\cite{KonoMagnonic}. For the free electron gas in cavity QED~\cite{rokaj2020} we showed that for non-interacting electrons and dipolar electromagnetic fields such a phase does not exist. However, the superradiant phase cannot be excluded for interacting electrons or for non-dipolar fields. Thus, as a continuation of the work on the free electron gas coupled to the cavity, it becomes highly interesting to investigate the role of electron correlations and of electromagnetic fields beyond the dipole approximation for the existence of the superradiant phase transition. Further, in the case that such a phase cannot be reached in equilibrium, it is worthwhile to explore whether this phase transition can be achieved by driving the system out of equilibrium with the use of external fields, currents, or lasers. 
    
\subsection*{Quantum Hall Effects Inside a Cavity}    
    
Two-dimensional electron systems at low temperatures when placed in a perpendicular homogeneous magnetic field exhibit quantization of the Hall conductance. The Hall conductance has been found to be either an integer or a fractional multiple of $e^2/h$. These effects are known as the integer~\cite{Klitzing} and the fractional~\cite{TsuifractionalQHE} quantum Hall effect respectively, and in most cases are described with help of the Landau-level-picture. In the third part of this thesis we studied two-dimensional Landau level systems under cavity-confinement and we were able to show analytically that the coupling to the quantized cavity field modifies the well-known Landau levels. In addition, we demonstrated that the strong light-matter coupling leads to the formation of hybrid quasiparticle states called Landau polaritons~\cite{rokaj2019}. Recently, the study of integer and fractional quantum Hall systems in cavity QED has attracted experimental interest and the Landau polaritons have been observed~\cite{Keller, ScalariScience, li2018, paravacini2019, SmolkaAtac}. 

Our work on the Landau polaritons opens the possibility to study integer and fractional quantum Hall systems strongly coupled to the quantized cavity field. Since the quantum Hall effects in most cases are described with the use of the Landau-level-picture, we are confident that our work can be used as the fundamental building block for the description of quantum Hall systems inside a cavity. The main phenomena to be investigated in the future, which are of current experimental interest~\cite{paravacini2019, SmolkaAtac}, are: (i) What is the effect of the cavity confinement on the Hall conductance? Does it remain still quantized or the electron-photon correlations modify this fundamental behavior? (ii) In the fractional regime, how does the Laughlin wavefunction~\cite{Laughlingfractional} and the incompressible properties of this quantum fluid change due to the photon field? (iii) Does the photon field induce novel correlations and many-body states?

Exploring these directions will advance and forward the field of cavity QED materials both theoretically and experimentally. We hope and believe that following the above research objectives will provide new insights on how to induce novel correlated states between matter and photons. Finally, by exploiting these hybrid states, the modification of material properties, like conduction or energy transfer, becomes possible and could potentially lead to technological applications.

\appendix


\chapter{Many-Body Hamiltonian without Dipole Self-Energy}\label{Many Mode DSE}

In section~\ref{No GS without DSE} we proved for the simple case of one electron in a binding, Coulombic-type of potential, coupled to a single mode of the photon field that the electron-photon system has no ground-state if the dipole self-energy given by Eq.~(\ref{Dipole self energy}) is neglected from the length-gauge Hamiltonian in Eq~(\ref{Lenght Hamiltonian}). The aim of this appendix is to generalize the latter proof to the generic case where we have $N$ electrons interacting via Coulomb forces, coupled to $M$ modes of the electromagnetic field in the length gauge, described by the Hamiltonian $\hat{H}_L$ given by Eq.~(\ref{Lenght Hamiltonian}). 

\begin{figure}
\begin{center}
 \includegraphics[width=3in, height=2.5in]{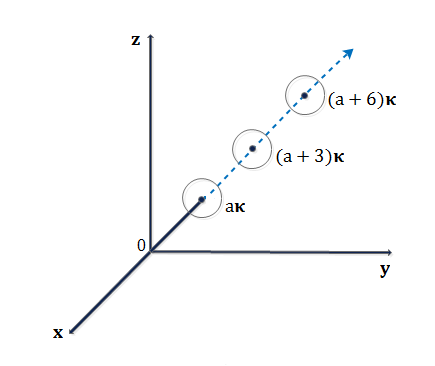}
\caption{\label{fig:three}Schematic depiction of the localization of the electronic wavefunction $\Psi_e$. The mollifiers are put on an equally spaced grid along the vector $\bi{w}$ such that there is no overlap between them.}   
\end{center}
\end{figure}

To investigate the importance of the dipole self-energy we will follow the strategy of section~\ref{No GS without DSE}, namely we will drop the dipole self-energy, which subsequently leads to the Hamiltonian
\begin{eqnarray}\label{eqa.5}
	\hat{H}^{\prime}&=&\hat{H}_L-\hat{\varepsilon}_{dip}=-\frac{\hbar^2}{2m}\sum\limits^{N}_{i=1}\bi{\nabla}^2_i+\frac{1}{4\pi \epsilon_0} \sum\limits^{N}_{i< j}\frac{e^2}{|\bi{r}_i-\bi{r}_j|}+\sum\limits^{N}_{i=1}v_{ext}(\bi{r}_{i})\nonumber\\
	&+&\sum\limits_{\bm{\kappa},\lambda}\left[ -\frac{\hbar\omega(\bm{\kappa})}{2}\frac{\partial^2}{\partial p^2_{\bm{\kappa},\lambda}}+\frac{\hbar\omega(\bm{\kappa})}{2}p^2_{\bm{\kappa},\lambda}+ \hat{V}_{int}\right],
\end{eqnarray}
where the bilinear interaction term between the photons and the charged particles is
\begin{equation}\
	\hat{V}_{int} = -\sum_{\bm{\kappa},\lambda}e \sqrt{\frac{\hbar\omega(\bm{\kappa})}{\epsilon_0V}} \bm{\varepsilon}_{\lambda}(\bm{\kappa})\cdot\bi{R} \;p_{\bm{\kappa},\lambda}=-\sum_{\bm{\kappa},\lambda}\bm{\zeta}_{\bm{\kappa},\lambda}\cdot \bi{R}p_{\bm{\kappa},\lambda}.
\end{equation}
Then, we will compute the energy of a test wavefunction with respect to $\hat{H}'$ and show that the energy can be lowered indefinitely. For the electronic part of the trial wavefunction we will consider a Slater determinant, and for simplicity we will assume a fully spin-polarized wavefunction such that we can separate the spin component of the wavefunction. Under these choices we have
\begin{eqnarray}\label{eqa.7}
	\Psi_e(\bi{r}_1,...,\bi{r}_N)=\frac{1}{\sqrt{N!}}\;\begin{tabular}{|c c c c|}
		$F_{1}(\bi{r}_1)$ &$ F_2(\bi{r}_1)$ & $\cdot\cdot\cdot$ & $F_{N}(\bi{r}_1)$\\
		$F_1(\bi{r}_2)$ & $F_2(\bi{r}_2)$ & $\cdot\cdot\cdot$ & $F_{N}(\bi{r}_2)$\\
		$\cdot$ & $\cdot$ & $\cdot\cdot\cdot$& $\cdot$\\
		$\cdot$ & $\cdot$ & $\cdot\cdot\cdot$& $\cdot$\\
		$F_1(\bi{r}_N)$ & $F_2(\bi{r}_N)$ & $\cdot\cdot\cdot$ & $F_N(\bi{r}_N)$
	\end{tabular}\quad .
	\label{tab:my_label}
\end{eqnarray}
For every component of the Slater determinant a normalized mollifier is used
\begin{eqnarray}\label{eqa.8}
	F_i(\bi{r}_j)=
	\begin{cases}
		\mathcal{N}\exp[-\frac{1}{1-|\bi{r}_j-\bi{a}_i|^2}]\qquad \textrm{if} \qquad|\bi{r}_j-\bi{a}_i|<1 \\
		0\qquad \textrm{if} \qquad |\bi{r}_j-\bi{a}_i|\geq 1  \\
		\textrm{where}\qquad \bi{a}_i=[a+3(i-1)]\bi{w}
	\end{cases}
\end{eqnarray}
where $\mathcal{N}$ is the normalization constant. The mollifiers are placed on a grid along an arbitrary direction $\bi{w}$ as depicted in Fig.~\ref{fig:three}. The mollifiers $F_i(\bi{r}_j)$ are non-zero within the unit ball $|\bi{r}_j-\bi{a}_i|< 1$, and zero outside it. Their supports are disjoint, such that they have no overlaps, and the vector $\bi{a}_i$ is the center of each of these unit balls. It is important to note that the position of mollifiers depends on an arbitrary parameter $a$.
For the photonic part we use
\begin{equation}\label{eqa.9}
	\Phi_p=\bigotimes^{M}_{\bm{\kappa},\lambda}\frac{1}{\sqrt{2}}\left(\phi_1(p_{\bm{\kappa},\lambda})+\phi_2(p_{\bm{\kappa},\lambda})\right),
\end{equation}
where $\phi_n(p_{\bm{\kappa},\lambda})$ are the normalized eigenfunctions of the corresponding harmonic oscillator with respect to the coordinate $p_{\bm{\kappa},\lambda}$. Thus, the full wavefunction is
\begin{equation}\label{eqa.10}
	\Psi=\Psi_e(\bi{r}_1,...,\bi{r}_N)\otimes\Phi_p.
\end{equation}
Due to the fact that $\langle p_{\bm{\kappa},\lambda}\Phi_p,p_{\bm{\kappa}^{\prime},\lambda^{\prime}}\Phi_p\rangle=2\delta_{\bm{\kappa}\bm{\kappa}^{\prime}}\delta_{\lambda\lambda^{\prime}}$ we have
\begin{eqnarray}\label{eqa.11}
	\langle\hat{V}_{int}\Psi|\hat{V}_{int}\Psi\rangle &=&\sum\limits^{M}_{\bm{\kappa},\bm{\kappa}^{\prime},\lambda,\lambda^{\prime}}\sum\limits^{N}_{i,j=1}\langle p_{\bm{\kappa},\lambda}\Phi_p|p_{\bm{\kappa}^{\prime},\lambda^{\prime}}\Phi_p\rangle \langle\bm{\zeta}_{\bm{\kappa},\lambda}\cdot\bi{r}_i\Psi_e|\bm{\zeta}_{\bm{\kappa}^{\prime},\lambda^{\prime}}\cdot\bi{r}_j\Psi_e\rangle
        \\
        &=&2\sum\limits^M_{\bm{\kappa},\lambda}\sum\limits^N_{i,j=1}\langle\bm{\zeta}_{\bm{\kappa}}\cdot\bi{r}_i\Psi_e|\bm{\zeta}_{\bm{\kappa}}\cdot\bi{r}_j\Psi_e\rangle<\infty. \nonumber
\end{eqnarray}
This means that the wavefunction $\Psi$ is part of the domain of $\hat{H}_L$ as well as $\hat{H}'$. The expression of the energy is
\begin{eqnarray}\label{eqa.13}
\langle\Psi|\hat{H}^{'}| \Psi\rangle=\langle\Psi_e|\hat{T}_e|\Psi_e\rangle+\langle\Psi_e|\hat{W}_e|\Psi_e\rangle+\langle\Psi_e|\hat{V}_{ext}|\Psi_e\rangle +\langle\Phi_p|\hat{H}_p|\Phi_p\rangle+\langle\Psi|\hat{V}_{int}|\Psi\rangle .
\end{eqnarray}
The kinetic energy of the electrons,
\begin{eqnarray}\label{eqa.14}
\langle\Psi_e|\hat{T}_e|\Psi_e\rangle&=&-\frac{\hbar^2}{2m}\sum\limits^{N}_{i=1}\langle\Psi_e|\bi{\nabla}^2_{i}|\Psi_e\rangle\nonumber\\
&=&-\frac{\hbar^2}{2m}\sum\limits^{N}_{i=1}\prod^{N}_{n=1}\;\int \limits_{|\bi{r}_{n}-\bi{a}_i|<1}d^3r_n \Psi_e(\bi{r}_1,...,\bi{r}_N) \bi{\nabla}^2_i \Psi_e(\bi{r}_1,...,\bi{r}_N),
\end{eqnarray}
is translationally invariant and thus we can perform the translation $\bi{r}_n \longrightarrow \bi{r}_n+a\bi{w}$. The integration volume after this transformation becomes
\begin{equation}\label{eqa.16}
	|\bi{r}_{n}-\bi{a}_i|<1 \longrightarrow |\bi{r}_n -3(i-1)\bi{w}|<1.
\end{equation}
As a consequence the result of the integral is a finite constant independent of the parameter $a$ 
\begin{eqnarray}\label{eqa.17b}
 \langle\Psi_e|\hat{T}_e|\Psi_e\rangle=-\frac{\hbar^2}{2m}\sum\limits^{N}_{i=1}\prod^{N}_{n=1}\;\;\int \limits_{|\bi{r}_{n}-3(i-1)\bm{\kappa}|<1}d^3r_n \Psi_e(\bi{r}_1,...,\bi{r}_N) \bi{\nabla}^2_{i} \Psi_e(\bi{r}_1,...,\bi{r}_N)=A.
\end{eqnarray}
After performing the same translation, the contribution of the Coulomb interaction becomes also independent of the parameter $a$
\begin{equation}\label{eqa.19}
	\langle\Psi_e|\hat{W}_e|\Psi_e\rangle=\sum\limits^{N}_{i<j}\prod^{N}_{n=1}\;\int \limits_{|\bi{r}_{n}-3(i-1)\bi{w}|<1}d^3r_n  W_e(\bi{r}_i-\bi{r}_j)|\Psi_e(\bi{r}_1,...,\bi{r}_N)|^2=D<\infty.
\end{equation}
Without loss of generality we choose the external potential to be negative such that
\begin{equation}\label{eqa.22}
	\langle\Psi_e|\hat{V}_{ext}|\Psi_e\rangle=\sum\limits^N_{i=1}\langle\Psi_e|v_{\textrm{ext}}(\bi{r}_i)|\Psi_e\rangle=-\tilde{V}_a\qquad \textrm{where}\qquad \tilde{V}_a\geq 0.
\end{equation}
The energy of the photons is
\begin{eqnarray}\label{eqa.23}
\langle\Phi_p|\hat{H}_p|\Phi_p\rangle&=&\frac{1}{2}\bigotimes^{M}_{\bm{\kappa},\bm{\kappa}^{\prime},\lambda,\lambda^{\prime}} \langle\psi_1(p_{\bm{\kappa},\lambda})+\psi_2(p_{\bm{\kappa},\lambda})|\hat{H}_p|\psi_1(p_{\bm{\kappa}^{\prime},\lambda^{\prime}})+\psi_{2}(p_{\bm{\kappa}^{\prime},\lambda^{\prime}})\rangle\nonumber\\
	&=&\sum^{M}_{\bm{\kappa}}\left(E_1(\bm{\kappa})+E_{2}(\bm{\kappa})\right),
\end{eqnarray}
where $E_n(\bm{\kappa})=\hbar\omega(\bm{\kappa})(n+1/2)$ are the eigenenergies of the harmonic oscillator. We note that to obtain the result above we also summed over the two polarizations of each mode, which are degenerate energetically. The contribution of the bilinear interaction between the electrons and the photon modes is
\begin{eqnarray}\label{eqa.24}
\langle\Psi|\hat{V}_{int}|\Psi\rangle&=&-\sum^{M}_{\bm{\kappa},\lambda}\langle\Phi_p|p_{\bm{\kappa},\lambda}|\Phi_p\rangle \langle\Psi_e|\bm{\zeta}_{\bm{\kappa},\lambda}\cdot\bi{R}|\Psi_e\rangle\nonumber\\
&=&-\sum\limits^{N}_{i=1}\sum\limits^M_{\bm{\kappa},\lambda}\bm{\zeta}_{\bm{\kappa},\lambda}\langle\Phi_p|p_{\bm{\kappa},\lambda}|\Phi_p\rangle\;\prod^{N}_{n=1}\;\;\int \limits_{|\bi{r}_{n}-\bi{a}_i|<1}d^3r_n \;\bi{r}_i|\Psi_e(\bi{r}_1,...,\bi{r}_N)|^2\nonumber\\
&=&-\sum^N_{i=1}\sum^M_{\bm{\kappa},\lambda}\bm{\zeta}_{\bm{\kappa},\lambda}\cdot\;\prod^{N}_{n=1}\;\;\int \limits_{|\bi{r}_{n}-\bi{a}_i|<1}d^3r_n \;\bi{r}_i|\Psi_e(\bi{r}_1,...,\bi{r}_N)|^2.
\end{eqnarray}
In equation (\ref{eqa.24}) we used that $\sum\limits^{M}_{\bm{\kappa},\lambda}\bm{\zeta}_{\bm{\kappa},\lambda}\langle\Phi_p|p_{\bm{\kappa},\lambda}|\Phi_p\rangle=\sum\limits^{M}_{\bm{\kappa},\lambda}\bm{\zeta}_{\bm{\kappa},\lambda}$. We perform once more the translation $\bi{r}_n \longrightarrow \bi{r}_n+a\bi{w}$ and we have
\begin{eqnarray}\label{eqa.25}
	 \langle\Psi|\hat{V}_{int}|\Psi\rangle&=&-\sum^N_{i=1}\sum^M_{\bm{\kappa},\lambda}\bm{\zeta}_{\bm{\kappa},\lambda}\cdot\prod^{N}_{n=1}\;\;\int \limits_{|\bi{r}_{n}-3(i-1)\bi{w}|<1}d^3r_n (\bi{r}_i+a\bi{w})|\Psi_e(\bi{r}_1,...,\bi{r}_N)|^2\nonumber\\
	 &=&-\sum^N_{i=1}\sum^M_{\bm{\kappa},\lambda}\bm{\zeta}_{\bm{\kappa},\lambda}\cdot\prod^{N}_{n=1}\;[\int \limits_{|\bi{r}_{n}-3(i-1)\bi{w}|<1}d^3r_n \;\bi{r}_i|\Psi_e(\bi{r}_1,...,\bi{r}_N)|^2 \\
	 &+&a\bi{w}\int\limits_{|\bi{r}_{n}-3(i-1)\bi{w}|<1}d^3r_n |\Psi_e(\bi{r}_1,...,\bi{r}_N)|^2 ].
\end{eqnarray}
The two integrals above do not depend in the parameter $a$. The result of the first integral is some finite constant and the result of the second integral is 1 because it is simply the norm of the electronic wavefunction. As a consequence we obtain
\begin{eqnarray}\label{eqa.26}
	\langle\Psi|\hat{V}_{int}|\Psi\rangle=-B-\sum^M_{\bm{\kappa},\lambda}\bm{\zeta}_{\bm{\kappa},\lambda}\cdot\sum^{N}_{i=1}a\bi{w}=-B-aN\sum^{M}_{\bm{\kappa},\lambda}\bm{\zeta}_{\bm{\kappa},\lambda}\cdot \bi{w}.
\end{eqnarray}
Then, by choosing $\bi{w}$ to be parallel to at least one of the coupling-strength polarization vectors $\bm{\zeta}_{\bm{\kappa},\lambda}$ the contribution of the bilinear interaction will be proportional to $-a$.
Finally, by summing up all five contributions entering the Hamiltonian $\hat{H}^{\prime}$, we obtain the following inequality for the energy of the system 
\begin{eqnarray}\label{eqa.28}
	\langle\Psi|\hat{H}^{'}|\Psi\rangle&=& A +D-\tilde{V}_a+\frac{1}{2}\sum^{M}_{\bm{\kappa}}\left(E_1(\bm{\kappa})+E_{2}(\bm{\kappa})\right)-B-aN\sum^M_{\bm{\kappa},\lambda}\bm{\zeta}_{\bm{\kappa},\lambda}\cdot\bi{w} \leq \nonumber\\
	&\leq& A +D +\sum^{M}_{\bm{\kappa}}\left(E_1(\bm{\kappa})+E_{2}(\bm{\kappa})\right) -B -aN\sum^M_{\bm{\kappa},\lambda}\bm{\zeta}_{\bm{\kappa},\lambda}\cdot\bi{w}\sim -a.
\end{eqnarray}
Since the parameter $a$ is arbitrary we can lower the energy indefinitely. This implies that the Hamiltonian $\hat{H}^{'}$ is unbounded from below and has no ground-state due to the elimination of the dipole self-energy~\cite{rokaj2017}.

\chapter{Free Electron Gas Coupled to Many-Modes}\label{Mode-Mode Interactions}

In this appendix we are interested in the case of the free electron gas coupled to an arbitrary finite amount of modes of the photon field. In chapter~\ref{Free Electron Gas in cavity QED} we provided the analytic solution for this system in the single-mode case. Here, we aim to generalize our solution to the many-mode case and demonstrate that the mode-mode interactions to do not fundamentally change the structure of the energy spectrum with respect to the single-mode case, and the spectrum of the effective field theory that we constructed in chapter~\ref{Effective QFT}. 
 
 The Pauli-Fierz Hamiltonian for $N$ non-interacting electrons coupled to the quantized photon field is
\begin{eqnarray}
\hat{H}&=&\frac{1}{2m_{\textrm{e}}}\sum\limits^{N}_{j=1}\left(\textrm{i}\hbar \mathbf{\nabla}_{j}+e \hat{\mathbf{A}}\right)^2 +\sum\limits_{\bm{\kappa},\lambda}\hbar\omega(\bm{\kappa})\left[\hat{a}^{\dagger}_{\bm{\kappa},\lambda}\hat{a}_{\bm{\kappa},\lambda}+\frac{1}{2}\right]\nonumber\\
\end{eqnarray}
where the quantized $\hat{\bi{A}}$-field is considered in the long-wavelength limit or dipole approximation. For the treatment of the many-mode case it is convenient to introduce for the description of the annihilation and creation operators the displacement coordinates $q_{\bm{\kappa},\lambda}$ and their conjugate momenta $\partial/\partial q_{\bm{\kappa},\lambda}$~\cite{rokaj2017}
\begin{eqnarray}
q_{\bm{\kappa},\lambda}=\frac{1}{\sqrt{2}}\left(\hat{a}_{\bm{\kappa},\lambda}+\hat{a}^{\dagger}_{\bm{\kappa},\lambda}\right)\;\; \& \;\;\frac{\partial}{\partial q_{\bm{\kappa},\lambda}}=\frac{1}{\sqrt{2}}\left(\hat{a}_{\bm{k},\lambda}-\hat{a}^{\dagger}_{\bm{\kappa},\lambda}\right).\nonumber\\
\end{eqnarray}
Then, the quantized vector potential in terms of the displacement coordinates is~\cite{rokaj2017}
\begin{eqnarray}
\hat{\bi{A}}=\sqrt{\frac{\hbar}{\epsilon_0V}}\sum_{\bm{\kappa},\lambda}\frac{\bm{\varepsilon}_{\lambda}(\bm{\kappa}) }{\sqrt{\omega(\bm{\kappa}})}q_{\bm{\kappa},\lambda}.
\end{eqnarray}
Furthermore, the Hamiltonian after expanding the covariant kinetic energy and writing the diamagnetic $\hat{\bi{A}}^2$ explicitly reads as
\begin{eqnarray}
\hat{H}&=&\sum\limits^{N}_{j=1}\left[-\frac{\hbar^2}{2m_{\textrm{e}}}\nabla^2_j +\frac{\textrm{i}e\hbar}{m_{\textrm{e}}} \hat{\mathbf{A}}\cdot\nabla_j\right]+\sum_{\bm{\kappa},\lambda}\frac{\hbar\omega(\bm{\kappa})}{2}\left(-\frac{\partial^2}{\partial q^2_{\bm{\kappa},\lambda}} +q^2_{\bm{\kappa},\lambda}\right)\nonumber\\
&+&\underbrace{\frac{\hbar \omega^2_p}{2}\sum_{\bm{\kappa},\bm{\kappa}^{\prime},\lambda,\lambda^{\prime}}\frac{\bm{\varepsilon}_{\lambda}(\bm{\kappa})\cdot \bm{\varepsilon}_{\lambda^{\prime}}(\bm{\kappa}^{\prime})}{\sqrt{\omega(\bm{\kappa})\omega(\bm{\kappa}^{\prime})}} q_{\bm{\kappa},\lambda}q_{\bm{\kappa}^{\prime},\lambda^{\prime}}}_{\frac{Ne^2}{2m_{\textrm{e}}}\hat{\bi{A}}^2}.
\end{eqnarray}
The part depending purely on the photonic degrees of freedom can be separated into a part being quadratic in the displacement coordinates $q^2_{\bm{\kappa},\lambda}$ and a part being bilinear $q_{\bm{\kappa},\lambda}q_{\bm{\kappa}^{\prime},\lambda^{\prime}}$ and we have
\begin{eqnarray}
\hat{H}&=&\sum\limits^{N}_{j=1}\left[-\frac{\hbar^2}{2m_{\textrm{e}}}\nabla^2_j +\frac{\textrm{i}e\hbar}{m_{\textrm{e}}} \hat{\mathbf{A}}\cdot\nabla_j\right]+\frac{\hbar \omega^2_p}{2}\sum_{\bm{\kappa}\neq \bm{\kappa}^{\prime},\lambda,\lambda^{\prime}}\frac{\bm{\varepsilon}_{\lambda}(\bm{\kappa})\cdot \bm{\varepsilon}_{\lambda^{\prime}}(\bm{\kappa}^{\prime})}{\sqrt{\omega(\bm{\kappa})\omega(\bm{\kappa}^{\prime})}} q_{\bm{\kappa},\lambda}q_{\bm{\kappa}^{\prime},\lambda^{\prime}}\nonumber\\
&+&\sum_{\bm{\kappa},\lambda}\frac{\hbar\omega(\bm{\kappa})}{2}\left(-\frac{\partial^2}{\partial q^2_{\bm{\kappa},\lambda}} +q^2_{\bm{\kappa},\lambda}\left(1+\frac{\omega^2_p}{\omega^2(\bm{\kappa})}\right)\right)
\end{eqnarray}
In addition, we introduce a new set of scaled photonic coordinates $u_{\bm{\kappa},\lambda}=q_{\bm{\kappa},\lambda}\sqrt{\hbar/\omega(\bm{\kappa})}$ and the dressed frequencies $\widetilde{\omega}^2(\bm{\kappa})=\omega^2(\bm{\kappa})+\omega^2_p$. Then, the Hamiltonian takes the form
\begin{eqnarray}
\hat{H}&=&\sum\limits^{N}_{j=1}\left[-\frac{\hbar^2}{2m_{\textrm{e}}}\nabla^2_j +\frac{\textrm{i}e\hbar}{m_{\textrm{e}}} \hat{\mathbf{A}}\cdot\nabla_j\right]+\sum_{\bm{\kappa},\lambda}\left(-\frac{\hbar^2}{2}\frac{\partial^2}{\partial u^2_{\bm{\kappa},\lambda}} +\frac{\widetilde{\omega}^2(\bm{\kappa})}{2}u^2_{\bm{\kappa},\lambda}\right)\nonumber\\ &+&\frac{\omega^2_p}{2}\sum_{\bm{\kappa}\neq \bm{\kappa}^{\prime},\lambda,\lambda^{\prime}} \bm{\varepsilon}_{\lambda}(\bm{\kappa})\cdot\bm{\varepsilon}_{\lambda^{\prime}}(\bm{\kappa}^{\prime}) u_{\bm{\kappa},\lambda} u_{\bm{\kappa}^{\prime},\lambda^{\prime}}.
\end{eqnarray}
In terms of the new set of coordinates the quantized field is 
\begin{eqnarray}
\hat{\mathbf{A}}=\sqrt{\frac{1}{\epsilon_0 V}}\sum_{\bm{\kappa},\lambda}\bm{\varepsilon}_{\lambda}(\bm{\kappa})u_{\bm{\kappa},\lambda}.
\end{eqnarray}
For simplicity we introduce the enlarged ``4-tuple'' variable $\alpha\equiv (\bm{\kappa},\lambda)=(\kappa_x,\kappa_y,\kappa_z,\lambda)$ and everything takes a much more compact form
\begin{eqnarray}
\hat{H}&=&\sum\limits^{N}_{j=1}\left[-\frac{\hbar^2}{2m_{\textrm{e}}}\nabla^2_j +\frac{\textrm{i}e\hbar}{m_{\textrm{e}}} \hat{\mathbf{A}}\cdot\nabla_j\right]-\frac{\hbar^2}{2}\sum^{M}_{\alpha=1}\frac{\partial^2}{\partial u^2_{\alpha}}+\frac{1}{2}\sum^{M}_{\alpha, \beta=1}W_{\alpha\beta}u_{\alpha}u_{\beta}\nonumber\\
&&\textrm{where}\;\;\; W_{\alpha\beta}=\widetilde{\omega}^2_{\alpha}\delta_{\alpha \beta}+\omega^2_p\mathcal{E}_{\alpha \beta}.
\end{eqnarray}
The matrix $\mathcal{E}_{\alpha \beta}$ is zero for $\alpha=\beta$, $\mathcal{E}_{\alpha \alpha}=0$, while for $\alpha \neq \beta$ this matrix is defined as the inner product of the polarization vectors $\mathcal{E}_{\alpha \beta}= \bm{\varepsilon}_{\alpha}\cdot \bm{\varepsilon}_{\beta}$.
\begin{eqnarray}
\mathcal{E}_{\alpha \beta}= \begin{cases}
		0 \;\;\;\;\textrm{for} \;\;\;\; \alpha =\beta  \\\\
		\bm{\varepsilon}_{\alpha}\cdot \bm{\varepsilon}_{\beta}\;\;\;\; \textrm{for} \;\;\;\; \alpha \neq \beta 
	\end{cases}
\end{eqnarray}
For the matrix $W$ it holds that it is real and symmetric and consequently can be brought into a diagonal form with the use of an orthogonal matrix $U$
\begin{eqnarray}\label{diagform}
\sum_{\gamma,\delta} U^{-1}_{\alpha\gamma}W_{\gamma\delta}U_{\delta\beta}= \Omega^2_{\alpha}\delta_{\alpha\beta}
\end{eqnarray}
where $\Omega^2_{\alpha}$ are the eigenvalues of the matrix $W_{\alpha\beta}$. Further, as the matrix $U$ is orthogonal, it means that is is also invertible, with its inverse $U^{-1}$ being equal to its transpose $U^{T}$. With the use of the matrix $U$ we can define the normal coordinates $z_{\gamma}$ and the canonical momenta $\partial/\partial z_{\gamma}$~\cite{faisal1987} 
\begin{eqnarray}
z_{\gamma}=\sum_{\alpha} U_{\alpha\gamma} u_{\alpha} \;\;\; \&\;\;\;\; \frac{\partial}{\partial z_{\gamma}}=\sum_{\alpha}U_{\alpha\gamma} \frac{\partial}{\partial u_{\alpha}} .
\end{eqnarray}
It is important to mention that the above coordinates and momenta are independent because they satisfy canonical commutation relations~\cite{faisal1987}. 
\begin{eqnarray}
[z_{\alpha},z_{\gamma}]=\left[\frac{\partial}{\partial z_{\alpha}},\frac{\partial}{\partial z_\gamma}\right]=0\;\;\; \textrm{and}\;\;\; \left[\frac{\partial}{\partial z_{\alpha}},z_{\gamma}\right]=\delta_{\alpha\gamma}.
\end{eqnarray}
The Hamiltonian in terms of the new coordinates and momenta is~\cite{faisal1987}
\begin{eqnarray}
\hat{H}=\sum\limits^{N}_{j=1}\left[-\frac{\hbar^2}{2m_{\textrm{e}}}\nabla^2_j +\frac{\textrm{i}e\hbar}{m_{\textrm{e}}} \hat{\mathbf{A}}\cdot\nabla_j\right]+\sum^M_{\gamma=1}\left(-\frac{\hbar^2}{2}\frac{\partial^2}{\partial z^2_{\gamma}}+\frac{\Omega^2_{\gamma}}{2}z^2_{\gamma}\right)\;\; \textrm{with}\;\; \hat{\mathbf{A}}=\sqrt{\frac{1}{\epsilon_0 V}}\sum_{\gamma=1}\widetilde{\bm{\varepsilon}}_{\gamma} z_{\gamma}.\nonumber\\
\end{eqnarray}
The new polarization vectors $\widetilde{\bm{\varepsilon}}_{\gamma}$ are defined as $\widetilde{\bm{\varepsilon}}_{\gamma}=\sum_{\alpha=1}\bm{\varepsilon}_{\alpha}U_{\alpha\gamma}$. As it was explained also in section~\ref{2DEG in QED} the free electron gas coupled to the quantized modes in the dipole approximation, is translationally invariant. As a consequence the wavefunctions of the electronic part are given by a Slater determinant $\Phi_{\bi{K}}$ constructed out of plane waves, as defined in Eq.~(\ref{Slater determinant}). Applying $\hat{H}$ on $\Phi_{\bi{K}}$  we obtain
\begin{eqnarray}
\hat{H}\Phi_{\bi{K}}=\left[\frac{\hbar^2}{2m_{\textrm{e}}}\sum\limits^{N}_{j=1}\mathbf{k}^2_j+\sum^M_{\gamma=1}\left(-\frac{\hbar^2}{2}\frac{\partial^2}{\partial z^2_{\gamma}}+\frac{\Omega^2_{\gamma}}{2}z^2_{\gamma}-\widetilde{g}z_{\gamma} \widetilde{\bm{\varepsilon}}_{\gamma}\cdot\mathbf{K}  \right)\right]\Phi_{\bi{K}}, \;\; \textrm{where}\;\; \widetilde{g}=\frac{e\hbar}{m_{\textrm{e}}\sqrt{\epsilon_0 V}}.\nonumber\\
\end{eqnarray}
By performing a square completion, the part of the Hamiltonian depending on $z_{\gamma}$ can be written as the sum of a set of displaced harmonic oscillators with frequencies $\Omega_{\gamma}$ 
\begin{eqnarray}
\hat{H}\Phi_{\bi{K}}=\left[\frac{\hbar^2}{2m_{\textrm{e}}}\sum\limits^{N}_{j=1}\mathbf{k}^2_j+\sum^M_{\gamma=1}\left(-\frac{\hbar^2}{2}\frac{\partial^2}{\partial z^2_{\gamma}}+\frac{\Omega^2_{\gamma}}{2}\left(z_{\gamma}-\frac{\widetilde{g}\widetilde{\bm{\varepsilon}}_{\gamma}\cdot\mathbf{K} }{\Omega^2_{\gamma}}\right)^2-\frac{\left(\widetilde{g}\widetilde{\bm{\varepsilon}}_{\gamma}\cdot\mathbf{K}\right)^2}{2\Omega^2_{\gamma}} \right)\right] \Phi_{\bi{K}}.\nonumber\\
\end{eqnarray}
The eigenfunctions of the part of the Hamiltonian depending on the coordizates $z_{\gamma}$ are Hermite functions~\cite{GriffithsQM} with argument $z_{\gamma}-\widetilde{g}\widetilde{\bm{\varepsilon}}_{\gamma}\cdot\mathbf{K} /\Omega^2_{\gamma}$
\begin{eqnarray}
H_{n_{\gamma}}\left(z_{\gamma}-\frac{\widetilde{g}\widetilde{\bm{\varepsilon}}_{\gamma}\cdot\mathbf{K} }{\Omega^2_{\gamma}}\right)
\end{eqnarray}
and eigenergies
\begin{eqnarray}
E_{n_{\gamma}}=\hbar\Omega_{\gamma}\left(n_{\gamma}+\frac{1}{2}\right) \;\;\; \textrm{with} \;\;\; n_{\gamma} \in \mathbb{N}\;\; \forall\; \gamma.
\end{eqnarray}
Then, the complete set of eigenfunctions of the electron-photon system is
\begin{eqnarray}
\Phi_{\bi{K}} \prod^M_{\gamma=1} H_{n_{\gamma}}\left(z_{\gamma}-\frac{\widetilde{g}\widetilde{\bm{\varepsilon}}_{\gamma}\cdot\mathbf{K} }{\Omega^2_{\gamma}}\right),
\end{eqnarray}
 and we find that the energy spectrum of the free electron gas coupled to an arbitrary amount of photon modes with the mode-mode interactions included is
\begin{eqnarray}
E_{\bi{k}}=\frac{\hbar^2}{2m_{\textrm{e}}}\sum\limits^{N}_{j=1}\mathbf{k}^2_j -\sum^M_{\gamma=1}\frac{\left(\widetilde{g}\widetilde{\bm{\varepsilon}}_{\gamma}\cdot\mathbf{K}\right)^2}{2\Omega^2_{\gamma}}+\sum^M_{\gamma=1}\hbar\Omega_{\gamma}\left(n_{\gamma}+\frac{1}{2}\right).
\end{eqnarray}
From the above result we conclude that the structure of the energy spectrum of the free electron gas in the many-mode case, and with the mode-mode interactions included, is the same with the one in the effective quantum field theory in Eq.~(\ref{effective energy}) and that the mode-mode interactions do not modify fundamentally the energy spectrum. 




\bibliographystyle{unsrt}
\cleardoublepage 
\phantomsection  
\renewcommand*{\bibname}{References}

\addcontentsline{toc}{chapter}{\textbf{References}}

\bibliography{Thesis}


\newpage

\section*{Eidesstattliche Versicherung/Declaration on oath}

Hiermit versichere ich an Eides statt, die vorliegende Dissertationsschrift selbst verfasst und
keine anderen als die angegebenen Hilfsmittel und Quellen benutzt zu haben.\\

Hamburg, 19.10.2021\\

\hfill  \noindent\rule{3cm}{0.4pt}

\hfill Vasil Rokaj

\end{document}